\documentstyle[preprint,aps]{revtex}
\tightenlines
\begin{document}
\draft

\title{Tetrad Gravity: II) Dirac's Observables.}

\author{Luca Lusanna}

\address{
Sezione INFN di Firenze\\
L.go E.Fermi 2 (Arcetri)\\
50125 Firenze, Italy\\
E-mail LUSANNA@FI.INFN.IT}

\author{and}

\author{Stefano Russo}

\address
{Condominio dei Pioppi 16\\
6916 Grancia (Lugano)\\
Svizzera}

\maketitle
\begin{abstract}

After a study of the Hamiltonian group of gauge transformations, whose
infinitesimal generators are the 14 first class constraints of a new
formulation of canonical tetrad gravity on globally hyperbolic, asymptotically
flat at spatial infinity, spacetimes with simultaneity spacelike
hypersurfaces $\Sigma_{\tau}$ diffeomorphic to $R^3$, the multitemporal
equations associated with the constraints generating space rotations and space 
diffeomorphisms on the cotriads are given. Their solutions give the dependence 
of the cotriads on $\Sigma_{\tau}$ and of their momenta on the six parameters
associated with such transformations. The choice of 3-coordinates on $\Sigma
_{\tau}$, namely the gauge fixing to the space diffeomorphisms constraints, is
equivalent to the choice of how to parametrize the dependence of the cotriad
on the last three degrees of freedom: namely to the choice of a parametrization
of the superspace of 3-geometries. The Shanmugadhasan canonical transformation,
corresponding to the choice of 3-orthogonal coordinates on $\Sigma_{\tau}$ and
adapted to 13 of the 14 first class constraints, is found, the
superhamiltonian constraint is rewritten in this canonical basis and the
interpretation of the gauge transformations generated by it is given. Some
interpretational problems connected with Dirac's observables are discussed. In
particular the gauge interpretation of tetrad gravity based on constraint
theory implies that a ``Hamiltonian kinematical gravitational field" is an 
equivalence class of pseudo-Riemannian spacetimes modulo the Hamiltonian group 
of gauge transformations: it includes a conformal 3-geometry and 
all the different 4-geometries (standard definition of a kinematical 
gravitational field, $Riem\, M^4/Diff\, M^4$) 
connected to it by the gauge transformations
generated by the constraints, in particular by the superhamiltonian constraint.
A ``Hamiltonian Einstein or dynamical gravitational field" is a kinematical 
one which satisfies the Hamilton-Dirac equations generated by the ADM energy: 
it coincides with the standard Einstein or dynamical gravitational field, 
namely a 4-geometry solution of Einstein's equations, since the
Hilbert and ADM actions both generate Einstein's equations so that the 
kinematical Hamiltonian gauge transformations are dynamically restricted to the
spacetime pseudodiffeomorphisms of the solutions of Einstein's equations.
Also the problem of the physical identification of the points of spacetime by
means of Komar-Bergmann individuating fields is discussed and some comments
on the theory of measurement are done.

\vskip 1truecm
\noindent \today
\vskip 1truecm
\noindent This work has been partially supported by the network ``Constrained 
Dynamical Systems" of the E.U. Programme ``Human Capital and Mobility".

\end{abstract}
\pacs{}

\newpage 

\vfill\eject

\section
{Introduction}

In the paper \cite{russo} [quoted as I in the following] a new formulation of
tetrad gravity was given and the 14 first class constraints of its Hamiltonian
description were found. In that paper there was a long Introduction about the 
research program whose aim is to find a unified description and a canonical 
reduction of the four interactions based on Dirac-Bergmann theory of 
constraints. Since the canonical reduction is based on the Shanmugadhasan 
canonical transformation, in which the original first class constraints are 
replaced by a subset of the new momenta (whose conjugate variables are
Abelianized gauge variables, in the terminology of gauge theories), its use in
generally covariant theories has not yet been studied, being associated with a
breaking of manifest general covariance. This second paper will explore this
approach, because it is the natural one from the point of view of constraint 
theory (namely presymplectic geometry), like the search of coordinate 
systems separating the variables is natural in the theory of partial 
differential equations. There will be a  presentation and a (often naive) 
solution of an ordered sequence of mathematical and interpretational problems, 
which have to be understood step by step to arrive at a final picture
(in a heuristic way at the first stage, when nothing better can be done)
and which will require an exact mathematical treatment in future
refinements of the theory.

First of all, after a discussion about the parametrization of lapse and shift
functions, following the treatment developed for Yang-Mills theories in
Ref.\cite{lusa}, in Section II we shall study the Hamiltonian group of gauge
transformations whose infinitesimal generators are the 14 first class 
constraints of tetrad gravity. We shall concentrate, in particular, on the
gauge transformations generated by the action of the rotation and space
pseudodiffeomorphism (passive diffeomorphisms) constraints on cotriads and
on the associated stability groups connected with Gribov ambiguity and
isometries.

Then, in Section III, we will define the multitemporal equations associated
with the constraints generating space rotations and space pseudodiffeomorphisms
[they form a Lie subalgebra of the algebra of gauge transformations]. Their
solution allows to find the dependence of cotriads and of their conjugate
momenta on the rotation angles and on the three parameters characterizing
space pseudodiffeomorphisms (changes of chart in the coordinate atlas of the
simultaneity spacelike hypersurface $\Sigma_{\tau}$). As a consequence a
generic cotriad, which has nine independent degrees of freedom, becomes a
function of three angles, of three pseudodiffeomorphims parameters and of
three unspecified functions.

In Section IV it is shown that the problem of the choice of the coordinates on
the simultaneity spacelike hypersurface $\Sigma_{\tau}$ is equivalent to the 
choice of the form of the functional dependence of the cotriad upon these three
unspecified functions. The functional dependence corresponding to 3-orthogonal
and to normal coordinates around a point on $\Sigma_{\tau}$ is explicitly
given. Since $\Sigma_{\tau}$ is assumed diffeomorphic to $R^3$, the 3-orthogonal
and normal (around a point)
coordinates are globally defined. Then we find the Shanmugadhasan canonical 
transformation Abelianizing the constraints generating space rotations and
space pseudodiffeomorphisms [other seven first class constraints are Abelian
from the beginning] in 3-orthogonal coordinates. This allows to get a 
parametrization of the superspace of 3-geometries in these coordinates.

In Section V a further canonical transformation on the superspace sector, plus
its conjugate momenta, allows to put the 3-metric on $\Sigma_{\tau}$ in a 
Misner form: the 3-metric in 3-orthogonal coordinates is parametrized by its
conformal factor $\phi (\tau ,\vec \sigma )=e^{q(\tau ,\vec \sigma )/2}=[det\,
{}^3g_{rs}(\tau ,\vec \sigma )]^{1/12}$ plus two other variables $r_{\bar
a}(\tau ,\vec \sigma )$, $\bar a=1,2,$ whose conjugate momenta are denoted
$\pi_{\phi}(\tau ,\vec \sigma )$, $\pi_{\bar a}(\tau ,\vec \sigma )$.
Since now 13 of the 14 first class constraints have been 
transformed in new momenta, we can write the last superhamiltonian constraint
in its final form in the 3-orthogonal gauge. This constraint  is no more an 
algebraic relation among the surviving canonical variables (the three parameters
labelling 3-geometries and their conjugate momenta: they are the Dirac
observables with respect to the gauge transformations generated by 13
constraints, superhamiltonian one excluded), but an integro-differential one 
for the conformal factor $\phi$ of the 3-metric
(namely the reduced Lichnerowicz equation),
because the momenta conjugate to the cotriads are related to the new momenta
conjugate to 3-geometries by an integral relation. The last gauge variable of
tetrad gravity is not a configurational quantity, but the momentum $\pi_{\phi}
(\tau ,\vec \sigma )=2\phi^{-1}(\tau ,\vec \sigma )\, \rho (\tau ,\vec \sigma 
)$ conjugate to the conformal factor $\phi (\tau ,\vec \sigma )$. This momentum 
describes a ``nonlocal" information on the extrinsic curvature of the spacelike
hypersurfaces $\Sigma_{\tau}$ and replaces the York internal extrinsic time
${}^3K(\tau ,\vec \sigma )$ [in the 3-orthogonal gauge ${}^3K$ is determined by 
an integral of $\pi_{\phi}(\tau ,\vec \sigma )$ over all $\Sigma_{\tau}$ with 
a nontrivial kernel]. The interpretation of the gauge transformations generated
by the superhamiltonian constraint is given.

Therefore, if we add the natural gauge-fixing $\pi_{\phi}(\tau ,\vec \sigma ) 
\approx 0$ [$\rho (\tau ,\vec \sigma )\approx 0$], instead of the maximal 
slicing condition ${}^3K(\tau ,\vec \sigma )\approx 0$ of the Lichnerowicz-York
conformal approach, we get an identification of the two pairs of canonical
variables $r_{\bar a}(\tau ,\vec \sigma )$, $\pi_{\bar a}(\tau ,\vec \sigma )$,
$\bar a=1,2$, describing a ``Hamiltonian kinematical gravitational field" (the 
equivalence class of spacetimes modulo the Hamiltonian group of gauge 
transformations, but not solution of the Hamilton-Dirac equations): it is
an equivalence class of 4-geometries (standard kinematical gravitational
fields as elements of $Riem\, M^4/Diff\, M^4$) containing a conformal 
3-geometry.
The reduced ADM energy is playing the role of the physical Hamiltonian for the
evolution in the mathematical time parameter labelling the leaves $\Sigma
_{\tau}$ of the foliation of spacetime associated with the chosen 3+1 splitting.

In Section VI there are some conclusions and a discussion of the
interpretational problems deriving from the two conflicting point of
views based on gauge invariant deterministic Dirac observables and on
generally covariant (but not gauge invariant) observables. It is shown that if 
we define a ``Hamiltonian Einstein or dynamical gravitational field" as a
kinematical one satisfying the Hamilton-Dirac (and therefore the Einstein)
equations, it coincides with the standard Einstein or dynamical gravitational
field, namely a 4-geometry solution of Einstein's equations. It is underlined
that this is a consequence of the fact that both the Hilbert and ADM actions 
(even if they have different Noether symmetries) generate the same Einstein 
equations, so that on the space of their solutions the Hamiltonian gauge
transformations are forced to be restricted to the dynamical symmetries of
Einstein's equations, namely the spacetime diffeomorphisms of the solutions.
Also a discussion of how to give a physical identification of the points of 
spacetime by means of the Komar-Bergmann individuating fields is given, and some
comments on the theory of measurement are done.

In Appendix A there are some notions on coordinate systems.
In Appendix B there are some notions concerning isometries and conformal
transformations.  
In Appendix C there is a review of the Lichnerowicz-York conformal approach.
In Appendices D and E  there is the expression of certain 3- and
4-tensors  in the final 3-orthogonal canonical basis of Section V.

\vfill\eject

\section
{Gauge Transformation Algebra and Group and the Stability Subgroups.}

As said in I, we shall consider only globally hyperbolic, asymptotically flat
at spatial infinity spacetimes $M^4$ with simultaneity spacelike hypersurfaces 
$\Sigma_{\tau}$ (the Cauchy surfaces) diffeomorphic to $R^3$. The configuration
variables of our approach to tetrad gravity are: i) lapse and shift functions
$N(\tau ,\vec \sigma )$, $N_{(a)}(\tau ,\vec \sigma )$ [the usual shift
functions are $N^r={}^3e^r_{(a)} N_{(a)}$]; ii) boost parameters $\varphi
_{(a)}(\tau ,\vec \sigma )$; iii) cotriads ${}^3e_{(a)r}(\tau ,\vec \sigma )$
on $\Sigma_{\tau}$. Their conjugate momenta are ${\tilde \pi}^N(\tau ,\vec 
\sigma )$, ${\tilde \pi}^{\vec N}_{(a)}(\tau ,\vec \sigma )$, ${\tilde \pi}
^{\vec \varphi}_{(a)}(\tau ,\vec \sigma )$, ${}^3{\tilde \pi}^r_{(a)}
(\tau ,\vec \sigma )$. The fourteen first class constraints and the Dirac 
Hamiltonian are [$\epsilon =\pm$ according to the chosen signature convention
for $M^4$: $\, \epsilon (+ - - - )$]

\begin{eqnarray}
{\tilde \pi}^N(\tau ,\vec \sigma )&\approx& 0,\nonumber \\
{\tilde \pi}^{\vec N}_{(a)}(\tau ,\vec \sigma )&\approx& 0,\nonumber \\
{\tilde \pi}^{\vec \varphi}_{(a)}(\tau ,\vec \sigma ) &\approx& 0,\nonumber \\
{}^3{\tilde M}_{(a)}(\tau ,\vec \sigma )&=&{1\over 2}\epsilon_{(a)(b)(c)}\,
{}^3{\tilde M}_{(b)(c)}(\tau ,\vec \sigma )=\epsilon_{(a)(b)(c)}\, {}^3e_{(b)r}
(\tau ,\vec \sigma )\, {}^3{\tilde \pi}^r_{(c)}(\tau ,\vec \sigma )
\approx 0,\nonumber \\
{}^3{\tilde \Theta}_r(\tau ,\vec \sigma )&=&{}^3{\tilde \pi}^s_{(a)}(\tau ,\vec 
\sigma ) \partial_r\, {}^3e_{(a)s}(\tau ,\vec \sigma )-\partial_s [{}^3e
_{(a)r}(\tau ,\vec \sigma )\, {}^3{\tilde \pi}^s_{(a)}(\tau ,\vec \sigma )]
=\nonumber \\
&=&-{}^3e_{(a)r}(\tau ,\vec \sigma ){\hat {\cal H}}_{(a)}(\tau ,\vec \sigma )-
{}^3\omega_{r(a)}(\tau ,\vec \sigma )\, {}^3{\tilde M}_{(a)}(\tau ,\vec \sigma )
\approx 0,\nonumber \\
{\hat {\cal H}}(\tau ,\vec \sigma )&=&\epsilon [ k\, {}^3e\, \epsilon
_{(a)(b)(c)}\, {}^3e^r_{(a)}\, {}^3e^s_{(b)}\, {}^3\Omega_{rs(c)}-\nonumber \\
&-&{1\over {8k\, {}^3e}} {}^3G_{o(a)(b)(c)(d)}\, {}^3e_{(a)r}\, {}^3{\tilde 
\pi}^r_{(b)}\, {}^3e_{(c)s}\, {}^3{\tilde \pi}^s_{(d)} ](\tau ,\vec \sigma )
\approx 0,\nonumber \\
&&{}\nonumber \\
H_{(D)}&=&\int d^3\sigma [N\, {\hat {\cal H}}-N_{(a)} {\hat {\cal H}}_{(a)}+
\lambda_N {\tilde \pi}^N+\lambda^{\vec N}_{(a)} {\tilde \pi}^{\vec N}_{(a)}+
\lambda^{\vec \varphi}_{(a)} {\tilde \pi}^{\vec \varphi}_{(a)}+{\mu}_{(a)} 
{}^3{\tilde M}_{(a)}](\tau ,\vec \sigma )=\nonumber \\
&=&\int d^3\sigma [N {\hat {\cal H}}+N^r\, {}^3{\tilde \Theta}_r+
\lambda_N {\tilde \pi}^N+\lambda^{\vec N}_{(a)} {\tilde \pi}^{\vec N}_{(a)}+
\lambda^{\vec \varphi}_{(a)} {\tilde \pi}^{\vec \varphi}_{(a)}+{\hat \mu}_{(a)} 
{}^3{\tilde M}_{(a)}](\tau ,\vec \sigma ).
\label{I0}
\end{eqnarray}

\noindent Here, $(a)=(1), (2), (3)$ is a flat index, while $\sigma^A=\{ \tau ;
\sigma^r \}$ [$A=(\tau , r)$] are $\Sigma_{\tau}$-adapted coordinates
for $M^4$. As shown in Section III of I, in each point of $\Sigma_{\tau}$
quantities like an internal
Euclidean vector $V_{(a)}(\tau ,\vec \sigma )$ transform as
Wigner spin 1 3-vectors under Lorentz transformations in $TM^4$ at that point.
The constraints ${}^3{\tilde \Theta}_r(\tau ,\vec \sigma )$ are the
generators of the extension of space pseudodiffeomorphisms (passive 
diffeomorphisms) in $Diff\, \Sigma_{\tau}$ to cotriads on $\Sigma_{\tau}$ [they 
replace the secondary constraints ${\hat {\cal H}}_{(a)}(\tau ,\vec \sigma 
)=\{ \partial_r\, {}^3{\tilde \pi}^r_{(a)}-\epsilon_{(a)(b)(c)}\, {}^3\omega
_{r(b)}\, {}^3{\tilde \pi}^r_{(c)} \} (\tau ,\vec \sigma )\approx 0$
(SO(3) Gauss laws), see I],
while ${}^3{\tilde M}_{(a)}(\tau ,\vec \sigma )$ is the generator of space 
rotations.

Since one of the most important motivations of our approach to tetrad gravity is
to arrive at a unified description of the four interactions \cite{russo,re}, we
need to find a solution to the deparametrization problem of general relativity 
\cite{isha}. This means that in the limit of vanishing Newton constant,
$G\, \rightarrow \, 0$, tetrad gravity plus any kind of matter should go in the
description of the given matter in Minkowski spacetime with a 3+1 decomposition 
based on its foliation with spacelike hypersurfaces, and, in particular, it
should be possible to recover the rest-frame Wigner-covariant instant form
description of such a matter \cite{lus1,re}. This was the main reason for the
restriction to the above class of spacetimes.

In the next paper \cite{russo3} we shall study the asymptotic behaviour for 
$|\vec \sigma | \, \rightarrow \infty$ of the fields of tetrad gravity
so that an asymptotic, at spatial infinity, Poincar\'e algebra of charges exists
without problems of supertranslations\cite{adm,reg,reg1,reg2,reg3} and the
asymptotic part of spacetime agrees as much as possible with Minkowski
spacetime [our definitions at this preliminary stage will be coordinate
dependent, because the Hamiltonian formulation of general relativity is not yet
so developed to be able to translate in it coordinate independent statements
about asymptotically flat spacetimes 
\cite{p0,p1,p2,p3,p4,p5,p6,p7,p8,p9,p12,chris,p13,p14,p16,hfr}].

For the time present, however, we must anticipate some of the results of that 
paper regarding the allowed class of lapse and shift functions: these
functions must be parametrized in a form allowing their identification at
spatial infinity, in a class of asymptotically Minkowskian coordinate systems,
with the flat lapse and shift functions which can be defined in the description 
of isolated systems in Minkowski spacetime on spacelike hyperplanes.

In Ref.\cite{lus1} 
scalar charged particles and electromagnetic fields in Minkowski
spacetime were described in parametrized form on an arbitrary foliation of it
(3+1 splitting) with spacelike hypersurfaces still denoted $\Sigma_{\tau}$, 
whose points $z^{(\mu )}(\tau ,\vec \sigma )$ [$(\mu )$ are flat Cartesian 
indices] are extra configuration variables with conjugate momenta $\rho
_{(\mu )}(\tau ,\vec \sigma )$: this is possible, because, contrary to curved
spacetimes, in Minkowski spacetime the transition coefficients $b^{(\mu )}_A=
z^{(\mu )}_A$ from arbitrary to $\Sigma_{\tau}$-adapted coordinates are flat
tetrads defining a holonomic basis of vector fields. 
Indeed, in each point of $\Sigma
_{\tau}$ the gradients $z^{(\mu )}_A(\tau ,\vec \sigma )=\partial z^{(\mu )}
(\tau ,\vec \sigma )/\partial \sigma^A$ (in Minkowski spacetime we use the
notation $A=(\tau ;\check r)$ to conform with Ref.\cite{lus1}) form a flat 
tetrad, i.e. ${}^4\eta^{(\mu )(\nu )}=z^{(\mu )}_A\, {}^4g^{AB} z^{(\nu )}_B$
with ${}^4g^{AB}$ being the inverse of the induced 4-metric ${}^4g_{AB}=
z^{(\mu )}_A\, {}^4\eta_{(\mu )(\nu )} z^{(\nu )}_B$ on $\Sigma_{\tau}$, with 
the evolution vector given by $z^{(\mu )}_{\tau}=N_{[z](flat)}l^{(\mu )}+
N^{\check r}_{[z](flat)}z^{(\mu )}_{\check r}$, 
where $l^{(\mu )}(\tau ,\vec \sigma )$ is 
the normal to $\Sigma_{\tau}$ in $z^{(\mu )}(\tau ,\vec \sigma )$ and
\hfill\break
\hfill\break
$N_{[z](flat)}(\tau ,\vec \sigma )=\sqrt{{}^4g_{\tau\tau}-{}^3\gamma^{\check 
r\check s}\, {}^4g_{\tau \check r}\, {}^4g_{\tau \check s}}=\sqrt{{}^4g
/{}^3\gamma}$,\hfill\break
 $N_{[z](flat) \check r}(\tau ,\vec \sigma )={}^3g
_{\check r\check s}(\tau ,\vec \sigma )N^{\check s}_{[z](flat)}(\tau ,\vec 
\sigma )={}^4g_{\tau \check r}$,\hfill\break
\hfill\break 
are the flat lapse and shift functions
defined through the metric like in general relativity
[here ${}^3\gamma^{\check r\check u}\, {}^4g_{\check u\check s}=\delta
^{\check r}_{\check s}$ with ${}^3\gamma^{\check r\check s}=-{}^3g^{\check 
r\check s}$ of signature (- - -) to conform with the notations of 
Ref.\cite{lus1}]; however, they are not independent
variables but functionals of $z^{(\mu )}(\tau ,\vec \sigma )$ in Minkowski 
spacetime. The independence of the description from the choice of the foliation 
is manifest 
due to the presence of four first class constraints whose structure is
independent from the system under investigation: \hfill\break
\hfill\break
${\cal H}_{(\mu )}(\tau ,\vec 
\sigma )=\rho_{(\mu )}(\tau ,\vec \sigma )-l_{(\mu )}(\tau ,\vec \sigma )
T_{system}^{\tau\tau}(\tau ,\vec \sigma )-z_{\check r (\mu )}(\tau ,\vec 
\sigma )T_{system}^{\tau \check r}(\tau ,\vec \sigma ) \approx 0$, \hfill\break
\hfill\break
where $T_{system}^{\tau\tau}(\tau ,\vec \sigma )$, $T_{system}^{\tau \check 
r}(\tau ,\vec \sigma )$, are the components of 
the energy-momentum tensor in the holonomic coordinate system on 
$\Sigma_{\tau}$ corresponding to the energy- and momentum-density of the
isolated system. These four constraints satisfy an Abelian Poisson algebra 
being solved in four momenta: $\lbrace {\cal H}_{(\mu )}(\tau ,\vec \sigma ),
{\cal H}_{(\nu )}(\tau ,{\vec \sigma}^{'}) \rbrace =0$.

The original Dirac Hamiltonian contains a piece given by
$\int d^3\sigma \lambda^{(\mu )}(\tau ,\vec \sigma ){\cal H}_{(\mu )}
(\tau ,\vec \sigma )$ with $\lambda^{(\mu )}(\tau ,\vec \sigma )$ arbitrary
Dirac multipliers. By using ${}^4\eta^{(\mu )(\nu )}=[l^{(\mu )}l^{(\nu )}-
z^{(\mu )}_{\check r}\, {}^3g^{\check r\check s} z^{(\nu )}_{\check s}](\tau ,
\vec \sigma )$ with ${}^3g^{\check r\check s}$ [inverse of ${}^3g_{\check 
r\check s}$] of signature (+++), we can write \hfill\break
\hfill\break
$\lambda_{(\mu )}(\tau ,\vec 
\sigma ){\cal H}^{(\mu )}(\tau ,\vec \sigma )=[(\lambda_{(\mu )}l^{(\mu )})
(l_{(\nu )}{\cal H}^{(\nu )})-(\lambda_{(\mu )}z^{(\mu )}_{\check r})({}^3g
^{\check r\check s} z_{\check s (\nu )}{\cal H}^{(\nu )})](\tau ,\vec \sigma )$
\hfill\break
${\buildrel {def} \over =}\,
N_{(flat)}(\tau ,\vec \sigma ) (l_{(\mu )}{\cal H}^{(\mu )})(\tau ,\vec \sigma )
-N_{(flat) \check r}(\tau ,\vec \sigma ) ({}^3g^{\check r\check s} z_{\check 
s (\nu )}{\cal H}^{(\nu )})(\tau ,\vec \sigma )$ \hfill\break
\hfill\break
with the (nonholonomic form of
the) constraints $(l_{(\mu )}{\cal H}^{(\mu )})(\tau ,\vec \sigma )\approx 0$,
$({}^3g^{\check r\check s} z_{\check s (\mu )} {\cal H}^{(\mu )})(\tau ,\vec 
\sigma )\approx 0$, satisfying the universal Dirac algebra 
[see the last three lines of Eqs.(\ref{I5})]. In this way we have defined
new  flat lapse and shift functions 

\begin{eqnarray}
N_{(flat)}(\tau ,\vec \sigma )&=& \lambda_{(\mu )}(\tau ,\vec \sigma ) 
l^{(\mu )}(\tau ,\vec \sigma ),\nonumber \\
N_{(flat) \check r}(\tau ,\vec \sigma )&=& \lambda_{(\mu )}(\tau ,\vec \sigma )
z^{(\mu )}_{\check r}(\tau ,\vec \sigma ).
\label{I1}
\end{eqnarray}

\noindent which have the same content of the arbitrary Dirac multipliers
$\lambda_{(\mu )}(\tau ,\vec \sigma )$, namely they multiply primary
first class constraints satisfying the Dirac algebra. In Minkowski spacetime
they are quite distinct from the previous lapse and shift functions 
$N_{[z](flat)}$, $N_{[z](flat) \check r}$, defined starting from the metric. 
In general relativity (where the coordinates $z^{\mu}(\tau ,\vec \sigma )$ do
not exist) the lapse and shift functions
defined starting from the 4-metric are also the coefficient (in the canonical
part of the Hamiltonian) of secondary first class constraints satisfying the
Dirac algebra [as shown in I, this is evident both for ADM canonical metric
gravity, see Eqs.(77) and (79) of I, and for canonical tetrad gravity, see
Eqs.(59), (60) and (62) of I]. 

Therefore, it is not clear how to arrive at the 
soldering of tetrad gravity with matter and of the parametrized Minkowski 
formulation for the same matter. However, when the parametrized Minkowski
formulation is restricted to spacelike hyperplanes, the two definitions of
lapse and shift functions coincide [and have the same linear grow in $\vec
\sigma$ as the asymptotic ones of tetrad gravity , in suitable asymptotic
Minkowski coordinates, according to existing literature on asymptotic
Poincar\'e charges at spatial infinity \cite{reg,reg1}]
and we get a consistent soldering with
canonical tetrad gravity if its 3+1 splittings are restricted to have leaves
$\Sigma_{\tau}$ approaching flat spacelike hyperplanes at spatial infinity in a
direction-independent way. 

Instead if we want to reduce the description of parametrized Minkowski theories
to one restricted to flat hyperplanes in Minkowski
spacetime, we have to add the gauge-fixings $z^{(\mu )}(\tau ,\vec \sigma )-
x^{(\mu )}_s(\tau )-b^{(\mu )}_{\check r}(\tau ) \sigma^{\check r} \approx 0$.
Here $x^{(\mu )}_s(\tau )$ denotes a point on the hyperplane $\Sigma_{\tau}$ 
chosen as an origin; the $b^{(\mu )}_{\check r}(\tau )$'s form an orthonormal
triad at $x^{(\mu )}_s(\tau )$ and the $\tau$-independent normal to the family 
of spacelike hyperplanes is $l^{(\mu )}=b^{(\mu )}_{\tau}=\epsilon^{(\mu )}{}
_{(\alpha )(\beta )(\gamma )}b^{(\alpha )}_{\check 1}(\tau )b^{(\beta )}
_{\check 2}(\tau )b^{(\gamma )}_{\check 3}(\tau )$. Each hyperplane is 
described by 10 configuration variables, $x^{(\mu )}_s(\tau )$, plus the 6 
independent degrees of freedom contained in the triad $b^{(\mu )}_{\check r}
(\tau )$, and by the 10 conjugate momenta: $p^{(\mu )}_s$ and 6 variables hidden
in a spin tensor $S^{(\mu )(\nu )}_s$\cite{lus1}. With these 20 canonical 
variables it is possible to build 10 Poincar\'e generators ${\bar p}^{(\mu )}_s
=p^{(\mu )}_s$, ${\bar J}^{(\mu )(\nu )}_s=x^{(\mu )}_sp^{(\nu )}_s-x^{(\nu )}
_sp^{(\mu )}_s+S^{(\mu )(\nu )}_s$.

After the restriction to spacelike hyperplanes the previous piece of the Dirac 
Hamiltonian is reduced to \hfill\break
\hfill\break
${\tilde \lambda}^{(\mu )}(\tau ){\tilde {\cal H}}_{(\mu )}(\tau )
-{1\over 2}{\tilde \lambda}^{(\mu )(\nu )}(\tau ){\tilde {\cal H}}_{(\mu 
)(\nu )}(\tau )$, \hfill\break
\hfill\break
because the time constancy of the gauge-fixings $z^{(\mu )}
(\tau ,\vec \sigma )-x^{(\mu )}_s(\tau )-b^{(\mu )}_{\check r}(\tau )\sigma
^{\check r}\approx 0$ implies $\lambda_{(\mu )}(\tau ,\vec \sigma )={\tilde
\lambda}_{(\mu )}(\tau )+{\tilde \lambda}_{(\mu )(\nu )}(\tau )b^{(\nu )}
_{\check r}\sigma^{\check r}$ with ${\tilde \lambda}^{(\mu )}(\tau )=-{\dot x}
^{(\mu )}_s(\tau )$, ${\tilde \lambda}^{(\mu )(\nu )}(\tau )=-{\tilde \lambda}
^{(\nu )(\mu )}(\tau )={1\over 2}\sum_{\check r}[{\dot b}^{(\mu )}_{\check r}
b^{(\nu )}_{\check r}-b^{(\mu )}_{\check r}{\dot b}^{(\nu )}_{\check r}](\tau )$
[$\, \, \, \dot {}$ means $d/d\tau$]. 
Since at this stage we have $z^{(\mu )}_{\check r}(\tau 
,\vec \sigma )\approx b^{(\mu )}_{\check r}(\tau )$, so that $z^{(\mu )}
_{\tau}(\tau ,\vec \sigma )\approx
N_{[z](flat)}(\tau ,\vec \sigma )l^{(\mu )}(\tau ,\vec \sigma )+N^{\check r}
_{[z](flat)}(\tau ,\vec \sigma )b^{(\mu )}_{\check r}(\tau ,\vec 
\sigma )\approx {\dot x}^{(\mu )}_s(\tau )+{\dot b}^{(\mu )}_{\check r}(\tau )
\sigma^{\check r}=-{\tilde \lambda}^{(\mu )}(\tau )-{\tilde
\lambda}^{(\mu )(\nu )}(\tau )b_{\check r (\nu )}(\tau )\sigma^{\check r}$,
it is only now that we get the coincidence of the two definitions of flat
lapse and shift functions, i.e. \hfill\break
\hfill\break
$N_{[z](flat)}\approx N_{(flat)}$, 
$N_{[z](flat) \check r}\approx  N_{(flat)\check r}$.\hfill\break
\hfill\break
The description on arbitrary families of spacelike hyperplanes is
independent from the choice of the family, due to the 10 first class 
constraints\hfill\break
\hfill\break
${\tilde {\cal H}}^{(\mu )}(\tau )=\int d^3\sigma {\cal H}^{(\mu )}
(\tau ,\vec \sigma )=p^{(\mu )}_s-$\hfill\break
$[total\, momentum\, of\, the\,
system\, inside\, the\, hyperplane]^{(\mu )}\approx 0$,\hfill\break 
\hfill\break
${\tilde {\cal H}}
^{(\mu )(\nu )}(\tau )=b^{(\mu )}_{\check r}(\tau ) \int d^3\sigma \, \sigma
^{\check r}{\cal H}^{(\nu )}(\tau ,\vec \sigma )-b^{(\nu )}_{\check r}(\tau ) 
\int d^3\sigma \, \sigma^{\check r} {\cal H}^{(\mu )}(\tau ,\vec \sigma )$
\hfill\break
$=S^{(\mu )(\nu )}_s-[intrinsic\, angular\, momentum\, of\, the\, system\, 
inside\, the\, hyperplane]^{(\mu )(\nu )}$\hfill\break
$=S^{(\mu )(\nu )}_s
-(b^{(\mu )}_{\check r}(\tau )l^{(\nu )}-b^{(\nu )}_{\check r}(\tau )l^{(\mu )}
)[boost\, part\, of\, system's\, angular\, momentum]^{\tau \check r}$\hfill\break
$-(b^{(\mu )}_{\check r}(\tau )b^{(\nu )}_{\check s}(\tau )-b^{(\nu )}_{\check
r}(\tau )b^{(\mu )}_{\check s}(\tau ))[spin\, part\, of\, system's\, angular\, 
momentum]^{\check r\check s}\approx 0$.\hfill\break 
\hfill\break
Therefore, on spacelike hyperplanes in Minkowski spacetime we have

\begin{eqnarray} 
N_{(flat)}(\tau ,\vec \sigma )&=&\lambda_{(\mu )}(\tau ,\vec \sigma )l^{(\mu )}
(\tau ,\vec \sigma ) \mapsto \nonumber \\
&\mapsto& N_{(flat)}(\tau ,\vec \sigma )=N_{[z](flat)}(\tau ,\vec \sigma )=
\nonumber \\
&&=-{\tilde \lambda}
_{(\mu )}(\tau )l^{(\mu )}-l^{(\mu )}{\tilde \lambda}_{(\mu )(\nu )}(\tau )b
^{(\nu )}_{\check s}(\tau ) \sigma^{\check s},\nonumber \\
N_{(flat)\, \check r}(\tau 
,\vec \sigma )&=&\lambda_{(\mu )}(\tau ,\vec \sigma )z^{(\mu )}_{\check r}
(\tau ,\vec \sigma ) \mapsto \nonumber \\
&\mapsto& N_{(flat )}(\tau ,\vec \sigma )=N_{[z](flat)\check r}
(\tau ,\vec \sigma )=\nonumber \\
&&=-{\tilde \lambda}
_{(\mu )}(\tau )b^{(\mu )}_{\check r}(\tau )-b^{(\mu )}_{\check r}(\tau ){\tilde
\lambda}_{(\mu )(\nu )}(\tau ) b^{(\nu )}_{\check s}(\tau ) \sigma^{\check s}.
\label{I2}
\end{eqnarray}

This is the main difference from the treatment of parametrized Minkowski
theories given in Refs.\cite{isha}: there, in the phase action (no 
configuration action is defined),  one uses $N_{[z](flat)}$, $N_{[z](flat)\check
r}$ in place of $N_{(flat)}$, $N_{(flat)\check r}$ also on arbitrary spalike 
hypersurfaces and not only on spacelike hyperplanes.

In Ref.\cite{p22} and in the book in Ref.\cite{dirac} (see also Ref.\cite{reg}),
Dirac introduced asymptotic Minkowski rectangular coordinates $z^{(\mu )}
_{(\infty )}(\tau ,\vec \sigma )=x^{(\mu )}_{(\infty )}(\tau )+b^{(\mu )}
_{(\infty )\, \check r}(\tau ) \sigma^{\check r}$ in $M^4$ at spatial infinity
[here $\{ \sigma^{\check r} \}$ are the  coordinates in an atlas 
of $\Sigma_{\tau}$, not matching the spatial coordinates $z^{(i)}_{(\infty )}
(\tau ,\vec \sigma )$]. For each value of $\tau$, the coordinates $x^{(\mu )}
_{(\infty )}(\tau )$ labels a point, near spatial infinity 
chosen as origin. On it there is a flat tetrad $b^{(\mu )}_{(\infty )\, A}
(\tau )= (\, l^{(\mu )}_{(\infty )}=b^{(\mu )}_{(\infty )\, \tau}=\epsilon
^{(\mu )}{}_{(\alpha )(\beta )(\gamma )} b^{(\alpha )}_{(\infty )\, \check 1}
(\tau )b^{(\beta )}_{(\infty )\, \check 2}(\tau )b^{(\gamma )}_{(\infty )\, 
\check 3}(\tau );\, b^{(\mu )}_{(\infty )\, \check r}(\tau )\, )$, with
$l^{(\mu )}_{(\infty )}$ $\tau$-independent, satisfying $b^{(\mu )}_{(\infty )\,
A}\, {}^4\eta_{(\mu )(\nu )}\, b^{(\nu )}_{(\infty )\, B}={}^4\eta_{AB}$ for
every $\tau$ and assumed to be tangent to the boundary $S^2_{\tau ,\infty}$
of $\Sigma_{\tau}$.

This suggests that, in a suitable class of coordinate systems asymptotic to
Minkowski coordinates (for the sake of simplifying the notation the indices
$\check r$ are replaced with $r$)
and with the general coordinate transformations
suitably restricted at spatial infinity so that it is not possible to go
outside this class, the lapse and shift functions of tetrad gravity should be
parametrized as

\begin{eqnarray}
N(\tau ,\vec \sigma )&=& N_{(as)}(\tau ,\vec \sigma )
+n(\tau ,\vec \sigma ),\quad\quad n(\tau ,\vec 
\sigma )\, \rightarrow_{|\vec \sigma | \, \rightarrow \infty} 0,\nonumber \\
N_{(a)}(\tau ,\vec \sigma )&=&N_{(as) (a)}(\tau ,\vec \sigma )
+n_{(a)}(\tau ,\vec \sigma )=,\nonumber \\
&=&{}^3e^r_{(a)}(\tau ,\vec \sigma )[N_{(as) r}(\tau ,\vec \sigma )+n_r(\tau
,\vec \sigma )],\quad\quad 
n_{(a)}(\tau ,\vec \sigma )\, \rightarrow_{|\vec \sigma | \, \rightarrow 
\infty} 0,\nonumber \\
&&{}\nonumber \\
N_{(as)}(\tau ,\vec \sigma )&=&-{\tilde \lambda}_{(\mu )}(\tau )l^{(\mu )}
_{(\infty )}-l^{(\mu )}_{(\infty )}{\tilde \lambda}_{(\mu )(\nu )}(\tau )
b^{(\nu )}_{(\infty) s}(\tau ) \sigma^{s}=\nonumber \\
&=&-{\tilde \lambda}_{\tau}(\tau )-{1\over 2}{\tilde \lambda}_{\tau s}(\tau )
\sigma^{s},\nonumber \\
N_{(as) r}(\tau ,\vec \sigma )&=&-b^{(\mu )}_{(\infty ) r}(\tau )
{\tilde \lambda}_{(\mu )}(\tau )-b^{(\mu )}_{(\infty ) r}(\tau ){\tilde 
\lambda}_{(\mu )(\nu )}(\tau ) b^{(\nu )}_{(\infty ) s}(\tau ) \sigma
^{s}=\nonumber \\
&=&-{\tilde \lambda}_r(\tau )-{1\over 2}{\tilde \lambda}_{rs}(\tau ) \sigma^s.
\label{I3}
\end{eqnarray}

This very strong assumption (which will be studied in more detail in
Ref.\cite{russo3}) implies that we are restricting the allowed 3+1 splittings
of $M^4$ to those whose leaves $\Sigma_{\tau}$ tend 
asymptotically at spatial infinity to Minkowski spacelike hyperplanes in a
direction-independent way and that only asymptotic coordinate systems are
allowed in which the lapse and shift functions have asymptotic behaviours 
similar to those of 
Minkowski spacelike hyperplanes; but this is coherent with Dirac's choice of 
asymptotic rectangular coordinates [modulo 3-diffeomorphisms not changing the 
nature of the coordinates] and with the assumptions used to define the 
asymptotic Poincar\'e charges. In a future paper \cite{russo4} it will be shown 
that in this way we can solve the deparametrization problem of general
relativity. It is  also needed to eliminate consistently supertranslations
and coordinate transformations not becoming the identity at spatial infinity
[they are not associated with the gravitational fields of isolated systems
\cite{ll}]. With these assumptions we have from Eqs.(6) of I: \hfill\break
\hfill\break
${}^4g_{\tau\tau}
(\tau ,\vec \sigma )=\epsilon \{ [N_{(as)}+n]^2-[N_{(as) (a)}+n_{(a)}]
[N_{(as) (a)}+n_{(a)}] \}(\tau ,\vec \sigma )=\epsilon \{ [N_{(as)}+n]^2-
[N_{(as) r}+n_r]{}^3e^r_{(a)}\, {}^3e^s_{(a)}[N_{(as) s}+n_s]\} (\tau ,\vec
\sigma )$,\hfill\break
${}^4g_{\tau r}(\tau ,\vec \sigma )=-\epsilon [{}^3e_{(a)r} 
(N_{(as) (a)}+n_{(a)})](\tau ,\vec \sigma )=-\epsilon [N_{(as) r}+n_r](\tau
,\vec \sigma )$ \hfill\break
\hfill\break
and the following form of the line element

\begin{eqnarray}
ds^2&=& \epsilon \Big( [N_{(as)}+n]^2 - [N_{(as) r}+n_r] {}^3e^r_{(a)}\, {}^3e^s
_{(a)} [N_{(as) s}+n_s] \Big) (d\tau )^2-\nonumber \\
&-&2\epsilon [N_{(as) r}+n_r] d\tau d\sigma^r -\epsilon \, {}^3e_{(a)r}\, 
{}^3e_{(a)s} d\sigma^r d\sigma^s=\nonumber \\
&=&\epsilon \Big( [N_{(as)}+n]^2(d\tau )^2-[{}^3e_{(a)r}d\sigma^r+(N_{(as)(a)}+
n_{(a)})d\tau ][{}^3e_{(a)s}d\sigma^s+(N_{(as)(a)}+n_{(a)})d\tau ]\Big) .
\label{I3b}
\end{eqnarray}

By using ${\tilde \lambda}_A(\tau )=\{ {\tilde \lambda}_{\tau}(\tau );
{\tilde \lambda}_r(\tau ) \}$, ${\tilde \lambda}_{AB}(\tau )=-{\tilde \lambda}
_{BA}(\tau )$, $n(\tau ,\vec \sigma )$, $n_{(a)}(\tau ,\vec \sigma )$
as new configuration variables [replacing $N(\tau ,\vec \sigma )$ and
$N_{(a)}(\tau ,\vec \sigma )$] in the
Lagrangian of I only produces the replacement of the first class constraints
${\tilde \pi}^N(\tau ,\vec \sigma )\approx 0$, ${\tilde \pi}^{\vec N}_{(a)}
(\tau ,\vec \sigma )\approx 0$, with the new first class constraints
${\tilde \pi}^n(\tau ,\vec \sigma )\approx 0$, ${\tilde \pi}^{\vec n}_{(a)}
(\tau ,\vec \sigma )\approx 0$, ${\tilde \pi}^A(\tau )\approx 0$, ${\tilde
\pi}^{AB}(\tau )=-{\tilde \pi}^{BA}(\tau )\approx 0$, 
corresponding to the vanishing of the canonical 
momenta conjugate to the new configuration variables [we assume the Poisson
brackets $\{ {\tilde \lambda}_A(\tau ),{\tilde \pi}^B(\tau )\} =\delta^B_A$,
$\{ {\tilde \lambda}_{AB}(\tau ), {\tilde \pi}^{CD}(\tau ) \} =\delta^C_A
\delta^D_B-\delta^D_A \delta^C_B$]. The only change in the
Dirac Hamiltonian is \hfill\break
\hfill\break
$\int d^3\sigma [\lambda_N {\tilde \pi}^N+\lambda
^{\vec N}_{(a)} {\tilde \pi}^{\vec N}_{(a)}](\tau ,\vec \sigma )\, \mapsto \,
\zeta_A(\tau ) {\tilde \pi}^A(\tau )+\zeta_{AB}(\tau ) {\tilde \pi}^{AB}(\tau )
+\int d^3\sigma [\lambda_n {\tilde \pi}^n+\lambda^{\vec n}_{(a)} {\tilde \pi}
^{\vec n}_{(a)}](\tau ,\vec \sigma )$ \hfill\break
\hfill\break
(the problem of its differentiability 
and of the needed surface terms \cite{witt} will be discussed in Ref.
\cite{russo3}).

A minimal set of (angle independent to avoid supertranslations \cite{reg1};
this is also in accord with what is needed to define color charges in
Yang-Mills theory\cite{lusa}) boundary conditions on the canonical 
variables of tetrad gravity, which will be justified in the next paper 
\cite{russo3}, is [$\,\, r=|\vec \sigma |\,\,$]

\begin{eqnarray}
&&{}^3e_{(a) r}(\tau ,\vec \sigma )\, {\rightarrow}_{r\, \rightarrow 
\infty }\, \delta_{(a)r}+{}^3w_{(a) r}(\tau ,\vec \sigma ),
\nonumber \\
&&{}^3w_{(a) r}(\tau ,\vec \sigma )={{{}^3w_{(as)(a) r}(\tau )}
\over r}+O(r^{-1}),\nonumber \\
&&{}^3g_{rs}(\tau ,\vec \sigma )=[{}^3e_{(a)r}\, {}^3e_{(a)s}]
(\tau ,\vec \sigma )\, {\rightarrow}_{r\, \rightarrow \infty}\, \delta
_{rs}+{}^3h_{rs}(\tau ,\vec \sigma ),\nonumber \\
&&{}^3h_{rs}(\tau ,\vec \sigma )={1\over r}  \Big[ \delta_{(a) 
r}\, {}^3w_{(as)(a)s}(\tau )+{}^3w_{(as)(a) r}(\tau )\delta
_{(a) s}\Big] +O(r^{-2}),\nonumber \\
&&{}^3g^{rs}(\tau ,\vec \sigma )=[{}^3e^{r}_{(a)}\, {}^3e
^{s}_{(a)}](\tau ,\vec \sigma )\, {\rightarrow}_{r\, 
\rightarrow \infty}\, \delta^{rs}+{}^3h^{rs}(\tau 
,\vec \sigma ),\nonumber \\
&&{}^3h^{rs}(\tau ,\vec \sigma )={1\over r}\Big[ \delta^{r}
_{(a)}\, {}^3w^{s}_{(as)(a)}(\tau )+{}^3w^{r}_{(as)(a)}(\tau ) 
\delta^{s}_{(a)}\Big] +O(r^{-2}),\nonumber \\
&&{}\nonumber \\
&&{}^3{\tilde \pi}^{r}_{(a)}(\tau ,\vec \sigma )\, {\rightarrow}_{r\, 
\rightarrow \infty}\, {}^3{\tilde p}^{r}_{(a)}(\tau ,\vec \sigma ),
\nonumber \\
&&{}^3{\tilde p}^{r}_{(a)}(\tau ,\vec \sigma )={{{}^3{\tilde p}
^{r}_{(as)(a)}(\tau )}\over {r^2}}+O(r^{-3}),\nonumber \\
&&{}^3{\tilde \Pi}^{rs}(\tau ,\vec \sigma )={1\over 4}\Big[ {}^3e^{r}_{(a)}\,
{}^3{\tilde \pi}^{s}_{(a)}+{}^3e^{s}_{(a)}\, {}^3{\tilde \pi}
^{r}_{(a)}\Big] (\tau ,\vec \sigma )\, {\rightarrow}_{r\, \rightarrow 
\infty}\, {}^3{\tilde k}^{rs}(\tau ,\vec \sigma ),\nonumber \\
&&{}^3{\tilde k}^{rs}={{{}^3{\tilde k}^{rs}
_{(as)}(\tau )}\over {r^2}}+O(r^{-3}),\nonumber \\
&&{}^3{\tilde k}^{rs}_{(as)}(\tau )={1\over 4}\Big[ \delta^{r}
_{(a)}\, {}^3{\tilde p}^{s}_{(as)(a)}+\delta^{s}_{(a)}\, 
{}^3{\tilde p}^{r}_{(as)(a)}\Big] (\tau ),\nonumber \\
&&{}\nonumber \\
&&n(\tau ,\vec \sigma )\, {\rightarrow}_{r\, \rightarrow \infty}\, 
O(r^{-\epsilon}),\nonumber \\
&&n_{(a)}(\tau ,\vec \sigma )=[{}^3e^{r}_{(a)}n_{r}](\tau ,\vec 
\sigma )\, {\rightarrow}_{r\, \rightarrow \infty}\, \delta^{r}_{(a)}n
_{r}(\tau ,\vec \sigma )+O(r^{-(1+\epsilon )}),\nonumber \\
&&n_r(\tau ,\vec \sigma )\, {\rightarrow}_{r\, \rightarrow \infty}\,
O(r^{-\epsilon}),\nonumber \\
&&{}\nonumber \\
&&{\tilde \pi}^n(\tau ,\vec \sigma )\, {\rightarrow}_{r\, \rightarrow \infty}\,
O(r^{-3}),\nonumber \\
&&{\tilde \pi}^{\vec n}_{(a)}(\tau ,\vec \sigma )\, {\rightarrow}_{r\, 
\rightarrow \infty}\, O(r^{-3}),\nonumber \\
\label{I4}
\end{eqnarray}

No special requirements are needed at this stage for the asymptotic behaviour 
of the configuration variables $\varphi_{(a)}(\tau ,\vec \sigma )$.

Let us momentarily forget the asymptotic variables ${\tilde \lambda}_A(\tau )$,
${\tilde \lambda}_{AB}(\tau )$ and their conjugate momenta ${\tilde \pi}^A
(\tau )\approx 0$, ${\tilde \pi}^{AB}(\tau )\approx 0$.
In the 32-dimensional functional phase space $T^{*}{\cal C}$ spanned by the 16 
variables $n(\tau ,\vec \sigma )$, $n_{(a)}(\tau ,\vec \sigma )$, $\varphi_{(a)}
(\tau ,\vec \sigma )$, ${}^3e_{(a)r}(\tau ,\vec \sigma )$ of the Lagrangian
configuration space ${\cal C}$ and by their 16 conjugate momenta, we have 14 
first class constraints ${\tilde \pi}^n(\tau ,\vec \sigma )\approx 0$, ${\tilde
\pi}^{\vec n}_{(a)}(\tau ,\vec \sigma )\approx 0$, ${\tilde \pi}^{\vec 
\varphi}_{(a)}(\tau ,\vec \sigma )\approx 0$, ${}^3{\tilde M}_{(a)}
(\tau ,\vec \sigma )\approx 0$, ${\hat {\cal H}}(\tau ,\vec 
\sigma )\approx 0$ and either ${}^3{\tilde \Theta}_r(\tau ,\vec \sigma )\approx
0$ or ${\hat {\cal H}}_{(a)}(\tau ,\vec \sigma )\approx 0$. Seven pairs of
conjugate canonical variables, $\lbrace n(\tau ,\vec \sigma ), {\tilde \pi}^n
(\tau ,\vec \sigma ); n_{(a)}(\tau ,\vec \sigma ), {\tilde \pi}^{\vec n}_{(a)}
(\tau ,\vec \sigma ); \varphi_{(a)}(\tau ,\vec \sigma ), {\tilde \pi}^{\vec
\varphi}_{(a)}(\tau ,\vec \sigma )\rbrace$, are already decoupled from the 
18-dimensional subspace spanned by $\lbrace {}^3e_{(a)r}(\tau ,\vec \sigma );
{}^3{\tilde \pi}^r_{(a)}(\tau ,\vec \sigma ) \rbrace$. The variables in
${\cal C}_g=\lbrace n(\tau ,\vec \sigma ), n_{(a)}(\tau ,\vec \sigma ),
\varphi_{(a)}(\tau ,\vec \sigma ) \rbrace$ are gauge variables, but due to the
decoupling there is no need to introduce gauge-fixing constraints to
eliminate them explicitly, at least at this stage.

Therefore, let us concentrate on the reduced 9-dimensional configuration 
function space ${\cal C}_e=\lbrace {}^3e_{(a)r}(\tau ,\vec \sigma )\rbrace$
[${\cal C}={\cal C}_g+{\cal C}_e$, $T^{*}{\cal C}=T^{*}{\cal C}_g+T^{*}{\cal
C}_e$] and on the 18-dimensional function phase space $T^{*}{\cal C}_e=
\lbrace {}^3e_{(a)r}(\tau ,\vec \sigma ), {}^3{\tilde \pi}^r_{(a)}(\tau ,\vec 
\sigma ) \rbrace$, on which we have 7 first class constraints ${}^3{\tilde M}
_{(a)}(\tau ,\vec \sigma )\approx 0$, ${}^3{\tilde \Theta}_r(\tau ,\vec \sigma )
\approx 0$, ${\hat {\cal H}}(\tau ,\vec \sigma )\approx 0$, whose Poisson 
brackets, defining an algebra $\bar g$, are given in Eqs.(63) of I

\begin{eqnarray}
\lbrace {}^3{\tilde M}_{(a)}(\tau ,\vec \sigma ),{}^3{\tilde M}_{(b)}
(\tau ,{\vec \sigma}^{'})\rbrace &=&\epsilon_{(a)(b)(c)}\, {}^3{\tilde M}_{(c)}
(\tau ,\vec \sigma ) \delta^3(\vec \sigma ,{\vec \sigma}^{'}),\nonumber \\
\lbrace {}^3{\tilde M}_{(a)}(\tau ,\vec \sigma ),{}^3{\tilde \Theta}_r
(\tau ,{\vec \sigma}^{'})\rbrace &=&{}^3{\tilde M}_{(a)}(\tau ,{\vec \sigma}
^{'})\, {{\partial \delta^3(\vec \sigma ,{\vec \sigma}^{'})}\over {\partial
\sigma^r}},\nonumber \\
\lbrace {}^3{\tilde \Theta}_r(\tau ,\vec \sigma ),{}^3{\tilde \Theta}_s
(\tau ,{\vec \sigma}^{'})\rbrace &=& \Big[
{}^3{\tilde \Theta}_r(\tau ,{\vec \sigma}
^{'}) {{\partial}\over {\partial \sigma^s}} +{}^3{\tilde \Theta}_s(\tau ,\vec 
\sigma ) {{\partial}\over {\partial \sigma^r}}\Big] \delta^3(\vec \sigma ,{\vec 
\sigma}^{'}),\nonumber \\
\lbrace {\hat {\cal H}}(\tau ,\vec \sigma ),{}^3{\tilde \Theta}_r(\tau ,{\vec 
\sigma}^{'})\rbrace &=& {\hat {\cal H}}(\tau ,{\vec \sigma}^{'}) {{\partial
\delta^3(\vec \sigma ,{\vec \sigma}^{'})}\over {\partial \sigma^r}},
\nonumber \\
\lbrace {\hat {\cal H}}(\tau ,\vec \sigma ),{\hat {\cal H}}(\tau ,{\vec 
\sigma}^{'})\rbrace &=&\Big[ {}^3e^r_{(a)}(\tau ,\vec \sigma )\, {\hat {\cal H}}
_{(a)}(\tau ,\vec \sigma ) +\nonumber \\
&+& {}^3e^r_{(a)}(\tau ,{\vec \sigma}^{'})\, {\hat {\cal H}}_{(a)}(\tau 
,{\vec \sigma}^{'}) \Big] {{\partial \delta^3(\vec \sigma
,{\vec \sigma}^{'})}\over {\partial \sigma^r}}=\nonumber \\
&=&\Big( \Big[ {}^3e^r_{(a)}\, {}^3e^s_{(a)}\, [{}^3{\tilde \Theta}_s+
{}^3\omega_{s(b)}\, {}^3{\tilde M}_{(b)}]\Big] (\tau ,\vec \sigma ) + 
\nonumber \\
&+&\Big[ {}^3e^r_{(a)}\, {}^3e^s_{(a)}\, [{}^3{\tilde \Theta}_s+{}^3\omega
_{s(b)}\, {}^3{\tilde M}_{(b)}]\Big] (\tau ,{\vec \sigma}^{'}) \Big) \, 
{{\partial \delta^3(\vec \sigma ,{\vec \sigma}^{'})}\over {\partial \sigma^r}}.
\label{I5}
\end{eqnarray}

Let us call ${\bar {\cal G}}$ the
(component connected to the identity of the) gauge group obtained from
successions of gauge transformations generated by these first class constraints.
Since ${}^3{\tilde M}_{(a)}(\tau ,\vec \sigma )$ [the generators of the inner 
gauge SO(3)-rotations] and  ${}^3{\tilde \Theta}_r(\tau ,\vec \sigma )$ [the 
generators of space pseudodiffeomorphisms (passive diffeomorphisms) in 
$Diff\, \Sigma_{\tau}$ extended to cotriads] form a Lie subalgebra ${\bar g}_R$ 
of $\bar g$ (the algebra of ${\bar {\cal G}}$),
let ${\bar {\cal G}}_R$ be the gauge group without the superhamiltonian 
constraint and ${\bar {\cal G}}_{ROT}$ its invariant subgroup containing only
SO(3) rotations. 
The addition to ${\bar g}_R$ of the superhamiltonian ${\hat {\cal H}}(\tau 
,\vec \sigma )$ introduces structure functions [the last of Eqs.(\ref{I5})] 
as in the ADM Hamiltonian formulation of metric gravity, so that $\bar g$ is 
not a Lie algebra.

The gauge group ${\bar {\cal G}}_R$ may be identified with the automorphism
group $Aut\, L\Sigma_{\tau}$ of the trivial principal SO(3)-bundle
$L\Sigma_{\tau} \approx \Sigma_{\tau}\times SO(3)$ of orthogonal coframes,
whose properties are studied in Ref.\cite{abba}. The automorphism group
$Aut\, L\Sigma_{\tau}$ contains the structure group SO(3) of $L\Sigma_{\tau}$ 
as a subgroup, and, moreover, $Aut\, L\Sigma_{\tau}$ is itself a principal 
bundle with base $Diff\, \Sigma_{\tau}$ (which acts on the base $\Sigma_{\tau}$
of $L\Sigma_{\tau}$) and structure group the group of gauge transformations
[$Gau\, L\Sigma_{\tau}$; see Ref.\cite{lusa} for a review of the notations] of
the principal bundle $L\Sigma_{\tau}$: therefore, locally $Aut\, L\Sigma
_{\tau}$ has the trivialization $[U\subset Diff\, \Sigma_{\tau}]\times SO(3)$
and we have

\begin{eqnarray}
\begin{array}{lcl}
Aut\, L\Sigma_{\tau} &\rightarrow & L\Sigma_{\tau}\approx \Sigma_{\tau}\times
SO(3)\\
\downarrow & {}& \downarrow \\
Diff\, \Sigma_{\tau} & \rightarrow & \Sigma_{\tau}
\end{array}
\label{aut1}
\end{eqnarray}

Since the geometric nature of the gauge transformations generated by the 
superhamiltonian constraint in the fixed time Hamiltonian description is
different from time diffeomorphisms, see for
instance Ref.\cite{wa,anton}, let us concentrate on the study of the non-Abelian
algebra ${\bar g}_R$ and of the associated group of gauge transformations
${\bar {\cal G}}_R$. Since ${\bar {\cal G}}_R$ contains $Diff\, \Sigma_{\tau}$
(or better its action on the cotriads), it is not a Hilbert-Lie group, at
least in standard sense\cite{diff,abba} 
(its differential structure is defined in an
inductive way); therefore, the standard technology from the
theory of Lie groups used for Yang-Mills theory 
[see Ref.\cite{lusa} and the appendix of Ref.\cite{bao}] is not directly
available. However this technology can be used for the invariant subgroup of 
gauge SO(3)-rotations. The main problem is that it is not clear how to 
parametrize the group 
manifold of $Diff\, \Sigma_{\tau}$: one only knows that its algebra (the 
infinitesimal diffeomorphisms) is isomorphic to the tangent bundle $T\Sigma
_{\tau}$\cite{diff}.

Moreover, while in a Lie (and also in a Hilbert-Lie) group the basic tool is the
group-theoretical exponential map (associated with the one-parameter
subgroups), which coincides with the geodesic exponential map when the group
manifold of a compact semisimple Lie group is regarded as a symmetric
Riemann manifold\cite{helga}, in $Diff\, \Sigma_{\tau}$
this map does not produce a diffeomorphism 
between a neighbourhood of zero in the algebra and a neighbourhood of the 
identity in $Diff\, \Sigma_{\tau}$\cite{diff,abba}. Therefore, to study the
Riemannian 3-manifold $\Sigma_{\tau}$ we have to use the geodesic exponential
map as the main tool\cite{oneil,kobay}, 
even if it is not clear its relationship
with the differential structure of $Diff\, \Sigma_{\tau}$.
The ``geodesic exponential map" at $p\in M^4$ sends each vector ${}^4V_p
={}^4V_p^{\mu}\partial_{\mu} \in T_pM^4$ at p to the point of unit parameter 
distance along the unique geodesic through p with tangent vector ${}^4V_p$ at 
p; in a small neighbourhood U of p the exponential map has an inverse:
$q\in U\subset M^4 \Rightarrow q=Exp\, {}^4V_p$ for some ${}^4V_p\in T_pM^4$. 
Then, ${}^4V^{\mu}_p$ are the ``normal coordinates" $x_2^{\mu}$ of $q$ and U is 
a ``normal neighbourhood" (see  Appendix A for a review of special coordinate 
systems). Let us remark that in this way one defines an inertial observer in 
free fall at $q$ in general relativity.

In Yang-Mills theory with trivial principal bundles $P(M,G)=M\times G$
\cite{lusa}, the abstract object behind the configuration space is the
connection 1-form $\omega$ on $P(M,G)=M\times G$ [G is a compact, semisimple,
connected, simply connected Lie group with compact, semisimple real Lie
algebra $g$]; instead Yang-Mills configuration space contains the gauge
potentials over the base M, ${}^{\sigma}A^{(\omega )}=\sigma^{*}\omega$, i.e.
the pull-backs to M of the connection 1-form through global cross sections
$\sigma :M\rightarrow P$. The group ${\cal G}$ of gauge transformations (its
component connected to the identity) acting on the gauge potentials on M is
interpreted in a passive sense as a change of global cross section at fixed
connection $\omega$, ${}^{\sigma_U}A^{(\omega )}=U^{-1}\, {}^{\sigma}A
^{(\omega )}\, U+U^{-1}dU$ (if $\sigma_U=\sigma \cdot U$ with $U:M\rightarrow
G$): this formula describes the gauge orbit associated with the given 
$\omega$. In this case, the
group manifold of ${\cal G}$ [which is the space of the cross sections of the
principal bundle P(M,G)] may be considered the principal bundle $P(M,G)=
M\times G$ itself  parametrized
with a special connection-dependent family of global cross sections, after
having chosen canonical coordinates of first kind on a reference fiber
(a copy of the group manifold of G) and having parallel (with respect to the 
given connection) transported them to the other fibers. In this way we avoid 
the overparametrization of ${\cal G}$ by means of the infinite-dimensional space
of all possible local and global cross sections from M to P (this would be the
standard description of ${\cal G}$). The infinitesimal gauge transformations
[the Lie algebra $g_{\cal G}$ of ${\cal G}$: it is a vector bundle whose
standard fiber is the Lie algebra $g$] in phase space are generated by the
first class constraints giving the Gauss laws $\Gamma_a\approx 0$. 
By Legendre pullback to
configuration space, we find  \hfill\break
\hfill\break
${}^{\sigma +\delta \sigma}A
^{(\omega )}={}^{\sigma}A^{(\omega )}+\delta_o\, {}^{\sigma}A^{(\omega )}=
{}^{\sigma}A^{(\omega )}+U^{-1}(dU+[{}^{\sigma}A^{(\omega )},U])={}^{\sigma}A
^{(\omega )}+{\hat D}^{(A)}\alpha ={}^{\sigma}A^{(\omega )}+\lbrace {}^{\sigma}A
^{(\omega )},\int \alpha_a\Gamma_a \rbrace$ if $U=I+\alpha$.

In our formulation of tetrad gravity the relevant configuration variables are
globally defined cotriads ${}^3e_{(a)r}(\tau ,\vec \sigma )$ on the
hypersurface $\Sigma_{\tau} \approx R^3$, which is a 
parallelizable Riemannian 3-manifold $(\Sigma_{\tau},
{}^3g_{rs}={}^3e_{(a)r}\, {}^3e_{(a)s})$ assumed asymptotically flat (therefore
noncompact) at spatial infinity and geodesically complete [so that, due to the 
Hopf-Rinow theorem \cite{oneil},
every two points of $\Sigma_{\tau}$ may be connected by a minimizing geodesic 
segment and there exists a point $p\in \Sigma_{\tau}$ from which $\Sigma_{\tau}$
is geodesically complete, that is the geodesic exponential map is defined on the
entire tangent space $T_p\Sigma_{\tau}$]; with these hypotheses we have 
$T\Sigma_{\tau}\approx \Sigma_{\tau}\times R^3$ and the coframe orthogonal
principal affine SO(3)-bundle is also trivial $L\Sigma_{\tau}\approx 
\Sigma_{\tau}\times SO(3)$ [its points are the abstract coframes ${}^3\theta
_{(a)}\, (={}^3e_{(a)r}d\sigma^r\,$ in global coordinates)]. 
In the phase space of tetrad gravity the rotations of the structure group SO(3) 
are generated 
by the first class constraints ${}^3{\tilde M}_{(a)}(\tau ,\vec \sigma )
\approx 0$. Therefore, in this case the abstract object behind the configuration
space is the so(3)-valued soldering 1-form ${}^3\theta ={\hat R}^{(a)}\, 
{}^3\theta_{(a)}$ [${\hat R}^{(a)}$ are the generators of the Lie algebra 
so(3)]. This shows that to identify the global cotriads ${}^3e_{(a)r}
(\tau ,\vec \sigma )$ we have to choose an atlas of coordinate charts on
$\Sigma_{\tau}$, so that in each chart ${}^3\theta \mapsto {\hat R}^{(a)}\,
{}^3e_{(a)r}(\tau ,\vec \sigma ) d\sigma^r$. Since $\Sigma_{\tau}$ is assumed 
diffeomorphic to $R^3$, global coordinate systems exist.

The general coordinate transformations or space pseudodiffeomorphisms of $Diff\,
\Sigma_{\tau}$ are denoted as $\vec \sigma \mapsto {\vec \sigma}^{'}(\vec 
\sigma )=\vec \xi (\vec \sigma )=\vec \sigma +{\hat {\vec \xi}}(\vec \sigma )$;
for infinitesimal pseudodiffeomorphisms, ${\hat {\vec \xi}}(\vec \sigma )=\delta
\vec \sigma (\vec \sigma )$ is an infinitesimal quantity and the inverse
infinitesimal pseudodiffeomorphism is 
$\vec \sigma ({\vec \sigma}^{'})={\vec \sigma}
^{'}-\delta \vec \sigma ({\vec \sigma}^{'})={\vec \sigma}^{'}-{\hat {\vec
\xi}}({\vec \sigma}^{'})$. The cotriads ${}^3e_{(a)r}(\tau ,\vec \sigma )$ and 
the 3-metric ${}^3g_{rs}(\tau ,\vec \sigma )={}^3e_{(a)r}(\tau ,\vec \sigma )\, 
{}^3e_{(a)s}(\tau ,\vec \sigma )$ transform as [see also Eqs.(30) and 
(31) of I; ${\hat V}(\vec \xi (\vec \sigma ))$ is the operator whose action
on functions is ${\hat V}(\vec \xi (\vec \sigma ))f(\vec \sigma )=f(\vec \xi
(\vec \sigma ))$]

\begin{eqnarray}
{}^3e_{(a)r}(\tau ,\vec \sigma ) &\mapsto& {}^3e^{'}_{(a)r}(\tau ,{\vec \sigma}
^{'}(\vec \sigma ))={{\partial \sigma^s}\over {\partial \sigma^{{'}r}}}\,
{}^3e_{(a)s}(\tau ,\vec \sigma ),\nonumber \\
{} &\Rightarrow& {}^3e_{(a)r}(\tau ,\vec \sigma )={{\partial \xi^s(\vec \sigma )
}\over {\partial \sigma^r}}\, {}^3e^{'}_{(a)s}(\tau ,\vec \xi (\vec \sigma ))=
{{\partial \xi^s(\vec \sigma )}\over {\partial \sigma^r}}\, {\hat V}(\vec \xi
(\vec \sigma ))\, {}^3e^{'}_{(a)s}(\tau ,\vec \sigma ),\nonumber \\
&&{}\nonumber \\
{}^3g_{rs}(\tau ,\vec \sigma ) &\mapsto& {}^3g^{'}_{rs}(\tau ,{\vec \sigma}^{'}
(\vec \sigma ))={{\partial \sigma^u}\over {\partial \sigma^{{'}r}}}
{{\partial \sigma^v}\over {\partial \sigma^{{'}s}}}\, {}^3g_{uv}(\tau ,\vec 
\sigma ),\nonumber \\
&&{}\nonumber \\
\delta \, {}^3e_{(a)r}(\tau ,\vec \sigma )&=&{}^3e^{'}_{(a)r}(\tau ,{\vec 
\sigma}^{'}(\vec \sigma ))-{}^3e_{(a)r}(\tau ,\vec \sigma )=\delta_o\, {}^3e
_{(a)r}(\tau ,\vec \sigma )+{\hat \xi}^s(\vec \sigma )\partial_s\, {}^3e_{(a)r}
(\tau ,\vec \sigma )=\nonumber \\
&=&{{\partial \sigma^s}\over {\partial \sigma^{{'}r}}}\, {}^3e_{(a)s}(\tau ,
\vec \sigma )-{}^3e_{(a)r}(\tau ,\vec \sigma )=-\partial_r{\hat \xi}^s(\vec 
\sigma )\, {}^3e_{(a)s}(\tau ,\vec \sigma ),\nonumber \\
\delta_o\, {}^3e_{(a)r}(\tau ,\vec \sigma )&=&{}^3e^{'}_{(a)r}(\vec \sigma )-
{}^3e_{(a)r}(\vec \sigma )=-[\partial_r{\hat \xi}^s(\vec \sigma )+\delta^s_r
{\hat \xi}^u(\vec \sigma )\partial_u]{}^3e_{(a)s}(\tau ,\vec \sigma )=
\nonumber \\
&=&[{\cal L}_{-{\hat \xi}^s\partial_s}\, {}^3e_{(a)u}(\tau ,\vec \sigma )
d\sigma^u]_r=-\lbrace {}^3e_{(a)r}(\tau ,\vec \sigma ), \int d^3\sigma_1 {\hat 
\xi}^s({\vec \sigma}_1)\, {}^3{\tilde \Theta}_s(\tau ,{\vec \sigma}_1)
\rbrace ,\nonumber \\
&&{}\nonumber \\
\delta \, {}^3g_{rs}(\tau ,\vec \sigma )&=&{}^3g^{'}_{rs}(\tau ,{\vec \sigma}
^{'}(\vec \sigma ))-{}^3g_{rs}(\tau ,\vec \sigma )=\delta_o\, {}^3g_{rs}(\tau ,
\vec \sigma )+{\hat \xi}^u(\vec \sigma )\partial_u\, {}^3g_{rs}(\tau ,
\vec \sigma )=\nonumber \\
&=&{{\partial \sigma^u}\over {\partial \sigma^{{'}r}}}{{\partial \sigma^v}\over
{\partial \sigma^{{'}s}}}\, {}^3g_{uv}(\tau ,\vec \sigma )-{}^3g_{rs}(\tau ,
\vec \sigma )=-[\delta^u_r\partial_s{\hat \xi}^v(\vec \sigma )+\delta^v_s
\partial_r{\hat \xi}^u(\vec \sigma )]{}^3g_{uv}(\tau ,\vec \sigma ),\nonumber \\
\delta_o\, {}^3g_{rs}(\tau ,\vec \sigma )&=&{}^3g^{'}_{rs}(\tau ,\vec \sigma )-
{}^3g_{rs}(\tau ,\vec \sigma )=\nonumber \\
&=&-\Big[ \delta^u_r\partial_s{\hat \xi}^v(\vec \sigma )
+\delta^v_s\partial_r{\hat \xi}^u(\vec \sigma )+\delta^u_r\delta^v_s{\hat \xi}
^w(\vec \sigma )\partial_w\Big] {}^3g_{uv}(\tau ,\vec \sigma )=\nonumber \\
&=&[{\cal L}_{-{\hat \xi}^w\partial_w}\, {}^3g_{uv}(\tau ,\vec \sigma )d\sigma^u
\otimes d\sigma^v]_{rs}=-\lbrace {}^3g_{rs}(\tau ,\vec \sigma ),\int 
d^3\sigma_1 {\hat \xi}^s({\vec \sigma}_1)\, {}^3{\tilde \Theta}_s(\tau ,{\vec 
\sigma}_1)\rbrace .\nonumber \\
&&
\label{IV1}
\end{eqnarray}

Instead the action of finite and infinitesimal gauge rotations of angles 
$\alpha_{(c)}(\vec \sigma )$ and $\delta \alpha_{(c)}(\vec \sigma )$ is 
respectively

\begin{eqnarray}
&&{}^3e_{(a)r}(\tau ,\vec \sigma )\mapsto \, {}^3R_{(a)(b)}(\alpha_{(c)}(\vec 
\sigma ))\, {}^3e_{(b)r}(\tau ,\vec \sigma ),\nonumber \\
&&\delta_o\, {}^3e_{(a)r}(\tau ,\vec \sigma )=\lbrace {}^3e_{(a)r}(\tau ,
\vec \sigma ), \int d^3\sigma_1 \delta \alpha_{(c)}({\vec \sigma}_1)\,
{}^3{\tilde M}_{(c)}(\tau ,{\vec \sigma}_1)\rbrace =\nonumber \\
&=&\epsilon_{(a)(b)(c)}\delta \alpha_{(b)}(\vec \sigma )\, {}^3e_{(c)r}(\tau ,
\vec \sigma ).
\label{IV2}
\end{eqnarray}

To identify the algebra ${\bar g}_R$ of 
${\bar {\cal G}}_R$, let us study its symplectic action on
$T^{*}{\cal C}_e$, i.e. the infinitesimal canonical transformations generated by
the first class constraints ${}^3{\tilde M}_{(a)}(\tau ,\vec \sigma )$,
${}^3{\tilde \Theta}_r(\tau ,\vec \sigma )$. Let us define the vector fields

\begin{eqnarray}
&&X_{(a)}(\tau ,\vec \sigma )=-\lbrace .,{}^3{\tilde M}_{(a)}(\tau ,\vec 
\sigma )\rbrace ,\nonumber \\
&&Y_r(\tau ,\vec \sigma )=-\lbrace .,{}^3{\tilde \Theta}_r(\tau ,\vec \sigma )
\rbrace .
\label{IV3}
\end{eqnarray}

\noindent Due to Eqs.(\ref{I5})  they close the algebra

\begin{eqnarray}
&&[X_{(a)}(\tau ,\vec \sigma ),X_{(b)}(\tau ,{\vec \sigma}^{'})]=\delta^3(\vec
\sigma ,{\vec \sigma}^{'}) \epsilon_{(a)(b)(c)} X_{(c)}(\tau ,\vec \sigma ),
\nonumber \\
&&[X_{(a)}(\tau ,\vec \sigma ),Y_r(\tau ,{\vec \sigma}^{'})]=-{{\partial
\delta^3(\vec \sigma ,{\vec \sigma}^{'})}\over {\partial \sigma^{{'}r}}}
X_{(a)}(\tau ,{\vec \sigma}^{'}),\nonumber \\
&&[Y_r(\tau ,\vec \sigma ),Y_s(\tau ,{\vec \sigma}^{'})]=-{{\partial \delta^3
(\vec \sigma ,{\vec \sigma}^{'})}\over {\partial \sigma^{{'}s}}}Y_r(\tau ,{\vec 
\sigma}^{'})-{{\partial \delta^3(\vec \sigma ,{\vec \sigma}^{'})}\over
{\partial \sigma^{{'}r}}}Y_s(\tau ,\vec \sigma ).
\label{IV4}
\end{eqnarray}

These six vector fields describe the symplectic action of rotation and
space pseudodiffeomorphism gauge transformations on the 
subspace of phase space containing cotriads ${}^3e_{(a)r}(\tau ,\vec \sigma )$ 
and their conjugate momenta ${}^3{\tilde \pi}^r_{(a)}(\tau ,\vec \sigma )$. The
non commutativity of rotations and space pseudodiffeomorphisms means that the 
action of a space pseudodiffeomorphism on a rotated cotriad produces a cotriad 
which differ by a rotation with modified angles from the action of the space
pseudodiffeomorphism
on the original cotriad: if $\vec \sigma \rightarrow {\vec \sigma}^{'}(\vec
\sigma )$ is a space-diffeomorphism and ${}^3R_{(a)(b)}(\alpha_{(c)}(\vec
\sigma ))$ is a rotation matrix parametrized with angles $\alpha_{(c)}(\vec
\sigma )$, then

\begin{eqnarray}
{}^3e_{(a)r}(\tau ,\vec \sigma )&\mapsto& {}^3e^{'}_{(a)r}(\tau ,{\vec \sigma}
^{'}(\vec \sigma ))={{\partial \sigma^s}\over {\partial \sigma^{{'}r}}}\,
{}^3e_{(a)s}(\tau ,\vec \sigma ),\nonumber \\
{}^3R_{(a)(b)}(\alpha_{(c)}(\vec \sigma )) &\,& {}^3e_{(b)r}(\tau ,\vec \sigma )
\mapsto {}^3R_{(a)(b)}(\alpha^{'}_{(c)}({\vec \sigma}^{'}(\vec \sigma )))\,
{}^3e^{'}_{(b)r}(\tau ,{\vec \sigma}^{'}(\vec \sigma ))=\nonumber \\
&=&{{\partial \sigma^s}\over {\partial \sigma^{{'}r}}}\Big[{}^3R_{(a)(b)}(\alpha
_{(c)}(\vec \sigma ))\, {}^3e_{(b)s}(\tau ,\vec \sigma )\Big] =\nonumber \\
&=&{}^3R_{(a)(b)}(\alpha_{(c)}(\vec \sigma ))\, {}^3e^{'}_{(b)r}({\vec \sigma}
^{'}(\vec \sigma )),\nonumber \\
\Rightarrow &{}& \alpha^{'}_{(c)}({\vec \sigma}^{'}(\vec \sigma ))=\alpha_{(c)}
(\vec \sigma ),
\label{IV5}
\end{eqnarray}

\noindent i.e. the rotation matrices, namely the angles $\alpha_{(c)}(\vec 
\sigma )$, behave as scalar fields under space pseudodiffeomorphisms. Under 
infinitesimal rotations ${}^3R_{(a)(b)}(\delta \alpha_{(c)}(\vec \sigma ))=
\delta_{(a)(b)}+\delta \alpha_{(c)}(\vec \sigma )({\hat R}^{(c)})_{(a)(b)}=
\delta_{(a)(b)}+\epsilon_{(a)(b)(c)} \delta \alpha_{(c)}(\vec \sigma )$ and
space pseudodiffeomorphisms ${\vec \sigma}^{'}(\vec \sigma )=\vec \sigma +\delta
\vec \sigma (\vec \sigma )$ [${\hat R}^{(c)}$ are the SO(3) generators in the
adjoint representation; $\delta \alpha_{(c)}(\vec \sigma )$, $\delta \vec
\sigma (\vec \sigma )$ are infinitesimal variations], we have

\begin{eqnarray}
\int d^3\sigma_1d^3\sigma_2 &{}& \delta \sigma^s({\vec \sigma}_2)\delta \alpha
_{(c)}({\vec \sigma}_1) [Y_s(\tau ,{\vec \sigma}_2),X_{(c)}(\tau ,{\vec 
\sigma}_1)] {}^3e_{(a)r}(\tau ,\vec \sigma )=\nonumber \\
&=&\int d^3\sigma_2 \delta \beta_{(c)}({\vec \sigma}_2) X_{(c)}(\tau ,{\vec 
\sigma}_2) {}^3e_{(a)r}(\tau ,\vec \sigma ),\nonumber \\
\delta \beta_{(c)}(\vec \sigma )&=&\delta \sigma^s(\vec \sigma ) {{\partial
\alpha_{(c)}(\vec \sigma )}\over {\partial \sigma^s}},\nonumber \\
\Rightarrow &{}& \alpha^{'}_{(c)}(\vec \sigma )=\alpha_{(c)}(\vec \sigma -
\delta \vec \sigma (\vec \sigma ))=\alpha_{(c)}(\vec \sigma )-\delta \beta_{(c)}
(\vec \sigma )\, \Rightarrow \delta_o\alpha_{(c)}(\vec \sigma )=-\delta
\beta_{(c)}(\vec \sigma ).\nonumber \\
&&{}
\label{IV6}
\end{eqnarray}

The group manifold of the group ${\bar {\cal G}}_R$ of gauge transformations 
[isomorphic to $Aut\, L\Sigma_{\tau}$] is 
locally parametrized by $\vec \xi (\vec \sigma )$ and by three angles $\alpha
_{(c)}(\vec \sigma )$ [which are also functions of $\tau$], 
which are scalar fields under 
pseudodiffeomorphisms, and contains an invariant subgroup ${\bar {\cal G}}
_{ROT}$ [the group of gauge transformations of $L\Sigma_{\tau}$; it is a 
splitting normal Lie subgroup of $Aut\, L\Sigma_{\tau}$ \cite{abba}$\,\,$],  
whose group manifold (in the passive interpretation) is the space
of the cross sections of the trivial principal bundle
$\Sigma_{\tau} \times SO(3)\approx L\Sigma_{\tau}$ over $\Sigma_{\tau}$, like 
in SO(3) Yang-Mills theory\cite{lusa}, if $\Sigma_{\tau}$ is ``topologically
trivial" (its homotopy groups $\pi_k(\Sigma_{\tau})$ all vanish); therefore, 
it may be parametrized as said above. As affine
function space of connections on this principal SO(3)-bundle we shall take
the space of spin connection 1-forms ${}^3\omega_{(a)}$, whose
pullback to $\Sigma_{\tau}$ by means of cross sections $\sigma :\Sigma_{\tau}
\rightarrow \Sigma_{\tau}\times SO(3)$ are the 
(Levi-Civita) spin connections ${}^3\omega
_{r(a)}(\tau ,\vec \sigma )d\sigma^r=\sigma^{*}\, {}^3\omega_{(a)}$ built
with the cotriads ${}^3e_{(a)r}(\tau ,\vec \sigma )$
[they and not the spin connections are the independent
variables of tetrad gravity]  such that
${}^3g_{rs}={}^3e_{(a)r}\, {}^3e_{(a)s}$.

Due to our hypotheses on $\Sigma_{\tau}$ (parallelizable, asymptotically
flat, topologically trivial, geodesically complete), the Hopf-Rinow theorem
implies the existence of (at least) one point $p\in \Sigma_{\tau}$ which can
be chosen as reference point and can be connected to every other point 
$q\in \Sigma_{\tau}$ with a minimizing geodesic segment $\gamma_{pq}$;
moreover, the theorem says that the geodesic exponential map $Exp_p$ is
defined on all $T_p\Sigma_{\tau}$. If $\Sigma_{\tau}$ is further restricted to
have sectional curvature ${}^3K_p(\Pi ) \leq 0$ for each $p\in \Sigma_{\tau}$
and each tangent plane $\Pi \subset T_p\Sigma_{\tau}$, the Hadamard theorem
\cite{oneil} says that for each $p\in \Sigma_{\tau}$ the geodesic exponential 
map $Exp_p:T_p\Sigma_{\tau}\rightarrow \Sigma_{\tau}$ is a diffeomorphism:
therefore, there is a unique geodesic joining any pair of points $p,q\in 
\Sigma_{\tau}$ and $\Sigma_{\tau}$ is diffeomorphic to $R^3$ as we have
assumed.

In absence of rotations, the group ${\bar {\cal G}}_R$ 
is reduced to the group $Diff\,
\Sigma_{\tau}$ of space-diffeomorphisms. In the active point of view,
diffeomorphisms are smooth mappings (with smooth inverse) $\Sigma_{\tau}
\rightarrow \Sigma_{\tau}$: under $Diff\, \Sigma_{\tau}$ a point $p\in \Sigma
_{\tau}$ is sent (in many ways) in every point of $\Sigma_{\tau}$. In the
passive point of view, the action of the elements of $Diff\, \Sigma_{\tau}$,
called pseudodiffeomorphisms,
on a neighbourhood of a point $p\in \Sigma_{\tau}$ is equivalent to all the
possible coordinatizations of the subsets of the neighbourhood of p [i.e. to
all possible changes of coordinate charts containing p]. 

A coordinate system (or chart) $(U,\sigma )$ in $\Sigma_{\tau}$ is a 
homeomorphism (which is also a diffeomorphism) $\sigma$ of an open set $U 
\subset \Sigma_{\tau}$ onto an open set $\sigma (U)$ of $R^3$: if $\sigma :U
\rightarrow \sigma (U)$ and $p\in U$, then $\sigma (p)=(\sigma
^r(p))$, where the functions $\sigma^r$ are called the coordinate functions of 
$\sigma$. An atlas on $\Sigma_{\tau}$ is a collection of charts in $\Sigma
_{\tau}$ such that: i) each point $p\in \Sigma_{\tau}$ is contained in the
domain of some chart; ii) any two charts overlap smoothly. Let ${\cal A}=
\lbrace (U_{\alpha},\sigma_{\alpha})\rbrace$ be the unique ``complete" atlas on
$\Sigma_{\tau}$, i.e. an atlas by definition containing each coordinate system
$(U_{\alpha},\sigma_{\alpha})$ in $\Sigma_{\tau}$ that overlaps smoothly with 
every coordinate system in ${\cal A}$.

Given a diffeomorphism $\phi :\Sigma_{\tau} \rightarrow \Sigma_{\tau}$ (i.e.
a smooth mapping with smooth inverse) and any chart $(U,\sigma )$ in ${\cal A}$,
then $(\phi (U), \sigma_{\phi}=\sigma \circ \phi^{-1})$ is another chart in 
${\cal A}$ with $\sigma (U)=\sigma_{\phi}(\phi (U))\subset R^3$ and,
if $p\in U$ and also $p\in \phi (U)$,
$\sigma_{\phi}(p)=(\sigma \circ \phi^{-1})(p)=\xi (p)$ [i.e. $\vec \sigma
\mapsto {\vec \sigma}_{\phi}(\vec \sigma )=\vec \xi (\vec \sigma )$]. Therefore,
each diffeomorphism $\phi :\Sigma_{\tau}\rightarrow \Sigma_{\tau}$ may be 
viewed as a mapping $\phi_{\cal A}:{\cal A}\rightarrow {\cal A}$. If we
consider a point $p\in \Sigma_{\tau}$ and the set ${\cal A}_p=\lbrace (U^p
_{\beta},\sigma^p_{\beta})\rbrace$ of all charts in ${\cal A}$ containing p,
then for each diffeomorphism $\phi :\Sigma_{\tau}\rightarrow \Sigma_{\tau}$ we 
will have $\phi_{\cal A}:{\cal A}_p\rightarrow {\cal A}_p$. This suggests that
a local parametrization of $Diff\, \Sigma_{\tau}$ around a point $p\in 
\Sigma_{\tau}$ (i.e. local diffeomorphisms defined on the open sets containing
p)  may be done by choosing an arbitrary chart $(U^p_o,
\sigma^p_o)$ as the local identity of diffeomorphisms [$\vec \xi (\vec \sigma )
=\vec \sigma$] and associating with every nontrivial diffeomorphism $\phi :
\Sigma_{\tau}\rightarrow \Sigma_{\tau}$ [$\vec \sigma \mapsto {\vec \sigma}^{'}
(\vec \sigma )=\vec \xi (\vec \sigma )$] the chart $(U^p_{\beta}=\phi (U^p_o),
\sigma^p_{\beta}=\sigma^p_o\circ \phi^{-1})$. Since 
$\Sigma_{\tau} \approx R^3$ admits global 
charts $\Xi$,  then the group manifold of $Diff\, \Sigma_{\tau}$ may 
be tentatively parametrized (in a nonredundant way)  with the space of smooth 
global cross sections (global coordinate systems) in a fibration $\Sigma_{\tau}
\times \Sigma_{\tau}\rightarrow \Sigma_{\tau}$ [each global cross section of 
this fibration is a copy $\Sigma^{(\Xi )}_{\tau}$ of $\Sigma_{\tau}$ with the
given coordinate system $\Xi$]: 
this is analogous to the parametrization of the gauge group 
of Yang-Mills theory with a family of global cross sections of the trivial 
principal bundle $P(M,G)=M\times G$ [see next Section for G=SO(3)]. 
The infinitesimal pseudodiffeomorphisms [the algebra $T\Sigma_{\tau}$ of $Diff\,
\Sigma_{\tau}$\cite{diff}; its generators in its symplectic action on $T^{*}
{\cal C}_e$ are the vector fields $Y_r(\tau ,\vec \sigma )$] would correctly 
correspond to the cross sections of the fibration $\Sigma_{\tau}\times 
T\Sigma_{\tau}\rightarrow \Sigma_{\tau}$. With more general $\Sigma_{\tau}$ 
the previous description would hold only locally.

By remembering Eq.(\ref{aut1}), the following picture emerges:\hfill\break
i) Choose a global coordinate system $\Xi$ on $\Sigma_{\tau} \approx R^3$
(for instance 3-orthogonal coordinates).\hfill\break
ii) In the description of $Diff\, \Sigma_{\tau}$ as $\Sigma_{\tau} \times 
\Sigma_{\tau} \rightarrow \Sigma_{\tau}$ this corresponds to the choice of a 
global cross section $\sigma_{\Xi}$ in $\Sigma_{\tau}\times \Sigma_{\tau}$,
chosen as conventional origin of the pseudodiffeomorphisms parametrized as
$\vec \sigma \mapsto \vec \xi (\vec \sigma )$.\hfill\break
iii) This procedure identifies a cross section ${\tilde \sigma}_{\Xi}$ of the
principal bundle $Aut\, L\Sigma_{\tau} \rightarrow Diff\, \Sigma_{\tau}$,
whose action on $L\Sigma_{\tau}$ will be the SO(3) gauge rotations in the
chosen coordinate system $\Xi$ on $\Sigma_{\tau}$.\hfill\break
iv) This will induce a $\Xi$-dependent trivialization of $L\Sigma_{\tau}$ to
$\Sigma_{\tau}^{(\Xi )}\times SO(3)$, in which $\Sigma_{\tau}$ has $\Xi$ as 
coordinate system and the identity cross section $\sigma^{(\Xi )}_I$ of 
$\Sigma_{\tau}^{(\Xi )}\times SO(3)$ corresponds to the origin of rotations in
the coordinate system $\Xi$ (remember that the angles are scalar fields under
pseudodiffeomorphisms in $Diff\, \Sigma_{\tau}$).\hfill\break
v) As
we will see in the next Section, it is possible to define new vector fields
${\tilde Y}_r(\tau ,\vec \sigma )$ which commute with the rotations ($[X_{(a)}
(\tau ,\vec \sigma ),{\tilde Y}_r(\tau ,{\vec \sigma}^{'})]=0$) and still 
satisfy the last line of Eqs.(\ref{IV4}). In this way the algebra ${\bar g}_R$
of the group ${\bar {\cal G}}_R$ is replaced (at least locally) by a new 
algebra ${\bar g}_R^{'}$, which  defines a  group ${\bar {\cal G}}_R^{'}$,
which is a (local) trivialization of $Aut\, L\Sigma_{\tau}$. 
It is at this level that the rotations in ${\bar {\cal G}}_{ROT}$ may be 
parametrized with a special family of cross sections of the trivial orthogonal
coframe bundle $\Sigma_{\tau}^{(\Xi )}\times SO(3) \approx L\Sigma_{\tau}$, as 
for SO(3) Yang-Mills theory, as said in iv).\hfill\break
We do not know whether these steps can be implemented rigorously in a global 
way for $\Sigma_{\tau} \approx R^3$; if this is possible, then the 
quasi-Shanmugadhasan canonical transformation of Section IV can be defined 
globally for global coordinate systems on $\Sigma_{\tau}$.

Both to study the singularity structure of DeWitt superspace
\cite{dew,fis,ing} for
the Riemannian 3-manifolds $\Sigma_{\tau}$ (the space of 3-metrics ${}^3g$
modulo $Diff\, \Sigma_{\tau}$), for instance the cone over cone singularities
of Ref.\cite{arms}, and the analogous phenomenon (called in this case Gribov
ambiguity) for the group ${\bar {\cal G}}_{ROT}$ of SO(3) gauge transformations,
we have to analyze the stability subgroups of the group ${\bar {\cal G}}_R$ of
gauge transformations for special cotriads ${}^3e_{(a)r}(\tau ,\vec \sigma )$,
the basic variables in tetrad gravity. In metric gravity, where the metric is
the basic variable and pseudodiffeomorphisms are the only gauge 
transformations, 
it is known that if the 3-metric ${}^3g$ over a noncompact 3-manifold like
$\Sigma_{\tau}$ satisfies boundary conditions compatible with being a
function in a Sobolev space $W^{2,s}$ with $s > 3/2$, then there exist special
metrics admitting ``isometries" [see Appendix B]. 
The group $Iso(\Sigma_{\tau},{}^3g)$ of isometries
of a 3-metric of a Riemann 3-manifold $(\Sigma_{\tau},{}^3g)$ is the subgroup
of $Diff\, \Sigma_{\tau}$ which leaves the functional form of the 3-metric
${}^3g_{rs}(\tau ,\vec \sigma )$ invariant [its Lie algebra is spanned by the
Killing vector fields]: the pseudodiffeomorphism $\vec \sigma \mapsto {\vec 
\sigma}^{'}(\vec \sigma )=\vec \xi (\vec \sigma )$ in $Diff\, \Sigma_{\tau}$ is 
an isometry in $Iso(\Sigma_{\tau},{}^3g)$ [see Eqs.(30) of I] if

\begin{equation}
{}^3g_{rs}(\tau ,{\vec \sigma}^{'}(\vec \sigma ))={}^3g^{'}_{rs}(\tau ,
{\vec \sigma}^{'}(\vec \sigma ))={{\partial \sigma^u}\over {\partial 
\sigma^{{'}r}}}{{\partial \sigma^v}\over {\partial \sigma^{{'}s}}}\,
{}^3g_{uv}(\tau ,\vec \sigma ).
\label{IV7}
\end{equation}

The function space of 3-metrics turns out to be a stratified manifold
\cite{fis} with singularities. Each stratum contains all metrics ${}^3g$
with the same subgroup $Iso(\Sigma_{\tau},{}^3g)\subset Diff\, \Sigma_{\tau}$
[isomorphic but not equivalent subgroups of $Diff\, \Sigma_{\tau}$ produce
different strata of 3-metrics]; each point in a stratum with n  
Killing vectors is the vertex of a cone, which is a stratum with n-1 Killing
vectors (the cone over cone structure of singularities\cite{arms}).

From \hfill\break
\hfill\break
${}^3g_{rs}(\tau ,{\vec \sigma}^{'}(\vec \sigma ))={}^3g_{rs}^{'}
(\tau ,{\vec \sigma}^{'}(\vec \sigma ))={}^3e^{'}_{(a)r}(\tau ,{\vec \sigma}
^{'}(\vec \sigma ))\, {}^3e^{'}_{(a)s}(\tau ,{\vec \sigma}^{'}(\vec \sigma ))=
{{\partial \sigma^u}\over {\partial \sigma^{{'}r}}}{{\partial \sigma^v}\over
{\partial \sigma^{{'}s}}}\, {}^3g_{uv}(\tau ,\vec \sigma )={{\partial
\sigma^u}\over {\partial \sigma^{{'}r}}}{{\partial \sigma^v}\over {\partial
\sigma^{{'}s}}}\, {}^3e_{(a)r}(\tau ,\vec \sigma )\, {}^3e_{(a)s}(\tau ,
\vec \sigma )$,\hfill\break
\hfill\break
${}^3e^{'}_{(a)r}(\tau ,{\vec \sigma}^{'}(\vec \sigma ))=
R_{(a)(b)}(\gamma (\tau ,{\vec \sigma}^{'}(\vec \sigma )))
{{\partial \sigma^u}\over {\partial \sigma^{{'}r}}}\, {}^3e_{(b)u}(\tau ,
\vec \sigma )$ \hfill\break
\hfill\break
[at the level of cotriads a pseudodiffeomorphism-dependent
rotation is allowed], 
it follows that also the functional form of the associated
cotriads is invariant under $Iso(\Sigma_{\tau},{}^3g)$

\begin{equation}
{}^3e_{(a)r}(\tau ,{\vec \sigma}^{'}(\vec \sigma ))={}^3e^{'}_{(a)r} (\tau ,
{\vec \sigma}^{'}(\vec \sigma ))=R_{(a)(b)}(\gamma (\tau ,{\vec \sigma}
^{'}(\vec \sigma )))
{{\partial \sigma^s}\over {\partial \sigma^{{'}r}}}\,
{}^3e_{(b)s}(\tau ,\vec \sigma ).
\label{IV8}
\end{equation}

Moreover, ${}^3g_{rs}(\tau ,{\vec \sigma}^{'}(\vec \sigma ))={}^3g^{'}_{rs}
(\tau ,{\vec \sigma}^{'}(\vec \sigma ))$ implies ${}^3\Gamma^{{'}u}_{rs}(\tau ,
{\vec \sigma}^{'}(\vec \sigma ))={}^3\Gamma^u_{rs}(\tau ,{\vec \sigma}^{'}
(\vec \sigma ))$ and 
${}^3R^{{'}u}{}_{rst}(\tau ,{\vec \sigma}^{'}(\vec \sigma ))
={}^3R^u{}_{rst}(\tau ,{\vec \sigma}^{'}(\vec \sigma ))$, so that $Iso(\Sigma
_{\tau},{}^3g)$ is also the stability group for the associated Christoffel 
symbols and Riemann tensor

\begin{eqnarray}
{}^3\Gamma^u_{rs}(\tau ,{\vec \sigma}^{'}(\vec \sigma ))&=&{}^3\Gamma^{{'}u}
_{rs}(\tau ,{\vec \sigma}^{'}(\vec \sigma ))=\nonumber \\
&=&{{\partial \sigma
^{{'}u}}\over {\partial \sigma^v}}{{\partial \sigma^m}\over {\partial
\sigma^{{'}r}}}{{\partial \sigma^n}\over {\partial \sigma^{{'}s}}}\,
{}^3\Gamma^v_{mn}(\tau ,\vec \sigma )+{{\partial^2\sigma^v}\over {\partial
\sigma^{{'}r}\partial \sigma^{{'}s}}}{{\partial \sigma^{{'}u}}\over
{\partial \sigma^v}},\nonumber \\
&&{}\nonumber \\
{}^3R^u{}_{rst}(\tau ,{\vec \sigma}^{'}(\vec \sigma ))&=&{}^3R^{{'}u}{}_{rst}
(\tau ,{\vec \sigma}^{'}(\vec \sigma ))=\nonumber \\
&=&{{\partial \sigma
^{{'}u}}\over {\partial \sigma^v}}{{\partial \sigma^l}\over {\partial
\sigma^{{'}r}}} {{\partial \sigma^m}\over {\partial \sigma^{{'}s}}}
{{\partial \sigma^n}\over {\partial \sigma^{{'}t}}}\, {}^3R^v{}_{lmn}(\tau ,
\vec \sigma ).
\label{IV10}
\end{eqnarray}

\noindent Let us remark that in the Yang-Mills case (see Ref.\cite{lusa} and
the end of this Section) the field strengths have generically a larger
stability group (the gauge copies problem) than the gauge potentials
(the gauge symmetry problem). Here, one expects that Riemann tensors 
(the field strengths) should
have a stability group ${\cal S}_R(\Sigma_{\tau},{}^3g)$ generically larger
of the one of the Christoffel symbols (the connection) ${\cal S}_{\Gamma}(\Sigma
_{\tau},{}^3g)$, which in turn should be larger of the isometry group of the
metric: ${\cal S}_R(\Sigma_{\tau},{}^3g) \supseteq {\cal S}_{\Gamma}(\Sigma
_{\tau},{}^3g) \supseteq Iso(\Sigma_{\tau},{}^3g)$. However, these stability
groups do not seem to have been explored in the literature.

Since the most general transformation in ${\bar {\cal G}}_R$ for
cotriads ${}^3e_{(a)r}(\tau ,\vec \sigma )$, spin connections
${\hat R}^{(a)}\, {}^3\omega_{r(a)}(\tau ,\vec \sigma )$ and field strengths
${\hat R}^{(a)}\, {}^3\Omega_{rs(a)}(\tau ,\vec \sigma )$ is [we send
$\Lambda \rightarrow \Lambda^{-1}$ in Eqs.(32) of I to conform 
with the notations of Ref.\cite{lusa}]

\begin{eqnarray}
{}^3e^{{'}R}_{(a)r}(\tau ,{\vec \sigma}^{'}(\vec \sigma ))&=&
{}^3R_{(a)(b)}(\alpha_{(c)}(\tau ,\vec \sigma )){{\partial \sigma^s}\over 
{\partial \sigma^{{'}r}}}\,
{}^3e_{(b)s}(\tau ,\vec \sigma ),\nonumber \\
&&{}\nonumber \\
{\hat R}^{(a)}\, {}^3\omega^{{'}R}_{r(a)}(\tau ,{\vec \sigma}^{'}(\vec \sigma )
&=&{{\partial \sigma^u}\over {\partial \sigma^{{'}r}}}
\Big[ {}^3R^{-1}(\alpha_{(e)}
(\tau ,\vec \sigma ))\, {\hat R}^{(a)}\, {}^3\omega_{u(a)}(\tau ,\vec \sigma )\,
{}^3R(\alpha_{(e)}(\tau ,\vec \sigma ))+\nonumber \\
&+&{}^3R^{-1}(\alpha_{(e)}(\tau ,\vec \sigma ))
\partial_u\, {}^3R(\alpha_{(e)}(\tau ,\vec \sigma ))\Big] =\nonumber \\
&=&{{\partial \sigma^u}\over {\partial \sigma^{{'}r}}}\Big[ {\hat R}^{(a)}\,
{}^3\omega_{u(a)}(\tau ,\vec \sigma )+{}^3R^{-1}(\alpha_{(e)}(\tau ,\vec 
\sigma ))\, {\hat D}^{(\omega )}_u\, {}^3R(\alpha_{(e)}(\tau ,\vec \sigma ))
\Big] =\nonumber \\
&=&{\hat R}^{(a)}\, {}^3\omega^{'}_{r(a)}(\tau ,{\vec \sigma}^{'}(\vec \sigma ))
+{}^3R^{-1}(\alpha_{(e)}(\tau ,\vec \sigma ))\, {\hat D}^{(\omega^{'})}_r\, 
{}^3R(\alpha^{'}_{(e)}({\tau ,\vec \sigma}^{'}(\vec \sigma ))),\nonumber \\
&&{}\nonumber \\
{\hat R}^{(a)}\, {}^3\Omega^{{'}R}_{rs(a)}(\tau ,{\vec \sigma}^{'}(\vec 
\sigma ))&=&{{\partial \sigma^u}\over {\partial \sigma^{{'}r}}}{{\partial
\sigma^v}\over {\partial \sigma^{{'}s}}}\, {}^3R^{-1}(\alpha_{(e)}(\tau ,\vec 
\sigma ))\, {\hat R}^{(a)}\, {}^3\Omega_{uv(a)}(\tau ,\vec \sigma )\, 
{}^3R(\alpha_{(e)}(\tau ,\vec \sigma ))=\nonumber \\
&=&{{\partial \sigma^u}\over {\partial \sigma^{{'}r}}}{{\partial
\sigma^v}\over {\partial \sigma^{{'}s}}}\, \Big( {\hat R}^{(a)}\, {}^3\Omega
_{uv(a)}(\tau ,\vec \sigma )+\nonumber \\
&+&{}^3R^{-1}(\alpha_{(e)}(\tau ,\vec 
\sigma ))\, \Big[ {\hat R}^{(a)}\, {}^3\Omega_{uv(a)}(\tau ,\vec \sigma ),{}^3R
(\alpha_{(e)}(\tau ,\vec \sigma ))\Big]\, \Big) =\nonumber \\
&=&{\hat R}^{(a)}\, {}^3\Omega^{'}_{rs(a)}(\tau ,{\vec \sigma}^{'}(\vec \sigma 
)) +\nonumber \\
&+&{}^3R^{-1}(\alpha_{(e)}^{'}(\tau ,{\vec \sigma}^{'}(\vec \sigma )))\,\Big[
{\hat R}^{(a)}\, {}^3\Omega^{'}_{rs(a)}(\tau ,{\vec \sigma}^{'}(\vec \sigma )),
{}^3R(\alpha_{(e)}^{'}(\tau ,{\vec \sigma}^{'}(\vec \sigma )))\Big] .
\label{IV13}
\end{eqnarray}

\noindent where $({\hat D}^{(\omega )}_r)_{(a)(b)}={\hat D}^{(\omega )}
_{(a)(b)r}(\tau ,\vec \sigma )=\delta_{(a)(b)}\partial_r+\epsilon_{(a)(c)(b)}\,
{}^3\omega_{r(c)}(\tau ,\vec \sigma )$ and ${}^3R(\alpha_{(e)})$ are $3\times
3$ rotation matrices, the behaviour of spin connections and field strengths 
under isometries can be studied.

Let us now briefly review the Gribov ambiguity for the spin connections
and the field strengths following Ref.\cite{lusa}.
All spin connections are invariant under gauge transformations belonging to the
center $Z_3$ of SO(3): ${}^3R\in Z_3$ $\Rightarrow \,\, {}^3\omega^R_{r(a)}=
{}^3\omega_{r(a)}$.

As shown in Ref.\cite{lusa}, there can be special spin connections ${}^3\omega
_{r(a)}(\tau ,\vec \sigma )$, which admit a stability subgroup ${\bar {\cal
G}}^{\omega}_{ROT}$ (``gauge symmetries") of ${\bar {\cal G}}_{ROT}$, leaving 
them fixed

\begin{equation}
{}^3R(\alpha_{(e)}(\tau ,\vec \sigma ))\in {\bar {\cal G}}^{\omega}_R
\Rightarrow {\hat D}^{(\omega )}_r\, {}^3R(\alpha_{(e)}(\tau ,\vec \sigma ))
=0\, \Rightarrow \, {}^3\omega^R_{r(a)}(\tau ,\vec \sigma )={}^3\omega_{r(a)}
(\tau ,\vec \sigma ).
\label{IV14}
\end{equation}

\noindent From Eq.(\ref{IV8}), it follows that under an isometry in $Iso\,
(\Sigma_{\tau},{}^3g)$ we have ${}^3\omega^{'}_{r(a)}(\tau ,{\vec \sigma}
^{'}(\vec \sigma ))= {}^3\omega_{r(a)}(\tau ,{\vec \sigma}^{'}(\vec \sigma ))$,
namely the rotations ${}^3R(\gamma (\tau ,{\vec \sigma}^{'}(\vec \sigma )))$
are gauge symmetries.

When there are gauge symmetries, the spin connection is 
``reducible": its holonomy 
group $\Phi^{\omega}$ is a subgroup of SO(3) [$\Phi^{\omega}\subset SO(3)$] and
${\bar {\cal G}}^{\omega}_R$ [which is always equal to the centralizer of the 
holonomy group in SO(3), $Z_{SO(3)}(\Phi^{\omega})$] satisfies 
${\bar {\cal G}}_{ROT}^{\omega}=Z_{SO(3)}(\Phi^{\omega})\supset Z_3$.

Moreover, there can be special field strengths ${}^3\Omega_{rs(a)}$
which admit a stability subgroup ${\bar {\cal G}}_{ROT}^{\Omega}$ of ${\bar 
{\cal G}}_{ROT}$ leaving them fixed

\begin{eqnarray}
{}^3R(\alpha_{(e)}(\tau ,\vec \sigma )&&\in {\bar {\cal G}}_R^{\Omega}\, 
\Rightarrow \,  [{\hat R}^{(a)}\, {}^3\Omega_{rs(a)}(\tau ,\vec \sigma ),{}^3R(
\alpha_{(e)}(\tau ,\vec \sigma )]=0\, \nonumber \\
&&\Rightarrow {}^3\Omega^R_{rs(a)}(\tau ,
\vec \sigma )={}^3\Omega_{rs(a)}(\tau ,\vec \sigma ).
\label{IV16}
\end{eqnarray}

\noindent We have ${\bar {\cal G}}_{ROT}^{\Omega}\supseteq {\bar {\cal G}}_{ROT}
^{\omega}=Z_{SO(3)}(\Phi^{\omega})\supset Z_3$ and there is the problem of 
``gauge copies":
there exist different spin connections ${}^3\omega_{r(a)}(\tau ,
\vec \sigma )$ giving rise to the same field strength ${}^3\Omega_{rs(a)}(\tau ,
\vec \sigma )$.

A spin connection is ``irreducible", when its holonomy group $\Phi^{\omega}$ 
is a ``not closed" irreducible matrix
subgroup of SO(3). In this case we have ${\bar {\cal G}}_{ROT}^{\Omega}\supset 
{\bar {\cal G}}_{ROT}^{\omega}=Z_{SO(3)}(\Phi^{\omega})=Z_3$ and
there are gauge copies, but not gauge symmetries.

Finally, a spin connection ${}^3\omega_{r(a)}(\tau ,\vec \sigma )$ is ``fully
irreducible" if $\Phi^{\omega}=SO(3)$: in this case there are neither gauge
symmetries nor gauge copies [${\bar {\cal G}}_{ROT}^{\Omega}={\bar {\cal G}}
_{ROT}
^{\omega}=Z_3$] and the holonomy bundle $P^{\omega}(p)$ of every point $p\in 
\Sigma_{\tau}\times SO(3)$ coincides with $\Sigma_{\tau}\times SO(3)$ itself,
so that every two points in $\Sigma_{\tau}\times SO(3)$ can be joined by a
$\omega$-horizontal curve. Only in this case  the covariant divergence is
an elliptic operator without zero modes (this  requires the use of special
weighted Sobolev spaces for the spin connections to exclude the irreducible and
reducible ones) and its Green function can be globally 
defined (absence of Gribov ambiguities).

In conclusion, the following diagram

\begin{eqnarray}
\begin{array}{ccccccc}
{}& {}& \rightarrow& {}& {}^3\omega_{r(a)}& \rightarrow& {}^3\Omega_{rs(a)}\\
{}^3e_{(a)r}& {}& {}& {}& {}& {}& \Updownarrow \\
{}& \rightarrow& {}^3g_{rs}& \rightarrow& {}^3\Gamma^u_{rs}& \rightarrow&
{}^3R^u{}_{vrs}, \end{array}
\label{IV17}
\end{eqnarray}

\noindent together with Eqs.(\ref{IV8}), (\ref{IV13}),  implies 
that, to avoid any kind of pathology associated with stability subgroups of
gauge transformations, one has to work with cotriads belonging to a function
space such that: i) there is no subgroup of isometries in the action of
$Diff\, \Sigma_{\tau}$ on the cotriads (no cone over cone structure of 
singularities in the lower branch of the diagram); ii) all the spin connections 
associated with the cotriads are fully irreducible (no type of Gribov
ambiguity in the upper branch of the diagram). Both these requirements point
towards the use of special weighted Sobolev spaces like in Yang-Mills
theory\cite{lusa,moncrief} (see Appendix C and its bibliography).

It would be useful to make a systematic study of the relationships between
the stability groups ${\cal S}_R(\Sigma_{\tau},{}^3g) \supseteq {\cal S}
_{\Gamma}(\Sigma_{\tau},{}^3g) \supseteq Iso\, (\Sigma_{\tau},{}^3g)$ and the
stability groups ${\bar {\cal G}}^{\Omega}_{ROT} \supseteq {\bar {\cal G}}
^{\omega}_{ROT}$ and to show rigorously that the presence of isometries (Gribov 
ambiguity) in the lower (upper) branch of the diagram implies the existence of
Gribov ambiguity (isometries) in the upper (lower) branch.

In Ref. \cite{russo3} there will be a complete discussion on the definition
of proper gauge transformations [ Eqs.(\ref{I4}) plus boundary conditions on
the parameters of gauge transformations (implying their angle-independent
approach to the identity at spatial infinity) will be needed]
, problem connected with the differentiability
of the Dirac Hamiltonian, with supertranslations, and with the asymptotic
behaviour of the constraints and of their gauge parameters. There will be also
the definition of the asymptotic Poincar\'e charges, which are the analogue of
the non-Abelian charges (generators of the improper `global or rigid' gauge 
transformations) of Yang-Mills theory; see Refs.\cite{p18,p21} for
interpretational problems. Instead,
see Ref.\cite{giul} for a treatment of large diffeomorphisms, the analogous of 
the large gauge transformations (due to winding number) of Yang-Mills theory
\cite{lusa}, not connected to the identity.

\vfill\eject

\section
{Multitemporal Equations and their Solution.}

In this Section we
study the multitemporal equations associated with the gauge 
transformations in ${\bar {\cal G}}_R$, to find a local parametrization of the 
cotriads ${}^3e_{(a)r}(\tau ,\vec \sigma )$ in terms of the parameters 
$\xi_r(\tau ,\vec \sigma )$ and $\alpha_{(a)}(\tau ,\vec \sigma )$ of
${\bar {\cal G}}_R$. We shall assume to have chosen a global coordinate system
$\Xi$ on $\Sigma_{\tau} \approx R^3$ to conform with the discussion of the
previous Section.

Let us start with the invariant subalgebra ${\bar g}_{ROT}$ [the algebra of 
${\bar {\cal G}}_{ROT}$] of rotations, whose generators are the vector fields 
$X_{(a)}(\tau ,\vec \sigma )$ of Eqs.(\ref{IV3}). Since the group manifold of 
${\bar {\cal G}}_{ROT}$ is a trivial principal bundle $\Sigma_{\tau}^{(\Xi )}
\times SO(3)\approx L\Sigma_{\tau}$ over $\Sigma_{\tau}$, endowed with the
coordinate system $\Xi$, with structure group SO(3) [to be
replaced by SU(2) when one studies the action of ${\bar {\cal G}}_{ROT}$ on 
fermion fields], we can use the results of Ref.\cite{lusa} for the case of SO(3)
Yang-Mills theory.

Let $\alpha_{(a)}$ be canonical coordinates of first kind on the group
manifold of SO(3). If $r^{(a)}$ are the generators of so(3), $[r^{(a)},r^{(b)}]
=\epsilon_{(a)(b)(c)}r^{(c)}$ [instead ${\hat R}^{(a)}$ are the generators in
the adjoint representation, $({\hat R}^{(a)})_{(b)(c)}=\epsilon_{(a)(b)(c)}$],
and if $\gamma_{\alpha}(s)=exp_{SO(3)}\, (s \alpha_{(a)}r^{(a)})$ is a 
one-parameter subgroup of SO(3) with tangent vector $\alpha_{(a)}r^{(a)}$
at the identity $I\in 
SO(3)$, then the element $\gamma_{\alpha}(1)=exp_{SO(3)}(\alpha_{(a)}r^{(a)})
\in N_I\subset SO(3)$ [$N_I$ is a neighbourhood of the identity such that
$exp_{SO(3)}$ is a diffeomorphism from a neighbourhood of $0\in so(3)$ to
$N_I$] is given coordinates $\lbrace \alpha_{(a)} \rbrace$. If ${\tilde Y}
_{(a)}$ and ${\tilde \theta}_{(a)}$ are dual bases ($i_{{\tilde Y}_{(a)}}
{\tilde \theta}_{(b)}=\delta_{(a)(b)}$) of left invariant vector fields and
left invariant (or Maurer-Cartan) 1-forms on SO(3), we have the standard
Maurer-Cartan structure equations \hfill\break
\hfill\break
$[{\tilde Y}_{(a)},{\tilde Y}_{(b)}]=
\epsilon_{(a)(b)(c)}{\tilde Y}_{(c)}$ \hfill\break
\hfill\break
[${\tilde Y}_{(a)}{|}_I=r^{(a)}\in so(3)$] and \hfill\break
\hfill\break
$d{\tilde \theta}_{(a)}=-{1\over 2}\epsilon_{(a)(b)(c)} {\tilde \theta}
_{(b)}\wedge {\tilde \theta}_{(c)}$ \hfill\break
\hfill\break
[${\tilde \theta}_{(a)}{|}_I=r_{(a)}\in
so(3)^{*}$, the dual Lie algebra; $TSO(3)\approx so(3)$, $T^{*}SO(3)
\approx so(3)^{*}$]. Then, from Lie theorems, in arbitrary coordinates on the 
group manifold we have \hfill\break
\hfill\break
${\tilde Y}_{(a)}=B_{(b)(a)}(\alpha ){{\partial}\over
{\partial \alpha_{(b)}}}$, ${\tilde \theta}_{(a)}=A_{(a)(b)}(\alpha )d\alpha
_{(b)}$, $A(\alpha )=B^{-1}(\alpha )$, $A(0)=B(0)=1$, \hfill\break
\hfill\break
and the Maurer-Cartan equations become \hfill\break
\hfill\break
${{\partial A_{(a)(c)}(\alpha )}\over {\partial \alpha_{(b)}}}
-{{\partial A_{(a)(b)}(\alpha )}\over {\partial \alpha_{(c)}}}=-\epsilon
_{(a)(u)(v)}A_{(u)(b)}(\alpha )A_{(v)(c)}(\alpha )$,\hfill\break
\hfill\break
${\tilde Y}_{(b)}
B_{(a)(c)}(\alpha )-{\tilde Y}_{(c)}B_{(a)(b)}(\alpha )=B_{(u)(b)}(\alpha )
{{\partial B_{(a)(c)}(\alpha )}\over {\partial \alpha_{(u)}}}-B_{(u)(c)}
(\alpha ){{\partial B_{(a)(b)}(\alpha )}\over {\partial \alpha_{(u)}}}=
B_{(a)(u)}(\alpha )\epsilon_{(u)(b)(c)}$. \hfill\break
\hfill\break
By definition these
coordinates are said canonical of first kind and
satisfy $A_{(a)(b)}(\alpha )\, \alpha_{(b)}=\alpha_{(a)}$ [compare
with Eq.(\ref{b5}) of Appendix A], 
so that we get $A(\alpha )=(e^{R\alpha}-1)/R\alpha$ with
$(R\alpha )_{(a)(b)}=({\hat R}^{(c)})_{(a)(b)}\alpha_{(c)}=\epsilon_{(a)(b)(c)}
\alpha_{(c)}$. The canonical 1-form on SO(3) is ${\tilde \omega}_{SO(3)}=
{\tilde \theta}_{(a)}r^{(a)}=A_{(a)(b)}(\alpha )d\alpha_{(b)}r^{(a)}$ [$=
a^{-1}(\alpha )d_{SO(3)}\, a(\alpha )$, $a(\alpha )\in SO(3)$; $d_{SO(3)}$ is
the exterior derivative on SO(3)]. Due to the Maurer-Cartan structure equations 
the 1-forms ${\tilde \theta}_{(a)}$ are not integrable on SO(3); however in the
neighbourhood $N_I\subset SO(3)$ we can integrate them along the preferred
defining line $\gamma_{\alpha}(s)$ defining the canonical coordinates of first
kind to get the phases \hfill\break
\hfill\break
$\Omega^{\gamma_{\alpha}}_{(a)}(\alpha (s))={}_{\gamma
_{\alpha}}\int_I^{\gamma_{\alpha}(s)}\, {\tilde \theta}_{(a)}{|}_{\gamma
_{\alpha}}={}_{\gamma_{\alpha}}\int_0^{\alpha (s)}\, A_{(a)(b)}(\bar \alpha )
d{\bar \alpha}_{(b)}$. \hfill\break
\hfill\break
If $d_{\gamma_{\alpha}}=ds{{d\alpha_{(a)}(s)}\over {ds}}
{{\partial}\over {\partial \alpha_{(a)}}}{|}_{\alpha =\alpha (s)}=d_{SO(3)}
{|}_{\gamma_{\alpha}(s)}$ is the directional derivative along $\gamma_{\alpha}$,
on $\gamma_{\alpha}$ we have $d_{\gamma_{\alpha}}\, {\Omega}^{\gamma
_{\alpha}}_{(a)}(\alpha (s))={\tilde \theta}_{(a)}(\alpha (s))$ and
$d_{\gamma_{\alpha}}\, {\tilde \theta}_{(a)}(\alpha (s))=0$ $\Rightarrow \,
d^2_{\gamma_{\alpha}}=0$. The analytic atlas ${\cal N}$ for the group manifold
of SO(3) is built by starting from the neighbourhood $N_I$ of the identity
with canonical coordinates of first kind by left multiplication by elements of
SO(3): ${\cal N}=\cup_{a\in SO(3)}\, \lbrace a \cdot N_I\rbrace$.

As shown in Ref.\cite{lusa} for $R^3\times SO(3)$, in a tubular neighbourhood
of the identity cross section $\sigma_I$ of the trivial principal bundle
$R^3\times SO(3)$ [in which each fiber is a copy of the SO(3) group
manifold] we can define generalized canonical coordinates of first kind on
each fiber so to build a coordinatization of $R^3\times SO(3)$. We
now extend this construction from the flat Riemannian manifold $(R^3,\delta
_{rs})$ to a Riemannian manifold $(\Sigma_{\tau},{}^3g_{rs})$ satisfying our
hypotheses, especially the Hadamard theorem, so
that the 3-manifold $\Sigma_{\tau}$, diffeomorphic to $R^3$, admits 
global charts.

Let us consider the fiber SO(3) over a point $p\in \Sigma_{\tau}$, chosen as
origin $\vec \sigma =0$ of the global chart $\Xi$ on $\Sigma_{\tau}$, with 
canonical coordinates of first kind $\alpha_{(a)}=\alpha_{(a)}(\tau ,\vec 0)$.
For a given spin connection ${}^3\omega_{(a)}$ on $\Sigma_{\tau}^{(\Xi )}\times
SO(3)$ let us consider the ${}^3\omega$-horizontal lift of the star of geodesics
of the Riemann 3-manifold $(\Sigma_{\tau},{}^3g_{rs}={}^3e_{(a)r}\, 
{}^3e_{(a)s})$ emanating from $p\in \Sigma_{\tau}$ through each point of the 
fiber SO(3). If the spin connection ${}^3\omega_{(a)}$ is fully irreducible, 
$\Sigma_{\tau}\times SO(3)$ is in this way foliated by a connection-dependent 
family of global cross sections defined by the ${}^3\omega$-horizontal lifts of
the star of geodesics [they are not ${}^3\omega$-horizontal cross sections, as
it was erroneously written in Ref.\cite{lusa}, since such cross sections do not 
exist when the holonomy groups in each point of $\Sigma_{\tau}\times SO(3)$ are
not trivial]. The canonical coordinates of first kind on the
reference SO(3) fiber may then be parallel (with respect to ${}^3\omega_{(a)}$)
transported to all the other fibers along these ${}^3\omega$-dependent global
cross sections. If $\tilde p=(p;\alpha_{(a)})=(\tau ,\vec 0;\alpha_{(a)}(\tau ,
\vec 0))$ is a point in $\Sigma_{\tau}\times SO(3)$ over $p\in \Sigma_{\tau}$,
if $\sigma_{(\tilde p)}:\Sigma_{\tau}\rightarrow \Sigma_{\tau}\times SO(3)$ is 
the ${}^3\omega$-dependent cross section through $\tilde p$ and if ${}^3\omega
^{(\tilde p)}_{r(a)}(\tau ,\vec \sigma )d\sigma^r=\sigma^{*}_{(\tilde p)}\,
{}^3\omega_{(a)}$, then the coordinates of the point intersected by $\sigma
_{(\tilde p)}$ on the SO(3) fiber over the point $p^{'}$ of $\Sigma_{\tau}$ with
coordinates $(\tau ,\vec \sigma )$ are

\begin{eqnarray}
\alpha_{(a)}(\tau ,\vec \sigma )&=&\alpha_{(b)}(\tau ,\vec 0)\, \zeta^{(\omega
^{(\tilde p)}_{(c)})}_{(b)(a)}(\vec \sigma ,\vec 0;\tau )=\nonumber \\
&=&\alpha_{(b)}
(\tau ,\vec 0)\Big( P_{\gamma_{pp^{'}}}\, 
e^{\int_{\vec 0}^{\vec \sigma}dz^r{\hat 
R}^{(c)}\, {}^3\omega^{(\tilde p)}_{r(c)}(\tau,\vec z)}\Big)_{(b)(a)},
\label{V1}
\end{eqnarray}

\noindent where $\zeta^{(\omega )}_{(b)(a)}(\vec \sigma ,\vec 0;\tau )$ is the
Wu-Yang nonintegrable phase with the path ordering evaluated along the
geodesic $\gamma_{pp^{'}}$ from p to $p^{'}$. The infinitesimal form is

\begin{eqnarray}
\alpha_{(a)}(\tau ,d\vec \sigma )&\approx& \alpha_{(a)}(\tau ,\vec 0)+{{\partial
\alpha_{(a)}(\tau ,\vec \sigma )}\over {\partial \sigma^r}}{|}_{\vec \sigma
=0} d\sigma^r \approx \nonumber \\
&\approx& \alpha_{(b)}(\tau ,\vec 0) 
\Big[ \delta_{(b)(a)}+({\hat R}^{(c)})_{(b)(a)}
\, {}^3\omega_{r(c)}(\tau ,\vec 0)d\sigma^r\Big] ,
\label{V2}
\end{eqnarray}

\noindent implying that the identity cross section $\sigma_I$ of $\Sigma_{\tau}
^{(\Xi )}\times SO(3)$ 
[$\alpha_{(a)}=\alpha_{(a)}(\tau ,\vec 0)=0$] is the origin for 
all SO(3) fibers: $\alpha_{(a)}(\tau ,\vec \sigma ){|}_{\sigma_I}=0$. As shown
in Ref.\cite{lusa}, on $\sigma_I$ we also have $\partial_r \alpha_{(a)}(\tau ,
\vec \sigma ){|}_{\sigma_I}=0$. The main property of this construction is that
these coordinates are such that a vertical infinitesimal increment $d\alpha
_{(a)}{|}_{\alpha =\alpha (\tau ,\vec \sigma )}$ of them along the defining
path (one-parameter subgroup) $\gamma_{\alpha (\tau ,\vec \sigma )}(s)$ in the
fiber SO(3) over $q\in \Sigma_{\tau}$ with coordinates $(\tau ,\vec \sigma )$ 
is numerically equal to the horizontal infinitesimal increment $\partial_r
\alpha_{(a)}(\tau ,\vec \sigma )d\sigma^r$ in going from $\vec \sigma$ to
$\vec \sigma +d\vec \sigma$ in $\Sigma_{\tau}$

\begin{equation}
d\alpha_{(a)}{|}_{\alpha =\alpha (\tau ,\vec \sigma )} = d\alpha_{(a)}(\tau ,
\vec \sigma )=\partial_r \alpha_{(a)}(\tau ,\vec \sigma ) d\sigma^r.
\label{V3}
\end{equation}

With this coordinatization of $\Sigma_{\tau}^{(\Xi )}\times SO(3)$,  
in the chosen global coordinate system $\Xi$ for $\Sigma_{\tau}$ in which the 
identity cross section $\sigma_I$ is chosen as the origin of the angles, as in 
Ref.\cite{lusa} we have
the following realization for the vector fields $X_{(a)}(\tau ,\vec \sigma )$
of Eqs.(\ref{IV3})

\begin{equation}
X_{(a)}(\tau ,\vec \sigma )=B_{(b)(a)}(\alpha_{(e)}(\tau ,\vec \sigma ))
{{\tilde \delta}\over {\delta \alpha_{(b)}(\tau ,\vec \sigma )}}\,
\Rightarrow {{\tilde \delta}\over {\delta \alpha_{(a)}(\tau ,\vec \sigma )}}=
A_{(b)(a)}(\alpha_{(e)}(\tau ,\vec \sigma )) X_{(b)}(\tau ,\vec \sigma ),
\label{V4}
\end{equation}

\noindent where the functional derivative is the directional functional 
derivative along the path $\gamma_{\alpha (\tau ,\vec \sigma )}(s)$ in $\Sigma
_{\tau}^{(\Xi )}\times SO(3)$ originating at the identity cross section 
$\sigma_I$ (the origin of all SO(3) fibers) in the SO(3) fiber over the point 
$p\in \Sigma_{\tau}$ with coordinates $(\tau ,\vec \sigma )$, 
corresponding in the
above construction to the path $\gamma_{\alpha}(s)$ defining the canonical
coordinates of first kind in the reference SO(3) fiber. It satisfies the 
commutator in Eq.(\ref{IV4}) due to the generalized Maurer-Cartan equations for
$\Sigma_{\tau}\times SO(3)$ [$A=B^{-1}$]

\begin{eqnarray}
B_{(u)(a)}(\alpha_{(e)}(\tau ,\vec \sigma ))&&{{\partial B_{(v)(b)}(\alpha
_{(e)})}\over {\partial \alpha_{(u)}}} {|}_{\alpha =\alpha (\tau ,\vec \sigma )}
-B_{(u)(b)}(\alpha_{(e)}(\tau ,\vec \sigma )){{\partial B_{(v)(a)}(\alpha_{(e)})
}\over {\partial \alpha_{(u)}}}
{|}_{\alpha =\alpha (\tau ,\vec \sigma )}=\nonumber \\
&=&B_{(v)(d)}(\alpha_{(e)}(\tau ,\vec \sigma )) \epsilon_{(d)(a)(b)},
\nonumber \\
{{\partial A_{(a)(c)}(\alpha_{(e)})}\over {\partial \alpha_{(b)}}}&{|}_{\alpha
=\alpha (\tau ,\vec \sigma )}&-{{\partial A_{(a)(b)}(\alpha_{(e)})}\over
{\partial \alpha_{(c)}}} {|}_{\alpha =\alpha (\tau ,\vec \sigma )}=\nonumber \\
&=&\epsilon_{(a)(u)(v)}A_{(u)(b)}(\alpha_{(e)}(\tau ,\vec \sigma ))A_{(v)(c)}
(\alpha_{(e)}(\tau ,\vec \sigma )),
\label{V5}
\end{eqnarray}

\noindent holding pointwise on each fiber of $\Sigma_{\tau}^{(\Xi )}\times 
SO(3)$ over $(\tau ,\vec \sigma )$ in a suitable tubular neighbourhood of the 
identity cross section.

By defining a generalized canonical 1-form for ${\bar {\cal G}}_{ROT}$,
\hfill\break
\hfill\break
$\tilde
\omega ={\hat R}^{(a)}\, {\tilde \theta}_{(a)}(\tau ,\vec \sigma )=H_{(a)}
(\alpha_{(e)}(\tau ,\vec \sigma )) d\alpha_{(a)}(\tau ,\vec \sigma )$, 
\hfill\break
where \hfill\break
${\tilde \theta}_{(a)}(\tau ,\vec \sigma )={\hat \theta}_{(a)}(\alpha_{(e)}
(\tau ,\vec \sigma ),\partial_r\alpha_{(e)}(\tau ,\vec \sigma ))={\tilde
\theta}_{(a)r}(\tau ,\vec \sigma )d\sigma^r=A_{(a)(b)}(\alpha_{(e)}(\tau ,
\vec \sigma )) d\alpha_{(b)}(\tau ,\vec \sigma )$ \hfill\break
\hfill\break
are the generalized
Maurer-Cartan 1-forms on the Lie algebra ${\bar g}_{ROT}$ of 
${\bar {\cal G}}_{ROT}$
and where we defined the matrices $H_{(a)}(\alpha_{(e)}(\tau ,\vec \sigma ))=
{\hat R}^{(b)}\, A_{(b)(a)}(\alpha_{(e)}(\tau ,\vec \sigma ))$, the previous
equations can be rewritten in the form of a zero curvature condition

\begin{equation}
{{\partial H_{(a)}(\alpha_{(e)})}\over {\partial \alpha_{(b)}}} {|}_{\alpha
=\alpha (\tau ,\vec \sigma )} - {{\partial H_{(b)}(\alpha_{(e)})}\over
{\partial \alpha_{(a)}}} {|}_{\alpha =\alpha (\tau ,\vec \sigma )} +
[H_{(a)}(\alpha_{(e)}(\tau ,\vec \sigma )),H_{(b)}(\alpha_{(e)}(\tau ,\vec 
\sigma ))]=0.
\label{V6}
\end{equation}

Eq.(64) of I and Eqs.(\ref{IV3}) and (\ref{V4}) give the following multitemporal
equations for the dependence of the cotriad ${}^3e_{(a)r}(\tau ,\vec \sigma )$ 
on the 3 gauge angles $\alpha_{(a)}(\tau ,\vec \sigma )$

\begin{eqnarray}
X_{(b)}(\tau ,{\vec \sigma}^{'})\, {}^3e_{(a)r}(\tau ,\vec \sigma )&=& 
B_{(c)(b)}(\alpha_{(e)}(\tau ,{\vec \sigma}^{'}) {{\tilde \delta \,
{}^3e_{(a)r}(\tau ,\vec \sigma )}\over {\delta \alpha_{(c)}(\tau ,{\vec 
\sigma}^{'})}} =\nonumber \\
&=&-\lbrace {}^3e_{(a)r}(\tau ,\vec \sigma ),{}^3{\tilde M}_{(b)}(\tau ,{\vec 
\sigma}^{'})\rbrace =-\epsilon_{(a)(b)(c)} \, {}^3e_{(c)r}(\tau ,\vec \sigma )
\delta^3(\vec \sigma ,{\vec \sigma}^{'}),\nonumber \\
&&{}\nonumber \\
\Rightarrow && {{\tilde \delta \, {}^3e_{(a)r}(\tau ,\vec \sigma )}\over
{\delta \alpha_{(b)}(\tau ,{\vec \sigma}^{'})}}=-\epsilon_{(a)(c)(d)}
A_{(c)(b)}(\alpha_{(e)}(\tau ,\vec \sigma ))\, {}^3e_{(a)r}(\tau ,\vec \sigma )
\delta^3(\vec \sigma ,{\vec \sigma}^{'})=\nonumber \\
&=&\Big[ {\hat R}^{(c)} A_{(c)(b)}(\alpha_{(e)}(\tau ,\vec \sigma ))
\Big]_{(a)(d)}\,
{}^3e_{(d)r}(\tau ,\vec \sigma ) \delta^3(\vec \sigma ,{\vec \sigma}^{'})=
\nonumber \\
&=& \Big[ H_{(b)}(\alpha_{(e)}(\tau ,\vec \sigma ))\Big]_{(a)(d)}\, {}^3e
_{(d)r}(\tau ,\vec \sigma ) \delta^3(\vec \sigma ,{\vec \sigma}^{'}).
\label{V7}
\end{eqnarray}

\noindent These equations are a functional multitemporal generalization of the
matrix equation ${d\over {dt}} U(t,t_o)= h\, U(t,t_o)$ , $U(t_o,t_o)=1$,
generating the concept of time-ordering. They are integrable (i.e. their
solution is path independent) due to Eqs.(\ref{V6}) and their solution is

\begin{equation}
{}^3e_{(a)r}(\tau ,\vec \sigma )={}^3R_{(a)(b)}(\alpha_{(e)}(\tau ,\vec 
\sigma ))\, {}^3{\bar e}_{(b)r}(\tau ,\vec \sigma ),
\label{V8}
\end{equation}

\noindent where [$l$ is an arbitrary path originating at the identity cross
section of $\Sigma_{\tau}^{(\Xi )}\times SO(3)$; due to the path independence 
it can be replaced with the defining path $\gamma_{(\alpha (\tau ,\vec 
\sigma )}(s)=\hat \gamma (\tau ,\vec \sigma ;s)$]

\begin{eqnarray}
{}^3R_{(a)(b)}(\alpha_{(e)}(\tau ,\vec \sigma ))&=& \Big( P\, e^{{}_{(l)}\int
_0^{\alpha_{(e)}(\tau ,\vec \sigma )}\, H_{(c)}({\bar \alpha}_{(e)}) {\cal
D}{\bar \alpha}_{(c)}} \Big)_{(a)(b)}=\nonumber \\
&=&\Big( P\, e^{{}_{(\hat \gamma )}\int_0^{\alpha_{(e)}(\tau ,\vec \sigma )}\, 
H_{(c)}({\bar \alpha}_{(e)}) {\cal D}{\bar \alpha}_{(c)}} \Big)_{(a)(b)}=
\nonumber \\
&=&\Big( P\, e^{\Omega^{\hat \gamma}(\alpha_{(e)}(\tau ,\vec \sigma ))} 
\Big)_{(a)(b)},
\label{V9}
\end{eqnarray}

\noindent is a point dependent rotation matrix [${}^3R^T_{(a)(b)}(\alpha )=
{}^3R^{-1}_{(a)(b)}(\alpha )$ since ${\hat R}^{(a)\dagger}=-{\hat R}^{(a)}$].

In Eq.(\ref{V9}) we introduced the generalized phase obtained by functional
integration along the defining path in $\Sigma_{\tau}^{(\Xi )}\times SO(3)$ 
of the generalized Maurer-Cartan 1-forms 

\begin{eqnarray}
\Omega^{\hat \gamma}(\alpha_{(e)}(\tau ,\vec \sigma ;s))&=& {}_{(\hat \gamma )}
\int_I^{\gamma_{\alpha (\tau ,\vec \sigma ;s)}}\, {\hat R}^{(a)} {\tilde \theta}
_{(a)} {|}_{\gamma_{\alpha (\tau ,\vec \sigma ;s)}}=\nonumber \\
&=&{}_{(\hat \gamma )}\int_0^{\alpha_{(e)}(\tau ,\vec \sigma ;s)}\, H_{(a)}
({\bar \alpha}_{(e)}) {\cal D}{\bar \alpha}_{(a)}=\nonumber \\
&=&{}_{(\hat \gamma )}\int_0^{\alpha_{(e)}(\tau ,\vec \sigma ;s)}\, {\hat R}
^{(a)} A_{(a)(b)}({\bar \alpha}_{(e)}) {\cal D} {\bar \alpha}_{(b)}.
\label{V10}
\end{eqnarray}

As shown in Ref.\cite{lusa}, we have \hfill\break
\hfill\break
$d_{\hat \gamma}\, \Omega^{\hat \gamma}
(\alpha_{(e)}(\tau ,\vec \sigma ;s))={\hat R}^{(a)} {\hat \theta}_{(a)}
(\alpha_{(e)}(\tau ,\vec \sigma ;s),\partial_r\alpha_{(e)}(\tau ,\vec \sigma ;
s))$, \hfill\break
\hfill\break
where $d_{\hat \gamma}$ is the restriction of the ``fiber or vertical"
derivative $d_V$ on $\Sigma_{\tau}^{(\Xi )}\times SO(3)$ [the BRST operator] to 
the defining path, satisfying $d^2_{\hat \gamma}=0$ due to the generalized
Maurer-Cartan equations.

In Eq.(\ref{V8}), ${}^3{\bar e}_{(a)r}(\tau ,\vec \sigma )$ are the cotriads
evaluated at $\alpha_{(a)}(\tau ,\vec \sigma )=0$ (i.e. on the identity
cross section). Being Cauchy data of Eqs.(\ref{V7}), they are independent from 
the angles $\alpha_{(a)}(\tau ,\vec \sigma )$, satisfy $\lbrace {}^3{\bar e}
_{(a)r}(\tau ,\vec \sigma ),{}^3{\tilde M}_{(b)}(\tau ,{\vec \sigma}^{'})
\rbrace =0$ and depend only on 6 independent functions [the $\alpha_{(a)}(\tau 
,\vec \sigma )$ are the 3 rotational degrees of freedom hidden in the 9
variables ${}^3e_{(a)r}(\tau ,\vec \sigma )$].

We have not found 3 specific conditions on cotriads implying their independency
from the angles $\alpha_{(a)}$.

Since from Eq.(34) of I we get for the spin connection [${\hat D}^{(\omega )}
_{(a)(b)r}(\tau ,\vec \sigma )$ is the SO(3) covariant derivative in the 
adjoint representation]

\begin{eqnarray}
X_{(b)}(\tau ,{\vec \sigma}^{'})\, {}^3\omega_{r(a)}(\tau ,\vec \sigma )&=&
B_{(c)(b)}(\alpha_{(e)}(\tau ,{\vec \sigma}^{'})) {{\tilde \delta \, {}^3\omega
_{r(a)}(\tau ,\vec \sigma )}\over {\delta \alpha_{(c)}(\tau ,{\vec \sigma}
^{'})}}  =\nonumber \\
&=&-\lbrace {}^3\omega_{r(a)}(\tau ,\vec \sigma ), {}^3{\tilde M}_{(b)}(\tau ,
{\vec \sigma}^{'})\rbrace =\nonumber \\
&=&\Big[ \delta_{(a)(b)}\partial_r+\epsilon_{(a)(c)(b)}
\, {}^3\omega_{r(c)}(\tau ,\vec \sigma )\Big] \delta^3(\vec \sigma ,{\vec 
\sigma}^{'})=\nonumber \\
&=&\Big[ \delta_{(a)(b)}\partial_r-({\hat R}^{(c)}\, {}^3\omega_{r(c)}(\tau 
,\vec \sigma ))_{(a)(b)}\Big] \delta^3(\vec \sigma ,{\vec \sigma}^{'})
={\hat D}^{(\omega 
)}_{(a)(b)r}(\tau ,\vec \sigma ) \delta^3(\vec \sigma ,{\vec \sigma}^{'}),
\nonumber \\
&&{}
\label{V11}
\end{eqnarray}

\noindent which is the same result as for the gauge potential of the SO(3)
Yang-Mills theory, we can use the results of Ref.\cite{lusa} to write the
solution of Eq.(\ref{V11})

\begin{eqnarray}
{}^3\omega_{r(a)}(\tau ,\vec \sigma )&=&A_{(a)(b)}(\alpha_{(e)}(\tau ,\vec 
\sigma )) \partial_r\alpha_{(b)}(\tau ,\vec \sigma )+{}^3\omega^{(T)}_{r(a)}
(\tau ,\vec \sigma ,\alpha_{(e)}(\tau ,\vec \sigma )),\nonumber \\
&&with\nonumber \\
{{\partial \, {}^3\omega^{(T)}_{r(a)}(\tau ,\vec \sigma ,\alpha_{(e)})}\over 
{\partial \alpha_{(b)}}}&&{|}_{\alpha =\alpha (\tau ,\vec \sigma )}=-\epsilon
_{(a)(d)(c)}A_{(d)(b)}(\alpha_{(e)}(\tau ,\vec \sigma ))\, {}^3\omega^{(T)}
_{r(c)}(\tau ,\vec \sigma ,\alpha_{(e)}(\tau ,\vec \sigma )).
\label{V12}
\end{eqnarray}

In ${}^3\omega_{r(a)}(\tau ,\vec \sigma )d\sigma^r={\tilde \theta}_{(a)}(\tau ,
\vec \sigma )+{}^3\omega^{(T)}_{r(a)}(\tau ,\vec \sigma ,\alpha_{(e)}(\tau ,
\vec \sigma )) d\sigma^r$, the first term is a pure gauge spin connection
(the BRST ghost), while the second one is the source of the field strength:
${}^3\Omega_{rs(a)}=\partial_r\, {}^3\omega^{(T)}_{s(a)}-\partial_s\, 
{}^3\omega^{(T)}_{r(a)}-\epsilon_{(a)(b)(c)}\, {}^3\omega^{(T)}_{r(b)}\, 
{}^3\omega^{(T)}_{s(c)}$. Moreover, the Hodge decomposition theorem 
[in the functional spaces where the spin connections are fully irreducible] 
implies that ${}^3\omega^{(\perp )}_{r(a)}(\tau ,\vec \sigma )={}^3\omega^{(T)}
_{r(a)}(\tau ,\vec \sigma ,\alpha_{(e)}(\tau ,\vec \sigma ))$ satisfies 
${}^3\nabla^r\, {}^3\omega^{(\perp )}_{r(a)}=0$.

Since we have $X_{(b)}(\tau ,{\vec \sigma}^{'})\, {}^3\omega^{(\perp )}_{r(a)}
(\tau ,\vec \sigma )=-\epsilon_{(a)(c)(b)}\, {}^3\omega^{(\perp )}_{r(c)}(\tau
,\vec \sigma ) \delta^3(\vec \sigma ,{\vec \sigma}^{'})$, we get

\begin{eqnarray}
{{\tilde \delta \, {}^3\omega^{(\perp )}_{r(a)}(\tau ,\vec \sigma )}\over
{\delta \alpha_{(b)}(\tau ,{\vec \sigma}^{'})}}&=&[H_{(b)}(\alpha_{(e)}(\tau ,
\vec \sigma ))]_{(a)(c)}\, {}^3\omega^{(\perp )}_{r(c)}(\tau ,\vec \sigma )
\delta^3(\vec \sigma ,{\vec \sigma }^{'}),\nonumber \\
\Rightarrow && {}^3\omega^{(\perp )}_{r(a)}(\tau ,\vec \sigma )=\Big( P\, 
e^{\Omega^{\hat \gamma}(\alpha_{(e)}(\tau ,\vec \sigma ))} \Big)_{(a)(b)}\, 
{}^3{\bar \omega}^{(\perp )}_{r(b)}(\tau ,\vec \sigma ),\nonumber \\
&&{}^3\nabla^r\, {}^3{\bar \omega}^{(\perp )}_{r(a)}(\tau ,\vec \sigma )=0.
\label{V13}
\end{eqnarray}

The transverse spin connection ${}^3{\bar \omega}^{(\perp )}_{r(a)}(\tau ,
\vec \sigma )$ is independent from the gauge angles $\alpha_{(a)}(\tau ,\vec 
\sigma )$ and is the source of the field strength ${}^3{\bar \Omega}_{rs(a)}=
\partial_r\, {}^3{\bar \omega}^{(\perp )}_{s(a)}-\partial_s\, {}^3{\bar \omega}
^{(\perp )}_{r(a)}-\epsilon_{(a)(b)(c)}\, {}^3{\bar \omega}^{(\perp )}_{r(b)}
\, {}^3{\bar \omega}^{(\perp )}_{s(c)}$ invariant under the rotation gauge
transformations. Clearly, ${}^3{\bar \omega}_{r(a)}^{(\perp )}$ is built with 
the reduced cotriads ${}^3{\bar e}_{(a)r}$.

Let us remark that for ${}^3\omega^F_{r(a)}(\tau , \vec \sigma )d\sigma^r=
{\tilde \theta}_{(a)}(\tau ,\vec \sigma )$ we get ${}^3\Omega_{rs(a)}(\tau ,
\vec \sigma )=0$ and then ${}^3R_{rsuv}=0$: in this case the Riemannian manifold
$(\Sigma_{\tau},{}^3g_{rs}={}^3e_{(a)r}\, {}^3e_{(a)s})$ becomes the Euclidean
manifold $(R^3,\, {}^3g^F_{rs})$ with ${}^3g^F_{rs}$ the flat 3-metric in 
curvilinear coordinates. Now Eq.(\ref{V8}) implies that ${}^3g_{rs}={}^3{\bar 
e}_{(a)r}\, {}^3{\bar e}_{(a)s}$ and then ${}^3g^F_{rs}(\tau ,\vec \sigma )
={{\partial {\tilde \sigma}^u}\over {\partial \sigma^r}}{{\partial {\tilde 
\sigma}^v}\over {\partial \sigma^s}}\delta_{uv}={}^3g^F_{rs}(\vec \sigma )$, if
${\tilde \sigma}^u(\vec \sigma )$ are Cartesian coordinates, so that\hfill\break
\hfill\break
${}^3\Gamma^{F\, u}_{rs}={{\partial \sigma^u}
\over {\partial {\tilde \sigma}^n}}{{\partial^2{\tilde \sigma}^n}\over {\partial
\sigma^r\partial \sigma^s}}={}^3e^u_{(a)}\partial_r\, {}^3e_{(a)s}=
{}^3\triangle^u_{rs}$ \hfill\break
\hfill\break
(see after Eqs.(21) of I); 
then, for an arbitrary 
${}^3g$ we have ${}^3\Gamma^u_{rs}={}^3\triangle^u_{rs}+{}^3{\bar \Gamma}^u
_{rs}$ with ${}^3{\bar \Gamma}^u_{rs}={}^3e^u_{(a)}\, {}^3e_{(b)s}\, {}^3\omega
_{r(a)(b)}$ the source of the Riemann tensor. This implies that\hfill\break 
\hfill\break
${}^3{\bar e}^F
_{(a)r}(\tau ,\vec \sigma )={{\partial {\tilde \sigma}^u}\over {\partial
\sigma^r}}\, {}^3{\tilde {\bar e}}^F_{(a)u}(\tau ,{\vec {\tilde \sigma}})=
\delta_{(a)u} {{\partial {\tilde \sigma}^u(\vec \sigma )}\over {\partial 
\sigma^r}}$.\hfill\break
\hfill\break
 Therefore, a flat cotriad on $R^3$ has the form

\begin{equation}
{}^3e^F_{(a)r}(\tau ,\vec \sigma )={}^3R_{(a)(b)}(\alpha_{(e)}(\tau ,\vec 
\sigma )) \delta_{(b)u} {{\partial {\tilde \sigma}^u(\vec \sigma )}\over
{\partial \sigma^r}}.
\label{V14}
\end{equation}

Eqs.(64) of I give the multitemporal equations for the momenta

\begin{eqnarray}
X_{(b)}(\tau ,{\vec \sigma }^{'})\, {}^3{\tilde \pi}^r_{(a)}(\tau ,\vec 
\sigma )&=&B_{(c)(b)}(\alpha_{(e)}(\tau ,{\vec \sigma }^{'}) {{\tilde
\delta \, {}^3{\tilde \pi}^r_{(a)}(\tau ,\vec \sigma )}\over {\delta
\alpha_{(c)}(\tau ,{\vec \sigma }^{'})}}=\nonumber \\
&=&-\epsilon_{(a)(b)(c)}\, {}^3{\tilde \pi}^r_{(c)}(\tau ,\vec \sigma ) \delta
^3(\vec \sigma ,{\vec \sigma }^{'}),
\label{V15}
\end{eqnarray}

\noindent whose solution is [${}^3{\bar {\tilde \pi}}^r_{(a)}(\tau ,\vec 
\sigma )$ depends only on 6 independent functions]

\begin{equation}
{}^3{\tilde \pi}^r_{(a)}(\tau ,\vec \sigma )={}^3R_{(a)(b)}(\alpha_{(e)}(\tau ,
\vec \sigma ))\, {}^3{\bar {\tilde \pi}}^r_{(b)}(\tau ,\vec \sigma ).
\label{V16}
\end{equation}

With the definition of SO(3) covariant derivative given in Eq.(\ref{V11}), the
constraints ${\hat {\cal H}}_{(a)}(\tau ,\vec \sigma )\approx 0$ of
Eqs.(61) and (62) of I, may be written as

\begin{eqnarray}
{\hat {\cal H}}_{(a)}(\tau ,\vec \sigma )&=&-{}^3e^r_{(a)}(\tau ,\vec \sigma )
\Big[ {}^3{\tilde \Theta}_r + {}^3\omega_{r(b)}\, {}^3{\tilde M}_{(b)}\Big]
(\tau ,\vec \sigma )=\nonumber \\
&=&{\hat D}^{(\omega )}_{(a)(b)r}(\tau ,\vec \sigma )\, {}^3{\tilde \pi}^r
_{(b)}(\tau ,\vec \sigma )\approx 0,
\label{V17}
\end{eqnarray}

\noindent so that we have [${}^3{\tilde \pi}^{(T)r}_{(a)}(\tau ,\vec \sigma )$
is a field with zero SO(3) covariant divergence]

\begin{eqnarray}
{}^3{\tilde \pi}^r_{(a)}(\tau ,\vec \sigma )&=&{}^3{\tilde \pi}^{(T)r}_{(a)}
(\tau ,\vec \sigma )-\int d^3\sigma^{'}\, \zeta^{(\omega )r}_{(a)(b)}(\vec 
\sigma ,{\vec \sigma }^{'};\tau )\, {\hat {\cal H}}_{(b)}(\tau ,{\vec \sigma }
^{'}),\nonumber \\
{}&&{\hat D}^{(\omega )}_{(a)(b)r}(\tau ,\vec \sigma )\, {}^3{\tilde \pi}
^{(T)r}_{(b)}(\tau ,\vec \sigma )\equiv 0.
\label{V18}
\end{eqnarray}

\noindent In this equation, we introduced the Green function of the SO(3)
covariant divergence, defined by

\begin{equation}
{\hat D}^{(\omega )}_{(a)(b)r}(\tau ,\vec \sigma )\, \zeta^{(\omega )r}_{(b)(c)}
(\vec \sigma ,{\vec \sigma }^{'};\tau )=-\delta_{(a)(c)} \delta^3(\vec \sigma ,
{\vec \sigma }^{'}).
\label{V19}
\end{equation}

In Ref.\cite{lusa}, this Green function was evaluated for $\Sigma_{\tau}=R^3$,
the flat Euclidean space, by using the Green function $\vec c(\vec \sigma -
{\vec \sigma }^{'})$ of the flat ordinary divergence [$\triangle =-{\vec 
\partial}^2_{\sigma}$] in Cartesian coordinates

\begin{eqnarray}
\vec c(\vec \sigma -{\vec \sigma }^{'})&=&{\vec \partial}_{\sigma}\, c(\vec 
\sigma -{\vec \sigma }^{'})={{{\vec \partial}_{\sigma}}\over {\triangle}}
\delta^3(\vec \sigma -{\vec \sigma }^{'})={{\vec \sigma -{\vec \sigma }^{'}}
\over {4\pi |\, \vec \sigma -{\vec \sigma }^{'}|{}^3}}= {{\vec n(\vec \sigma -
{\vec \sigma }^{'})}\over {4\pi (\vec \sigma -{\vec \sigma }^{'})^2}},
\nonumber \\
&&{}\nonumber \\
{\vec \partial}_{\sigma}\cdot \vec c(\vec \sigma -{\vec \sigma }^{'})&=&
-\delta^3(\vec \sigma -{\vec \sigma }^{'}),
\label{V20}
\end{eqnarray}

\noindent where $\vec n(\vec \sigma -{\vec \sigma }^{'})$ is the tangent to the
flat geodesic (straight line segment) joining the point of coordinates $\vec 
\sigma $ and ${\vec \sigma }^{'}$, so that $\vec n(\vec \sigma -{\vec \sigma }
^{'})\cdot {\vec \partial}_{\sigma}$ is the directional derivative along the
flat geodesic.

With our special family of Riemannian 3-manifolds $(\Sigma_{\tau},{}^3g)$, we
would use Eq.(\ref{V20}) in the special global normal chart in which the star of
geodesics originating from the reference point p becomes a star of straight 
lines. In non normal coordinates, the Green function $\vec c(\vec \sigma -
{\vec \sigma }^{'})$ will be replaced with the gradient of the Synge world 
function\cite{synge} or DeWitt geodesic interval bitensor\cite{dew} $\sigma
_{DW}(\vec \sigma ,{\vec \sigma }^{'})$ [giving the arc length of the geodesic
from $\vec \sigma $ to ${\vec \sigma }^{'}$] adapted from the Lorentzian $M^4$
to the Riemannian $\Sigma_{\tau}$, i.e. \hfill\break
\hfill\break
$d^r_{\gamma_{pp^{'}}}(\vec \sigma ,
{\vec \sigma }^{'})={1\over 3}\sigma_{DW}^r(\vec \sigma ,{\vec \sigma }^{'})=
{1\over 3}\, {}^3\nabla^r_{\sigma}\, \sigma_{DW}(\vec \sigma ,{\vec \sigma }
^{'})={1\over 3} \partial^r_{\sigma}\, \sigma_{DW}(\vec \sigma ,{\vec \sigma }
^{'})$ \hfill\break
\hfill\break
giving in each point $\vec \sigma $ the tangent to the geodesic
$\gamma_{pp^{'}}$ joining the points p and $p^{'}$ of coordinates $\vec \sigma $
and ${\vec \sigma }^{'}$ in the direction from $p^{'}$ to p. Therefore, the
Green function is [$\partial_rd^r_{\gamma_{pp^{'}}}(\vec \sigma ,{\vec
\sigma}^{'})=-\delta^3(\vec \sigma ,{\vec \sigma}^{'})$; $d^r_{\gamma_{pp^{'}}}
(\vec \sigma ,{\vec \sigma}^{'})\, \partial_r$ is the directional derivative
along the geodesic $\gamma_{pp^{'}}$ at $p$ of coordinates $\vec \sigma$]

\begin{equation}
\zeta^{(\omega )r}_{(a)(b)}(\vec \sigma ,{\vec \sigma }^{'};\tau )=d^r_{\gamma
_{pp^{'}}}(\vec \sigma ,{\vec \sigma }^{'})
\Big( P_{\gamma_{pp^{'}}}\, e^{\int^{\vec \sigma }_{{\vec \sigma }^{'}}
d\sigma_1^s\, {\hat R}^{(c)}\, {}^3\omega_{s(c)}(\tau ,{\vec \sigma }_1)}
\Big)_{(a)(b)},
\label{V21}
\end{equation}

\noindent with the path ordering done
along the geodesic $\gamma_{pp^{'}}$. This path ordering
(Wu-Yang nonintegrable phase or geodesic Wilson line) is defined on all
$\Sigma_{\tau}\times SO(3)$ only if the spin connection is fully irreducible;
it is just the parallel transporter of Eq.(\ref{V1}).

Eqs.(\ref{V8}) show the dependence of the cotriad on the 3 angles $\alpha_{(a)}
(\tau ,\vec \sigma )$, which therefore must be expressible only in terms of the
cotriad itself and satisfy $\lbrace \alpha_{(a)}(\tau ,\vec \sigma ),\alpha
_{(b)}(\tau ,{\vec \sigma }^{'})\rbrace =0$. They are the rotational gauge
variables, canonically conjugate to  Abelianized rotation constraints
${\tilde \pi}^{\vec \alpha}_{(a)}(\tau ,\vec \sigma )\approx 0$. From
Eqs.(\ref{V4}), since the functional derivatives commute, we see that we
have\cite{lusa,hwang}

\begin{eqnarray}
{\tilde \pi}^{\vec \alpha}_{(a)}(\tau ,\vec \sigma )&=&{}^3{\tilde M}_{(b)}
(\tau ,\vec \sigma ) A_{(b)(a)}(\alpha_{(e)}(\tau ,\vec \sigma ))\approx 0,
\nonumber \\
&&{}\nonumber \\
\lbrace {\tilde \pi}^{\vec \alpha}_{(a)}(\tau ,\vec \sigma ),{\tilde \pi}
^{\vec \alpha}_{(b)}(\tau ,{\vec \sigma }^{'})\rbrace &=&0,\nonumber \\
\lbrace \alpha_{(a)}(\tau ,\vec \sigma ),{\tilde \pi}^{\vec \alpha}_{(b)}
(\tau ,{\vec \sigma }^{'})\rbrace &=&-A_{(c)(b)}(\alpha_{(e)}(\tau ,{\vec 
\sigma }^{'})) X_{(c)}(\tau ,{\vec \sigma }^{'}) \alpha_{(a)}(\tau ,
\vec \sigma )=\nonumber \\
&=&-\delta_{(a)(b)} \delta^3(\vec \sigma ,{\vec \sigma }^{'}).
\label{V23}
\end{eqnarray}

The functional equation determining $\alpha_{(a)}(\tau ,\vec \sigma )$ in terms
of ${}^3e_{(a)r}(\tau ,\vec \sigma )$ is

\begin{eqnarray}
-\delta_{(a)(b)} \delta^3(\vec \sigma ,{\vec \sigma }^{'})&=&\lbrace \alpha
_{(a)}(\tau ,\vec \sigma ),{}^3{\tilde M}_{(c)}(\tau ,{\vec \sigma }^{'})
\rbrace A_{(c)(b)}(\alpha_{(e)}(\tau ,{\vec \sigma }^{'})=\nonumber \\
&=&\epsilon_{(c)(u)(v)}\, {}^3e_{(u)r}(\tau ,{\vec \sigma }^{'}) \lbrace
\alpha_{(a)}(\tau ,\vec \sigma ),{}^3{\tilde \pi}^r_{(v)}(\tau ,{\vec \sigma }
^{'})\rbrace A_{(c)(b)}(\alpha_{(e)}(\tau ,{\vec \sigma }^{'}))=\nonumber \\
&=&\epsilon_{(c)(u)(v)}\, A_{(c)(b)}(\alpha_{(e)}(\tau ,{\vec \sigma }^{'}))
\, {}^3e_{(u)r}(\tau ,{\vec \sigma }^{'}) {{\delta \alpha_{(a)}(\tau ,\vec 
\sigma )}\over {\delta \, {}^3e_{(v)r}(\tau ,{\vec \sigma }^{'})}},\nonumber \\
&&{}\nonumber \\
\Rightarrow && \epsilon_{(b)(u)(v)}\, {}^3e_{(u)r}(\tau ,{\vec \sigma }^{'})
{{\delta \alpha_{(a)}(\tau ,\vec \sigma )}\over {\delta \, {}^3e_{(v)r}(\tau ,
{\vec \sigma }^{'})}}=-B_{(a)(b)}(\alpha_{(e)}(\tau ,\vec \sigma ))\delta^3
(\vec \sigma ,{\vec \sigma }^{'}),\nonumber \\
{}&&\epsilon_{(b)(u)(v)}\, {}^3e_{(u)r}(\tau ,{\vec \sigma }^{'}) 
\Big[ A_{(a)(c)}
(\alpha_{(e)}(\tau ,\vec \sigma )){{\delta \alpha_{(a)}(\tau ,\vec \sigma )}
\over {\delta \, {}^3e_{(v)r}(\tau ,{\vec \sigma }^{'})}}\Big] =-\delta_{(a)(b)}
\delta^3(\vec \sigma ,{\vec \sigma }^{'}),\nonumber \\
{}&&\epsilon_{(b)(u)(v)}\, {}^3e_{(u)r}(\tau ,{\vec \sigma }^{'}) {{\delta
\Omega^{\hat \gamma}_{(a)}(\alpha_{(e)}(\tau ,\vec \sigma ))}\over {\delta
\, {}^3e_{(v)r}(\tau ,{\vec \sigma }^{'})}}=-\delta_{(a)(b)}\delta^3(\vec 
\sigma ,{\vec \sigma }^{'}),\nonumber \\
\Rightarrow && \epsilon_{(b)(u)(v)} {{\delta
\Omega^{\hat \gamma}_{(a)}(\alpha_{(e)}(\tau ,\vec \sigma ))}\over {\delta
\, {}^3e_{(v)r}(\tau ,{\vec \sigma }^{'})}}=-{1\over 3}\delta_{(a)(b)}
{}^3e^r_{(u)}(\tau ,{\vec \sigma}^{'}) \delta^3(\vec 
\sigma ,{\vec \sigma }^{'}),\nonumber \\
{}&& {{\delta
\Omega^{\hat \gamma}_{(a)}(\alpha_{(e)}(\tau ,\vec \sigma ))}\over {\delta
\, {}^3e_{(b)r}(\tau ,{\vec \sigma }^{'})}}={1\over 6} ({\hat R}^{(u)})
_{(a)(b)}\, {}^3e^r_{(u)}(\tau ,\vec \sigma ) \delta^3(\vec 
\sigma ,{\vec \sigma }^{'}).
\label{V24}
\end{eqnarray}

This equation is not integrable like the corresponding one in the Yang-Mills
case \cite{lusa}. Having chosen a global coordinate system $\Xi$ on $\Sigma
_{\tau}$ as the conventional origin of pseudodiffeomorphisms, the discussion in 
Section II allows to define the trivialization $\Sigma^{(\Xi )}_{\tau} \times 
SO(3)$ of the coframe bundle $L\Sigma_{\tau}$. If: \hfill\break
i) $\sigma^{(\Xi )}_I$ is
the identity cross section of $\Sigma^{(\Xi )}_{\tau} \times SO(3)$,
corresponding to the coframe ${}^3\theta^I_{(a)}={}^3e^I_{(a)r} d\sigma^r$ in
$L\Sigma_{\tau}$ [$\sigma^r$ are the coordinate functions of $\Xi$]; 
\hfill\break
ii) $\sigma^{(\Xi )}$ is an arbitrary global cross section of $\Sigma^{(\Xi )}
_{\tau}\times SO(3)$, corresponding to a coframe ${}^3\theta_{(a)}={}^3e
_{(a)r} d\sigma^r$ in $L\Sigma_{\tau}$, in a tubolar neighbourhood of the
identity cross section where the generalized canonical coordinates of first kind
 on the fibers of $\Sigma^{(\Xi )}_{\tau}\times SO(3)$ (discussed at the 
beginning of this Section) are defined; \hfill\break
iii) $\sigma^{(\Xi )}(s)$ is the family
of global cross sections of $\Sigma^{(\Xi )}_{\tau}\times SO(3)$ connecting
$\Sigma^{(\Xi )}_I=\sigma^{(\Xi )}(s=0)$ and $\Sigma^{(\Xi )}=\sigma^{(\Xi )}
(s=1)$ so that on each fiber the point on $\sigma^{(\Xi )}_I$ is connected
with the point on $\Sigma^{(\Xi )}$ by the defining path $\hat \gamma$ of
canonical coordinates of first kind;\hfill\break
then the formal solution of the previous equation is

\begin{equation}
\Omega^{\hat \gamma}_{(a)}(\alpha_{(e)}(\tau ,\vec \sigma ))={1\over 6}
\, {}_{\hat \gamma}\int^{{}^3e_{(a)r}(\tau ,\vec \sigma )}
_{{}^3e^I_{(a)r}(\tau ,\vec \sigma )}\, ({\hat R}^{(u)}
)_{(a)(b)}\, {}^3e^r_{(u)}\, {\cal D}\, {}^3e_{(b)r},
\label{V24a}
\end{equation}

\noindent where the path integral is made along the path of coframes connecting
${}^3\theta^I_{(a)}$ with ${}^3\theta_{(a)}$ just described. As in 
Ref.\cite{lusa} , to get the angles $\alpha_{(a)}(\tau ,\vec \sigma )$ from
$\Omega^{\hat \gamma}_{(a)}(\alpha_{(e)}(\tau ,\vec \sigma ))$, we essentially 
have to invert the equation $\Omega^{\hat \gamma}_{(a)}(\alpha_{(e)})=
{}_{\hat \gamma} \int^{\alpha_{(e)}}_{0}\, A_{(a)(b)}(\bar \alpha )
d{\bar \alpha}_{(b)}$ with $A=(e^{R\alpha}-1)/R\alpha$.

Let us now study the multitemporal equations associated with 
pseudodiffeomorphisms
to find the dependence of ${}^3e_{(a)r}(\tau ,\vec \sigma )$ on the parameters
$\xi^r(\tau ,\vec \sigma )$. Disregarding momentarily rotations, let us look for
a realization of vector fields ${\tilde Y}_r(\tau ,\vec \sigma )$ satisfying 
the last line of Eqs.(\ref{IV4}). If we put

\begin{equation}
{\tilde Y}_r(\tau ,\vec \sigma )=-{{\partial \xi^s(\tau ,\vec \sigma )}\over
{\partial \sigma^r}} {{\delta}\over {\delta \xi^s(\tau ,\vec \sigma )}},
\label{V25}
\end{equation}

\noindent we find 

\begin{eqnarray}
[{\tilde Y}_r(\tau ,\vec \sigma ),{\tilde Y}_s(\tau ,{\vec 
\sigma }^{'})]&=&[{{\partial \xi^u(\tau ,\vec \sigma )}\over {\partial \sigma
^r}}{{\delta}\over {\delta \xi^u(\tau ,\vec \sigma )}},{{\partial \xi^v(\tau ,
{\vec \sigma}^{'})}\over {\partial \sigma^{{'}s}}}{{\delta}\over {\delta
\xi^v(\tau ,{\vec \sigma }^{'})}}]=\nonumber \\
&=&{{\partial \xi^u(\tau ,\vec \sigma )}\over {\partial \sigma^r}}{{\partial
\delta^3(\vec \sigma ,{\vec \sigma }^{'})}\over {\partial \sigma^{{'}s}}}
{{\delta}\over {\delta \xi^u(\tau ,{\vec \sigma }^{'})}}-{{\partial \xi^u(\tau 
,{\vec \sigma }^{'})}\over {\partial \sigma^{{'}s}}}{{\partial \delta^3(\vec 
\sigma ,{\vec \sigma }^{'})}\over {\partial \sigma^r}}{{\delta}\over {\delta
\xi^u(\tau ,\vec \sigma )}}=\nonumber \\
&=&-{{\partial \xi^u(\tau ,\vec \sigma )}\over {\partial \sigma^r}}{{\partial
\delta^3(\vec \sigma ,{\vec \sigma }^{'})}\over {\partial \sigma^{s}}}
{{\delta}\over {\delta \xi^u(\tau ,{\vec \sigma }^{'})}}+{{\partial \xi^u(\tau 
,{\vec \sigma }^{'})}\over {\partial \sigma^{{'}s}}}{{\partial \delta^3(\vec 
\sigma ,{\vec \sigma }^{'})}\over {\partial \sigma^{{'}r}}}{{\delta}\over 
{\delta \xi^u(\tau ,\vec \sigma )}}=\nonumber \\
&=&\Big[ -{{\partial}\over {\partial \sigma^s}}\Big( {{\partial \xi^u(\tau ,\vec
\sigma )}\over {\partial \sigma^r}}\delta^3(\vec \sigma ,{\vec \sigma }^{'})
\Big) +{{\partial^2\xi^u(\tau ,\vec \sigma )}\over {\partial \sigma^r\partial
\sigma^s}}\delta^3(\vec \sigma ,{\vec \sigma }^{'})\Big] {{\delta}\over {\delta
\xi^u(\tau ,{\vec \sigma}^{'})}}+\nonumber \\
&+&\Big[ {{\partial}\over {\partial \sigma^{{'}r}}}\Big( {{\partial \xi^u(\tau 
,{\vec \sigma}^{'})}\over {\partial \sigma^{{'}s}}}\delta^3(\vec \sigma ,{\vec 
\sigma }^{'})\Big) -{{\partial^2\xi^u(\tau ,{\vec \sigma}^{'})}\over {\partial 
\sigma^{{'}r}\partial \sigma^{{'}s}}}\delta^3(\vec \sigma ,{\vec \sigma }^{'})
\Big] {{\delta}\over {\delta \xi^u(\tau ,\vec \sigma )}}=\nonumber \\
&=&-{{\partial \delta^3(\vec \sigma ,{\vec \sigma }^{'})}\over
{\partial \sigma^s}}{{\partial \xi^u(\tau ,{\vec \sigma}^{'})}\over {\partial
\sigma^{{'}r}}}{{\delta}\over {\delta \xi^u(\tau ,{\vec \sigma}^{'})}}+
{{\partial \delta^3(\vec \sigma ,{\vec \sigma }^{'})}\over
{\partial \sigma^{{'}r}}}{{\partial \xi^u(\tau ,\vec \sigma )}\over {\partial
\sigma^s}}{{\delta}\over {\delta \xi^u(\tau ,\vec \sigma )}}=
\nonumber \\
&=&-{{\partial \delta^3(\vec \sigma ,{\vec \sigma }^{'})}\over
{\partial \sigma^{{'}s}}}{\tilde Y}_r(\tau ,{\vec \sigma }^{'})-{{\partial
\delta^3(\vec \sigma ,{\vec \sigma }^{'})}\over {\partial \sigma^{{'}r}}}
{\tilde Y}_s(\tau ,\vec \sigma ),
\label{V26}
\end{eqnarray}

\noindent in accord with the last of Eqs.(\ref{IV4}).
Therefore, the role of the Maurer-Cartan matrix B for rotations is taken by 
minus the Jacobian matrix of the pseudodiffeomorphism $\vec \sigma \mapsto \vec
\xi (\vec \sigma )$. To take into account the noncommutativity of rotations
and pseudodiffeomorphisms [the second line of Eqs.(\ref{IV4})], we need the 
definition

\begin{equation}
Y_r(\tau ,\vec \sigma )=-\lbrace .,{}^3{\tilde \Theta}_r(\tau ,\vec \sigma )
\rbrace =-{{\partial \xi^s(\tau ,\vec \sigma )}\over {\partial \sigma^r}}
{{\delta}\over {\delta \xi^s(\tau ,\vec \sigma )}}-{{\partial \alpha_{(a)}
(\tau ,\vec \sigma )}\over {\partial \sigma^r}} {{\tilde \delta}\over 
{\delta \alpha_{(a)}(\tau ,\vec \sigma )}}.
\label{V27}
\end{equation}

Clearly the last line of Eqs.(\ref{IV4}) is satisfied, while regarding the 
second line we have consistently

\begin{eqnarray}
[X_{(a)}(\tau ,\vec \sigma ),Y_r(\tau ,{\vec \sigma }^{'})]&=&-[B_{(b)(a)}
(\alpha_{(e)}(\tau ,\vec \sigma )){{\tilde \delta}\over {\delta \alpha_{(b)}
(\tau ,\vec \sigma )}},{{\partial \alpha_{(c)}(\tau ,{\vec \sigma }^{'})}\over
{\partial \sigma^{{'}r}}}{{\tilde \delta}\over {\delta \alpha_{(c)}(\tau ,
{\vec \sigma }^{'})}}]=\nonumber \\
&=&-B_{(b)(a)}(\alpha_{(e)}(\tau ,\vec \sigma )){{\partial \delta^3(\vec \sigma 
,{\vec \sigma }^{'})}\over {\partial \sigma^{{'}r}}}{{\tilde \delta}\over 
{\delta \alpha_{(b)}(\tau ,{\vec \sigma }^{'})}}+\nonumber \\
&+&{{\partial \alpha_{(c)}(\tau 
,\vec \sigma )}\over {\partial \sigma^r}}\delta^3(\vec \sigma ,{\vec \sigma }
^{'}){{\partial B_{(b)(a)}(\alpha_{(e)})}\over {\partial \alpha_{(c)}}}{|}
_{\alpha =\alpha (\tau ,\vec \sigma )}{{\tilde \delta}\over {\delta \alpha
_{(b)}(\tau ,\vec \sigma )}}=\nonumber \\
&=&-{{\partial \delta^3(\vec \sigma ,{\vec \sigma }^{'})}\over {\partial 
\sigma^{{'}r}}} X_{(a)}(\tau ,{\vec \sigma }^{'}).
\label{V28}
\end{eqnarray}

From Eqs.(\ref{V27}) and (\ref{V4}) we get

\begin{eqnarray}
{{\delta}\over {\delta \xi^r(\tau ,\vec \sigma )}}&=&-{{\partial \sigma^s(\vec 
\xi )}\over {\partial \xi^r}}{|}_{\vec \xi =\vec \xi (\tau ,\vec \sigma )}
\Big[ Y_s(\tau ,\vec \sigma )+A_{(a)(b)}(\alpha_{(e)}(\tau ,\vec \sigma ))
{{\partial \alpha_{(b)}(\tau ,\vec \sigma )}\over {\partial \sigma^s}} X_{(a)}
(\tau ,\vec \sigma )\Big] =\nonumber \\
&=&{{\partial \sigma^s(\vec \xi )}\over {\partial \xi^r}}{|}_{\vec \xi =\vec 
\xi (\tau ,\vec \sigma )}\Big[
\lbrace .,{}^3{\tilde \Theta}_s(\tau ,\vec \sigma )
\rbrace +{\tilde \theta}_{(a)s}(\tau ,\vec \sigma )\lbrace .,{}^3{\tilde M}
_{(a)}(\tau ,\vec \sigma )\rbrace \Big]\, {\buildrel {def}\over =}\nonumber \\
&{\buildrel {def} \over =}&\lbrace .,{\tilde \pi}^{\vec \xi}_r(\tau ,\vec 
\sigma )\rbrace ,\nonumber \\
{}&&\nonumber \\
\Rightarrow && \lbrace \xi^r(\tau ,\vec \sigma ),{\tilde \pi}^{\vec \xi}_s(\tau
,{\vec \sigma }^{'})\rbrace =\delta^r_s \delta^3(\vec \sigma ,{\vec \sigma }
^{'}),\nonumber \\
{}&&\lbrace {\tilde \pi}^{\vec \xi}_r(\tau ,\vec \sigma ),{\tilde \pi}^{\vec 
\xi}_s(\tau ,{\vec \sigma }^{'})\rbrace =0,
\label{V29}
\end{eqnarray}

\noindent where ${\tilde \pi}^{\vec \xi}_r(\tau ,\vec \sigma )$ is the momentum 
conjugate to the 3 gauge variables $\xi^r(\tau ,\vec \sigma )$, which will be 
functions only of the cotriads. On the space of cotriads the Abelianized form of
the pseudodiffeomorphism constraints is

\begin{eqnarray}
{\tilde \pi}^{\vec \xi}_r(\tau ,\vec \sigma )&=&{{\partial \sigma^s(\vec \xi )}
\over {\partial \xi^r}}{|}_{\vec \xi =\vec \xi (\tau ,\vec \sigma )} \Big[ {}^3
{\tilde \Theta}_s(\tau ,\vec \sigma )+{\hat {\tilde \theta}}_{(a)s}(\alpha_{(e)}
(\tau ,\vec \sigma ),\partial_u\alpha_{(e)}(\tau ,\vec \sigma ))\, {}^3{\tilde
M}_{(a)}(\tau ,\vec \sigma )\Big] =\nonumber \\
&=&{{\partial \sigma^s(\vec \xi )}
\over {\partial \xi^r}}{|}_{\vec \xi =\vec \xi (\tau ,\vec \sigma )} \Big[ {}^3
{\tilde \Theta}_s+{{\partial \alpha_{(a)}}\over {\partial \sigma^s}} {\tilde
\pi}^{\vec \alpha}_{(a)}\Big] (\tau ,\vec \sigma )\approx 0,
\label{V30}
\end{eqnarray}

\noindent and both $\xi^r(\tau ,\vec \sigma )$
and ${\tilde \pi}^{\vec \xi}_r(\tau ,\vec \sigma )$
have zero Poisson bracket with $\alpha_{(a)}(\tau ,\vec \sigma )$,
${\tilde \pi}^{\vec \alpha}_{(a)}(\tau ,\vec \sigma )$.

Therefore, the 6 gauge variables $\xi^r(\tau ,\vec \sigma )$ and $\alpha_{(a)}
(\tau ,\vec \sigma )$ and their conjugate momenta form 6 canonical pairs of a 
new canonical basis adapted to the rotation and pseudodiffeomorphisms 
constraints and replacing 6 of the 9 conjugate pairs ${}^3e_{(a)r}(\tau ,\vec 
\sigma )$, ${}^3{\tilde \pi}^r_{(a)}(\tau ,\vec \sigma )$.

From Eqs.(64) of I and from Eqs.(\ref{V27}) and (\ref{V8}), we get

\begin{eqnarray}
Y_s(\tau ,{\vec \sigma }^{'})\, {}^3e_{(a)r}(\tau ,\vec \sigma )&=&
-\Big( {{\partial
\xi^u(\tau ,{\vec \sigma }^{'})}\over {\partial \sigma^{{'}s}}}{{\delta}
\over {\delta \xi^u(\tau ,{\vec \sigma }^{'})}}+{{\partial \alpha_{(c)}(\tau ,
{\vec \sigma }^{'})}\over {\partial \sigma^{{'}r}}}{{\tilde \delta}\over 
{\delta \alpha_{(c)}(\tau ,{\vec \sigma }^{'})}}\Big) \cdot \nonumber \\
&\cdot & \Big[
{}^3R_{(a)(b)}(\alpha_{(e)}(\tau ,\vec \sigma ))\, {}^3{\bar e}_{(b)r}
(\tau ,\vec \sigma )\Big] =\nonumber \\
&=&-{}^3R_{(a)(b)}(\alpha_{(e)}(\tau ,\vec \sigma )){{\partial \xi^u(\tau ,
{\vec \sigma }^{'})}\over {\partial \sigma^{{'}s}}}{{\delta \, {}^3{\bar e}
_{(b)r}(\tau ,\vec \sigma )}\over {\delta \xi^u(\tau ,{\vec \sigma }^{'})}}-
\nonumber \\
&-&{{\partial \alpha_{(c)}(\tau ,{\vec \sigma }^{'})}\over {\partial \sigma
^{{'}s}}}{{\tilde \delta {}^3R_{(a)(b)}(\alpha_{(e)}(\tau ,\vec \sigma ))}\over
{\delta \alpha_{(c)}(\tau ,{\vec \sigma }^{'})}}\, {}^3{\bar e}_{(b)r}(\tau ,
\vec \sigma )=\nonumber \\
&=&-{}^3R_{(a)(b)}(\alpha_{(e)}(\tau ,\vec \sigma )){{\partial \xi^u(\tau ,
{\vec \sigma}^{'})}\over {\partial \sigma^{{'}s}}}{{\delta \, {}^3{\bar e}
_{(b)r}(\tau ,\vec \sigma )}\over {\delta \xi^u(\tau ,{\vec \sigma }^{'})}}-
\nonumber \\
&-&{{\partial \, {}^3R_{(a)(b)}(\alpha_{(e)}(\tau ,\vec \sigma ))}\over {\partial
\sigma^s}}\delta^3(\vec \sigma ,{\vec \sigma }^{'})\, {}^3{\bar e}_{(b)r}(\tau ,
\vec \sigma )=\nonumber \\
&&{}\nonumber \\
&=&-\lbrace {}^3e_{(a)r}(\tau ,\vec \sigma ),{}^3{\tilde \Theta}_s(\tau ,{\vec 
\sigma }^{'})\rbrace =\nonumber \\
&=&-{{\partial \, {}^3e_{(a)r}(\tau ,\vec \sigma )}\over
{\partial \sigma^s}}\delta^3(\vec \sigma ,{\vec \sigma }^{'})+{}^3e_{(a)s}
(\tau ,\vec \sigma ){{\partial \delta^3(\vec \sigma ,{\vec \sigma }^{'})}\over
{\partial \sigma^r}}=\nonumber \\
&=&-{}^3R_{(a)(b)}(\alpha_{(e)}(\tau ,\vec \sigma )){{\partial \, {}^3{\bar e}
_{(b)r}(\tau ,\vec \sigma )}\over {\partial \sigma^s}}\delta^3(\vec \sigma ,
{\vec \sigma }^{'})-\nonumber \\
&-&{{\partial {}^3R_{(a)(b)}(\alpha_{(e)}(\tau ,\vec \sigma ))}
\over {\partial \sigma^s}}\, {}^3{\bar e}_{(b)r}(\tau ,\vec \sigma )\delta^3
(\vec \sigma ,{\vec \sigma }^{'})+\nonumber \\
&+&{}^3R_{(a)(b)}(\alpha_{(e)}(\tau ,\vec \sigma ))\, {}^3{\bar e}_{(b)r}
(\tau ,\vec \sigma ){{\partial \delta^3(\vec \sigma ,{\vec \sigma }^{'})}\over
{\partial \sigma^r}},
\label{V31}
\end{eqnarray}

\noindent so that the pseudodiffeomorphism multitemporal equations for 
${}^3{\bar e}_{(a)r}(\tau ,\vec \sigma )$ are

\begin{eqnarray}
-{\tilde Y}_s(\tau ,{\vec \sigma }^{'})\, {}^3{\bar e}_{(a)r}(\tau ,\vec 
\sigma )&=&{{\partial \xi^u(\tau ,{\vec \sigma }^{'})}\over {\partial \sigma
^{{'}s}}} {{\delta \, {}^3{\bar e}_{(a)r}(\tau ,\vec \sigma )}\over {\delta
\xi^u(\tau ,{\vec \sigma }^{'})}}=\nonumber \\
&=&{{\partial \, {}^3{\bar e}_{(a)r}(\tau ,\vec \sigma )}\over {\partial 
\sigma^s}} \delta^3(\vec \sigma ,{\vec \sigma }^{'})-{}^3{\bar e}_{(a)s}(\tau ,
\vec \sigma ) {{\partial \delta^3(\vec \sigma ,{\vec \sigma }^{'})}\over
{\partial \sigma^{{'}r}}}.
\label{V32}
\end{eqnarray}

Analogously, from Eqs.(64) of I and Eqs.(\ref{V25}) and (\ref{V16}) we have

\begin{eqnarray}
Y_s(\tau ,{\vec \sigma }^{'})\, {}^3{\tilde \pi}^r_{(a)}(\tau ,\vec \sigma )&=&
-\Big( {{\partial
\xi^u(\tau ,{\vec \sigma }^{'})}\over {\partial \sigma^{{'}s}}}{{\delta}
\over {\delta \xi^u(\tau ,{\vec \sigma }^{'})}}+{{\partial \alpha_{(c)}(\tau ,
{\vec \sigma }^{'})}\over {\partial \sigma^{{'}r}}}{{\tilde \delta}\over 
{\delta \alpha_{(c)}(\tau ,{\vec \sigma }^{'})}}\Big) \cdot \nonumber \\
&\cdot & \Big[ {}^3R_{(a)(b)}(\alpha_{(e)}(\tau ,\vec \sigma ))\, {}^3{\bar 
{\tilde \pi}}^r_{(b)}(\tau ,\vec \sigma )\Big] =\nonumber \\
&=&-\Big[
{}^3R_{(a)(b)}(\alpha_{(e)}(\tau ,{\vec \sigma }^{'}))\, {}^3{\bar {\tilde
\pi}}^r_{(b)}(\tau ,{\vec \sigma }^{'})\Big] {{\partial \delta^3(\vec \sigma ,
{\vec \sigma }^{'})}\over {\partial \sigma^{{'}s}}}+\nonumber \\
&+&\delta^r_s {}^3R_{(a)(b)}
(\alpha_{(e)}(\tau ,\vec \sigma ))\, {}^3{\bar {\tilde \pi}}^u_{(b)}(\tau ,
\vec \sigma ){{\partial \delta^3(\vec \sigma ,{\vec \sigma }^{'})}
\over {\partial \sigma^{{'}u}}},
\label{V33}
\end{eqnarray}

\noindent and we get the pseudodiffeomorphism multitemporal equation for 
${}^3{\bar {\tilde \pi}}^r_{(a)}(\tau ,\vec \sigma )$

\begin{eqnarray}
-{\tilde Y}_s(\tau ,{\vec \sigma }^{'})\, {}^3{\bar {\tilde \pi}}^r_{(a)}
(\tau ,\vec \sigma )&=&{{\partial \xi^u(\tau ,{\vec \sigma }^{'})}\over
{\partial \sigma^{{'}s}}}{{\delta \, {}^3{\bar {\tilde \pi}}^r_{(a)}(\tau ,
\vec \sigma )}\over {\delta \xi^u(\tau ,{\vec \sigma }^{'})}}=\nonumber \\
&=&-{}^3{\bar {\tilde \pi}}^r_{(a)}(\tau ,{\vec \sigma }^{'}){{\partial \delta
^3(\vec \sigma ,{\vec \sigma }^{'})}\over {\partial \sigma^{{'}s}}}-\delta^r_s
\, {}^3{\bar {\tilde \pi}}^u_{(a)}(\tau ,\vec \sigma ){{\partial \delta^3
(\vec \sigma ,{\vec \sigma }^{'})}\over {\partial \sigma^{{'}u}}}.
\label{V34}
\end{eqnarray}

Let us remark that the Jacobian matrix satisfies an equation like (\ref{V32})

\begin{eqnarray}
-{\tilde Y}_s(\tau ,{\vec \sigma }^{'})&&{{\partial \xi^u(\tau ,\vec \sigma )}
\over {\partial \sigma^r}}={{\partial \xi^v(\tau ,{\vec \sigma }^{'})}\over
{\partial \sigma^{{'}s}}}{{\delta}\over {\delta \xi^v(\tau ,{\vec \sigma }^{'})
}}{{\partial \xi^u(\tau ,\vec \sigma )}\over {\partial \sigma^r}}=\nonumber \\
&=&{{\partial \xi^u(\tau ,{\vec \sigma }^{'})}\over {\partial \sigma^{{'}s}}}
{{\partial \delta^3(\vec \sigma ,{\vec \sigma }^{'})}\over {\partial \sigma^r}}
=-{{\partial^u(\tau ,{\vec \sigma }^{'})}\over {\partial \sigma^{{'}s}}}
{{\partial \delta^3(\vec \sigma ,{\vec \sigma }^{'})}\over {\partial
\sigma^{{'}r}}}=\nonumber \\
&=&-{{\partial \xi^u(\tau ,\vec \sigma )}\over {\partial \sigma^s}}
{{\partial \delta^3(\vec \sigma ,{\vec \sigma }^{'})}\over {\partial \sigma
^{{'}r}}}+{{\partial^2\xi^u(\tau ,\vec \sigma )}\over {\partial \sigma^r
\partial \sigma^s}}\delta^3(\vec \sigma ,{\vec \sigma }^{'})=\nonumber \\
&=&{{\partial}\over {\partial \sigma^s}}({{\partial \xi^u(\tau ,\vec \sigma )}
\over {\partial \sigma^r}})\delta^3(\vec \sigma ,{\vec \sigma }^{'})-{{\partial
\xi^u(\tau ,\vec \sigma )}\over {\partial \sigma^s}}{{\partial \delta^3
(\vec \sigma ,{\vec \sigma }^{'})}\over {\partial \sigma^{{'}r}}}.
\label{V35}
\end{eqnarray}

\noindent so that the identity ${{\partial \xi^u(\tau ,{\vec \sigma }^{'})}
\over {\partial \sigma^{{'}s}}}{{\delta f(\tau ,\vec \xi (\tau ,\vec \sigma ))}
\over {\delta \xi^u(\tau ,{\vec \sigma }^{'})}}={{\partial f(\tau ,\vec \xi
(\tau ,\vec \sigma ))}\over {\partial \sigma^s}}\delta^3(\vec \sigma ,
{\vec \sigma }^{'})$, implies the following solutions of the multitemporal
equations [again $\hat V(\vec \xi (\tau ,\vec \sigma ))$ is the operator with
the action $\hat V(\vec \xi (\tau ,\vec \sigma ))f(\tau ,\vec \sigma )=
f(\tau ,\vec \xi (\tau ,\vec \sigma ))$; and Eqs.(\ref{IV5}) is used]

\begin{eqnarray}
{}^3{\bar e}_{(a)r}(\tau ,\vec \sigma )&=&{{\partial \xi^s(\tau ,\vec \sigma )}
\over {\partial \sigma^r}}\, {}^3{\hat e}_{(a)s}(\tau ,\vec \xi (\tau ,
\vec \sigma ))={{\partial \xi^s(\tau ,\vec \sigma )}\over {\partial \sigma^r}}
\hat V(\vec \xi (\tau ,\vec \sigma ))\, {}^3{\hat e}_{(a)s}(\tau ,\vec \sigma )
,\nonumber \\
{}&&{{\delta \, {}^3{\hat e}_{(a)r}(\tau ,\vec \sigma )}\over {\delta \xi^s
(\tau ,{\vec \sigma }^{'})}}=0,\nonumber \\
{}^3e_{(a)r}(\tau,\vec \sigma )&=&{}^3R_{(a)(b)}(\alpha_{(e)}(\tau ,\vec 
\sigma )) {{\partial \xi^s(\tau ,\vec \sigma )}\over {\partial \sigma^r}}\,
{}^3{\hat e}_{(b)s}(\tau ,\vec \xi (\tau ,\vec \sigma ))=\nonumber \\
&=&{{\partial \xi^s(\tau ,\vec \sigma )}\over {\partial \sigma^r}}\, {}^3R
_{(a)(b)}(\alpha^{'}_{(e)}(\tau ,\vec \xi (\tau ,\vec \sigma )))\, {}^3{\hat e}
_{(b)s}(\tau ,\vec \xi (\tau ,\vec \sigma )),\nonumber \\
&&{}\nonumber \\
{}^3g_{rs}(\tau ,\vec \sigma )&=&{{\partial \xi^u(\tau ,\vec \sigma )}\over
{\partial \sigma^r}} {{\partial \xi^v(\tau ,\vec \sigma )}\over {\partial
\sigma^s}}\, {}^3{\hat e}_{(a)u}(\tau ,\vec \xi (\tau ,\vec \sigma ))\,
{}^3{\hat e}_{(a)v}(\tau ,\vec \xi (\tau ,\vec \sigma )).
\label{V36}
\end{eqnarray}

Here the cotriads ${}^3{\hat e}_{(a)r}(\tau ,\vec \sigma )$ depend only on 3
degrees of freedom and are Dirac observables with respect to both
Abelianized rotations and pseudodiffeomorphisms. 
Again, like in the case of rotations,
we have not found 3 specific conditions on the cotriads implying this final 
reduction. This is due to the fact that, even if one has a trivial coframe
bundle, one does not know the group manifold of $Diff\, \Sigma_{\tau}$ and that
there is no canonical identity for pseudodiffeomorphisms and therefore also for 
rotations inside the gauge group ${\bar {\cal G}}_R$.

Eqs.(\ref{V36}) are the counterpart in tetrad gravity of the solutions of the 3
elliptic equations for the gravitomagnetic vector potential ${\check W}^r$ of 
the conformal approach (see the end of Appendix C).

If ${{\partial \sigma^r(\vec \xi )}\over {\partial \xi^s}}{|}_{\vec \xi =\vec 
\xi (\tau ,\vec \sigma )}$ is the inverse Jacobian matrix and $|\, {{\partial 
\xi (\tau ,\vec \sigma )}\over {\partial \sigma}}\, |$ the determinant of the
Jacobian matrix, the following identities

\begin{eqnarray}
\delta^r_s&=&{{\partial \sigma^r(\vec \xi )}\over {\partial \xi^u}}{|}_{\vec \xi
=\vec \xi (\tau ,\vec \sigma )}\, {{\partial \xi^u(\tau ,\vec \sigma )}\over
{\partial \sigma^s}},\nonumber \\
\Rightarrow && {{\partial }\over {\partial \sigma^v}}{{\partial \sigma^r(\vec
\xi )}\over {\partial \xi^u}}{|}_{\vec \xi =\vec \xi (\tau ,\vec \sigma )}=-
{{\partial \sigma^s(\vec \xi )}\over {\partial \xi^u}}{|}_{\vec \xi =\vec \xi
(\tau ,\vec \sigma )}\, {{\partial \sigma^r(\vec \xi )}\over {\partial \xi^w}}
{|}_{\vec \xi =\vec \xi (\tau ,\vec \sigma )}\, {{\partial^2\xi^w(\tau ,\vec 
\sigma )}\over {\partial \sigma^v\partial \sigma^s}},\nonumber \\
\Rightarrow && {{\delta}\over {\delta \xi^v(\tau ,{\vec \sigma }^{'})}}
{{\partial \sigma^r(\vec \xi )}\over {\partial \xi^u}}{|}_{\vec \xi =\vec \xi
(\tau ,\vec \sigma )}=-{{\partial \sigma^s(\vec \xi )}\over {\partial \xi^u}}
{|}_{\vec \xi =\vec \xi (\tau ,\vec \sigma )} {{\partial \sigma^r(\vec \xi )}
\over {\partial \xi^s}}{|}_{\vec \xi =\vec \xi (\tau ,\vec \sigma )}
{{\partial \delta^3(\vec \sigma ,{\vec \sigma }^{'})}\over {\partial \sigma^v}},
\nonumber \\
{}&&\downarrow \nonumber \\
-{\tilde Y}_s(\tau ,{\vec \sigma }^{'})&&{{\partial \sigma^r(\vec \xi )}\over
{\partial \xi^u}}{|}_{\vec \xi =\vec \xi (\tau ,\vec \sigma  )}=\nonumber \\
&=&{{\partial}\over {\partial \sigma^s}}
\Big( {{\partial \sigma^r(\vec \xi )}\over 
{\partial \xi^u}}{|}_{\vec \xi =\vec \xi (\tau,\vec \sigma  )}\Big) \delta^3
(\vec \sigma ,{\vec \sigma }^{'})+\delta^r_s{{\partial \sigma^v(\vec \xi )}
\over {\partial \xi^u}}{|}_{\vec \xi -\vec \xi (\tau ,\vec \sigma )}\,
{{\partial \delta^3(\vec \sigma ,{\vec \sigma }^{'})}\over {\partial
\sigma^{{'}v}}},
\label{V37}
\end{eqnarray}

\noindent and [use is done of $\delta ln\, det\, M=Tr\, (M^{-1}\delta M)$]

\begin{eqnarray}
{{\partial}\over {\partial \sigma^r}}&& |\, {{\partial \xi (\tau ,\vec \sigma )}
\over {\partial \sigma }}\, | = |\, {{\partial \xi (\tau ,\vec \sigma )}\over
{\partial \sigma}}\, |\, {{\partial \sigma^s(\vec \xi )}\over {\partial \xi^u}}
{|}_{\vec \xi =\vec \xi (\tau ,\vec \sigma )}\, {{\partial^2\xi^u(\tau ,\vec 
\sigma )}\over {\partial \sigma^r\partial \sigma^s}},\nonumber \\
{{\delta}\over {\delta \xi^r(\tau ,{\vec \sigma }^{'})}} && |\, {{\partial \xi
(\tau ,\vec \sigma )}\over {\partial \sigma}}\, |=|\, {{\partial \xi (\tau ,
\vec \sigma )}\over {\partial \sigma}}\, |\, {{\partial \sigma^s(\vec \xi )}
\over {\partial \xi^r}}{|}_{\vec \xi =\vec \xi (\tau ,\vec \sigma )}\,
{{\partial \delta^3(\vec \sigma ,{\vec \sigma }^{'})}\over {\partial \sigma^s}},
\nonumber \\
{}&&\downarrow \nonumber \\
-{\tilde Y}_s(\tau ,{\vec \sigma }^{'})&& |\, {{\partial \xi (\tau ,\vec \sigma 
)}\over {\partial \sigma}}\, |=-|\, {{\partial \xi (\tau ,\vec \sigma )}\over
{\partial \sigma }}\, |\, {{\partial \delta^3(\vec \sigma ,{\vec \sigma }^{'})}
\over {\partial \sigma^{{'}s}}},
\label{V38}
\end{eqnarray}

\noindent allow to get

\begin{eqnarray}
{}^3{\tilde \pi}^r_{(a)}(\tau ,\vec \sigma )&=&{}^3R_{(a)(b)}(\alpha_{(a)}
(\tau ,\vec \sigma ))\, {}^3{\bar {\tilde \pi}}^r_{(b)}(\tau ,\vec \sigma )=
\nonumber \\
&=&{}^3R_{(a)(b)}(\alpha_{(e)}(\tau ,\vec \sigma ))\, |\, {{\partial \xi 
(\tau ,\vec \sigma )}\over {\partial \sigma}}\, |\, {{\partial \sigma^r(\vec
\xi )}\over {\partial \xi^s}}{|}_{\vec \xi =\vec \xi (\tau ,\vec \sigma )}\,
{}^3{\hat {\tilde \pi}}^s_{(b)}(\tau ,\vec \xi (\tau ,\vec \sigma ))=
\nonumber \\
&=&{}^3R_{(a)(b)}(\alpha_{(e)}(\tau .\vec \sigma ))\, |\, {{\partial \xi (\tau ,
\vec \sigma )}\over {\partial \sigma}}\, |\, {{\partial \sigma^r(\vec \xi )}
\over {\partial \xi^s}}{|}_{\vec \xi =\vec \xi (\tau ,\vec \sigma )}\,
\hat V(\vec \xi (\tau ,\vec \sigma ))\, {}^3{\hat {\tilde \pi}}^s_{(b)}(\tau .
\vec \sigma ),
\label{V39}
\end{eqnarray}

\noindent where ${}^3{\hat {\tilde \pi}}^r_{(a)}(\tau ,\vec \sigma )$ are Dirac
observables with respect to both Abelianized rotations and 
pseudodiffeomorphisms. In a similar way we get

\begin{equation}
{}^3e^r_{(a)}(\tau ,\vec \sigma )={}^3R_{(a)(b)}(\alpha_{(e)}(\tau ,\vec 
\sigma )) {{\partial \sigma^r(\vec \xi )}\over {\partial \xi^s}}{|}_{\vec \xi =
\vec \xi (\tau ,\vec \sigma )}\, {}^3{\hat e}^r_{(b)}(\tau ,\vec \xi (\tau ,
\vec \sigma )),
\label{V40}
\end{equation}

\noindent with ${}^3{\hat e}^r_{(a)}(\tau ,\vec \sigma )$ the Dirac observables 
for triads dual to ${}^3e_{(a)r}(\tau ,\vec \sigma )$. The line element becomes

\begin{eqnarray}
ds^2&=&\epsilon \Big( 
[N_{(as)}+n]^2 - [N_{(as) r}+n_r] {{\partial \sigma^r(\vec 
\xi )}\over {\partial \xi^u}} {}^3{\hat e}^u_{(a)}(\vec \xi )\, {}^3{\hat e}^v
_{(a)}(\vec \xi ) {{\partial \sigma^s(\vec \xi )}\over {\partial \xi^v}}
[N_{(as) s}+n_s] \Big) (d\tau )^2-\nonumber \\
&-&2\epsilon [N_{(as) r}+n_r] d\tau d\sigma^r-\epsilon {{\partial \xi^u}\over
{\partial \sigma^r}}\, {}^3{\hat e}_{(a)u}(\vec \xi )\, {}^3{\hat e}
_{(a)v}(\vec \xi ) {{\partial \xi^v}\over {\partial \sigma^s}} 
d\sigma^r d\sigma^s=\nonumber \\
&=&\epsilon \Big( [N_{(as)}+n]^2(d\tau )^2- [{}^3{\hat e}_{(a)u}(\vec \xi )
{{\partial \xi^u}\over {\partial \sigma^r}}d\sigma^r+{}^3{\hat e}^u_{(a)}(\vec 
\xi ){{\partial \sigma^r(\vec \xi )}\over {\partial \xi^u}}(N_{(as)r}+n_r)
d\tau ]\nonumber \\
&&[{}^3{\hat e}_{(a)v}(\vec \xi )
{{\partial \xi^v}\over {\partial \sigma^s}}d\sigma^s+{}^3{\hat e}^v_{(a)}(\vec 
\xi ){{\partial \sigma^s(\vec \xi )}\over {\partial \xi^v}}(N_{(as)s}+n_s)
d\tau ] \Big) .
\label{V40b}
\end{eqnarray}

To get $\xi^r(\tau ,\vec \sigma )$ in terms of the cotriads we have to solve
the equations [use is done of Eq.(\ref{V29}), of (62) of I and of $\lbrace
\xi^r(\tau ,\vec \sigma ),{}^3{\tilde M}_{(a)}(\tau ,{\vec \sigma }^{'})
\rbrace =0$]

\begin{eqnarray}
\delta^r_s\delta^3(\vec \sigma ,{\vec \sigma }^{'})&=&\lbrace \xi^r(\tau ,\vec 
\sigma ),{\tilde \pi}^{\vec \xi}_s(\tau ,{\vec \sigma }^{'})\rbrace =
\nonumber \\
&=&{{\partial \sigma^u(\vec \xi )}\over {\partial \xi^s}}{|}_{\vec \xi =\vec \xi
(\tau ,{\vec \sigma}^{'} )}\, 
\lbrace \xi^r(\tau ,\vec \sigma ),{}^3{\tilde \Theta}_u
(\tau ,{\vec \sigma }^{'})\rbrace =\nonumber \\
&=&{{\partial \sigma^u(\vec \xi )}\over {\partial \xi^s}}{|}_{\vec \xi =\vec \xi
(\tau ,{\vec \sigma}^{'} )} 
\Big[ \Big( {{\partial \, {}^3e_{(a)v}(\tau ,{\vec \sigma }
^{'})}\over {\partial \sigma^{{'}u}}}-{{\partial \, {}^3e_{(a)u}(\tau ,
{\vec \sigma }^{'})}\over {\partial \sigma^{{'}v}}}\Big) {{\delta \xi^r(\tau ,
\vec \sigma )}\over {\delta \, {}^3e_{(a)v}(\tau ,{\vec \sigma }^{'})}}-
\nonumber \\
&-&{}^3e_{(a)u}(\tau ,{\vec \sigma }^{'}){{\partial}\over {\partial 
\sigma^{{'}v}}}\, {{\delta \xi^r(\tau ,\vec \sigma )}\over {\delta \, {}^3e
_{(a)v}(\tau ,{\vec \sigma }^{'})}}\Big] ,\nonumber \\
&&\Downarrow \nonumber \\
\Big( \Big[ \delta_{(a)(b)}\partial^{'}_v&-& {}^3e^u_{(a)} (\partial^{'}_u\, 
{}^3e_{(b)v}-\partial^{'}_v\, {}^3e_{(b)u})\Big] (\tau ,{\vec \sigma}^{'})
{{\delta}\over {\delta \, {}^3e_{(b)v}(\tau ,{\vec \sigma}^{'})}}+\nonumber \\
&+&\delta^3(\vec \sigma ,{\vec \sigma }^{'})
{}^3e^u_{(a)}(\tau ,\vec \sigma ){{\partial}\over {\partial \sigma^u}}
\Big) \xi^r(\tau ,\vec \sigma ) =0.
\label{V41}
\end{eqnarray}

We do not know how to solve these equations along some privileged path in the
group manifold of $Diff\, \Sigma_{\tau}$ after having chosen a global
coordinate system $\Xi$ as a conventional origin of pseudodiffeomorphisms
[this identifies a conventional identity cross section $\Sigma_{\tau}^{(\Xi )}$
in the proposed description of $Diff\, \Sigma_{\tau}$ with the fibration
$\Sigma_{\tau}\times \Sigma_{\tau} \rightarrow \Sigma_{\tau}$ for the case
$\Sigma_{\tau}\approx R^3$], due to the poor understanding of the geometry and
differential structure of this group manifold. Presumably, since the fibers of
$\Sigma_{\tau}\times \Sigma_{\tau}$ are also copies of $\Sigma_{\tau}$, on each
one of them one can try to define an analogue of canonical coordinates of
first kind by using the geodesic exponential map:\hfill\break
i) choose a reference fiber $\Sigma_{\tau ,0}$ in $\Sigma_{\tau}\times \Sigma
_{\tau}$ over a point $p=(\tau ,\vec 0)$ chosen as origin in the base (and then
connected to all the points in base with geodesics; for $\Sigma_{\tau}\approx 
R^3$ this is well defined; the global cross sections corresponding to global
coordinate systems should be horizontal lifts of this geodesic star with
respect to some notion of connection on the fibration);\hfill\break
ii) if $q_o$ is the point in $\Sigma_{\tau}\times \Sigma_{\tau}$ at the 
intersection of $\Sigma_{\tau ,0}$ with the conventional identity cross
section $\Sigma_{\tau}^{(\Xi )}$ and $q_1$ the point where $\Sigma_{\tau ,0}$
intersects a nearby global cross section $\Sigma_{\tau}^{(\Xi^{'})}$ [$\Xi^{'}$
is another global coordinate system on $\Sigma_{\tau}$], we can consider the 
geodesic $\gamma_{q_oq_1}$ on $\Sigma_{\tau ,0}$;\hfill\break
iii) use the geodesic exponential map along the geodesic $\gamma_{q_oq_1}$ to
define ``pseudodiffeomorphism coordinates $\vec \xi (\tau ,\vec 0)$" describing 
the transition from the global coordinate system $\Xi$ to $\Xi^{'}$ over the
base point $p=(\tau ,\vec 0)$;\hfill\break
iv) parallel transport these coordinates on the fiber $\Sigma_{\tau ,0}$ to the
other fibers along the geodesics of the cross sections $\Sigma_{\tau}
^{(\Xi^{'})}$.

If this coordinatization of the group manifold of $Diff\, \Sigma_{\tau}$ for
$\Sigma_{\tau}\approx R^3$ can be justified, then one could try to solve the 
previous equations.

Instead, we are able to give a formal expression for the operator $\hat V(\vec 
\xi (\vec \sigma ))$ [for the sake of simplicity we do not consider the 
$\tau$-dependence], whose action on functions $f(\vec \sigma )$ is $\hat V
(\vec \xi (\vec \sigma )) f(\vec \sigma )=f(\vec \xi (\vec \sigma ))$. We have

\begin{equation}
\hat V(\vec \xi (\vec \sigma ))=P_{\gamma}\, e^{(\int_{\vec \sigma }^{\vec \xi
(\vec \sigma )} {{\partial \sigma^r(u)}\over {\partial u^s}} {\cal D}u^s)
{{\partial}\over {\partial \sigma^r}} },
\label{V42}
\end{equation}

\noindent where the path ordering is along the geodesic $\gamma$ in $\Sigma
_{\tau}$ joining the points with coordinates $\vec \sigma $ and ${\vec 
\sigma }^{'}=\vec \xi (\vec \sigma )$. For infinitesimal pseudodiffeomorphisms 
$\vec \sigma \mapsto {\vec \sigma }^{'}(\vec \sigma )=\vec \xi (\vec \sigma )=
\vec \sigma +\delta \vec \sigma (\vec \sigma )$ [with inverse ${\vec \sigma }
^{'}=\vec \xi \mapsto \vec \sigma (\vec \xi )=\vec \xi -\delta \vec \sigma 
(\vec \xi )$], we have

\begin{eqnarray}
\hat V(\vec \sigma +\delta \vec \sigma )&\approx& 
1+\Big[ \delta \sigma^s(\vec \sigma )
{{\partial \sigma^r(\vec \xi )}\over {\partial \xi^s}}{|}_{\vec \xi (\vec 
\sigma )-\delta \vec \sigma (\vec \xi (\vec \sigma ))}\Big] {{\partial}\over
{\partial \sigma^r}}\approx 1+\delta \sigma^s(\vec \sigma ){{\partial}\over 
{\partial \sigma^s}}:\nonumber \\
&:& f(\vec \sigma )\mapsto f(\vec \sigma )+\delta \sigma^s(\vec \sigma )
{{\partial f(\vec \sigma )}\over {\partial \sigma^s}}\approx f(\vec \sigma +
\delta \vec \sigma (\vec \sigma )).
\label{V43}
\end{eqnarray}

Formally we have [if $\delta /\delta \xi^r(\vec \sigma )$ is interpreted as the
directional functional derivative along $\gamma$]

\begin{eqnarray}
{{\delta}\over {\delta \xi^r({\vec \sigma }^{'})}} [\hat V(\vec \xi (\vec 
\sigma )) f(\vec \sigma )]&=&\delta^3(\vec \sigma ,{\vec \sigma }^{'})
{{\partial \sigma^s(\vec \xi )}\over {\partial \xi^r}}{|}_{\vec \xi =\vec \xi 
(\vec \sigma )}\, {{\partial}\over {\partial \sigma^s}}[\hat V(\vec \xi (\vec 
\sigma ))f(\vec \sigma )]=\nonumber \\
&=&\delta^3(\vec \sigma ,{\vec \sigma }^{'}) {{\partial \sigma^s(\vec \xi )}
\over {\partial \xi^r}}{|}_{\vec \xi =\vec \xi (\vec \sigma )}{{\partial
f(\vec \xi (\vec \sigma ))}\over {\partial \sigma^s}}=\delta^3(\vec \sigma ,
{\vec \sigma }^{'}){{\partial f(\vec \xi )}\over {\partial \xi^r}}{|}_{\vec \xi
=\vec \xi (\vec \sigma )}=\nonumber \\
&=&{{\delta f(\vec \xi (\vec \sigma ))}\over {\delta \xi^r({\vec \sigma }^{'})}}
.
\label{V44}
\end{eqnarray}

By using Eqs.(\ref{V19}) and (64) of I, we get

\begin{eqnarray}
{\hat D}^{(\omega )}_{(a)(b)r}(\tau ,\vec \sigma )&& \lbrace \zeta^{(\omega )r}
_{(b)(c)}(\vec \sigma ,{\vec \sigma}_1;\tau ),{}^3{\tilde M}_{(g)}(\tau ,
{\vec \sigma}_2)\rbrace =\nonumber \\
&=&-\epsilon_{(a)(d)(b)} \lbrace {}^3\omega_{s(d)}(\tau ,
\vec \sigma ),{}^3{\tilde M}_{(g)}(\tau ,{\vec \sigma}_2)\rbrace
\zeta^{(\omega )s}_{(b)(c)}(\vec \sigma ,{\vec \sigma}_1;\tau ),\nonumber \\
&&{}\nonumber \\
{\hat D}^{(\omega )}_{(a)(b)r}(\tau ,\vec \sigma )&& \lbrace \zeta^{(\omega )r}
_{(b)(c)}(\vec \sigma ,{\vec \sigma}_1;\tau ),{}^3{\tilde \Theta}_u(\tau ,{\vec
\sigma}_2)\rbrace =\nonumber \\
&=&-\epsilon_{(a)(d)(f)} \lbrace {}^3\omega_{s(d)}(\tau ,\vec
\sigma ),{}^3{\tilde \Theta}_u(\tau ,{\vec \sigma}_2)\rbrace \zeta^{(\omega )s}
_{(f)(c)}(\vec \sigma ,{\vec \sigma}_1;\tau ).
\label{V44a}
\end{eqnarray}

\noindent Then Eqs.(64) of I, (\ref{V11}) and (\ref{V19}) imply the
following transformation properties under rotations and space 
pseudodiffeomorphisms
of the Green function of the SO(3) covariant divergence (which we do not know
how to verify explicitly due to the path-ordering contained in it)

\begin{eqnarray}
\lbrace \zeta^{(\omega )r}_{(a)(b)}(\vec \sigma ,{\vec \sigma}_1;\tau ),&&
{}^3{\tilde M}_{(g)}(\tau ,{\vec \sigma}_2)\rbrace ={{\partial}\over {\partial 
\sigma_2^s}}\Big[
\zeta^{(\omega )r}_{(a)(e)}(\vec \sigma ,{\vec \sigma}_2;\tau )
\epsilon_{(e)(g)(f)} \zeta^{(\omega )s}_{(f)(b)}({\vec \sigma}_2,{\vec \sigma}
_1;\tau )\Big] +\nonumber \\
&+&\zeta^{(\omega )r}_{(a)(e)}(\vec \sigma ,{\vec \sigma}_2)\, {}^3\omega_{s(e)}
(\tau ,{\vec \sigma}_2)\, \zeta^{(\omega )s}_{(g)(b)}({\vec \sigma}_2,{\vec 
\sigma}_1;\tau )-\nonumber \\
&-&\zeta^{(\omega )r}_{(a)(g)}(\vec \sigma ,{\vec \sigma}_2;\tau )
\, {}^3\omega_{s(f)}(\tau ,{\vec \sigma}_2)\, \zeta^{(\omega )s}_{(f)(b)}({\vec
\sigma}_2,{\vec \sigma}_1;\tau ),\nonumber \\
&&{}\nonumber \\
\lbrace \zeta^{(\omega )r}_{(a)(b)}(\vec \sigma ,{\vec \sigma}_1;\tau ),&&
{}^3{\tilde \Theta}_u(\tau ,{\vec \sigma}_2)\rbrace =\nonumber \\
&=&\int d^3\sigma_3 \zeta^{(\omega )r}
_{(a)(e)}(\vec \sigma ,{\vec \sigma}_3;\tau )\, \epsilon_{(e)(d)(f)}
\lbrace {}^3\omega_{s(d)}(\tau ,{\vec \sigma}_3), {}^3{\tilde \Theta}_u(\tau ,
{\vec \sigma}_2)\rbrace \zeta^{(\omega )s}_{(f)(b)}({\vec \sigma}_3,{\vec 
\sigma}_1;\tau ).\nonumber \\
&&{}
\label{V44b}
\end{eqnarray}

Collecting all previous results, we obtain the following form for the
Dirac Hamiltonian 

\begin{eqnarray}
{\hat H}_{(D)ADM}&=&\int d^3\sigma \Big[ n {\hat {\cal H}}-
n_{(a)}\, {}^3e_{(a)^r}\, {}^3{\tilde \Theta}_r+\nonumber \\
&+&\lambda_n\, {\tilde \pi}^n+\lambda^{\vec n}_{(a)}{\tilde \pi}^{\vec n}_{(a)}
+\lambda_{(a)}^{\vec \varphi}{\tilde \pi}^{\vec \varphi}_{(a)}+{\hat \mu}_{(a)}
\, {}^3{\tilde M}_{(a)}\Big] (\tau ,\vec \sigma )+\nonumber \\
&+&{\zeta}_A(\tau ) {\tilde \pi}^A(\tau )+{\zeta}_{AB}(\tau )
{\tilde \pi}^{AB}(\tau )=\nonumber \\
&=&\int d^3\sigma \Big[ n {\hat {\cal H}}- n_{(a)}\, {}^3e^r_{(a)}
{{\partial \xi^s}\over {\partial \sigma^r}}\, {\tilde \pi}_s^{\vec \xi}+
\lambda_n{\tilde \pi}^n+\nonumber \\
&+&\lambda^{\vec n}_{(a)}{\tilde \pi}^{\vec n}_{(a)}
+\lambda_{(a)}^{\vec \varphi}{\tilde \pi}^{\vec \varphi}_{(a)}+({\hat \mu}_{(b)}
B_{(b)(a)}(\alpha_{(e)})+ \nonumber \\
&+&n_{(b)}\, {}^3e^r_{(b)} {{\partial \alpha
_{(a)}}\over {\partial \sigma^r}}) {\tilde \pi}^{\vec \alpha}_{(a)}\Big] (\tau ,
\vec \sigma )+{\zeta}_A(\tau ) {\tilde \pi}^A(\tau )+{\zeta}
_{AB}(\tau ) {\tilde \pi}^{AB}(\tau )=\nonumber \\
&=&\int d^3\sigma \Big[ n {\hat {\cal H}}- n_{(a)}\, {}^3e^r_{(a)}
{{\partial \xi^s}\over {\partial \sigma^r}}\, {\tilde \pi}_s^{\vec \xi}+
\nonumber \\
&+&\lambda_n{\tilde \pi}^n+\lambda^{\vec n}_{(a)}{\tilde \pi}^{\vec n}_{(a)}
+\lambda_{(a)}^{\vec \varphi}{\tilde \pi}^{\vec \varphi}_{(a)}+{\tilde \mu}
_{(a)} {\tilde \pi}^{\vec \alpha}_{(a)}\Big] (\tau ,\vec \sigma )+\nonumber \\
&+&{\zeta}_A(\tau ) {\tilde \pi}^A(\tau )+{\zeta}_{AB}(\tau )
{\tilde \pi}^{AB}(\tau ),
\label{V45}
\end{eqnarray}

\noindent where ${\tilde \mu}_{(a)}$ are new Dirac multipliers.

The phase space action, which usually is incorrectly written without the primary
constraints, is

\begin{eqnarray}
\bar S&=& 
\int d\tau d^3\sigma \Big[ {}^3{\tilde \pi}^r_{(a)} \partial_{\tau}\, 
{}^3e_{(a)r}-n {\hat {\cal H}}+ n_{(a)}{\cal H}_{(a)}-\nonumber \\
&-&\lambda_n {\tilde \pi}^n-\lambda^{\vec n}_{(a)} {\tilde \pi}^{\vec n}_{(a)}-
\lambda^{\vec \varphi}_{(a)} {\tilde \pi}^{\vec \varphi}_{(a)}-\mu_{(a)}\,
{}^3{\tilde M}_{(a)}\Big] (\tau ,\vec \sigma )-\nonumber \\
&-&{\zeta}_A(\tau ) {\tilde \pi}^A(\tau )-{\zeta}_{AB}(\tau )
{\tilde \pi}^{AB}(\tau )=\nonumber \\
&=&\int d\tau d^3\sigma \Big[
{}^3{\tilde \pi}^r_{(a)}\partial_{\tau}\, {}^3e_{(a)r}-
n {\hat {\cal H}}+ n_{(a)}\, {}^3e^r_{(a)}\, {}^3{\tilde
\Theta}_r-\nonumber \\
&-&\lambda_n {\tilde \pi}^n-\lambda^{\vec n}_{(a)} {\tilde \pi}^{\vec n}_{(a)}-
\lambda^{\vec \varphi}_{(a)} {\tilde \pi}^{\vec \varphi}_{(a)}-{\hat \mu}_{(a)}
\, {}^3{\tilde M}_{(a)}\Big] (\tau ,\vec \sigma )-\nonumber \\
&-&{\zeta}_A(\tau ) {\tilde \pi}^A(\tau )-{\zeta}_{AB}(\tau )
{\tilde \pi}^{AB}(\tau )=\nonumber \\
&=&\int d\tau d^3\sigma [{}^3{\tilde \pi}^r_{(a)} \partial_{\tau}\, {}^3e
_{(a)r}-n {\hat {\cal H}}+ n_{(a)}\, {}^3e^r_{(a)}
{{\partial \xi^s}\over {\partial \sigma^r}} {\tilde \pi}^{\vec \xi}_s-
\nonumber \\
&-&\lambda_n {\tilde \pi}^n-\lambda^{\vec n}_{(a)} {\tilde \pi}^{\vec n}_{(a)}-
\lambda^{\vec \varphi}_{(a)} {\tilde \pi}^{\vec \varphi}_{(a)}-{\tilde \mu}
_{(a)}\, {\tilde \pi}^{\vec \alpha}_{(a)}] (\tau ,\vec \sigma )-\nonumber \\
&-&{\zeta}_A(\tau ) {\tilde \pi}^A(\tau )-{\zeta}_{AB}(\tau )
{\tilde \pi}^{AB}(\tau ).
\label{V46}
\end{eqnarray}

In conclusion the 18-dimensional phase space spanned by ${}^3e_{(a)r}$ and 
${}^3{\tilde \pi}^r_{(a)}$ has a global [since $\Sigma_{\tau}\approx R^3$]  
canonical
basis, in which 12 variables are $\alpha_{(a)}$, ${\tilde \pi}^{\vec \alpha}
_{(a)}\approx 0$, $\xi^r$, ${\tilde \pi}^{\vec \xi}_r\approx 0$. The remaining
6 variables, hidden in the reduced quantities ${}^3{\hat e}_{(a)r}$, ${}^3{\hat
{\tilde \pi}}^r_{(a)}$, are 3 pairs of conjugate Dirac's observables with 
respect to the gauge transformations in ${\bar {\cal G}}_R$, namely they are
invariant under Abelianized rotations and space 
pseudodiffeomorphisms [and, therefore, weakly invariant under the original 
rotations and space pseudodiffeomorphisms]
connected with the identity and obtainable as a succession of infinitesimal 
gauge transformations. However, since space pseudodiffeomorphisms connect 
different charts in the atlas of $\Sigma_{\tau}$ and since $\xi^r(\tau ,\vec 
\sigma )=\sigma^r$ means to choose as origin of space 
pseudodiffeomorphisms an arbitrary chart, the functional
form of the Dirac's observables will depend on the chart chosen as origin.
This will reflect itself in the freedom of how to parametrize the reduced
cotriad ${}^3{\hat e}_{(a)r}(\tau ,\vec \sigma )$ in terms of only 3 independent
functions: in each chart `c' they will be denoted $Q^{(c)}_r(\tau ,\vec \sigma 
)$ and, if `c+dc' is a new chart connected to `c' by an infinitesimal space
pseudodiffeomorphism of parameters $\vec \xi (\tau ,\vec \sigma )$, then we 
will have $Q^{(c+dc)}_r(\tau ,\vec \sigma )={{\partial \xi^s(\tau ,\vec 
\sigma )}\over  {\partial \sigma^r}} Q^{(c)}_s(\tau ,\vec \xi (\tau ,\vec 
\sigma ))$.

The real invariants under pseudodiffeomorphisms of a Riemannian
3-manifold $(\Sigma_{\tau},{}^3g)$ [for which no explicit basis is known], can
be expressed in every chart `c' as functionals of the 3 independent functions
$Q^{(c)}_r(\tau ,\vec \sigma )$. Therefore, these 3 functions give a local
(chart-dependent) coordinatization of the space of 3-geometries (superspace or
moduli space) $Riem\, \Sigma_{\tau}/Diff\, \Sigma_{\tau}$ \cite{witt,fis}.

By using Eqs.(\ref{V36}) and (\ref{V39}) in the Hamiltonian expressions of the
4-tensors of Appendix B of I, we can get the most important 4-tensors on the
pseudo-Riemannian 4-manifold $(M^4,\, {}^4g)$ expressed in terms of
${\tilde \lambda}_A$, ${\tilde \pi}^A\approx 0$, ${\tilde \lambda}_{AB}$,
${\tilde \pi}^{AB}\approx 0$, $n$, ${\tilde \pi}^n\approx 0$, $n_{(a)}$, 
${\tilde \pi}^{\vec n}_{(a)}\approx 0$, 
$\alpha_{(a)}$, ${\tilde \pi}^{\vec \alpha}_{(a)}\approx 0$,
$\xi^r$, ${\tilde \pi}^{\vec \xi}_r\approx 0$, and of the (non canonically
conjugate) Dirac's observables with respect to the action of ${\bar {\cal G}}_R$
, i.e. ${}^3{\hat e}_{(a)r}$, ${}^3{\hat {\tilde \pi}}^r_{(a)}$. If we could
extract from ${}^3{\hat e}_{(a)r}$, ${}^3{\hat {\tilde \pi}}^r_{(a)}$, the
Dirac observables with respect to the gauge transformations generated by the
superhamiltonian constraint ${\hat {\cal H}}(\tau ,\vec \sigma )\approx 0$,
then we could express all 4-tensors in terms of these final Dirac
observables (the independent Cauchy data of tetrad gravity), of the gauge
variables $n$, $n_{(a)}$, $\alpha_{(a)}$, $\xi^r$ and of the gauge variable
associated with ${\hat {\cal H}}(\tau ,\vec \sigma )\approx 0$
(see Section V), when all the
constraints are satisfied. Therefore, we would get not only a chart-dependent
expression of the 4-metrics ${}^4g\in Riem\, M^4$, but also of the 4-geometries
in $Riem\, M^4/Diff\, M^4$.

In the next Section we shall study the simplest charts of the atlas of $\Sigma
_{\tau}$, namely the 3-orthogonal ones. See Appendix A for more information
about special coordinate charts.

\vfill\eject

\section
{The Quasi-Shanmugadhasan Canonical Transformation in 3-Orthogonal Coordinates.}

The quasi-Shanmugadhasan canonical transformation 
\cite{sha} [``quasi-" because we are not 
including the superhamiltonian constraint ${\hat {\cal H}}(\tau ,\vec \sigma )
\approx 0$] Abelianizing the rotation and pseudodiffeomorphism constraints
${}^3{\tilde M}_{(a)}(\tau ,\vec \sigma )\approx 0$, ${}^3{\tilde \Theta}_r
(\tau ,\vec \sigma )\approx 0$, will send the canonical basis ${}^3e_{(a)r}
(\tau ,\vec \sigma )$, ${}^3{\tilde \pi}^r_{(a)}(\tau ,\vec \sigma )$, of 
$T^{*}{\cal C}_e$ in a new basis whose conjugate pairs are $\Big( \alpha_{(a)}
(\tau ,\vec \sigma ), {\tilde \pi}^{\vec \alpha}_{(a)}(\tau ,\vec \sigma )
\approx 0\Big)$, $\Big( \xi^r(\tau ,\vec \sigma ),{\tilde \pi}^{\vec \xi}
_r(\tau ,\vec \sigma )\approx 0)$ for the gauge sector and $\Big( Q_r(\tau 
,\vec \sigma ), 
{\tilde \Pi}^r(\tau ,\vec \sigma )\Big)$ for the sector of Dirac observables.

Therefore, we must parametrize the Dirac observables ${}^3{\hat e}_{(a)r}(\tau ,
\vec \sigma )$ in terms of three functions $Q_r(\tau ,\vec \sigma )$,
${}^3{\hat e}_{(a)r}(\tau ,\vec \sigma )={}^3{\hat e}_{(a)r}[Q_s(\tau ,
\vec \sigma )]$, and then find how the Dirac observables ${}^3{\hat {\tilde
\pi}}^r_{(a)}(\tau ,\vec \sigma )$ are expressible in terms of $Q_r(\tau ,
\vec \sigma )$, ${\tilde \Pi}^r(\tau ,\vec \sigma )$, ${\tilde \pi}^{\vec
\xi}_r(\tau ,\vec \sigma )$, ${\tilde \pi}^{\vec \alpha}_{(a)}(\tau ,
\vec \sigma )$ [they cannot depend on $\alpha_{(a)}(\tau ,\vec \sigma )$, 
$\xi_r(\tau ,\vec \sigma )$, because they are Dirac observables]. Since from 
Eqs.(\ref{V36}) we get

\begin{eqnarray}
{}^3g_{rs}(\tau ,\vec \sigma )&=&{}^3e_{(a)r}(\tau ,\vec \sigma )\, {}^3e_{(a)s}
(\tau ,\vec \sigma )={}^3{\bar e}_{(a)r}(\tau ,\vec \sigma )\, {}^3{\bar e}
_{(a)s}(\tau ,\vec \sigma )=\nonumber \\
&=&{{\partial \xi^u(\tau ,\vec \sigma )}\over {\partial \sigma^r}}{{\partial
\xi^v(\tau ,\vec \sigma )}\over {\partial \sigma^s}}\, {}^3{\hat e}_{(a)u}
[Q_w(\tau ,\vec \xi (\tau ,\vec \sigma ))]\, {}^3{\hat e}_{(a)v}[Q_w(\tau ,
\vec \xi (\tau ,\vec \sigma ))]=\nonumber \\
&=&{{\partial \xi^u(\tau ,\vec \sigma )}\over {\partial \sigma^r}}{{\partial
\xi^v(\tau ,\vec \sigma )}\over {\partial \sigma^s}}\, {}^3{\hat g}_{uv}
[Q_w(\tau ,\vec \xi (\tau ,\vec \sigma ))],
\label{VI1}
\end{eqnarray}

\noindent the new metric ${}^3{\hat g}_{uv}(\tau ,\vec \xi )$ must depend only 
on the functions $Q_w(\tau ,\vec \xi )$. This shows that the parametrization of
${}^3{\hat e}_{(a)r}(\tau ,\vec \sigma )$ will depend on the chosen system
of coordinates, which will be declared the origin $\vec \xi (\tau ,\vec \sigma 
)=\vec \sigma $ of pseudodiffeomorphisms from the given chart. Therefore, each 
Dirac observable 3-metric ${}^3{\hat g}_{uv}$ is an element of DeWitt superspace
\cite{dew} for Riemannian 3-manifolds: it defines a 3-geometry on $\Sigma
_{\tau}$.

The simplest global system of coordinates on $\Sigma_{\tau}\approx R^3$, 
where to learn how to construct the 
quasi-Shanmugadhasan canonical transformation, is the 3-orthogonal one 
, in which ${}^3{\hat g}_{uv}$ is diagonal. In it
we have the parametrization

\begin{eqnarray}
{}^3{\hat e}_{(a)r}(\tau ,\vec \sigma )&=&\delta_{(a)r} Q_r(\tau ,\vec \sigma )
\, \Rightarrow {}^3{\hat e}^r_{(a)}(\tau ,\vec \sigma )={{\delta^r_{(a)}}
\over {Q_r(\tau ,\vec \sigma )}},\nonumber \\
\Rightarrow && {}^3{\hat g}_{rs}(\tau ,\vec \sigma )=\delta_{rs} Q^2_r(\tau ,
\vec \sigma ),\nonumber \\
ds^2&=&\epsilon \Big( [N_{(as)}+n]^2-
[N_{(as) r}+n_r] \sum_u{{\partial \sigma
^r(\vec \xi )}\over {\partial \xi^u}}{1\over {Q^2_u(\vec \xi )}}{{\partial
\sigma^s(\vec \xi )}\over {\partial \xi^u}}[N_{(as) s}+n_s] \Big) (d\tau )^2-
\nonumber \\
&-&2\epsilon [N_{(as) r}+n_r]d\tau d\sigma^r-\epsilon 
\sum_u {{\partial \xi^u}\over
{\partial \sigma^r}} Q^2_u(\vec \xi ){{\partial \xi^u}\over {\partial
\sigma^s}} d\sigma^r d\sigma^s=\nonumber \\
&=&\epsilon \Big( [N_{(as)}+n]^2(d\tau )^2 -\delta_{uv}[Q_u{{\partial \xi^u}
\over {\partial \sigma^r}}d\sigma^r+{1\over {Q_u}}{{\partial \sigma^r(\vec 
\xi )}\over {\partial \xi^u}}(N_{(as)r}+n_r)d\tau ]\nonumber \\
&&[Q_v{{\partial \xi^v}
\over {\partial \sigma^s}}d\sigma^s+{1\over {Q_v}}{{\partial \sigma^s(\vec 
\xi )}\over {\partial \xi^v}}(N_{(as)s}+n_s)d\tau ],
\label{VI2}
\end{eqnarray}

\noindent with $Q_r(\tau ,\vec \sigma )=1+h_r(\tau ,\vec \sigma ) > 0$ to avoid
singularities. The 3 functions $Q^2_r(\tau ,\vec \sigma )$ give a local 
parametrization of superspace; the presence of singularities in superspace
depends on the boundary conditions for $Q_r(\tau ,\vec \sigma )$, i.e. on the
possible existence of stability subgroups (isometries) of the group ${\bar
{\cal G}}$ of gauge transformations, which we assume to be absent if a
suitable weighted Sobolev space is chosen for cotriads.

Let us remark that if we change the parametrization of ${}^3{\hat e}_{(a)r}$, 
giving it as a different function of 3 ${\check Q}_r$, this amounts to a
canonical transformation $Q_r, {\tilde \Pi}^r \mapsto {\check Q}_r, {\check 
{\tilde \Pi}}^r$ with $\delta_{(a)r}Q_r={}^3{\hat e}_{(a)r}[{\check Q}_s]$
together with a redefinition of the origin of space 
pseudodiffeomorphism [the new global
chart is the new  origin defined as ${\vec \xi}{}^{\,\, '}(\tau ,\vec \sigma )=
\vec \sigma$ with ${\vec \xi}{}^{\,\, '}$ that functional of 
$\vec \xi$ dictated by the pseudodiffeomorphism connecting the two global
charts; however, ${\vec \xi}{}^{\,\, '}$ can be renamed $\vec \xi$ being a 
canonical variable of our basis].
In the quasi-Shanmugadhasan canonical transformation we will study in this
Section, this will be reflected in the change of the expression giving
${}^3{\tilde \pi}^r_{(a)}$ in terms of the new variables. 

The choice of the
parametrization of ${}^3{\hat e}_{(a)r}$ is equivalent to the coordinate
conditions of Refs.\cite{dirr,isha}.
See Eqs.(\ref{w1}) of Appendix A for a parametrization of the cotriads
${}^3{\hat e}_{(a)r}$ corresponding to normal coordinates around the point 
$\lbrace \tau ,\vec \sigma =0 \rbrace \in \Sigma_{\tau}$.

Since the rotation constraints ${}^3{\tilde M}_{(a)}=\epsilon_{(a)(b)(c)}\, 
{}^3e_{(b)r}\, {}^3{\tilde \pi}^r_{(c)}={1\over 2}\epsilon_{(a)(b)(c)}\,
{}^3{\tilde M}_{(b)(c)}$ may be written as \hfill\break
\hfill\break
${}^3{\tilde M}_{(a)(b)}={}^3e
_{(a)r}\, {}^3{\tilde \pi}^r_{(b)}-{}^3e_{(b)r}\, {}^3{\tilde \pi}^r_{(a)}=
\epsilon_{(a)(b)(c)}\, {}^3{\tilde M}_{(c)}=\epsilon_{(a)(b)(c)}\, {\tilde 
\pi}^{\vec \alpha}_{(d)} B_{(d)(c)}(\alpha_{(e)})$ \hfill\break
\hfill\break
due to Eqs.(\ref{V18}),
we may extract the dependence of ${}^3{\tilde \pi}^r_{(a)}(\tau ,\vec \sigma )$
from ${\tilde \pi}^{\vec \alpha}_{(a)}(\tau ,\vec \sigma )$

\begin{eqnarray}
{}^3{\tilde \pi}^r_{(a)}&=&{}^3e^r_{(b)}\, {}^3e_{(b)s}\, {}^3{\tilde \pi}^s
_{(a)}=\nonumber \\
&=&{1\over 2}\, {}^3e^r_{(b)}
\Big( {}^3e_{(b)s}\, {}^3{\tilde \pi}^s_{(a)}+{}^3e
_{(a)s}\, {}^3{\tilde \pi}^s_{(b)}\Big) +{1\over 2}\, 
{}^3e^r_{(b)}\Big( {}^3e_{(b)s}\, {}^3{\tilde \pi}^s_{(a)}-{}^3e_{(a)s}\, 
{}^3{\tilde \pi}^s_{(b)}\Big) =\nonumber \\
&=&{1\over 2}\, {}^3e^r_{(b)}
\Big( {}^3e_{(b)s}\, {}^3{\tilde \pi}^s_{(a)}+{}^3e
_{(a)s}\, {}^3{\tilde \pi}^s_{(b)}\Big) -\nonumber \\
&-&{1\over 2}\, {}^3e^r_{(b)}\, \epsilon_{(a)(b)(c)} {\tilde \pi}
^{\vec \alpha}_{(d)} B_{(d)(c)}(\alpha_{(e)})\, {\buildrel {def} \over =},
\nonumber \\
&{\buildrel {def} \over =}& {1\over 2}\, {}^3e^r_{(b)} \Big[ Z_{(a)(b)}-\epsilon
_{(a)(b)(c)} {\tilde \pi}^{\vec \alpha}_{(d)} B_{(d)(c)}(\alpha_{(e)})\Big]
\nonumber \\
{}&&\nonumber \\
Z_{(a)(b)}&=&Z_{(b)(a)}={}^3e_{(a)s}\, {}^3{\tilde \pi}^s_{(b)}+{}^3e_{(b)s}\,
{}^3{\tilde \pi}^s_{(a)}=Z_{(a)(b)}[\alpha_{(e)}, \xi^r, {\tilde \pi}^{\vec \xi}
_r, Q_r, {\tilde \Pi}^r].
\label{VI3}
\end{eqnarray}

To extract the dependence of ${}^3{\tilde \pi}^r_{(a)}(\tau ,\vec \sigma )$ on
${\tilde \pi}^{\vec \xi}_r(\tau ,\vec \sigma )$, let us recall 
Eqs.(62) of I, (\ref{V18}) and (\ref{V30})

\begin{eqnarray}
{\hat {\cal H}}_{(a)}(\tau ,\vec \sigma )&=&{\hat D}^{(\omega )}_{(a)(b)r}
(\tau ,\vec \sigma )\, {}^3{\tilde \pi}^r_{(b)}(\tau ,\vec \sigma )=\nonumber \\
&=&-{}^3e^r_{(a)}(\tau ,\vec \sigma )
\Big[ {}^3{\tilde \Theta}_r+{}^3\omega_{r(b)}\, 
{}^3{\tilde M}_{(b)}\Big] (\tau ,\vec \sigma )=\nonumber \\
&=&-{}^3e^r_{(a)}(\tau ,\vec \sigma )
\Big[ {{\partial \xi^s}\over {\partial \sigma^r}}
{\tilde \pi}^{\vec \xi}_s+(B_{(b)(c)}(\alpha_{(e)})\, {}^3\omega_{r(c)}-
{{\partial \alpha_{(b)}}\over {\partial \sigma^r}})\, {\tilde \pi}^{\vec
\alpha}_{(b)}\Big] (\tau ,\vec \sigma ),
\label{VI4}
\end{eqnarray}

\noindent and Eqs.(\ref{V18})

\begin{equation}
{}^3{\tilde \pi}^r_{(a)}(\tau ,\vec \sigma )={}^3{\tilde \pi}^{(T)r}_{(a)}
(\tau ,\vec \sigma )-\int d^3\sigma_1\, \zeta^{(\omega )r}_{(a)(b)}(\vec \sigma 
,{\vec \sigma }_1;\tau )\, {\hat {\cal H}}_{(b)}(\tau ,{\vec \sigma }_1).
\label{VI5}
\end{equation}

Then we have

\begin{eqnarray}
\Big[
Z_{(a)(b)}&-&\epsilon_{(a)(b)(c)} {\tilde \pi}^{\vec \alpha}_{(d)} B_{(d)(c)}
(\alpha_{(e)})\Big] (\tau ,\vec \sigma )=2\Big[ 
{}^3e_{(b)r}\, {}^3{\tilde \pi}^r_{(a)}
\Big] (\tau ,\vec \sigma )=\nonumber \\
&=&2\Big[
{}^3e_{(b)r}\, {}^3{\tilde \pi}^{(T)r}_{(a)}\Big] (\tau ,\vec \sigma )-2
{}^3e_{(b)r}(\tau ,\vec \sigma ) \int d^3\sigma_1\, \zeta^{(\omega )r}_{(a)(c)}
(\vec \sigma ,{\vec \sigma }_1;\tau ) {\hat {\cal H}}_{(c)}(\tau ,{\vec \sigma 
}_1)=\nonumber \\
&=&S_{(a)(b)}(\tau ,\vec \sigma )-\nonumber \\
&-&\int d^3\sigma_1\Big[ {}^3e_{(b)r}(\tau ,\vec 
\sigma )\, \zeta^{(\omega )r}_{(a)(c)}(\vec \sigma ,{\vec \sigma }_1;\tau )+
{}^3e_{(a)r}(\tau ,\vec \sigma )\, \zeta^{(\omega )r}_{(b)(c)}(\vec \sigma ,
{\vec \sigma }_1;\tau )\Big] {\hat {\cal H}}_{(c)}(\tau ,{\vec \sigma }_1)+
\nonumber \\
&+&\Big[ {}^3e_{(b)r}\, {}^3{\tilde \pi}^{(T)r}_{(a)}-{}^3e_{(a)r}\, 
{}^3{\tilde \pi}^{(T)r}_{(b)}\Big] (\tau ,\vec \sigma )-\nonumber \\
&-&\int d^3\sigma_1 \Big[ {}^3e_{(b)r}(\tau ,\vec 
\sigma )\, \zeta^{(\omega )r}_{(a)(c)}(\vec \sigma ,{\vec \sigma }_1;\tau )-
{}^3e_{(a)r}(\tau ,\vec \sigma )\, \zeta^{(\omega )r}_{(b)(c)}(\vec \sigma ,
{\vec \sigma }_1;\tau )\Big] {\hat {\cal H}}_{(c)}(\tau ,{\vec \sigma }_1), 
\nonumber \\
&&{}
\label{VI6}
\end{eqnarray}

\noindent with

\begin{equation}
S_{(a)(b)}(\tau ,\vec \sigma )=S_{(b)(a)}(\tau ,\vec \sigma )=\Big[
{}^3e_{(a)r}\,
{}^3{\tilde \pi}^{(T)r}_{(b)}+{}^3e_{(b)r}\, {}^3{\tilde \pi}^{(T)r}_{(a)}\Big]
(\tau ,\vec \sigma ).
\label{VI7}
\end{equation}

By equating the terms symmetric and antisymmetric in $(a)\, \Leftrightarrow
\, (b)$, we get

\begin{eqnarray}
&&Z_{(a)(b)}(\tau ,\vec \sigma )=S_{(a)(b)}(\tau ,\vec \sigma )-\nonumber \\
&-&\int d^3\sigma_1 \Big[ {}^3e_{(b)r}(\tau ,\vec 
\sigma )\, \zeta^{(\omega )r}_{(a)(c)}(\vec \sigma ,{\vec \sigma }_1;\tau )+
{}^3e_{(a)r}(\tau ,\vec \sigma )\, \zeta^{(\omega )r}_{(b)(c)}(\vec \sigma ,
{\vec \sigma }_1;\tau )\Big] {\hat {\cal H}}_{(c)}(\tau ,{\vec \sigma }_1),
\nonumber \\
&&{}\nonumber \\
&&\Big[
{}^3e_{(a)r} {}^3{\tilde \pi}^{(T)r}_{(b)}-{}^3e_{(b)r}\, {}^3{\tilde \pi}
^{(T)r}_{(a)}](\tau ,\vec \sigma )=\epsilon_{(a)(b)(c)}[{\tilde \pi}_{(d)}
^{\vec \alpha}B_{(d)(c)}(\alpha_{(e)})\Big] (\tau ,\vec \sigma )+\nonumber \\
&+&\int d^3\sigma_1 \Big[ {}^3e_{(b)r}(\tau ,\vec 
\sigma )\, \zeta^{(\omega )r}_{(a)(c)}(\vec \sigma ,{\vec \sigma }_1;\tau )-
{}^3e_{(a)r}(\tau ,\vec \sigma )\, \zeta^{(\omega )r}_{(b)(c)}(\vec \sigma ,
{\vec \sigma }_1;\tau )\Big] {\hat {\cal H}}_{(c)}(\tau ,{\vec \sigma }_1),
\label{VI8}
\end{eqnarray}

\noindent so that we obtain the following dependence of ${}^3{\tilde \pi}^r
_{(a)}$ on ${\tilde \pi}^{\vec \alpha}_{(a)}$ and ${\tilde \pi}^{\vec \xi}_r$

\begin{eqnarray}
{}^3{\tilde \pi}^r_{(a)}(\tau ,\vec \sigma )&=&{1\over 2}{}^3e^r_{(b)}(\tau ,
\vec \sigma )\Big( S_{(a)(b)}(\tau ,\vec \sigma )-\epsilon_{(a)(b)(c)}
{\tilde \pi}^{\vec \alpha}_{(d)}(\tau ,\vec \sigma )B_{(d)(c)}(\alpha_{(e)}
(\tau ,\vec \sigma ))-\nonumber \\
&-&\int d^3\sigma_1 \Big[ {}^3e_{(b)u}(\tau ,\vec 
\sigma )\, \zeta^{(\omega )u}_{(a)(c)}(\vec \sigma ,{\vec \sigma }_1;\tau )+
{}^3e_{(a)u}(\tau ,\vec \sigma )\, \zeta^{(\omega )u}_{(b)(c)}(\vec \sigma ,
{\vec \sigma }_1;\tau )\Big] \cdot \nonumber \\
&\cdot& {}^3e^w_{(c)}(\tau ,{\vec \sigma }_1)
\Big[ {{\partial \xi^s}\over {\partial 
\sigma_1^w}}{\tilde \pi}^{\vec \xi}_s+(B_{(d)(f)}(\alpha_{(e)})\, {}^3\omega
_{w(f)}-{{\partial \alpha_{(d)}}\over {\partial \sigma_1^w}}){\tilde \pi}
^{\vec \alpha}_{(d)}\Big] (\tau ,{\vec \sigma }_1)\Big) .
\label{VI9}
\end{eqnarray}

Therefore, all the dependence of ${}^3{\tilde \pi}^r_{(a)}$ on ${\tilde \Pi}^r$
is hidden in $S_{(a)(b)}$. To find it, let us impose the canonicity of the
transformation \hfill\break
\hfill\break
${}^3e_{(a)r}, {}^3{\tilde \pi}^r_{(a)} \mapsto \alpha_{(a)}, 
{\tilde \pi}^{\vec \alpha}_{(a)}, \xi^r, {\tilde \pi}^{\vec \xi}_r, Q_r, 
{\tilde \Pi}^r$ \hfill\break
\hfill\break
by taking into account that \hfill\break
\hfill\break
$\lbrace \alpha_{(a)}(\tau ,\vec
\sigma ),{\tilde \pi}^{\vec \alpha}_{(b)}(\tau ,{\vec \sigma }^{'})\rbrace
=\delta_{(a)(b)}\delta^3(\vec \sigma ,{\vec \sigma }^{'})$, \hfill\break
$\lbrace \xi^r
(\tau ,\vec \sigma ),{\tilde \pi}^{\vec \xi}_s(\tau ,{\vec \sigma }^{'})\rbrace
=\lbrace Q_s(\tau ,\vec \sigma ),{\tilde \Pi}^r(\tau ,{\vec \sigma }^{'})
\rbrace =\delta^r_s\delta^3(\vec \sigma ,{\vec \sigma }^{'})$ \hfill\break
and that
$\lbrace {}^3e_{(a)r}(\tau ,\vec \sigma ),\alpha_{(b)}(\tau ,{\vec \sigma }
^{'})\rbrace =\lbrace {}^3e_{(a)r}(\tau ,\vec \sigma ),\xi_s(\tau ,{\vec 
\sigma }^{'})\rbrace =\lbrace {}^3e_{(a)r}(\tau ,\vec \sigma ),Q_s(\tau ,{\vec 
\sigma }^{'})\rbrace =0$

\begin{eqnarray}
\delta^s_r\delta_{(a)(b)}\delta^3(\vec \sigma ,{\vec \sigma }^{'})&=&\lbrace 
{}^3e_{(a)r}(\tau ,\vec \sigma ),{}^3{\tilde \pi}^s_{(b)}(\tau ,{\vec \sigma }
^{'})\rbrace =\nonumber \\
&=&\int d^3\sigma_1\Big[
\lbrace {}^3e_{(a)r}(\tau ,\vec \sigma ),{\tilde \pi}^{\vec
\alpha}_{(c)}(\tau ,{\vec \sigma }_1)\rbrace \lbrace \alpha_{(c)}(\tau ,
{\vec \sigma }_1),{}^3{\tilde \pi}^s_{(b)}(\tau ,{\vec \sigma }^{'})\rbrace +
\nonumber \\
&+&\lbrace {}^3e_{(a)r}(\tau ,\vec \sigma ),{\tilde \pi}^{\vec \xi}_u(\tau ,
{\vec \sigma }_1)\rbrace \lbrace \xi^u(\tau ,{\vec \sigma }_1),{}^3{\tilde \pi}
^s_{(b)}(\tau ,{\vec \sigma }^{'})\rbrace +\nonumber \\
&+&\lbrace {}^3e_{(a)r}(\tau ,\vec \sigma ),{\tilde \Pi}^u(\tau ,{\vec \sigma }
_1)\rbrace \lbrace Q_u(\tau ,{\vec \sigma }_1),{}^3{\tilde \pi}^s_{(b)}(\tau ,
{\vec \sigma }^{'})\rbrace \Big] =\nonumber \\
&=&\int d^3\sigma_1\Big[
{{\tilde \delta {}^3e_{(a)r}(\tau ,\vec \sigma )}\over
{\delta \alpha_{(c)}(\tau ,{\vec \sigma }_1)}}{{\delta {}^3{\tilde \pi}^s_{(b)}
(\tau ,{\vec \sigma }^{'})}\over {\delta {\tilde \pi}^{\vec \alpha}_{(c)}(\tau ,
{\vec \sigma }_1)}}+{{\delta {}^3e_{(a)r}(\tau ,\vec \sigma )}\over {\delta
\xi^u(\tau ,{\vec \sigma }_1)}}{{\delta {}^3{\tilde \pi}^s_{(b)}(\tau ,
{\vec \sigma }^{'})}\over {\delta {\tilde \pi}^{\vec \xi}_u(\tau ,{\vec \sigma 
}_1)}}+\nonumber \\
&+&{{\delta {}^3e_{(a)r}(\tau ,\vec \sigma )}\over {\delta Q_u(\tau ,{\vec 
\sigma }_1)}}{{\delta {}^3{\tilde \pi}^s_{(b)}(\tau ,{\vec \sigma }^{'})}\over
{\delta {\tilde \Pi}^u(\tau ,{\vec \sigma }_1)}}\Big] .
\label{VI10}
\end{eqnarray}

\noindent [we could replace $\alpha_{(a)}(\tau ,\vec \sigma )$ with
$\alpha_{(a)}(\tau ,\vec \xi (\tau ,\vec \sigma ))$, since the angles are scalar
fields under pseudodiffeomorphims].

Since ${}^3e_{(a)r}(\tau ,\vec \sigma )={}^3R_{(a)(b)}(\alpha_{(e)}(\tau ,
\vec \sigma )){{\partial \xi^u}\over {\partial \sigma^r}}\delta_{(b)u}Q_u(\tau 
,\vec \xi (\tau ,\vec \sigma ))$, 
by using Eqs.(\ref{V9}) and $H_{(a)}(\alpha_{(e)})={\hat R}
^{(d)}A_{(d)(a)}(\alpha_{(e)})$ we get

\begin{eqnarray}
{{\tilde \delta {}^3e_{(a)r}(\tau ,\vec \sigma )}\over {\delta \alpha_{(c)}
(\tau ,{\vec \sigma }_1)}}&=&\delta^3(\vec \sigma ,{\vec \sigma }_1)\Big[
H_{(c)}(\alpha_{(e)}(\tau ,\vec \sigma )){}^3R(\alpha_{(e)}(\tau ,\vec \sigma ))
\Big]_{(a)(b)}\nonumber \\
&&\sum_u{{\partial \xi^u(\tau ,\vec \sigma )}\over {\partial \sigma^r}}
\delta_{(b)u}Q_u(\tau ,\vec \xi (\tau ,\vec \sigma ))=\nonumber \\
&=&\delta^3(\vec \sigma ,{\vec \sigma }_1)\epsilon_{(a)(n)(d)}A_{(d)(c)}
(\alpha_{(e)}(\tau ,\vec \sigma )){}^3R_{(n)(m)}(\alpha_{(e)}(\tau ,\vec 
\sigma ))\cdot \nonumber \\
&\cdot& \sum_u{{\partial \xi^u(\tau ,\vec \sigma )}\over {\partial \sigma^r}}
\delta_{(m)u}Q_u(\tau ,\vec \xi (\tau ,\vec \sigma )),\nonumber \\
{{\delta {}^3e_{(a)r}(\tau ,\vec \sigma )}\over {\delta \xi^u(\tau ,{\vec 
\sigma }_1)}}&=&{}^3R_{(a)(n)}(\alpha_{(e)}(\tau ,\vec \sigma ))
\sum_v\delta_{(n)v}\Big[
{{\partial \xi^v(\tau ,\vec \sigma )}\over {\partial \sigma^r}}{{\partial
Q_v(\tau ,\vec \xi )}\over {\partial \xi^u}}{|}_{\vec \xi =\vec \xi (\tau ,
\vec \sigma )}\delta^3(\vec \sigma ,{\vec \sigma }_1)+\nonumber \\
&+&\delta^v_u
Q_v(\tau ,\vec \xi (\tau ,\vec \sigma )){{\partial \delta^3(\vec \sigma ,
{\vec \sigma }_1)}\over {\partial \sigma^r}}\Big] ,\nonumber \\
{{\delta {}^3e_{(a)r}(\tau ,\vec \sigma )}\over {\delta Q_u(\tau ,{\vec 
\sigma }_1)}}&=&{}^3R_{(a)(n)}(\alpha_{(e)}(\tau ,\vec \sigma ))\sum_v
{{\partial \xi^v
(\tau ,\vec \sigma )}\over {\partial \sigma^r}}\delta_{(n)v}
\delta^u_v\delta^3(\vec \xi
(\tau ,\vec \sigma ),{\vec \sigma }_1).
\label{VI11}
\end{eqnarray}

Then, Eqs.(\ref{VI9}) give

\begin{eqnarray}
{{\delta {}^3{\tilde \pi}^s_{(b)}(\tau ,{\vec \sigma }^{'})}\over {\delta 
{\tilde \pi}^{\vec \alpha}_{(c)}(\tau ,{\vec \sigma }_1)}}&=&{1\over 2}
{}^3e^s_{(h)}(\tau ,{\vec \sigma }^{'})\Big( -\epsilon_{(b)(h)(k)}
B_{(c)(k)}(\alpha_{(e)}(\tau ,{\vec \sigma }^{'}))\delta^3({\vec \sigma }_1,
{\vec \sigma }^{'})-\nonumber \\
&-&\Big[ {}^3e_{(b)t}(\tau ,{\vec \sigma }^{'})\, \zeta^{(\omega )t}_{(h)(k)}
({\vec \sigma }^{'},{\vec \sigma }_1;\tau )+{}^3e_{(h)t}(\tau ,{\vec \sigma }
^{'})\, \zeta^{(\omega )t}_{(b)(k)}({\vec \sigma }^{'},{\vec \sigma }_1;\tau )
\Big] \cdot \nonumber \\
&\cdot& {}^3e^w_{(k)}(\tau ,{\vec \sigma }_1)\Big[ B_{(c)(f)}(\alpha_{(e)})\,
{}^3\omega_{w(f)}-{{\partial \alpha_{(c)}}\over {\partial \sigma^w_1}}\Big]
(\tau ,{\vec \sigma }_1) \Big) ,\nonumber \\
{{\delta {}^3{\tilde \pi}^s_{(b)}(\tau ,{\vec \sigma }^{'})}\over {\delta 
{\tilde \pi}^{\vec \xi}_u(\tau ,{\vec \sigma }_1)}}&=&-{1\over 2} {}^3e^s_{(h)}
(\tau ,{\vec \sigma }^{'}) \Big[ {}^3e_{(b)t}(\tau ,{\vec \sigma }^{'})\, 
\zeta^{(\omega )t}_{(h)(k)}({\vec \sigma }^{'},{\vec \sigma }_1;\tau )+
\nonumber \\
&+&{}^3e_{(h)t}(\tau ,{\vec \sigma }
^{'})\, \zeta^{(\omega )t}_{(b)(k)}({\vec \sigma }^{'},{\vec \sigma }_1;
\tau )\Big] {}^3e^w_{(k)}(\tau ,{\vec \sigma }_1){{\partial \xi^u(\tau ,{\vec 
\sigma }_1)}\over {\partial \sigma_1^w}},\nonumber \\
{{\delta {}^3{\tilde \pi}^s_{(b)}(\tau ,{\vec \sigma }^{'})}\over {\delta
{\tilde \Pi}^u(\tau ,{\vec \sigma }_1)}}&=&{1\over 2}\, {}^3e^s_{(h)}(\tau ,
{\vec \sigma }^{'}) {{\delta S_{(b)(h)}(\tau ,{\vec \sigma }^{'})}\over
{\delta {\tilde \Pi}^u(\tau ,{\vec \sigma }_1)}}.
\label{VI12}
\end{eqnarray}

\noindent so that Eqs.(\ref{VI10}) become

\begin{eqnarray}
\delta^s_r\delta_{(a)(b)}&&\delta^3(\vec \sigma  ,{\vec \sigma }^{'})=
\nonumber \\
&=&-{1\over 2}\epsilon_{(a)(n)(d)}\, {}^3R_{(n)(m)}(\alpha_{(e)}(\tau ,\vec 
\sigma ))\sum_v{{\partial \xi^v(\tau ,\vec \sigma )}\over {\partial \sigma^r}}
\delta_{(m)v}Q_v(\tau ,\vec \xi (\tau ,\vec \sigma ))\cdot \nonumber \\
&\cdot& {}^3e^s_{(h)}(\tau ,{\vec \sigma }^{'}) \Big( \epsilon_{(b)(h)(d)}
\delta^3(\vec \sigma ,{\vec \sigma }^{'})+{}^3e^w_{(k)}(\tau ,\vec \sigma )
\Big[ {}^3\omega_{w(d)}(\tau ,\vec \sigma )-A_{(d)(c)}
(\alpha_{(e)}(\tau ,\vec \sigma ))\cdot \nonumber \\
&\cdot& {{\partial \alpha_{(c)}(\tau ,\vec \sigma )}\over {\partial \sigma^w}}
\Big] T_{(b)(h)(k)}({\vec \sigma }^{'},\vec \sigma ;\tau )\, \Big)-\nonumber \\
&-&{1\over 2}\, {}^3R_{(a)(n)}(\alpha_{(e)}(\tau ,\vec \sigma )) 
\sum_v\delta_{(n)v} \Big( \,
{{\partial \xi^v(\tau ,\vec \sigma )}\over {\partial \sigma^r}}
\, {}^3e^w_{(k)}(\tau ,\vec \sigma )\cdot \nonumber \\
&\cdot& {{\partial Q_v(\tau ,\vec \xi )}\over {\partial \sigma^w}}
{}^3e^s_{(h)}(\tau ,{\vec \sigma }^{'}) T_{(b)(h)(k)}({\vec \sigma }^{'},
\vec \sigma ;\tau )+\nonumber \\
&+&Q_v(\tau ,\vec \xi (\tau ,\vec \sigma ))\, {}^3e^s_{(h)}
(\tau ,{\vec \sigma }^{'}) \nonumber \\
&\cdot& {{\partial}\over {\partial \sigma^r}} \Big[ {}^3e^w_{(k)}(\tau ,
\vec \sigma ){{\partial \xi^v(\tau ,\vec \sigma )}\over {\partial \sigma^w}}
T_{(b)(h)(k)}({\vec \sigma }^{'},\vec \sigma ;\tau )
\Big]\,\, \Big) +\nonumber \\
&+&{1\over 2}\, {}^3R_{(a)(n)}(\alpha_{(e)}(\tau ,\vec \sigma ))\sum_v{{\partial
\xi^v(\tau ,\vec \sigma )}\over {\partial \sigma^r}}\delta_{(n)v}\, {}^3e^s
_{(h)}(\tau ,{\vec \sigma }^{'}) {{\delta S_{(h)(b)}(\tau ,{\vec \sigma }^{'})}
\over {\delta {\tilde \Pi}^v(\tau ,\vec \xi (\tau ,\vec \sigma ))}},
\label{VI13}
\end{eqnarray}

\noindent where we introduced the notation

\begin{eqnarray}
T_{(b)(h)(k)}({\vec \sigma }^{'},\vec \sigma ;\tau )&=&{}^3e_{(b)t}(\tau ,{\vec 
\sigma }^{'})\, \zeta^{(\omega )t}_{(h)(k)}({\vec \sigma }^{'},\vec \sigma ;
\tau )+{}^3e_{(h)t}(\tau ,{\vec \sigma }^{'})\, \zeta^{(\omega )t}_{(b)(k)}
({\vec \sigma }^{'},\vec \sigma ;\tau )=\nonumber \\
&=&\sum_rQ_r(\tau ,{\vec \sigma}^{'})
\Big[ \delta_{(b)r} \zeta^{(\omega )r}_{(h)(k)}
({\vec \sigma}^{'},\vec \sigma ;\tau )+\delta_{(h)r} \zeta^{(\omega )r}_{(b)(k)}
({\vec \sigma}^{'},\vec \sigma ;\tau )\Big] =\nonumber \\
&=&\sum_r Q_r(\tau ,\vec \sigma ) d^r_{\gamma_{P^{'}P}}({\vec \sigma}^{'},\vec 
\sigma )\nonumber \\
&&\Big[ \delta_{(b)r} \Big(
P_{\gamma_{P^{'}P}}\, e^{\int^{{\vec \sigma}^{'}}_{\vec \sigma}d\sigma_1^s\,
{\hat R}^{(c)}\, {}^3\omega_{s(c)}(\tau ,{\vec \sigma}_1)}\, \Big)_{(h)(k)}+
\nonumber \\
&+&\delta_{(h)r}\, \Big( P_{\gamma_{P^{'}P}}\, e^{\int^{{\vec \sigma}^{'}}
_{\vec \sigma}d\sigma_1^s\, {\hat R}^{(c)}\, 
{}^3\omega_{s(c)}(\tau ,{\vec \sigma}_1)}\, \Big)_{(b)(k)}\, \Big] .
\label{VI14}
\end{eqnarray}

By multiplying this equation by ${}^3R^{-1}_{(g)(a)}(\alpha_{(e)}(\tau ,
\vec \sigma ))={}^3R^T_{(g)(a)}(\alpha_{(e)}(\tau ,\vec \sigma ))$ and then
by sending $(g)\mapsto (a)$, we get

\begin{eqnarray}
\sum_v{{\partial \xi^v(\tau ,\vec \sigma )}\over {\partial \sigma^r}}
\delta_{(a)v}&& {}^3e^s_{(h)}(\tau ,{\vec 
\sigma }^{'}){{\delta S_{(h)(b)}(\tau ,{\vec \sigma }^{'})}\over {\delta
{\tilde \Pi}^v(\tau ,\vec \xi (\tau ,\vec \sigma ))}}=\nonumber \\
&=&2 \delta^s_r\, {}^3R_{(b)(a)}(\alpha_{(e)}(\tau ,\vec \sigma )) \delta^3
(\vec \sigma ,{\vec \sigma }^{'})+\nonumber \\
&+&{}^3R_{(f)(a)}(\alpha_{(e)}(\tau ,\vec \sigma )) \epsilon_{(f)(n)(d)}\,
{}^3R_{(n)(m)}(\alpha_{(e)}(\tau ,\vec \sigma )) \sum_v
{{\partial \xi^v(\tau ,\vec 
\sigma )}\over {\partial \sigma^r}}\cdot \nonumber \\
&\cdot& \delta_{(m)v}Q_v(\tau ,\vec \xi (\tau ,\vec 
\sigma ))\, {}^3e^s_{(h)}(\tau ,{\vec \sigma }^{'})
\Big[ \epsilon_{(b)(h)(d)}\delta^3(\vec \sigma ,{\vec \sigma }^{'})+\nonumber \\
&+&{}^3e^w_{(k)}(\tau ,\vec \sigma )({}^3\omega_{w(d)}(\tau ,\vec \sigma )-
A_{(d)(c)}(\alpha_{(e)}(\tau ,\vec \sigma )){{\partial \alpha_{(c)}(\tau ,\vec 
\sigma )}\over {\partial \sigma^w}}) \nonumber \\
&\cdot& T_{(b)(h)(k)}({\vec \sigma }^{'},
\vec \sigma ;\tau )\Big] +\nonumber \\
&+&\sum_v
\delta_{(a)v}{{\partial \xi^v(\tau ,\vec \sigma )}\over {\partial \sigma^r}}
{{\partial Q_v(\tau ,\vec \xi (\tau ,\vec \sigma ))}\over {\partial \sigma^w}}
\, {}^3e^w_{(k)}(\tau ,\vec \sigma )\, {}^3e^s_{(h)}(\tau ,{\vec \sigma }^{'})
\nonumber \\
&\cdot& T_{(b)(h)(k)}({\vec \sigma }^{'},\vec \sigma ;\tau )+\nonumber \\
&+&\sum_v
\delta_{(a)v}Q_v(\tau ,\vec \xi (\tau ,\vec \sigma ))\, {}^3e^s_{(h)}(\tau ,
{\vec \sigma }^{'})\nonumber \\
&\cdot& {{\partial}\over {\partial \sigma^r}} \Big[ {}^3e^w_{(k)}(\tau ,
\vec \sigma ){{\partial \xi^v(\tau ,\vec \sigma )}\over {\partial \sigma^w}}
T_{(b)(h)(k)}({\vec \sigma }^{'},\vec \sigma ;\tau )\Big] .\nonumber \\
&&{}
\label{VI15}
\end{eqnarray}

From Eqs.(\ref{V9}) and (\ref{V10}) and from Eq.(5.32) of the second paper in 
Ref.\cite{lusa}, we have \hfill\break
\hfill\break
${}^3R_{(f)(a)}\epsilon_{(f)(n)(d)}\, {}^3R_{(n)(m)}=
[{}^3R^{-1}\, {\hat R}^{(d)}\, {}^3R]_{(a)(m)}=({\hat R}^{(n)})_{(a)(m)}\,
{}^3R_{(n)(d)}=\epsilon_{(a)(m)(n)}\, {}^3R_{(n)(d)}$ \hfill\break
\hfill\break
and, then,
$\epsilon_{(a)(m)(n)}\, {}^3R_{(n)(d)} \epsilon_{(b)(g)(d)}=-
[{\hat R}^{(a)}\, {}^3R\, {\hat R}^{(b)}]_{(m)(g)}$. Then, by
multiplying the previous equation by ${}^3e_{(g)s}(\tau ,{\vec \sigma }^{'})$
we obtain [also using Eq.(\ref{V12})]

\begin{eqnarray}
\sum_v
{{\partial \xi^v(\tau ,\vec \sigma )}\over {\partial \sigma^r}}&& \delta_{(a)v}
{{\delta S_{(g)(b)}(\tau ,{\vec \sigma }^{'})}\over {\delta {\tilde \Pi}^v
(\tau ,\vec \sigma ))}}=\nonumber \\
&=&\Big( 
2\, {}^3e_{(g)r}(\tau ,\vec \sigma )\, {}^3R_{(b)(a)}(\alpha_{(e)}(\tau ,
\vec \sigma ))-\Big[ {\hat R}^{(a)}\, {}^3R(\alpha_{(e)}(\tau 
,\vec \sigma ))\, {\hat R}^{(b)} \Big]_{(m)(g)}\cdot \nonumber \\
&\cdot& \sum_v{{\partial \xi^v(\tau ,\vec \sigma )}\over {\partial \sigma^r}}
\delta_{(m)v}Q_v(\tau ,\vec \xi (\tau ,\vec \sigma ))\Big) \, 
\delta^3(\vec \sigma ,{\vec \sigma }^{'})+\nonumber \\
&+&\sum_v{{\partial \xi^v(\tau ,\vec \sigma )}\over {\partial \sigma^r}}
{}^3e^w_{(k)}(\tau ,\vec \sigma ) \Big(
\, \Big[ {\hat R}^{(a)}\, {}^3R(\alpha_{(e)}(\tau ,\vec \sigma ))\Big]_{(m)(d)}
\delta_{(m)v}Q_v(\tau ,\vec \xi (\tau ,\vec \sigma ))\cdot \nonumber \\
&\cdot& {}^3\omega^{(T)}_{w(d)}(\tau ,\vec \sigma ,\alpha_{(e)}(\tau ,
\vec \sigma ))+\nonumber \\
&+&\sum_v\delta_{(a)v}\, {{\partial Q_v(\tau ,\vec \xi (\tau ,\vec \sigma 
))}\over {\partial \sigma^w}}\, \Big)
T_{(b)(g)(k)}({\vec \sigma }^{'},\vec \sigma ;\tau )+\nonumber \\
&+&\sum_v
\delta_{(a)v}Q_v(\tau ,\vec \xi (\tau ,\vec \sigma )){{\partial}\over
{\partial \sigma^r}} 
\Big[ {}^3e^w_{(k)}(\tau ,\vec \sigma ){{\partial \xi^v(\tau ,
\vec \sigma )}\over {\partial \sigma^w}} T_{(b)(g)(k)}({\vec \sigma }^{'},\vec 
\sigma ;\tau )\Big] ,\nonumber \\
&&{}
\label{VI16}
\end{eqnarray}

\noindent and by multiplication by ${{\partial \sigma^r(\vec \xi )}\over 
{\partial \xi^u}} {|}_{\vec \xi =\vec \xi (\tau ,\vec \sigma )}$ we have
[there is no sum over $u$]

\begin{eqnarray}
&&{{\delta S_{(g)(b)}(\tau ,{\vec \sigma }^{'})}\over {\delta
[\delta_{(a)u}\, {\tilde \Pi}^u(\tau ,\vec \xi (\tau ,\vec \sigma ))]}}=
\delta_{(a)u} {{\delta S_{(g)(b)}(\tau ,{\vec \sigma }^{'})}\over {\delta
{\tilde \Pi}^u(\tau ,\vec \xi (\tau ,\vec \sigma ))}}=\nonumber \\
&&{}\nonumber \\
&=&\Big( 2\, {}^3e_{(g)r}(\tau ,\vec \sigma ){{\partial 
\sigma^r(\vec \xi )}\over {\partial \xi^u}}{|}_{\vec \xi =\vec \xi (\tau ,
\vec \sigma )}\, {}^3R_{(b)(a)}(\alpha_{(e)}(\tau ,\vec \sigma ))-\nonumber \\
&-&\Big[ {\hat R}^{(a)}\, {}^3R(\alpha_{(e)}(\tau ,\vec \sigma ))\, {\hat R}
^{(b)}\Big]_{(m)(g)} \delta_{(m)u}Q_u(\tau ,\vec \xi (\tau ,\vec \sigma ))
\Big) \delta^3(\vec \sigma ,{\vec \sigma }^{'})+\nonumber \\
&+&\Big( \, \Big[ {\hat R}^{(a)}\, {}^3R(\alpha_{(e)}(\tau ,\vec \sigma ))
\Big]_{(m)(d)}\delta_{(m)u}Q_u(\tau ,\vec \xi (\tau ,\vec \sigma ))\, 
{}^3\omega^{(T)}_{w(d)}(\tau ,\vec \sigma ,\alpha_{(e)}(\tau ,
\vec \sigma ))+\nonumber \\
&+&\delta_{(a)u}{{\partial Q_u(\tau ,\vec \xi (\tau ,\vec \sigma ))}\over
{\partial \sigma^w}}\, \Big) {}^3e^w_{(k)}(\tau ,\vec \sigma )\,  T_{(b)(g)(k)}
({\vec \sigma }^{'},\vec \sigma ;\tau )+\nonumber \\
&+&\sum_v
\delta_{(a)v}Q_v(\tau ,\vec \xi (\tau ,\vec \sigma )){{\partial \sigma^r
(\vec \xi)}\over {\partial \xi^u}}{|}_{\vec \xi =\vec \xi (\tau ,\vec \sigma )}
{{\partial}\over {\partial \sigma^r}} \Big[ {}^3e^w_{(k)}(\tau ,\vec \sigma )
{{\partial \xi^v(\tau ,\vec \sigma )}\over {\partial \sigma^w}} T_{(b)(g)(k)}
({\vec \sigma }^{'},\vec \sigma ;\tau )\Big] =\nonumber \\
&&{}\nonumber \\
&=&\sum_v\Big( 2\, {}^3e_{(g)r}(\tau ,\vec \sigma ){{\partial 
\sigma^r(\vec \xi )}\over {\partial \xi^v}}{|}_{\vec \xi =\vec \xi (\tau ,
\vec \sigma )}\, {}^3R_{(b)(c)}(\alpha_{(e)}(\tau ,\vec \sigma ))-\nonumber \\
&-&\Big[ {\hat R}^{(c)}\, {}^3R(\alpha_{(e)}(\tau ,\vec \sigma ))\, {\hat R}
^{(b)}\Big]_{(m)(g)} \delta_{(m)v}Q_v(\tau ,\vec \xi (\tau ,\vec \sigma ))
\Big) \,\,\,
\delta_{(c)(a)}\delta_{(a)u}\delta^v_u\delta^3(\vec \sigma ,{\vec \sigma }^{'})+
\nonumber \\
&+&\int d^3\sigma_1\, \delta_{(a)(c)}\sum_v\delta^v_u\delta^3(\vec \sigma ,{\vec 
\sigma }_1) \delta_{(a)u} \cdot \nonumber \\
&\cdot& \Big( \, {}^3e^w_{(k)}(\tau ,{\vec \sigma}_1)\,  T_{(b)(g)(k)}
({\vec \sigma }^{'},{\vec \sigma}_1;\tau )\cdot \nonumber \\
&\cdot& 
\Big[ \, [ {\hat R}^{(c)}\, {}^3R(\alpha_{(e)}(\tau ,{\vec \sigma}_1))
]_{(m)(d)}\delta_{(m)v}Q_v(\tau ,\vec \xi (\tau ,{\vec \sigma}_1))\, 
{}^3\omega^{(T)}_{w(d)}(\tau ,{\vec \sigma}_1,\alpha_{(e)}
(\tau ,{\vec \sigma}_1))+\nonumber \\
&+& \delta_{(c)v}{{\partial Q_v(\tau ,\vec \xi (\tau ,{\vec \sigma}_1))}\over
{\partial \sigma_1^w}}\, \Big]  +\nonumber \\
&+&\sum_t\delta_{(c)t}Q_t(\tau ,\vec \xi (\tau ,{\vec \sigma}_1)){{\partial 
\sigma_1^r(\vec \xi)}\over {\partial \xi^v}}{|}_{\vec \xi 
=\vec \xi (\tau ,{\vec \sigma}_1)}{{\partial}\over 
{\partial \sigma_1^r}} \Big[ {}^3e^w_{(k)}(\tau ,{\vec \sigma}_1)
{{\partial \xi^t(\tau ,{\vec \sigma}_1)}\over {\partial \sigma_1^w}} 
T_{(b)(g)(k)}({\vec \sigma }^{'},{\vec \sigma}_1;\tau )\Big]\,\, 
\Big) .\nonumber \\
&&{}
\label{VI17}
\end{eqnarray}

This is the final equation for $S_{(a)(b)}$ in terms of ${\tilde \Pi}^u$. 
Since we have [no sum over u,v]

\begin{eqnarray}
{{\delta \, [\delta_{(c)v}{\tilde \Pi}^v(\tau ,\vec \xi (\tau ,{\vec \sigma }
^{'}))]}\over {\delta \, [\delta_{(a)u}{\tilde \Pi}^u(\tau ,\vec \xi (\tau ,
\vec \sigma ))]}}&=&\delta_{(a)u}\delta_{(c)v}\delta^v_u\delta^3(\vec \xi
(\tau ,\vec \sigma ),\vec \xi (\tau ,{\vec \sigma }^{'}))=\nonumber \\
&=&\delta_{(a)(c)}\delta_{(a)u}\delta^v_u
{{\delta^3(\vec \sigma ,{\vec \sigma }^{'})}\over 
{ |{{\partial \xi}\over {\partial \sigma}}(\tau ,\vec \sigma )| }},
\label{VI18}
\end{eqnarray}

\noindent the final solution for $S_{(a)(b)}$ is

\begin{eqnarray}
&&S_{(a)(b)}(\tau ,\vec \sigma )=\nonumber \\
&&{}\nonumber \\
&=&|\, {{\partial \xi}\over {\partial \sigma}}(\tau ,\vec \sigma )\, |\cdot
\Big(\,\sum_v \Big[ 2\, {}^3e_{(a)r}(\tau ,\vec \sigma ){{\partial \sigma^r(\vec
\xi )}\over {\partial \xi^v}}{|}_{\vec \xi =\vec \xi(\tau ,\vec \sigma )}\,
{}^3R_{(b)(c)}(\alpha_{(e)}(\tau ,\vec \sigma ))-\nonumber \\
&-&[{\hat R}^{(c)}\, {}^3R(\alpha_{(e)}(\tau ,\vec \sigma ))\, {\hat R}^{(b)}]
_{(m)(a)}\delta_{(m)v}Q_v(\tau ,\vec \xi (\tau ,\vec \sigma ))\Big] \,
\delta_{(c)v} {\tilde \Pi}^v(\tau ,\vec \xi (\tau ,\vec \sigma ))+\nonumber \\
&+&\int d^3\sigma_1\, \sum_v\delta_{(c)v}
{\tilde \Pi}^v(\tau ,\vec \xi (\tau ,{\vec 
\sigma }_1))\, { {|{{\partial \xi}\over {\partial \sigma_1}}(\tau ,{\vec 
\sigma }_1)|} \over {|{{\partial \xi}\over {\partial \sigma}}(\tau ,\vec \sigma 
)|} }\cdot \nonumber \\
&\cdot& \Big( \,\, {}^3e^w_{(k)}(\tau ,{\vec \sigma }_1)\, T_{(b)(a)(k)}
(\vec \sigma ,{\vec \sigma }_1;\tau )\cdot \nonumber \\
&\cdot& \Big[ [{\hat R}^{(c)}\, {}^3R(\alpha_{(e)}(\tau ,{\vec \sigma }_1))]
_{(m)(d)}\delta_{(m)v}Q_v(\tau ,\vec \xi (\tau ,{\vec \sigma }_1))\, 
{}^3\omega^{(T)}_{w(d)}(\tau ,{\vec \sigma 
}_1,\alpha_{(e)}(\tau ,{\vec \sigma }_1))+\nonumber \\
&+&\delta_{(c)v}{{\partial Q_v(\tau ,\vec \xi (\tau ,{\vec \sigma }_1))}\over
{\partial \sigma_1^w}}\, \Big] +\nonumber \\
&+&\sum_t\delta_{(c)t}Q_t(\tau ,\vec \xi (\tau ,{\vec \sigma }_1)){{\partial
\sigma^r_1(\vec \xi )}\over {\partial \xi^v}}{|}_{\vec \xi =\vec \xi (\tau ,
{\vec \sigma }_1)}\, \cdot \nonumber \\
&\cdot& {{\partial}\over {\partial \sigma_1^r}}\, [{}^3e^w_{(k)}(\tau ,{\vec 
\sigma }_1) {{\partial \xi^t(\tau ,{\vec \sigma }_1)}\over {\partial
\sigma^w_1}}\, T_{(b)(a)(k)}(\vec \sigma ,{\vec \sigma }_1;\tau ) \,\, 
\Big) \, \Big).
\label{VI19}
\end{eqnarray}

We have put equal to zero an arbitrary integration constant, namely an 
arbitrary function $f_{(a)(b)}(\alpha_{(c)}, \xi^r, Q_r)$, which would 
contribute a term $g^r_{(a)}={1\over 2}f_{(a)(b)}\, {}^3e^r_{(b)}={1\over 
{2Q_r}}f_{(a)(b)}(\alpha_{(c)}, \xi^s, Q_s)\delta^r_{(b)}$ to 
${}^3{\tilde \pi}^r_{(a)}$. Let us remark 
that the canonical transformation \hfill\break
\hfill\break
${}^3e_{(a)r}$, ${}^3{\tilde \pi}^r_{(a)}$
$\mapsto$ $\alpha_{(a)}$, $\xi^r$, $Q_r$, ${\tilde \pi}^{\vec \alpha}_{(a)}$,
${\tilde \pi}^{\vec \xi}_r$, ${\tilde \Pi}^r$ \hfill\break
\hfill\break
is a point transformation
$q^i, p_i$ $\mapsto$ ${\tilde q}^i$, ${\tilde p}_i$ with $q^i=q^i({\tilde q}^j)$
, $p_i({\tilde q}^j,{\tilde p}_k)$, which is defined modulo a so called trival
phase canonical transformation $q^i({\tilde q}^j)$, $p_i({\tilde q}^j,
{\tilde p}_k)+f_i({\tilde q}^j)$ with $f_i({\tilde q}^j)=\partial f({\tilde q}
^j)/\partial {\tilde q}^i$. Therefore, even if we cannot check explicitly the 
validity of $\{ \, {}^3{\tilde \pi}^r_{(a)}(\tau ,\vec \sigma ),\, {}^3{\tilde 
\pi}^s_{(b)}(\tau ,{\vec \sigma}^{'}\} =0$ due to the presence of the 
path-orderings in the expression of the momenta in terms of the new variables, 
these Poisson brackets imply that $g^r_{(a)}$ is the gradient of a function of
$\alpha_{(a)}$, $\xi^r$, $Q_r$, so that our choice $f_{(a)(b)}=0$ amounts to
a trivial phase canonical transformation.

Therefore the cotriad and its momentum have the following expression in terms
of the new canonical variables [Eqs. (\ref{V12}) and (\ref{VI9}) are used] 

\begin{eqnarray}
{}^3e_{(a)r}(\tau ,\vec \sigma )&=&{}^3R_{(a)(b)}(\alpha_{(e)}(\tau ,\vec 
\sigma ))\, {{\partial \xi^s(\tau ,\vec \sigma )}\over {\partial \sigma^r}}
\, \delta_{(b)s}\, Q_s(\tau ,\vec \xi (\tau ,\vec \sigma )),\nonumber \\
&&{}\nonumber \\
{}^3{\tilde \pi}^r_{(a)}(\tau ,\vec \sigma )&=&{1\over 2}\, {{\delta_{(b)}^r}
\over {Q_r(\tau ,\vec \sigma )}}\,\,\,
|\, {{\partial \xi}\over {\partial \sigma}}(\tau ,\vec \sigma )\, |\cdot
\nonumber \\
&\cdot& \Big(\,\sum_v \Big[ 2\, \sum_s \delta_{(a)s}Q_s(\tau ,\vec \sigma )
{{\partial \sigma^s(\vec \xi )}\over {\partial \xi^v}}{|}_{\vec \xi =\vec 
\xi(\tau ,\vec \sigma )}\,
{}^3R_{(b)(c)}(\alpha_{(e)}(\tau ,\vec \sigma ))-\nonumber \\
&-&[{\hat R}^{(c)}\, {}^3R(\alpha_{(e)}(\tau ,\vec \sigma ))\, {\hat R}^{(b)}]
_{(m)(a)}\delta_{(m)v}Q_v(\tau ,\vec \xi (\tau ,\vec \sigma ))\Big] \,
\delta_{(c)v}{\tilde \Pi}^v(\tau ,\vec \xi (\tau ,\vec \sigma ))+\nonumber \\
&&{}\nonumber \\
&+&\int d^3\sigma_1\, \sum_v\delta_{(c)v}
{\tilde \Pi}^v(\tau ,\vec \xi (\tau ,{\vec 
\sigma }_1))\, { {|{{\partial \xi}\over {\partial \sigma_1}}(\tau ,{\vec 
\sigma }_1)|} \over {|{{\partial \xi}\over {\partial \sigma}}(\tau ,\vec \sigma 
)|} }\cdot \nonumber \\
&\cdot& \Big( \,\, \sum_w{{\delta_{(k)w}}\over {Q_w(\tau ,{\vec \sigma}_1)}}
T_{(b)(a)(k)}(\vec \sigma ,{\vec \sigma }_1;\tau ) \Big[ \delta_{(c)v}
{{\partial Q_v(\tau ,\vec \xi (\tau ,{\vec \sigma }_1))}\over
{\partial \sigma_1^w}}+\nonumber \\
&+&[{\hat R}^{(c)}\, {}^3R(\alpha_{(e)}(\tau ,{\vec \sigma }_1))]
_{(m)(d)}\delta_{(m)v}Q_v(\tau ,\vec \xi (\tau ,{\vec \sigma }_1))
{}^3\omega^{(T)}_{w(d)}(\tau ,{\vec \sigma}_1,\alpha_{(e)}(\tau ,
{\vec \sigma }_1)) \Big] +\nonumber \\
&+&\sum_t \delta_{(c)t}Q_t(\tau ,\vec \xi (\tau ,{\vec \sigma }_1)){{\partial
\sigma_1^s(\vec \xi )}\over {\partial \xi^v}}{|}_{\vec \xi =\vec \xi (\tau ,
{\vec \sigma }_1)}\, \cdot \nonumber \\
&\cdot& {{\partial}\over {\partial \sigma_1^s}}\, \Big[ \sum_w {{\delta_{(k)w}}
\over {Q_w(\tau ,{\vec \sigma }_1)}}
{{\partial \xi^t(\tau ,{\vec \sigma }_1)}\over {\partial
\sigma^w_1}}\, T_{(b)(a)(k)}(\vec \sigma ,{\vec \sigma }_1;\tau )\Big]\, 
\Big) \, \Big) -\nonumber \\
&&{}\nonumber \\
&-&{1\over 2}\, {{\delta^r_{(b)}}\over {Q_r(\tau ,\vec \sigma )}}\, 
\Big( \,\,\epsilon_{(a)(b)(c)}
{\tilde \pi}^{\vec \alpha}_{(d)}(\tau ,\vec \sigma )B_{(d)(c)}(\alpha_{(e)}
(\tau ,\vec \sigma ))+\nonumber \\
&+&\int d^3\sigma_1\, T_{(b)(a)(c)}(\vec \sigma ,{\vec \sigma }_1;\tau )
\sum_w{{\delta^w_{(c)}}\over {Q_w(\tau ,{\vec \sigma }_1)}}\cdot \nonumber \\
&\cdot& \Big[ {{\partial \xi^s}\over {\partial 
\sigma_1^w}}\, {\tilde \pi}^{\vec \xi}_s +B_{(d)(f)}(\alpha_{(e)})\,
{}^3\omega^{(T)}_{w(f)}(.,\alpha_{(e)})\, 
{\tilde \pi}^{\vec \alpha}_{(d)}\Big] (\tau ,{\vec \sigma 
}_1)\,\, \Big) =\nonumber \\
&&{}\nonumber \\
&=&{1\over 2}\, {{\delta_{(b)}^r}
\over {Q_r(\tau ,\vec \sigma )}}\,\,\,
|\, {{\partial \xi}\over {\partial \sigma}}(\tau ,\vec \sigma )\, |\cdot
\nonumber \\
&\cdot& \Big(\, 2\, {}^3R_{(b)(c)}(\alpha_{(e)}(\tau ,\vec \sigma ))
\sum_{s,v}\delta_{(a)s}Q_s(\tau ,\vec \sigma )
{{\partial \sigma^s(\vec \xi )}\over {\partial \xi^v}}{|}_{\vec \xi =\vec 
\xi(\tau ,\vec \sigma )}\, \delta_{(c)v}{\tilde \Pi}^v(\tau ,\vec \xi (\tau ,
\vec \sigma ))+\nonumber \\
&+&\int d^3\sigma_1\, 
{ {|{{\partial \xi}\over {\partial \sigma_1}}(\tau ,{\vec 
\sigma }_1)|} \over {|{{\partial \xi}\over {\partial \sigma}}(\tau ,\vec \sigma 
)|} } \sum_v\delta_{(c)v} {\tilde \Pi}^v(\tau ,\vec \xi (\tau ,{\vec 
\sigma }_1))\cdot \nonumber \\
&\cdot& \Big( \sum_w{{\delta_{(c)v}\delta_{(k)w}}\over {Q_w(\tau ,{\vec \sigma}
_1)}} T_{(b)(a)(k)}(\vec \sigma ,{\vec \sigma}_1;\tau ) {{\partial Q_v(\tau ,
\vec \xi (\tau ,{\vec \sigma}_1))}\over {\partial \sigma^w_1}}+\nonumber \\
&+&\sum_t\delta_{(c)t}Q_t(\tau ,\vec \xi (\tau ,{\vec \sigma}_1))
{{\partial \sigma_1^s(\vec \xi )}\over {\partial \xi^v}}{|}_{\vec \xi =\vec 
\xi(\tau ,{\vec \sigma}_1)}\cdot \nonumber \\
&\cdot& {{\partial}\over {\partial \sigma_1^s}}\, \Big[ \sum_w {{\delta_{(k)w}}
\over {Q_w(\tau ,{\vec \sigma }_1)}}
{{\partial \xi^t(\tau ,{\vec \sigma }_1)}\over {\partial
\sigma^w_1}}\, T_{(b)(a)(k)}(\vec \sigma ,{\vec \sigma }_1;\tau )\Big]\, 
\Big) \, \Big) -\nonumber \\
&&{}\nonumber \\
&-&{1\over 2}\, {{\delta^r_{(b)}}\over {Q_r(\tau ,\vec \sigma )}}\, 
\Big( \,\,\epsilon_{(a)(b)(c)}
{\tilde \pi}^{\vec \alpha}_{(d)}(\tau ,\vec \sigma )B_{(d)(c)}(\alpha_{(e)}
(\tau ,\vec \sigma ))+\nonumber \\
&+&\int d^3\sigma_1\, T_{(b)(a)(c)}(\vec \sigma ,{\vec \sigma }_1;\tau )
\sum_w{{\delta^w_{(c)}}\over {Q_w(\tau ,{\vec \sigma }_1)}}\cdot \nonumber \\
&\cdot& \Big[ {{\partial \xi^s}\over {\partial 
\sigma_1^w}}\, {\tilde \pi}^{\vec \xi}_s +B_{(d)(f)}(\alpha_{(e)})\,
{}^3\omega^{(T)}_{w(f)}(.,\alpha_{(e)})\, 
{\tilde \pi}^{\vec \alpha}_{(d)}\Big] (\tau ,{\vec \sigma 
}_1)\,\, \Big) =\nonumber \\
&&{}\nonumber \\
&=&{|}_{\alpha_{(a)}=0, \xi^r=\sigma^r, {\tilde \pi}^{\vec \alpha}_{(a)}=
{\tilde \pi}^{\vec \xi}_r=0}\quad {}^3{\hat {\tilde \pi}}^r
_{(a)}(\tau ,\vec \sigma ).
\label{VI20}
\end{eqnarray}

\noindent and from Eqs.(57) and (84)  of I we get

\begin{eqnarray}
{}^3K_{rs}&=&\sum_u {{\epsilon Q_rQ_sQ_u}\over {4kQ_1Q_2Q_3}} \, 
{}^3G_{o(a)(b)(c)(d)} \delta_{(a)r}\delta_{(b)s}\delta_{(c)u}\,
{}^3{\tilde \pi}^u_{(d)},\nonumber \\
&&{}\nonumber \\
{}^3K&=&-{{\epsilon}\over {2kQ_1Q_2Q_3}}\, {}^3{\tilde \Pi}=-{{\epsilon}\over
{4kQ_1Q_2Q_3}}\sum_r\delta_{(a)r}Q_r\, {}^3{\tilde \pi}^r_{(a)},\nonumber \\
&&{}\nonumber \\
{}^2{\tilde \Pi}^{rs}&=&\epsilon kQ_1Q_2Q_3({}^3K^{rs}-Q_r^2\delta^{rs}\, {}^3K)
={1\over 4}\Big[ {{\delta^r_{(a)}\, {}^3{\tilde \pi}^s_{(a)}}\over {Q_r}}+
{{\delta^s_{(a)}\, {}^3{\tilde \pi}^r_{(a)}}\over {Q_s}}\Big] .
\label{VI21}
\end{eqnarray}

Due to the presence of the Green function it is not possible to rewrite the
final expression of ${}^3{\tilde \pi}^r_{(a)}$ explicitly in the form of
Eq.(\ref{V39}).

However, the functions ${}^3\Gamma^u_{rs}$, ${}^3R_{rsuv}$, ${}^3\omega_{r(a)}$,
${}^3\Omega_{rs(a)}$ and ,by using Eqs.(57) of I, ${}^3K_{rs}$ [and also the
metric ADM momentum ${}^3{\tilde \Pi}^{rs}$ of Eq.(84) of I and the 
Weyl-Schouten ${}^3C_{rsu}$ and Cotton-York ${}^3{\cal H}_{rs}$ tensors defined
after Eq.(9) of I] may now be 
expressed in terms of $\alpha_{(a)}$, ${\tilde \pi}^{\vec \alpha}_{(a)}$, 
$\xi^r$, ${\tilde \pi}^{\vec \xi}_r$, $Q_r$, ${\tilde \Pi}^r$, and then
Eqs.(38), (39), (40), (43), (46), (47), (A1),
(A4), (A5), (A6) of I, allow to reconstruct the functions
${}^4g_{AB}$, ${}^4E^{(\alpha )}_{\mu}$, ${}^4\Gamma^{\alpha}_{\beta\gamma}$,
${}^4R_{\mu\nu\alpha\beta}$, ${}^4\omega_{\mu (\alpha )(\beta )}$,
${}^4\Omega_{\mu\nu (\alpha )(\beta )}$, ${}^4C_{\mu\nu\alpha\beta}$,
in terms of the canonical basis ${\tilde \lambda}_A$, ${\tilde \pi}^A\approx 0$,
${\tilde \lambda}_{AB}$, ${\tilde \pi}^{AB}\approx 0$,
$n$, ${\tilde \pi}^n\approx 0$, $n_{(a)}$, ${\tilde \pi}^{\vec n}_{(a)}\approx
0$, $\varphi_{(a)}$, ${\tilde \pi}^{\vec \varphi}_{(a)}\approx 0$, $\alpha
_{(a)}$, ${\tilde \pi}^{\vec \alpha}_{(a)}\approx 0$, $\xi^r$, ${\tilde \pi}
^{\vec \xi}_r\approx 0$, $Q_r$, ${\tilde \Pi}^r$.
In the new basis only the superhamiltonian constraint of Eq.(61) of I
is left. Instead the inverse canonical transformation cannot be computed
explicitly till when one does not understand how to solve Eqs.(\ref{V41}).

\vfill\eject

\section{A New Canonical Basis and the Superhamiltonian Constraint.}

Let us study the reduced phase space spanned by the canonical coordinates
${\tilde \lambda}_A$, ${\tilde \pi}^A
\approx 0$, ${\tilde \lambda}_{AB}$, ${\tilde \pi}^{AB}\approx 0$,
$n$, ${\tilde \pi}^n\approx 0$, $n_{(a)}$, ${\tilde \pi}^{\vec n}
_{(a)}\approx 0$, $\varphi_{(a)}$, ${\tilde \pi}^{\vec \varphi}_{(a)}\approx 0$
(for the spacetime description), and $Q_r$, ${\tilde \Pi}^r$ (for the
superspace of 3-geometries) obtained by adding the gauge-fixing constraints
$\varphi_{(a)}(\tau ,\vec \sigma )\approx 0$
[their time constancy implies $\lambda^{\vec \varphi}_{(a)}(\tau ,
\vec \sigma ) \approx 0$],
$\alpha_{(a)}(\tau ,\vec \sigma )\approx 0$, $\xi^r(\tau ,\vec \sigma )
\approx \sigma^r$ and by going to Dirac brackets. This means to restrict
the Cauchy data of cotriads on $\Sigma_{\tau}$ by eliminating the gauge
degrees of freedom of boosts, rotations and space
pseudodiffeomorphisms, i.e. by restricting
ourselves to 3-orthogonal coordinates on $\Sigma_{\tau}$ and by having made the 
choice of the $\Sigma_{\tau}$-adapted tetrads ${}^4_{(\Sigma_{\tau})}{\check
{\tilde E}}^A_{(\alpha )}$ [see Eqs.(39), (40) of I rewritten in terms of
the Dirac observables ${}^3{\hat e}^r_{(a)}$ dual to ${}^3{\hat e}_{(a)r}$] as
the reference nongeodesic congruence of timelike ``nonrotating" observers
with 4-velocity field $l^A(\tau ,\vec \sigma )$.

By remembering Eqs.(\ref{V23}) and (\ref{V29}), the Dirac brackets are
strongly equal to

\begin{eqnarray}
\lbrace A(\tau ,\vec \sigma )&,&B(\tau ,{\vec \sigma}^{'})\rbrace {}^{*}=
\lbrace A(\tau ,\vec \sigma ),B(\tau ,{\vec \sigma}^{'})\rbrace +\nonumber \\
&+&\int d^3\sigma_1\Big[ \lbrace A(\tau ,\vec \sigma ),\alpha_{(a)}(\tau ,{\vec 
\sigma}_1)\rbrace \lbrace {\tilde \pi}^{\vec \alpha}_{(a)}(\tau ,{\vec 
\sigma}_1),B(\tau ,{\vec \sigma}^{'})\rbrace -\nonumber \\
&-&\lbrace A(\tau ,\vec \sigma ),
{\tilde \pi}^{\vec \alpha}_{(a)}(\tau ,{\vec \sigma}_1)\rbrace \lbrace \alpha
_{(a)}(\tau ,{\vec \sigma}_1),B(\tau ,{\vec \sigma}^{'})\rbrace +\nonumber \\
&+&\lbrace A(\tau ,\vec \sigma ),\xi^r(\tau ,{\vec \sigma}_1)\rbrace \lbrace
{\tilde \pi}^{\vec \xi}_r(\tau ,{\vec \sigma}_1),B(\tau ,{\vec \sigma}^{'})
\rbrace -\nonumber \\
&-&\lbrace A(\tau ,\vec \sigma ),{\tilde \pi}^{\vec \xi}_r(\tau ,{\vec 
\sigma}_1)\rbrace \lbrace\xi^r(\tau ,{\vec \sigma}_1),B(\tau ,{\vec \sigma}^{'})
\rbrace \Big] \equiv \nonumber \\
&\equiv& \lbrace A(\tau ,\vec \sigma ),B(\tau ,{\vec \sigma}^{'})\rbrace +
\nonumber \\
&+& \int d^3\sigma_1 \Big( \, \Big[\, \lbrace A(\tau ,\vec \sigma ),\alpha
_{(a)}(\tau ,{\vec \sigma}_1)\rbrace \lbrace {}^3{\tilde M}_{(b)}(\tau ,{\vec 
\sigma}_1),B(\tau ,{\vec \sigma}^{'})\rbrace -\nonumber \\
&-&\lbrace A(\tau ,\vec \sigma ),{}^3{\tilde M}
_{(b)}(\tau ,{\vec \sigma}_1)\rbrace \lbrace \alpha_{(a)}(\tau ,{\vec \sigma}_1)
B(\tau ,{\vec \sigma}^{'})\rbrace \Big] \cdot \nonumber \\
&\cdot& A_{(b)(a)}(\alpha_{(e)}(\tau ,{\vec \sigma}_1)) + {{\partial 
\sigma^s_1(\vec \xi )}\over {\partial \xi^r}}{|}_{\vec \xi =\vec \xi (\tau ,
{\vec \sigma}_1)} \cdot \nonumber \\
&\cdot& \Big[ \lbrace A(\tau ,\vec \sigma ),\xi^r(\tau ,{\vec \sigma}_1)\rbrace
\lbrace {}^3{\tilde \Theta}_s(\tau ,{\vec \sigma}_1),B(\tau ,{\vec \sigma}^{'})
\rbrace -\nonumber \\
&-&\lbrace A(\tau ,\vec \sigma ),{}^3{\tilde \Theta}_s(\tau ,{\vec 
\sigma}_1)\rbrace \lbrace \xi^r(\tau ,{\vec \sigma}_1),B(\tau ,{\vec 
\sigma}^{'}) \rbrace \Big] +\nonumber \\
&+&{{\partial \sigma^s_1(\vec \xi )}\over {\partial \xi^r}}{|}_{\vec \xi =\vec 
\xi (\tau ,{\vec \sigma}_1)} A_{(b)(a)}(\alpha_{(e)}(\tau ,{\vec \sigma}_1))
{{\partial \alpha_{(a)}(\tau ,{\vec \sigma}_1)}\over {\partial \sigma^s_1}}
\cdot \nonumber \\
&\cdot& \Big[ \lbrace A(\tau ,\vec \sigma ),\xi^r(\tau ,{\vec \sigma}_1)\rbrace
\lbrace {}^3{\tilde M}_{(b)}(\tau ,{\vec \sigma}_1),B(\tau ,{\vec \sigma}^{'})
\rbrace -\nonumber \\
&-&\lbrace A(\tau ,\vec \sigma ),{}^3{\tilde M}_{(b)}(\tau ,{\vec 
\sigma}_1)\rbrace \lbrace \xi^r(\tau ,{\vec \sigma}_1),B(\tau ,{\vec \sigma}
^{'})\rbrace \Big]\, \Big) .
\label{VII0}
\end{eqnarray}

\noindent Since the variables $\alpha_{(a)}(\tau ,\vec \sigma )$, $\xi^r(\tau ,
\vec \sigma )$, are not known as explicit functions of the cotriads, these
Dirac brackets can be used only implicitly. As it will be shown in
Ref.\cite{russo3}, we must have $\alpha_{(a)}(\tau ,\vec \sigma )\, 
\rightarrow \, O(r^{-(1+\epsilon )})$ and $\xi^r(\tau ,\vec \sigma )\, 
\rightarrow \, \sigma^r+O(r^{-\epsilon})$ for $r\, \rightarrow \infty$ 
to preserve Eqs.(\ref{I4}).

We have seen in Section III
that the differential geometric description for rotations already
showed that the restriction to the identity cross section $\alpha_{(a)}(\tau ,
\vec \sigma )=0$ implied also $\partial_r\alpha_{(a)}(\tau ,\vec \sigma )=0$;
we also have $A_{(a)(b)}(\alpha_{(e)}(\tau ,\vec \sigma )){|}_{\alpha =0}=0$.
When we add the gauge-fixings $\alpha_{(a)}(\tau ,\vec \sigma )\approx 0$,
the derivatives of all orders of $\alpha_{(a)}(\tau ,\vec \sigma )$ weakly
vanish at $\alpha_{(a)}(\tau ,\vec \sigma )=0$. Similarly, it can be shown that,
if we have the pseudodiffeomorphism $\vec \xi (\tau ,\vec \sigma )=\vec \sigma +
{\hat {\vec \xi}}(\tau ,\vec \sigma )$ so that for $\vec \xi (\tau ,\vec 
\sigma )\rightarrow \vec \sigma $ we have ${\hat {\vec \xi}}(\tau ,\vec \sigma )
\rightarrow \delta \vec \sigma  (\tau ,\vec \sigma )$, then the quantities 
${}^3e_{(a)r}(\tau ,\vec \sigma )$, $\partial_r\, {}^3e_{(a)s}(\tau ,\vec 
\sigma )$, ${}^3\omega_{r(a)}(\tau ,\vec \sigma )$, ${}^3\Omega_{rs(a)}(\tau ,
\vec \sigma )$, become functions only of $Q_r(\tau ,\vec \sigma )$ for
$\vec \xi (\tau ,\vec \sigma )\rightarrow \vec \sigma $ and $\alpha_{(a)}
(\tau ,\vec \sigma )\rightarrow 0$ only if we have the following behaviour
of the parameters $\xi^r(\tau ,\vec \sigma )$

\begin{eqnarray}
{{\partial \delta \sigma^r(\tau ,\vec \sigma )}\over {\partial \sigma^s}}
{|}_{\vec \xi =\vec \sigma}=0 &\Rightarrow& {{\partial \xi^r(\tau ,\vec \sigma 
)}\over {\partial \sigma^s}}{|}_{\vec \xi =\vec \sigma }=\delta^r_s,\quad
{{\partial^2 \xi^r(\tau ,\vec \sigma )}\over {\partial \sigma^s\partial 
\sigma^u}}{|}_{\vec \xi =\vec \sigma}=0,\nonumber \\
{{\partial^2\delta \sigma^r(\tau ,\vec \sigma )}\over {\partial \sigma^u\partial
\sigma^v}}{|}_{\vec \xi =\vec \sigma }=0 
&\Rightarrow& [{{\partial}\over {\partial
\sigma^u}}{{\partial \sigma^r(\tau ,\vec \sigma )}\over {\partial \xi^v}}
{|}_{\vec \xi =\vec \xi (\tau ,\vec \sigma )}]{|}_{\vec \xi =\vec \sigma }=0,
\nonumber \\
{{\partial^3\delta \sigma^r(\tau ,\vec \sigma )}\over {\partial \sigma^s
\partial \sigma^u\partial \sigma^v}}{|}_{\vec \xi =\vec \sigma }=0 
&\Rightarrow& [{{\partial^2}\over
{\partial \sigma^u\partial \sigma^v}}{{\partial \sigma^r(\vec \xi)}\over
{\partial \xi^s}}{|}_{\vec \xi =\vec \xi (\tau ,\vec \sigma )}]{|}_{\vec \xi =
\vec \sigma }=0.
\label{VII1}
\end{eqnarray}

\noindent These conditions should be satisfied by the parameters of 
pseudodiffeomorphisms near the identity, i.e. near the chart chosen as 
reference chart (the 3-orthogonal one in this case).
With the gauge-fixings $\xi^r(\tau ,\vec \sigma )\approx \vec \sigma$ all these
properties are satisfied.

By using the Dirac Hamiltonian (\ref{V45}) \hfill\break
\hfill\break
${\hat H}_{(D)ADM}=\int d^3\sigma
[n {\hat {\cal H}}-{\tilde n}^r{\tilde \pi}^{\vec \xi}_r+\lambda_n{\tilde \pi}
^n+\lambda^{\vec n}_{(a)}{\tilde \pi}^{\vec \varphi}_{(a)}
+{\tilde \mu}_{(a)}{\tilde \pi}^{\vec \alpha}_{(a)}](\tau
,\vec \sigma )+\zeta_A(\tau ){\tilde \pi}^A(\tau )+\zeta_{AB}(\tau ){\tilde
\pi}^{AB}(\tau )$ \hfill\break
\hfill\break
with ${\tilde n}^r=n_{(a)}\, {}^3e^s_{(a)}{{\partial \xi^r}
\over {\partial \sigma^s}}=n_u\, {}^3g^{uv} {{\partial \xi^r}\over {\partial 
\sigma^v}}$, the time constancy of the
gauge fixings gives \hfill\break
\hfill\break
$\partial_{\tau}\alpha_{(a)}\, {\buildrel \circ \over =}\,
{\tilde \mu}_{(a)}\approx 0$ and $\partial_{\tau}[\xi^r-\sigma^r]\,
{\buildrel \circ \over =}\, {\tilde n}^r=n_{(a)}\, {}^3e^s_{(a)} {{\partial
\xi^r}\over {\partial \sigma^s}} \approx 0$, \hfill\break
\hfill\break
so that we get the three new
constraints $n_{(a)}(\tau ,\vec \sigma )\approx 0$ implying the vanishing
of the part of the shift vector associated with proper gauge transformations
[and $N_{(a)}=N_{(as)\, (a)}$, see Eq.(\ref{I3})]. 
Then we have $\partial_{\tau} n_{(a)}\, {\buildrel \circ
\over =}\, \lambda^{\vec n}_{(a)} \approx 0$. Now $n_{(a)}(\tau ,\vec \sigma )
\approx 0$ implies $n^r(\tau ,\vec \sigma )\approx 0$ and, from 
Eqs.(\ref{VI2}), \hfill\break
\hfill\break
$ds^2=\epsilon ([N_{(as)}+n]^2-\sum_u N^2_{(as) u}/Q^2_u)
(d\tau )^2-2\epsilon N_{(as) r} d\tau d\sigma^r -\epsilon \sum_u Q^2_u 
(d\sigma^u)^2$.\hfill\break
\hfill\break
If we would add the extra gauge-fixings $N_{(as) r}\approx 0$,
this would be the definition of ``synchronous" 
coordinates in $M^4$, whose problem is the tendency to develop coordinate
singularities in short times\cite{misner,lifsh} [see Ref.\cite{yoyo}
for the problems of the fixation of N and $N^r$ 
in ADM metric gravity (coordinate conditions to
rebuild spacetime) and for the origin of the coordinates used in numerical
gravity (see Ref.\cite{nume} for a recent review of it)].

However, as we shall see in the next paper\cite{russo3}, these results will 
not be valid after the addition to the Dirac Hamiltonian of the surface terms 
needed to make it differentiable.

Since, as already said, a change of coordinate chart with a space 
pseudodiffeomorphism implies the redefinition of the functions $\vec \xi (\tau ,
\vec \sigma )$, we should explore systematically the effect of other
gauge-fixings of the type $\vec \xi (\tau ,\vec \sigma )-\vec f(\tau ,\vec 
\sigma )\approx 0$ 
for arbitrary vector functions $\vec f$ [so that $\partial_{\tau}
(\xi^r-f^r)\, {\buildrel \circ \over =}\, n_{(a)}\, {}^3e^s_{(a)}
{{\partial \xi^r}\over {\partial \sigma^s}}-\partial_{\tau}f^r\approx 0$,
which implies $n_{(a)} \approx  {}^3e_{(a)r}{{\partial \sigma^r}\over
{\partial \xi^s}} \partial_{\tau}f^s$ or $n_u\, {}^3g^{uv}\partial_vf^r\approx 
\partial_{\tau}f^r$], which describes the ``residual gauge freedom" of going
from a 3-orthogonal gauge ``in $M^4$" to another one [the allowed canonical 
transformations $Q_r$, ${\tilde \Pi}^r$ $\mapsto$ ${\check Q}^r(Q)$, ${\check
{\tilde \Pi}}^r(Q,{\tilde \Pi})$ which leave ${}^3{\hat g}_{rs}$ diagonal].
Let us remark that a redefinition of the functions $\vec \xi (\tau ,\vec 
\sigma )$ also implies a redefinition of the angles $\alpha_{(a)}(\tau ,\vec 
\sigma )$ [rotations and pseudodiffeomorphisms do not commute]: therefore this
``residual gauge freedom" allows to go from a 3-orthogonal gauge $A_1$ to 
another one $A_2$ ``rotating" with respect to $A_1$ [so that for instance we 
could get $n_r(\tau ,\vec \sigma )=2\epsilon_{rst} \sigma^s \omega^t$: for the
observer in $(\tau ,\vec \sigma )$ in $A_1$ the triad ${}^3{\hat e}^r_{(a)}
(\tau ,\vec \sigma )$ would be Fermi-Walker transported along his worldline,
while in $A_2$ it would rotate with angular velocity $\vec \omega$ with respect
to the one in $A_1$ \cite{stephani,mtw}].

The Dirac Hamiltonian reduces to \hfill\break
\hfill\break
$H_{(D)ADM,R}\equiv \int d^3\sigma [ n{\hat 
{\cal H}}_R+\lambda_n{\tilde \pi}^n](\tau ,
\vec \sigma )+{\zeta}_A(\tau ){\tilde \pi}^A(\tau )+{\zeta}_{AB}
(\tau ){\tilde \pi}^{AB}(\tau )$, \hfill\break
\hfill\break
where ${\hat {\cal H}}_R$ is the reduced superhamiltonian constraint.

This amounts to the Schwinger time gauge: Eqs. (46), (40) of I imply for the
cotetrad ${}^4E^{(\alpha )}_A={}^4_{(\Sigma )}{\check {\tilde E}}^{(\alpha )}_A$
with ${}^4_{(\Sigma )}{\check {\tilde E}}^{(o)}_{\tau}=N_{(as)}+n$, 
${}^4_{(\Sigma )}{\check {\tilde E}}^{(o)}_r=0$, ${}^4_{(\Sigma )}{\check 
{\tilde E}}^{(a)}_{\tau}=N_{(as)(a)}$, ${}^4_{(\Sigma )}{\check {\tilde E}}
^{(a)}_r={}^3{\hat e}_{(a)r}$.

At the level of Dirac brackets the constraints ${\hat {\cal H}}_{(a)}\approx 0$
(or ${}^3{\tilde \Theta}_r\approx 0$) and ${}^3{\tilde M}_{(a)}\approx 0$
[and also the derived ADM constraints ${}^3{\tilde \Pi}^{rs}{}_{|s}\approx 0$
as shown in Section V of I]
hold strongly, so that the reduced quantities ${}^3{\hat e}_{(a)r}$ and
${}^3{\hat {\tilde \pi}}^r_{(a)}$ [which now describe only three pairs of 
conjugate variables in each point of $\Sigma_{\tau}$] must obey\hfill\break
\hfill\break
${}^3{\hat e}_{(a)r}\, {}^3{\hat {\tilde \pi}}^r_{(b)}-{}^3{\hat e}_{(b)r}\,
{}^3{\hat {\tilde \pi}}^r_{(a)}\equiv 0,\quad\quad
\partial_r\, {}^3{\hat {\tilde \pi}}^r_{(a)}-\epsilon_{(a)(b)(c)}\, {}^3{\hat
\omega}_{r(b)}\, {}^3{\hat {\tilde \pi}}^r_{(c)} \equiv 0,$\hfill\break
$or\quad {}^3{\hat {\tilde \pi}}^s_{(a)}\partial_r\, {}^3{\hat e}_{(a)s}-
\partial_s({}^3{\hat e}_{(a)r}\, {}^3{\hat {\tilde \pi}}^s_{(a)})\equiv 0,$
\hfill\break
$or\quad \partial_s\, {}^3{\hat {\tilde \Pi}}^{rs}+{}^3{\hat \Gamma}^r_{su}\,
{}^3{\hat {\tilde \Pi}}^{su}\equiv 0.$\hfill\break
\hfill\break
Therefore, the ADM momentum 
${}^3{\hat {\tilde \Pi}}^{rs}$ is strongly transverse, ${}^3{\hat
{\tilde \Pi}}^{rs}\equiv {}^3{\hat {\tilde \Pi}}^{rs}_t$, and, according to
the result (\ref{tt4}) of Appendix C, can be written as ${}^3{\hat {\tilde 
\Pi}}^{rs}_t={}^3{\hat {\tilde \Pi}}^{rs}_{TT}+{}^3{\hat {\tilde \Pi}}^{rs}
_{Tr,t}$ with both the terms transverse and the first one traceless. Since
now ${}^3{\hat {\tilde \Pi}}^{rs}_t$ contains only 3 independent degrees of
freedom [the ${}^3{\tilde \Pi}^r(\tau ,\vec \sigma )$], 
we see that ${}^3{\hat {\tilde \Pi}}^{rs}_{TT}$ should describe the
spin-two wave part of the ADM momentum, while ${}^3{\hat {\tilde \Pi}}^{rs}
_{Tr,t}$ should describe the mean extrinsic curvature through its trace.
However, Eq.(84) of I does not imply $\lbrace {}^3{\hat {\tilde \Pi}}^{rs}
(\tau ,\vec \sigma ),{}^3{\hat {\tilde \Pi}}^{uv}(\tau ,{\vec \sigma}_1)\rbrace
{}^{*}=0$ at the level of these
Dirac brackets, since ${}^3{\tilde \Pi}^{rs}(\tau ,
\vec \sigma )$ does not commute with the supermomentum constraints [one would 
get $\lbrace {}^3{\hat {\tilde \Pi}}^{rs}
(\tau ,\vec \sigma ),{}^3{\hat {\tilde \Pi}}^{uv}(\tau ,{\vec \sigma}_1)\rbrace
{}^{*}=0$ if these would be the Dirac brackets only with respect to the second
class pairs ${\tilde \pi}^{\vec \alpha}_{(a)}\approx 0$, $\alpha_{(a)}\approx 0$
, in accord with Section V of I].

Some algebraic calculations for $\xi^r(\tau ,\vec \sigma )\rightarrow
\sigma^r$ give [to get the expressions with $n_r\not= 0$, replace $N_{(as)r}$
with $N_{(as)r}+n_r$]

\begin{eqnarray}
&&{}^3e_{(a)r} \mapsto {}^3{\hat e}_{(a)r}=\delta_{(a)r} Q_r,\nonumber \\
&&{}^3e^r_{(a)} \mapsto {}^3{\hat e}^r_{(a)} ={{\delta^r_{(a)}}\over {Q_r}},
\nonumber \\
&&{}^3e=det\, |{}^3e_{(a)r}| = \sqrt{\gamma} 
\mapsto {}^3{\hat e}=\sqrt{\hat \gamma}=Q_1Q_2Q_3,\nonumber \\
&&{}^3g_{rs}={}^3e_{(a)r}\, {}^3e_{(a)s} \mapsto {}^3{\hat g}_{rs}=\delta_{rs} 
Q^2_r,\nonumber \\
&&{}^3g^{rs}={}^3e^r_{(a)}\, {}^3e^s_{(a)} \mapsto {}^3{\hat g}^{rs}=
\delta^{rs}/ Q^2_r,\nonumber \\
&&ds^2 \mapsto d{\hat s}^2=\epsilon \Big[ (N_{(as)}+n)^2-\sum_u{{N^2_{(as)u}}
\over {Q^2_u}}\Big] (d\tau )^2-\nonumber \\
&-&2\epsilon N_{(as)r} d\tau d\sigma^r -\epsilon \sum_uQ^2_u (d\sigma^u)^2=
\epsilon \Big[ (N_{(as)}+n)^2(d\tau )^2-\nonumber \\
&&-\delta_{uv}(Q_ud\sigma^u+{{N_{(as)u}}\over {Q_u}}d\tau )(Q_vd\sigma^v+
{{N_{(as)v}}\over {Q_v}}d\tau )\Big] ,\nonumber \\
&&{}\nonumber \\
&&{}^3\Gamma^r_{uv} \mapsto {}^3{\hat \Gamma}^r_{uv}=\delta^r_u {{\partial_v
Q_u}\over {Q_u}}+\delta^r_v {{\partial_uQ_v}\over {Q_v}}-\delta_{uv}\delta^r_s
{{Q_u\partial_sQ_u}\over {Q^2_s}}=\nonumber \\
&&=-\delta_{uv}\delta^r_s({{Q_u}\over {Q_s}})^2 \partial_sln\, Q_u+\delta^r_u 
\partial_vln\, Q_u+\delta^r_v \partial_uln\, Q_v,\nonumber \\
&&{}\nonumber \\
&&{}^3\omega_{r(a)} \mapsto {}^3{\hat \omega}_{r(a)}=\epsilon_{(a)(b)(c)}
\delta_{(b)r}\delta_{(c)u}{{\partial_uQ_r}\over {Q_u}}=\epsilon_{(a)(b)(c)}
\delta_{(b)r}\delta_{(c)u}{{Q_r}\over {Q_u}} \partial_uln\, Q_r.
\label{VII2}
\end{eqnarray}

\noindent
The expressions for ${}^3\Omega_{rs(a)}$, ${}^3R_{rsuv}$, ${}^3R_{rs}$,
${}^3R$, will be given in Appendix D after a final canonical transformation.

Moreover, from Eq.(\ref{VI20}) we have

\begin{eqnarray}
{}^3{\tilde \pi}^r_{(a)}(\tau ,\vec \sigma ) &\mapsto& {}^3{\hat {\tilde \pi}}
^r_{(a)}(\tau ,\vec \sigma )=\int d^3\sigma_1\, {\cal K}^r_{(a)s}(\vec \sigma ,
{\vec \sigma }_1;\tau |Q]\, {\tilde \Pi}^s(\tau ,{\vec \sigma }_1),
\nonumber \\
&&{}\nonumber \\
{\cal K}^r_{(a)s}(\vec \sigma ,{\vec \sigma }_1;\tau |Q]&=&\delta^r_{(a)}\delta
^r_s\delta^3(\vec \sigma ,{\vec \sigma }_1)+{\cal T}^r_{(a)s}(\vec \sigma ,
{\vec \sigma}_1;\tau |Q],\nonumber \\
&&{}\nonumber \\
{\cal T}^r_{(a)s}(\vec \sigma ,{\vec \sigma}_1;\tau |Q]&=&{{\delta^r_{(b)}}\over
{2Q_r(\tau ,\vec \sigma )}}
\Big[ \sum_{w\not= s} {{\delta_{(k)w}}\over {Q_w(\tau ,
{\vec \sigma}_1)}} {{\partial Q_s(\tau ,{\vec \sigma}_1)}\over {\partial
\sigma_1^w}} T_{(b)(a)(k)}(\vec \sigma ,{\vec \sigma}_1;\tau )+\nonumber \\
&+& \delta_{(k)s} {{\partial}\over {\partial \sigma_1^s}} T_{(b)(a)(k)}(\vec 
\sigma ,{\vec \sigma}_1;\tau ) \Big] ,\nonumber \\
&&{}\nonumber \\
\delta^r_{(b)} T_{(b)(a)(k)}(\vec \sigma ,{\vec \sigma}_1;\tau )&=& Q_r(\tau ,
\vec \sigma ) d^r_{\gamma_{PP_1}}(\vec \sigma ,{\vec \sigma}_1) \Big( P_{\gamma
_{PP_1}}\, e^{\int^{\vec \sigma}_{{\vec \sigma}_1} d\sigma_2^w\, {}^3{\hat 
\omega}_{w(c)} {\hat R}^{(c)} }\, \Big)_{(a)(k)} +\nonumber \\
&+&\sum_u \delta_{(a)u}Q_u(\tau ,\vec \sigma ) d^u_{\gamma_{PP_1}}(\vec \sigma ,
{\vec \sigma}_1) \delta^r_{(b)} \Big( P_{\gamma_{PP_1}}\,
e^{\int^{\vec \sigma}_{{\vec \sigma}_1} d\sigma_2^w\, {}^3{\hat 
\omega}_{w(c)} {\hat R}^{(c)} }\, \Big)_{(b)(k)}.\nonumber \\
&&{}
\label{VII3}
\end{eqnarray}

\noindent so that we have 

\begin{eqnarray}
&&{}^3K_{rs} \mapsto {}^3{\hat K}_{rs}={{\epsilon}\over {4k}} 
{{Q_rQ_s}\over {Q_1Q_2Q_3}} \sum_u
\Big( \delta_{ru}\delta_{(a)s}+\delta_{su}\delta_{(a)r}-\delta_{rs}\delta
_{(a)u}\Big) Q_u\, {}^3{\hat {\tilde \pi}}^u_{(a)},\nonumber \\
&&{}^3K \mapsto {}^3\hat K={}^3{\hat g}^{rs}\, {}^3{\hat K}_{rs}
=-{{\epsilon}\over {4k}}\sum_u \delta_{(a)u}{{Q_u}\over {Q_1Q_2Q_3}}\,
{}^3{\hat {\tilde \pi}}^u_{(a)},\nonumber \\
&&{}^3{\tilde \Pi}^{rs} \mapsto {}^3{\hat {\tilde \Pi}}^{rs}=
{{1}\over 4} \Big[ {{\delta^r
_{(a)}}\over {Q_r}}\, {}^3{\hat {\tilde \pi}}^s_{(a)}+{{\delta^s_{(a)}}\over 
{Q_s}}\, {}^3{\hat {\tilde \pi}}^r_{(a)}\Big] \equiv {}^3{\hat {\tilde \Pi}}
^{rs}_t={}^3{\hat {\tilde \Pi}}^{rs}_{TT}+{}^3{\hat {\tilde \Pi}}^{rs}_{Tr,t}
,   \nonumber \\
&&{}^3{\tilde \Pi}={}^3g_{rs}\, {}^3{\tilde \Pi}^{rs} \mapsto {}^3{\hat {\tilde
\Pi}}=-2\epsilon k\, Q_1Q_2Q_3\, {}^3\hat K={{1}\over 2}\, \sum_r Q_r\delta
_{(a)r}\, {}^3{\hat {\tilde \pi}}^r_{(a)},\nonumber \\
&&{}^3{\tilde \Pi}_{(a)(b)}={}^3e_{(a)r}\, {}^3e_{(b)s}\, {}^3{\tilde \Pi}^{rs}
\mapsto {}^3{\hat {\tilde \Pi}}_{(a)(b)}={{1}\over 4} \sum_rQ_r 
\Big[ \delta_{(a)r}
\, {}^3{\hat {\tilde \pi}}^r_{(b)}+\delta_{(b)r}\, {}^3{\hat {\tilde \pi}}^r
_{(a)}\Big] .
\label{VII4}
\end{eqnarray}

\noindent The determination of the gravitomagnetic potential $W^r(\tau ,\vec 
\sigma )$, see Appendix C, by solving the elliptic equations associated with the
supermomentum constraints in the conformal approach to metric gravity, has been 
replaced here by the determination of the kernel ${\cal K}^r_{(a)s}(\vec \sigma 
,{\vec \sigma}^{'};\tau |Q]$ connecting the old momenta ${}^3{\hat {\tilde 
\pi}}^r_{(a)}(\tau ,\vec \sigma )$ to the new canonical ones ${\tilde \Pi}
^r(\tau ,\vec \sigma )$.

The reduced superhamiltonian constraint becomes [in the last line Eq(\ref{I5})
is used; $k=c^3/16\pi G$ with G the Newton constant]

\begin{eqnarray}
&&{\hat {\cal H}}(\tau ,\vec \sigma )=\epsilon
\Big[ k{}^3e\, \epsilon_{(a)(b)(c)}\, {}^3e^r
_{(a)}\, {}^3e^s_{(b)}\, {}^3\Omega_{rs(c)}-\nonumber \\
&-&{1\over {8k\, {}^3e}}\, {}^3G_{o(a)(b)(c)(d)}\, {}^3e_{(a)r}\, {}^3{\tilde
\pi}^r_{(b)}\, {}^3e_{(c)s}\, {}^3{\tilde \pi}^s_{(d)}\Big] (\tau ,\vec \sigma )
\nonumber \\
&&{}\nonumber \\
&&\mapsto {\hat {\cal H}}_R(\tau ,\vec \sigma )=\epsilon
\Big[ kQ_1Q_2Q_3 \epsilon_{(a)(b)(c)}
\sum_{r,s}{{\delta_{(a)r}\delta_{(b)s}}\over {Q_rQ_s}}\, {}^3{\hat \Omega}
_{rs(c)}-\nonumber \\
&-&{1\over {8k\, Q_1Q_2Q_3}}\sum_{r,s} \Big( \delta_{rs}\delta_{(a)(b)}+\delta
_{(a)r}\delta_{(b)s}-\delta_{(a)s}\delta_{(b)r}\Big)\, Q_r\, {}^3{\hat {\tilde 
\pi}}^r_{(b)}\, Q_s\, {}^3{\hat {\tilde \pi}}^s_{(a)}\Big] (\tau ,\vec \sigma )=
\nonumber \\
&&{}\nonumber \\
&=&-\sum_{r,s} \epsilon \Big( {{Q_rQ_s}\over {8k\, Q_1Q_2Q_3}}\Big) (\tau ,\vec 
\sigma )\, \int d^3\sigma_1
d^3\sigma_2\, {\tilde \Pi}^u(\tau ,{\vec \sigma }_1) {\cal K}^r_{(b)u}
(\vec \sigma ,{\vec \sigma }_1;\tau |Q]\cdot \nonumber \\
&\cdot& \Big( \delta_{rs}\delta_{(a)(b)}+\delta_{(a)r}\delta_{(b)s}-\delta
_{(a)s}\delta_{(b)r}\Big)\, {\cal K}^s_{(a)v}(\vec \sigma ,{\vec \sigma }_2;
\tau |Q]\,{\tilde \Pi}^v(\tau ,{\vec \sigma }_2)+\nonumber \\
&+&\epsilon \Big( kQ_1Q_2Q_3\, \epsilon_{(a)(b)(c)}\sum_{r,s}
{{\delta_{(a)r}\delta_{(b)s}}\over
{Q_rQ_s}}\, {}^3{\hat \Omega}_{rs(c)}\Big) (\tau ,\vec \sigma )=\nonumber \\
&&{}\nonumber \\
&=&-\epsilon \Big({1\over {8k\, Q_1Q_2Q_3}}\Big) (\tau ,\vec \sigma )\, \Big( 
\,\, \Big[ 2\sum_u (Q_u{\tilde \Pi}
^u)^2-(\sum_uQ_u{\tilde \Pi}^u)^2\Big] (\tau ,\vec \sigma )+\nonumber \\
&+&2\sum_{r,s}
Q_r(\tau ,\vec \sigma )Q_s(\tau ,\vec \sigma ){\tilde \Pi}^r(\tau ,\vec 
\sigma ) (2\delta_{rs}-1)\delta_{(a)s} \cdot \nonumber \\
&\cdot & \int d^3\sigma_1\, {\cal T}^s_{(a)v}
(\vec \sigma ,{\vec \sigma }_1;\tau |Q] {\tilde \Pi}^v(\tau ,{\vec \sigma }_1)+
\nonumber \\
&+&\sum_{r,s}
Q_r(\tau ,\vec \sigma )Q_s(\tau ,\vec \sigma ) \int d^3\sigma_1d^3\sigma_2\, 
{\tilde \Pi}^u(\tau ,{\vec \sigma }_1) {\cal T}^r_{(b)u}(\vec \sigma ,{\vec 
\sigma}_1;\tau |Q]\cdot \nonumber \\
&\cdot& \Big( \delta_{rs}\delta_{(a)(b)}+\delta_{(a)r}\delta_{(b)s}-\delta
_{(a)s}\delta_{(b)r}\Big) {\cal T}^s_{(a)v}(\vec \sigma ,{\vec \sigma }_2;\tau 
|Q]\, {\tilde \Pi}^v(\tau ,{\vec \sigma }_2)\,\, \Big) +\nonumber \\
&+&\epsilon \Big[ kQ_1Q_2Q_3 \epsilon_{(a)(b)(c)}\sum_{r,s}
{{\delta_{(a)r}\delta_{(b)s}}\over {Q_rQ_s}}\, {}^3{\hat \Omega}_{rs(c)}[Q]\, 
\Big] (\tau ,\vec \sigma )\approx 0,\nonumber \\
&&{}\nonumber \\
&&\lbrace {\hat {\cal H}}_R(\tau ,\vec \sigma ),{\hat {\cal H}}_R(\tau ,
{\vec \sigma }^{'})\rbrace {}^{*} \equiv \lbrace {\hat {\cal H}}(\tau ,\vec 
\sigma ),{\hat {\cal H}}(\tau ,{\vec \sigma}^{'})\rbrace {}^{*}\equiv 
\nonumber \\
&&\equiv -{{\partial}\over {\partial \sigma^s}} \Big[\, {{\partial \sigma^s(\vec
\xi)}\over {\partial \xi^r}}{|}_{\vec \xi =\vec \xi (\tau ,\vec \sigma )}\,
\lbrace \xi^r(\tau ,\vec \sigma ),{\hat {\cal H}}(\tau ,{\vec \sigma}^{'})
\rbrace \, {\hat {\cal H}}_R(\tau ,\vec \sigma ) \Big] +\nonumber \\
&&+{{\partial}\over {\partial \sigma^{{'}s}}} 
\Big[\, {{\partial \sigma^{{'}s}(\vec
\xi)}\over {\partial \xi^r}}{|}_{\vec \xi =\vec \xi (\tau ,{\vec \sigma}^{'})}\,
\lbrace \xi^r(\tau ,{\vec \sigma}^{'}),{\hat {\cal H}}(\tau ,\vec \sigma )
\rbrace \, {\hat {\cal H}}_R(\tau ,{\vec \sigma}^{'}) \Big] \approx 0.
\label{VII5}
\end{eqnarray}

The constraint is no more an algebraic relation among the final variables, but
rather an integrodifferential equation for one of them.

Let us now consider a new canonical transformation from the basis
$Q_u(\tau ,\vec \sigma )$, ${\tilde \Pi}^u(\tau ,\vec \sigma )$
to a new basis $q_u(\tau ,\vec \sigma )$, $\rho_u(\tau ,\vec \sigma )$
defined in the following way

\begin{eqnarray}
q_u(\tau ,\vec \sigma )&=&ln\, Q_u(\tau ,\vec \sigma ),\nonumber \\
\rho_u(\tau ,\vec \sigma )&=&Q_u(\tau ,\vec \sigma )\, {\tilde \Pi}^u(\tau ,
\vec \sigma ),\nonumber \\
&&{}\nonumber \\
&&\lbrace q_u(\tau ,\vec \sigma ),\rho_v(\tau ,{\vec \sigma }^{'})\rbrace =
\delta_{uv}\delta^3(\vec \sigma ,{\vec \sigma }^{'}),\nonumber \\
&&Q_u(\tau ,\vec \sigma )=e^{q_u(\tau ,\vec \sigma )},\quad\quad
{\tilde \Pi}^u(\tau ,\vec \sigma )=e^{-q_u(\tau ,\vec \sigma )}\, 
\rho_u(\tau ,\vec \sigma ).
\label{VII6}
\end{eqnarray}

It is convenient to make one more canonical transformation, like for the
determination of the center of mass of a particle system\cite{lus1}, to the
following new set

\begin{eqnarray}
&&q(\tau ,\vec \sigma )={1\over 3}\sum_u q_u(\tau ,\vec \sigma )=
{1\over 3} \sum_u ln\, Q_u(\tau ,\vec \sigma ) ,\nonumber \\
&&r_{\bar a}(\tau ,\vec \sigma )=\sqrt{3} \sum_u \gamma_{\bar au} 
q_u(\tau ,\vec \sigma )=\sqrt{3}\sum_u\gamma_{\bar au} ln\, Q_u(\tau ,\vec 
\sigma ),\quad\quad \bar a=1,2, \nonumber \\
&&\rho (\tau ,\vec \sigma )=\sum_u \rho_u(\tau ,\vec \sigma )=\sum_u
[Q_u{\tilde \Pi}^u](\tau ,\vec \sigma ),\nonumber \\
&&\pi_{\bar a}(\tau ,\vec \sigma )={1\over {\sqrt{3}}} \sum_u \gamma_{\bar au}
\rho_u(\tau ,\vec \sigma )={1\over {\sqrt{3}}}\sum_u\gamma_{\bar au}
[Q_u{\tilde \Pi}^u](\tau ,\vec \sigma ),\quad\quad \bar a=1,2,\nonumber \\
&&{}\nonumber \\
&&\lbrace q(\tau ,\vec \sigma ),\rho (\tau ,{\vec \sigma }^{'})\rbrace =
\delta^3(\vec \sigma ,{\vec \sigma }^{'}),\quad\quad
\lbrace r_{\bar a}(\tau ,\vec \sigma ),\pi_{\bar b}(\tau ,{\vec \sigma }^{'})
\rbrace =\delta_{\bar a\bar b}\delta^3(\vec \sigma ,{\vec \sigma }^{'}),
\nonumber \\
&&{}\nonumber \\
&&q_u(\tau ,\vec \sigma )=q(\tau ,\vec \sigma )+{1\over {\sqrt{3}}} \sum_{\bar 
a}\gamma_{\bar au} r_{\bar a}(\tau ,\vec \sigma ),\quad\quad
Q_u(\tau ,\vec \sigma )=e^{q_u(\tau ,\vec \sigma )},\nonumber \\
&&\rho_u(\tau ,\vec \sigma )={1\over 3} \rho(\tau ,\vec \sigma )+\sqrt{3}
\sum_{\bar a} \gamma_{\bar au} \pi_{\bar a}(\tau ,\vec \sigma ),\quad\quad
{\tilde \Pi}^u(\tau ,\vec \sigma )=[e^{-q_u} \rho_u](\tau ,\vec \sigma ),
\nonumber \\
&&{}\nonumber \\
&&{}^3{\hat {\tilde \pi}}^r_{(a)}(\tau ,\vec \sigma )=
\sum_s \int d^3\sigma_1 {\cal K}
^r_{(a)s}(\vec \sigma ,{\vec \sigma}_1,\tau |q,r_{\bar a}]\cdot \nonumber \\
&&\cdot (e^{-q-{1\over {\sqrt{3}}}\sum_{\bar a}\gamma_{\bar as}r_{\bar a}})
(\tau ,{\vec \sigma}_1) \Big[ {1\over 3}\rho +\sqrt{3} \sum_{\bar b}
\gamma_{\bar bs} \pi_{\bar b}\Big] (\tau ,{\vec \sigma}_1),\nonumber \\
&&{}\nonumber \\
&&{\cal K}^r_{(a)s}(\vec \sigma ,{\vec \sigma}_1,\tau |q,r_{\bar a}]=
\delta^r_{(a)}\delta^r_s\delta^3(\vec \sigma ,{\vec \sigma}_1)+{\cal T}^r
_{(a)s}(\vec \sigma ,{\vec \sigma}_1,\tau |q,r_{\bar a}],\nonumber \\
&&{}\nonumber \\
&&{\cal T}^r_{(a)s}(\vec \sigma ,{\vec \sigma}_1;\tau |q,r_{\bar a}]
={1\over 2}e^{-{1\over {\sqrt{3}}}\sum_{\bar c}\gamma_{\bar cr}r_{\bar c}
(\tau ,\vec \sigma )}\Big[ \sum_{w\not= s} \delta_{(k)w} e^{{1\over {\sqrt{3}}}
\sum_{\bar c}(\gamma_{\bar cw}-\gamma_{\bar cs})r_{\bar c}(\tau ,{\vec
\sigma}_1)}\cdot \nonumber \\
&&\cdot \Big(\, {{\partial q(\tau ,{\vec \sigma}_1)}\over {\partial \sigma_1
^w}}+{1\over {\sqrt{3}}} \sum_{\bar c}\gamma_{\bar cs}{{\partial r_{\bar 
c}(\tau ,{\vec \sigma}_1)}\over {\partial \sigma_1^w}}\Big) 
e^{-q(\tau ,\vec \sigma )}
\delta^r_{(b)} T_{(b)(a)(k)}(\vec \sigma ,{\vec \sigma}_1;\tau )+\nonumber \\
&&+\delta_{(k)s} {{\partial}\over {\partial \sigma_1^s}} e^{-q(\tau, \vec
\sigma )} \delta^r_{(b)} T_{(b)(a)(k)}(\vec \sigma ,{\vec \sigma}_1;\tau ) 
\Big] ,\nonumber \\
&&{}\nonumber \\
&&e^{-q(\tau ,\vec \sigma )} \delta^r_{(b)} T_{(b)(a)(k)}(\vec \sigma ,{\vec
\sigma}_1;\tau ) =\nonumber \\
&&=e^{ {1\over {\sqrt{3}}}\sum_{\bar c}\gamma_{\bar cr}r_{\bar 
c}(\tau ,\vec \sigma )} d^r_{\gamma_{PP_1}} \Big( P_{\gamma_{PP_1}}\, e^{\int
^{\vec \sigma}_{{\vec \sigma}_1}d\sigma_2^w\, {}^3{\hat \omega}_{w(c)}(\tau ,
{\vec \sigma}_2){\hat R}^{(c)} }\, \Big)_{(a)(k)}+\nonumber \\
&&+\sum_u\delta_{(a)u} e^{ {1\over {\sqrt{3}}} \sum_{\bar c}\gamma_{\bar cu}
r_{\bar c}(\tau ,\vec \sigma )} d^u_{\gamma_{PP_1}}(\vec \sigma ,{\vec 
\sigma}_1) \delta^r_{(b)}\Big( P_{\gamma_{PP_1}}\, e^{\int
^{\vec \sigma}_{{\vec \sigma}_1}d\sigma_2^w\, {}^3{\hat \omega}_{w(c)}(\tau ,
{\vec \sigma}_2){\hat R}^{(c)} }\, \Big)_{(b)(k)},\nonumber \\
&&{}\nonumber \\
&&{}^3{\hat K}_{rs}={{\epsilon}\over {4k}} e^{ {1\over {\sqrt{3}}}\sum_{\bar c}
(\gamma_{\bar cr}+\gamma_{\bar cs})r_{\bar c} } \sum_u
\Big( \delta_{ru}\delta_{(a)s}
+\delta_{su}\delta_{(a)r}-\delta_{rs}\delta_{(a)u}\Big)
e^{ {1\over {\sqrt{3}}} \sum_{\bar c}\gamma_{\bar cu}r_{\bar c}}\, {}^3{\hat 
{\tilde \pi}}^u_{(a)},\nonumber \\
&&{}^3{\hat K}=-{{\epsilon}\over {4k}} e^{-2q} \sum_u \delta_{(a)u} e^{ {1\over 
{\sqrt{3}}}\sum_{\bar c}\gamma_{\bar cu}r_{\bar c}}\, {}^3{\hat {\tilde \pi}}
^u_{(a)},\nonumber \\
&&{}\nonumber \\
&&{}^3{\hat {\tilde \Pi}}^{rs}={{1}\over 4}e^{-q} \Big[ e^{-{1\over 
{\sqrt{3}}}\sum_{\bar a}\gamma_{\bar ar}r_{\bar a}}\, \delta^r_{(a)}\, 
{}^3{\hat {\tilde \pi}}^s_{(a)}+e^{-{1\over {\sqrt{3}}}\sum_{\bar a}\gamma
_{\bar as}r_{\bar a}}\, \delta^s_{(a)}\,
{}^3{\hat {\tilde \pi}}^r_{(a)}\Big] ,\nonumber \\
&&{}^3{\hat {\tilde \Pi}}=-2\epsilon k e^{3q}\, {}^3\hat K={1\over 2}
\, \sum_r\, e^{q+{1\over {\sqrt{3}}}\sum_{\bar a}\gamma_{\bar ar}r_{\bar a}}\, 
\delta^r_{(a)}\, {}^3{\hat {\tilde \pi}}^r_{(a)},\nonumber \\
&&{}
\label{VII8}
\end{eqnarray}

\noindent where $\gamma_{\bar au}$ are numerical constants satisfying\cite{lus1}

\begin{equation}
\sum_u \gamma_{\bar au}=0,\quad\quad \sum_u\, \gamma_{\bar au}\gamma_{\bar bu}
=\delta_{\bar a\bar b},\quad\quad \sum_{\bar a}\, \gamma_{\bar au}\gamma
_{\bar av}=\delta_{uv}-{1\over 3}.
\label{VII9}
\end{equation}

In terms of these variables we have (we reintroduce $n_r\not= 0$ to take into 
account more general situations)

\begin{eqnarray}
{}^3{\hat g}_{rs}&=&e^{2q}\, \left( \begin{array}{ccc} e^{ {2\over {\sqrt{3}}}
\sum_{\bar a}\gamma_{\bar a1}r_{\bar a}}&0&0\\ 0& e^{ {2\over {\sqrt{3}}}
\sum_{\bar a}\gamma_{\bar a2}r_{\bar a}}&0\\ 0&0& e^{ {2\over {\sqrt{3}}}
\sum_{\bar a}\gamma_{\bar a3}r_{\bar a}} \end{array} \right) =  
e^{2q}\,\, {}^3{\hat g}^{diag}_{rs},
\nonumber \\
&&{}\nonumber \\
\hat \gamma &=&{}^3\hat g= {}^3{\hat e}^2=e^{6q},\quad\quad
det\, |{\hat g}^{diag}_{rs}| =1,\nonumber \\
&&{}\nonumber \\
d{\hat s}^2&=&\epsilon \Big( [N_{(as)}+n]^2-e^{-2q} \sum_u e^{ -{2\over 
{\sqrt{3}}}\sum_{\bar a} \gamma_{\bar au} r_{\bar a}} [N_{(as) u}+n_u]^2\Big)
(d\tau )^2 -\nonumber \\
&&-2\epsilon [N_{(as)r}+n_r]d\tau d\sigma^r-\epsilon e^{2q} \sum_u e^{ {2\over 
{\sqrt{3}}}\sum_{\bar a} \gamma_{\bar au} r_{\bar a}} (d\sigma^u)^2=\nonumber \\
&&=\epsilon \Big( [N_{(as)}+n]^2(d\tau )^2-\nonumber \\
&&-\delta_{uv}[e^qe^{ {2\over 
{\sqrt{3}}}\sum_{\bar a} \gamma_{\bar au} r_{\bar a}} d\sigma^u+e^{-q}e^{-
{2\over {\sqrt{3}}}\sum_{\bar a} \gamma_{\bar au} r_{\bar a}}(N_{(as)u}+n_u)
d\tau ]\nonumber \\
&&[e^qe^{ {2\over 
{\sqrt{3}}}\sum_{\bar a} \gamma_{\bar av} r_{\bar a}} d\sigma^v+e^{-q}e^{-
{2\over {\sqrt{3}}}\sum_{\bar a} \gamma_{\bar av} r_{\bar a}}(N_{(as)v}+n_v)
d\tau ],\nonumber \\
&&{}\nonumber \\
q&=& {1\over 6}\, ln\, {}^3\hat g,\nonumber \\
r_{\bar a}&=&{{\sqrt{3}}\over 2} \sum_r \gamma_{\bar ar}\, ln\, {{{}^3{\hat g}
_{rr}}\over {{}^3\hat g}}.
\label{VII10}
\end{eqnarray}

\noindent The momenta ${}^3{\hat {\tilde \pi}}^r_{(a)}$ and ${}^3{\hat {\tilde
\Pi}}^{rs}$ and the mean extrinsic curvature ${}^3\hat K$ are linear functions 
of the new momenta $\rho$ and $r_{\bar c}$, but with a coordinate-dependent
integral kernel. The variables $\rho$ and $r_{\bar a}$ replace ${}^3\hat K$ and 
${}^3{\hat K}^{rs}_{TT}\,\,$ [or $\, {}^3{\hat {\tilde \Pi}}$ and ${}^3{\hat
{\tilde \Pi}}^{rs}_{TT}\,\,$] of the conformal approach respectively (see
Appendix C) after the solution of the supermomentum constraints (i.e. after the
determination of the gravitomagnetic potential) in the 3-orthogonal gauges. It
would be important to find the expression of $\rho$ and $r_{\bar a}$ in terms
of ${}^3{\hat g}_{rs}$ and ${}^3{\hat K}_{rs}\,\,$ [or ${}^3{\hat {\tilde
\Pi}}^{rs}$].

In terms of the variables $q, r_{\bar a}$, we have the following results

\begin{eqnarray}
&&{}^3{\hat e}_{(a)r}=\delta_{(a)r} e^{q_r}=\delta_{(a)r} e^{q+
{1\over {\sqrt{3}}}\sum_{\bar a}\gamma_{\bar ar}r_{\bar a}} \nonumber \\
&&{\rightarrow}_{r_{\bar a}\rightarrow 0}\, \delta_{(a)r} e^q {\rightarrow}
_{q\, \rightarrow 0}\, \delta_{(a)r},\quad
{\rightarrow}_{q\, \rightarrow 0}\, \delta_{(a)r} e^{{1\over {\sqrt{3}}}\sum
_{\bar a}\gamma_{\bar ar}r_{\bar a}},\nonumber \\
&&{}^3{\hat e}^r_{(a)}=\delta_{(a)}^r e^{-q_r}=\delta_{(a)}^r e^{-q-
{1\over {\sqrt{3}}}\sum_{\bar a}\gamma_{\bar ar}r_{\bar a}} \nonumber \\
&&{\rightarrow}_{r_{\bar a}\rightarrow 0}\, \delta_{(a)}^r e^{-q}
{\rightarrow}_{q\, \rightarrow 0}\, \delta^r_{(a)},\quad
{\rightarrow}_{q\, \rightarrow 0}\, \delta^r_{(a)} e^{-{1\over {\sqrt{3}}}\sum
_{\bar a}\gamma_{\bar ar}r_{\bar a}},\nonumber \\
&&{}\nonumber \\
&&{}^3{\hat g}_{rs}=\delta_{rs} e^{2q_r}=\delta_{rs} e^{2q+{2\over {\sqrt{3}}}
\sum_{\bar a}\gamma_{\bar ar}r_{\bar a}}\nonumber \\ 
&&{\rightarrow}_{r_{\bar a}\rightarrow 0}\, \delta_{rs}e^{2q}\, {\rightarrow}
_{q\, \rightarrow 0}\, \delta_{rs},\quad {\rightarrow}_{q\, \rightarrow 0}\,
\delta_{rs} e^{{2\over {\sqrt{3}}}\sum_{\bar a}\gamma_{\bar ar}r_{\bar a}}
,\nonumber \\
&&{}^3{\hat g}^{rs}=\delta^{rs} e^{-2q_r}=\delta^{rs} e^{-2q-{2\over {\sqrt{3}}}
\sum_{\bar a}\gamma_{\bar ar}r_{\bar a}}\nonumber \\ 
&&{\rightarrow}_{r_{\bar a}\rightarrow 0}\, \delta^{rs}e^{-2q}\, {\rightarrow}
_{q\, \rightarrow 0}\, \delta^{rs},\quad {\rightarrow}_{q\, \rightarrow 0}\, 
\delta^{rs} e^{-{2\over {\sqrt{3}}}\sum_{\bar a}\gamma_{\bar ar}r_{\bar a}}
,\nonumber \\
&&{}\nonumber \\
&&{}^3\hat e=\sqrt{\hat \gamma}=e^{\sum_rq_r}=e^{3q}\, {\rightarrow}
_{q\, \rightarrow 0}\, 1,\nonumber \\
&&{}\nonumber \\
&&{}^3{\hat \Gamma}^r_{uv}=-\delta_{uv} 
\sum_s \delta^r_s e^{2(q_u-q_s)} \partial_sq_u
+\delta^r_u \partial_vq_u +\delta^r_v \partial_uq_v=\nonumber \\
&&=-\delta_{uv}\sum_s\delta^r_s e^{{2\over {\sqrt{3}}}\sum_{\bar a}(\gamma
_{\bar au}-\gamma_{\bar as})r_{\bar a}} \Big[ \partial_sq+{1\over {\sqrt{3}}}
\sum_{\bar b}\gamma_{\bar bu}\partial_sr_{\bar b}\Big] +\nonumber \\
&&+\delta^r_u\Big[ \partial_vq+{1\over {\sqrt{3}}}\sum_{\bar a}\gamma_{\bar au}
\partial_vr_{\bar a}\Big] +\delta^r_v\Big[ \partial_uq+{1\over {\sqrt{3}}}
\sum_{\bar a}\gamma_{\bar av}\partial_ur_{\bar a}\Big]\nonumber \\
&&{\rightarrow}_{r_{\bar a}\rightarrow 0}\, -\delta_{uv}\sum_s\delta^r_s 
\partial_sq+\delta^r_u \partial_vq +\delta^r_v \partial_uq\, {\rightarrow}_{q\, 
\rightarrow 0}\, 0,\nonumber \\
&&{\rightarrow}_{q\, \rightarrow 0}\, {1\over {\sqrt{3}}}
\Big( -\delta_{uv}\sum_s
\delta^r_s e^{{2\over {\sqrt{3}}}\sum_{\bar a}(\gamma_{\bar au}-\gamma
_{\bar as})r_{\bar a}}\sum_{\bar b}\gamma_{\bar bu}\partial_sr_{\bar b}+\sum
_{\bar a}\Big[ \delta^r_u\gamma_{\bar au}\partial_vr_{\bar a}+\delta^r_v\gamma
_{\bar av}\partial_ur_{\bar a}\Big] \Big), \nonumber \\
&&\sum_u\, {}^3{\hat \Gamma}^u_{uv} = \partial_v \sum_u q_u = 3\partial_v q,
\nonumber \\
&&{}\nonumber \\
&&{}^3{\hat \omega}_{r(a)}=\epsilon_{(a)(b)(c)}\delta_{(b)r}\delta_{(c)u}
e^{q_r-q_u} \partial_uq_r=\nonumber \\
&&=\epsilon_{(a)(b)(c)}\delta_{(b)r}\delta_{(c)u} e^{{1\over {\sqrt{3}}}\sum
_{\bar a}(\gamma_{\bar ar}-\gamma_{\bar au})r_{\bar a}}\Big[ \partial_uq+{1
\over {\sqrt{3}}}\sum_{\bar b}\gamma_{\bar br} \partial_ur_{\bar b}\Big]
\nonumber \\
&&{\rightarrow}_{r_{\bar a}\rightarrow 0}\, \epsilon_{(a)(b)(c)}\delta_{(b)r}
\delta_{(c)u} \partial_uq\, {\rightarrow}_{q\, \rightarrow 0}\, 0,\nonumber \\
&&{\rightarrow}_{q\, \rightarrow 0}\, {1\over {\sqrt{3}}}\epsilon_{(a)(b)(c)}
\sum_u\, \delta_{(b)r}\delta_{(c)u} e^{{1\over {\sqrt{3}}}\sum_{\bar a}
(\gamma_{\bar ar}-\gamma_{\bar au})r_{\bar a}}\sum_{\bar b}\gamma_{\bar br}
\partial_ur_{\bar b}.
\label{VIIa}
\end{eqnarray}

See Appendix D for the expression of other 3-tensors and Appendix E for the
corresponding expression of 4-tensors.

Since the proper gauge transformations go to the identity at spatial infinity,
Eqs.(\ref{VII6}), (\ref{VII8}), (\ref{VII3}), (\ref{VII2}) and
(\ref{I4}) imply the following boundary conditions

\begin{eqnarray}
&&{}^3{\hat e}_{(a)r}(\tau ,\vec \sigma )=\delta_{(a)r}Q_r(\tau ,\vec \sigma )
\, {\rightarrow}_{r\, \rightarrow \infty}\, \delta_{(a)r}+{{{}^3{\hat w}
_{(as)(a)r}(\tau )}\over r}+O(r^{-2}),\nonumber \\
&&Q_r(\tau ,\vec \sigma )\, {\rightarrow}_{r\, \rightarrow \infty}\, 
1+{{Q_{(as)r}(\tau )}\over r}+O(r^{-2}),\quad
{}^3{\hat w}_{(as)(a)r}(\tau )=\delta_{(a)r}Q_{(as)r}(\tau ),\nonumber \\
&&q(\tau ,\vec \sigma )\, {\rightarrow}_{r\, \rightarrow \infty}\, {1\over 
{3r}}\sum_uQ_{(as)u}(\tau )+O(r^{-2}),\nonumber \\
&&r_{\bar a}(\tau ,\vec \sigma )\, {\rightarrow}_{r\, \rightarrow \infty}\, 
{{ \sqrt{3}}\over r}\sum_u\gamma_{\bar au}Q_{(as)u}(\tau )+O(r^{-2}),
\nonumber \\
&&{}\nonumber \\
&&{}^3{\hat {\tilde \pi}}^r_{(a)}(\tau ,\vec \sigma )\, {\rightarrow}
_{r\, \rightarrow \infty}\,
{{{}^3{\hat {\tilde p}}^r_{(as)(a)}(\tau )}\over {r^2}}+O(r^{-3}),\nonumber \\
&&{\tilde \Pi}^r(\tau ,\vec \sigma )\, {\rightarrow}_{r\, \rightarrow \infty}\,
{{{\tilde \Pi}^r_{(as)}(\tau )}\over {r^2}} +O(r^{-3}),\nonumber \\
&&\rho (\tau ,\vec \sigma )\, {\rightarrow}_{r\, \rightarrow \infty}\,
{1\over {r^2}}\sum_u\gamma_{\bar au}{\tilde \Pi}^u_{(as)}(\tau )+O(r^{-3}),
\nonumber \\
&&\pi_{\bar a}(\tau ,\vec \sigma )\, {\rightarrow}_{r\, \rightarrow \infty}\,
{1\over {\sqrt{3}\, r^2}}\sum_u\gamma_{\bar au}{\tilde \Pi}^u_{(as)}(\tau )+
O(r^{-3}),\nonumber \\
&&{}^3{\hat \omega}_{r(a)}(\tau ,\vec \sigma )\, {\rightarrow}_{r\, 
\rightarrow \infty}\,
{{{}^3{\hat \omega}_{(as)r(a)}(\tau )}\over {r^2}}+O(r^{-3}).
\label{VIIb}
\end{eqnarray}

\noindent By using Appendix D, we find that 
${}^3{\hat R}_{rsuv}(\tau ,\vec \sigma )$ and
${}^3{\hat \Omega}_{rs(a)}(\tau ,\vec \sigma )$ go as $O(r^{-3})$ for
$r\, \rightarrow \infty$.

The superhamiltonian constraint takes the following final reduced form

\begin{eqnarray}
{\hat {\cal H}}_R(\tau ,\vec \sigma )&=&-\epsilon
{1\over {8k}}e^{-\sum_uq_u(\tau ,\vec \sigma )} \Big( \,\, [2\sum_u\, \rho_u^2
-(\sum_u\rho_u)^2](\tau ,\vec \sigma )+\nonumber \\
&+&2\sum_{r,s}e^{q_s(\tau ,\vec \sigma )}\rho_r(\tau ,\vec \sigma )
(2\delta_{rs}-1)\delta_{(a)s} \int d^3\sigma_1\, {\cal T}^s_{(a)v}(\vec \sigma ,
{\vec \sigma }_1;\tau |e^{q_t}]\, e^{-q_v(\tau ,{\vec \sigma }_1)}
\rho_v(\tau ,{\vec \sigma }_1)+\nonumber \\
&+&\sum_{r,s}e^{q_r(\tau ,\vec \sigma )+q_s(\tau ,\vec \sigma )} \int d^3\sigma
_1d^3\sigma_2 \sum_{uv}
e^{-q_u(\tau ,{\vec \sigma }_1)-q_v(\tau ,{\vec \sigma }_2)}
\, \rho_u(\tau ,{\vec \sigma }_1) {\cal T}^r_{(b)u}(\vec \sigma ,{\vec \sigma }
_1;\tau |e^{q_t}]\nonumber \\
&\cdot& \Big( \delta_{rs}\delta_{(a)(b)}+\delta_{(a)r}\delta_{(b)s}-\delta
_{(a)s}\delta_{(b)r}\Big) {\cal T}^s_{(a)v}(\vec \sigma ,{\vec \sigma }_2;
\tau |e^{q_t}]\, \rho_v(\tau ,{\vec \sigma }_2)\,\, \Big) +\nonumber \\
&+&\epsilon k\, \Big[ e^{\sum_uq_u-q_r-q_s}\Big] (\tau ,\vec \sigma ) 
\epsilon_{(a)(b)(c)} \delta
_{(a)r}\delta_{(b)s}\, {}^3{\hat \Omega}_{rs(c)}[e^{q_t}](\tau ,\vec \sigma )
=\nonumber \\
&&{}\nonumber \\
&=&-\epsilon {{e^{-q(\tau ,\vec \sigma )}}\over {8k}}\Big[ (e^{-2q}[6 \sum_{\bar a}\pi^2
_{\bar a}-{1\over 3}\rho^2])(\tau ,\vec \sigma )+\nonumber \\
&+&2 (e^{-q}\sum_ue^{{1\over {\sqrt{3}}}\sum_{\bar a}\gamma_{\bar au}r_{\bar a}}
[2\sqrt{3}\sum_{\bar b}\gamma_{\bar bu}\pi_{\bar b}-{1\over 3}\rho ])(\tau
,\vec \sigma )\times \nonumber \\
&&\int d^3\sigma_1 \sum_r \delta^u_{(a)} {\cal T}^r_{(a)r}(\vec \sigma ,{\vec
\sigma}_1,\tau |q,r_{\bar a}] \Big( e^{-q-{1\over {\sqrt{3}}}\sum_{\bar a}
\gamma_{\bar ar}r_{\bar a}}[{{\rho}\over 3}+\sqrt{3}\sum_{\bar b}\gamma_{\bar 
br} \pi_{\bar b}]\Big) (\tau ,{\vec \sigma}_1)+\nonumber \\
&+&\int d^3\sigma_1d^3\sigma_2 \Big( \sum_u e^{{2\over {\sqrt{3}}}\sum_{\bar a}
\gamma_{\bar au}+r_{\bar a}(\tau ,\vec \sigma )} \times \nonumber \\
&&\sum_r{\cal T}^u_{(a)r}(\vec \sigma ,{\vec
\sigma}_1,\tau |q,r_{\bar a}] \Big( e^{-q-{1\over {\sqrt{3}}}\sum_{\bar a}
\gamma_{\bar ar}r_{\bar a}}[{{\rho}\over 3}+\sqrt{3}\sum_{\bar b}\gamma_{\bar 
br} \pi_{\bar b}]\Big) (\tau ,{\vec \sigma}_1)\times \nonumber \\
&&\sum_s {\cal T}^u_{(a)s}(\vec \sigma ,{\vec
\sigma}_2,\tau |q,r_{\bar a}] \Big( e^{-q-{1\over {\sqrt{3}}}\sum_{\bar a}
\gamma_{\bar as}r_{\bar a}}[{{\rho}\over 3}+\sqrt{3}\sum_{\bar c}\gamma_{\bar 
cs} \pi_{\bar c}]\Big) (\tau ,{\vec \sigma}_2)+\nonumber \\
&+&\sum_{uv} e^{{1\over {\sqrt{3}}}\sum_{\bar a}(\gamma_{\bar au}+\gamma_{\bar 
av})r_{\bar a}(\tau ,\vec \sigma )} (\delta^u_{(b)}\delta^v_{(a)}-\delta^u_{(a)}
\delta^v_{(b)})\times \nonumber \\
&&\sum_r {\cal T}^u_{(a)r}(\vec \sigma ,{\vec
\sigma}_1,\tau |q,r_{\bar a}] \Big( e^{-q-{1\over {\sqrt{3}}}\sum_{\bar a}
\gamma_{\bar ar}r_{\bar a}}[{{\rho}\over 3}+\sqrt{3}\sum_{\bar b}\gamma_{\bar 
br} \pi_{\bar b}]\Big) (\tau ,{\vec \sigma}_1)\nonumber \\
&&\sum_s {\cal T}^v_{(b)s}(\vec \sigma ,{\vec
\sigma}_2,\tau |q,r_{\bar a}] \Big( e^{-q-{1\over {\sqrt{3}}}\sum_{\bar a}
\gamma_{\bar as}r_{\bar a}}[{{\rho}\over 3}+\sqrt{3}\sum_{\bar c}\gamma_{\bar 
cs} \pi_{\bar c}]\Big) (\tau ,{\vec \sigma}_2)\, \Big)\, \Big]\, +\nonumber \\
&+&\epsilon k \sum_{r,s}\Big[ e^{q-{1\over {\sqrt{3}}}\sum_{\bar a}(\gamma
_{\bar ar}+\gamma_{\bar as})r_{\bar a}}\Big] (\tau, \vec 
\sigma ) \epsilon_{(a)(b)(c)}\delta_{(a)r}\delta_{(b)s} \,
{}^3{\hat \Omega}_{rs(c)}[q,r_{\bar c}](\tau ,\vec \sigma )\approx 0.
\label{VII11}
\end{eqnarray}

\noindent The last line is equal to $\epsilon k e^{3q}\, {}^3\hat 
R[q,r_{\bar a}]$.

The conformal factor $q(\tau ,\vec \sigma )$ of the 3-metric has been
interpreted as an ``intrinsic internal time" [it is not a scalar and is 
proportional to Misner's time $\Omega=-{1\over 3}\, ln\,
\sqrt{\hat \gamma}$ \cite{misner} for asymptotically flat spacetimes
(see Appendix C):  $q=-{1\over 2} \Omega$], to be contrasted with York's
``extrinsic internal time" ${\cal T}=-{4\over 3}\epsilon k \, {}^3K={2\over 
{3\sqrt{\gamma}}}{}^3{\tilde \Pi}$ 
[see Ref.\cite{beig} for a review of the known
results with York's extrinsic internal time, 
Ref.\cite{qadir} for York cosmic time versus proper time
and Refs.\cite{ish,kuchar1} for more
general reviews about the problem of time in general relativity].

Let us also remark that if we would have added only the gauge-fixing $\alpha
_{(a)}(\tau ,\vec \sigma )\approx 0$ [so that the associated Dirac brackets 
would coincide with the ADM Poisson brackets for metric gravity as already 
said], the four variables $\vec \xi (\tau ,\vec \sigma )$, $q(\tau ,\vec 
\sigma )$ [with conjugate momenta ${\tilde \pi}^{\vec \xi}_r(\tau ,\vec 
\sigma )\approx 0$, $\rho (\tau ,\vec \sigma )$] would correspond to the 
variables used in Ref.\cite{ish} to label the points of the spacetime $M^4$
(assumed compact), following the suggestion of Ref.\cite{dc11}, if $q(\tau 
,\vec \sigma )$ is interpreted as a time variable (see Section VI
for a different identification of points). However, the example of the
foliation of Minkowski spacetime with rectangular coordinates by means of
spacelike hyperplanes, shows that both internal intrinsic [$q(\tau ,\vec \sigma
)$] and extrinsic [${}^3K(\tau ,\vec \sigma )$] times cannot be used as time
labels to identify the leaves: i) ${}^3K=0$ on every leaf; ii) $q=0$ on
every leaf. Therefore, we shall not accept $q(\tau ,\vec \sigma )$ as a time
variable for $M^4$: the problem of time in the Hamiltonian formulation will be 
discussed in Ref.\cite{russo3} (see also Section VI). A related problem 
(equivalent to the transition from a Cauchy problem to a Dirichlet one and
requiring a definition of which time parameter has to be used) is the
validity of the ``full sandwich conjecture" \cite{dc11,mtw} [given two nearby
3-metrics on Cauchy surfaces $\Sigma_{\tau_1}$ and $\Sigma_{\tau_2}$, there is
a unique spacetime $M^4$, satisfying Einstein's equations, with these 3-metrics 
on those Cauchy surfaces] and of the ``thin sandwich conjecture" [given ${}^3g$ 
and $\partial_{\tau}\, {}^3g$ on $\Sigma_{\tau}$, there is a unique spacetime 
$M^4$ with these initial data satisfying Einstein's equations]: see Ref.
\cite{bafo} for the non validity of the ``full" case and for the restricted 
validity (and its connection with constraint theory) of the ``thin" case.

See Appendix C for some
notions on mean extrinsic curvature slices, for the
TT-decomposition, for more comments about internal intrinsic and
extrinsic times and for a review of the Lichnerowicz-York
conformal approach to the reduction of metric gravity.
In this approach the superhamiltonian constraint (namely the elliptic 
Lichnerowicz equation) is solved in the variable $\phi (\tau ,\vec \sigma )=
e^{{1\over 2}q(\tau ,\vec \sigma )}$, namely in 
the conformal factor $q(\tau ,\vec \sigma )$ rather than in its
conjugate momentum $\rho (\tau ,\vec \sigma )$.
In the conformal approach one uses York's variables
\cite{york}, because most of the work on the Cauchy problem for Einstein's
equations in metric gravity
[see the reviews\cite{cho,beig} with their rich bibliography and
Ref.\cite{ise}, where it is shown (see the end of Appendix C and Eq.(\ref{c7})
for the notations) that if one extracts the transverse traceless
part ${}^3{\tilde \Pi}^{rs}_{TT}$ of ${}^3{\tilde \Pi}_A^{rs}$, one may define 
a local canonical basis with variables ${\cal T}$, ${\cal P}_{\cal T}$, 
${}^3\sigma_{rs}$,  ${}^3{\tilde \Pi}^{rs}_{TT}$: it is called the York map] 
is done by using spacelike hypersurfaces $\Sigma$ of constant
mean extrinsic curvature in the compact case [see Refs.\cite{cho,i1,i2}]
and with the maximal slicing condition
${\cal T}(\tau ,\vec \sigma )=0$ (it may be extended to non constant 
${\cal T}$) in the asymptotically free case [see also Ref.\cite{im} for recent 
work in the compact case with non constant ${\cal T}$ and Ref.\cite{brill1}
for solutions of Einstein's equations in presence of matter which do not
admit constant mean extrinsic curvature slices]. 
Let us remark that
in Minkowski spacetime ${}^3K(\tau ,\vec \sigma )=0$ are the hyperplanes, 
while ${}^3K(\tau ,\vec \sigma )=const.$ are the mass hyperboloids, 
corresponding to the instant and point form of the dynamics
according to Dirac\cite{dd} 
respectively [see Refs.\cite{gaida} for other types of foliations].
It would be extremely important to have some ideas how to find explicitly the 
canonical transformation from our canonical basis in the 3-orthogonal gauges
to the canonical basis whose existence is assured by the York map\cite{ise}.

Instead in Ref.\cite{beig} in the case of compact spacetimes the 
superhamiltonian constraint is interpreted as a ``time-dependent Hamiltonian" 
for general relativity in the intrinsic internal time $q$ .

We shall see in Ref.\cite{russo3}, that  in asymptotically flat (at
spatial infinity) spacetimes the canonically reduced superhamiltonian
constraint ${\hat {\cal H}}_R(\tau ,\vec \sigma )\approx 0$ in 3-orthogonal
coordinates has to be interpreted (like in the conformal approach) as an 
integrodifferential equation, the reduced
Lichnerowicz equation,  for the conformal factor $\phi (\tau ,\vec \sigma )=
e^{{1\over 2}q(\tau ,\vec \sigma )}$ whose solution gives it as a functional 
of the canonical variables $r_{\bar a}(\tau ,\vec \sigma )$, $\pi_{\bar 
a}(\tau ,\vec \sigma )$, and of the last gauge variable: the momentum $\pi
_{\phi}(\tau ,\vec \sigma )=2\phi^{-1}(\tau \vec \sigma ) \rho (\tau ,\vec 
\sigma )$ conjugate to the
conformal factor. The evolution in $\tau$ (the time parameter labelling the
leaves $\Sigma_{\tau}$ of the foliation associated with the 3+1 splitting of
$M^4$) will be shown to be generated by the ADM energy [absent in closed 
spacetimes]. The solution $\phi =e^{q/2}\approx e^{F[r_{\bar a},
\pi_{\bar a}, \rho ]}$ of the reduced Lichnerowicz equation
determines an equivalence class of 3-geometries (or conformal 3-geometry)
parametrized by the gauge variable $\rho (\tau ,\vec \sigma )$
[conformal gauge orbit]; the natural representative of an
equivalence class is obtained with the gauge-fixing $\rho (\tau ,\vec \sigma )
\approx 0$: ${}^3g_{rs}=e^{4F[r_{\bar a}, \pi_{\bar a}, 0]}\, {}^3{\hat g}
^{diag}_{rs}[r_{\bar a}, \pi_{\bar a}]$.

The functions $r_{\bar a}(\tau ,\vec \sigma )$, $\bar a=1,2$, give a 
parametrization of the Hamiltonian physical degrees of
freedom of the gravitational field and of the space of conformal
3-geometries [the quotient of superspace 
by the group $Weyl\, \Sigma_{\tau}$, if by varying $\rho$ the solution 
$\phi =e^{q/2} \approx e^{F[r_{\bar a},\pi_{\bar a}, \rho ]}$ 
of the Lichnerowicz equation spans all the Weyl rescalings]: 
it turns out that a point (a 3-geometry)
in this space, i.e. a ${}^3{\hat g}^{diag}_{rs}$ [it is simultaneously
the York\cite{york} reduced metric and the Misner's one\cite{misner} in
3-orthogonal coordinates; see the end of Section VI],
is a class of conformally related 3-metrics (conformal gauge orbit).

When we add the
gauge-fixing $\rho (\tau ,\vec \sigma )\approx 0$ to the superhamiltonian
constraint and go to Dirac brackets eliminating the conjugate variables
$q(\tau ,\vec \sigma )$, $\rho(\tau ,\vec \sigma )$, the functions $r_{\bar 
a}(\tau ,\vec \sigma )$ and $\pi_{\bar a}(\tau ,\vec \sigma )$ become the
physical canonical variables for the gravitational field in this special
3-orthogonal gauge; this does not happens 
with the gauge-fixing ${}^3K(\tau ,\vec \sigma )\approx 0\, (or\, const.)$.
The ADM energy, which, in this gauge, depends only on $r_{\bar a}$, $\pi_{\bar 
a}$ is the Hamiltonian generating the $\tau$-evolution of the physical
(non covariant) gravitational field degrees of freedom
[this corresponds to the two dynamical
equations contained in the 10 Einstein equations in this gauge].
In Ref.\cite{russo3} there will be a more complete discussion of these points.

Since there are statements [see Ref.\cite{conf}; Ref.\cite{ciuf} contains a 
recent review with a rich bibliography] on the existence and unicity of
solutions of the 5 equations of ADM metric gravity (the Lichnerowicz equation
or superhamiltonian constraint, the 3 supermomentum constraints and the gauge
fixing (maximal slicing condition) ${}^3K(\tau ,\vec \sigma )=0$) 
and since our approach to tetrad
gravity contains metric gravity, it is reasonable that this demonstration can
be extended to the reduced Lichnerowicz equation [obtained by putting into it
a solution of the supermomentum constraints possible only in tetrad gravity
since it uses the Green function (\ref{V21})] with the gauge fixing ${}^3K(\tau
,\vec \sigma )=0$ replaced with $\rho (\tau ,\vec \sigma )=0$.

Let us remark that  Minkowski spacetime in Cartesian coordinates is a solution
of Einstein equations, which in the 3-orthogonal gauges corresponds to
$q=\rho =r_{\bar a}=\pi_{\bar a}=0$ [$\phi =1$] and $n=n_r=N_{(as)r}=0$, 
$N_{(as)}
=\epsilon$ [for $q=\rho =r_{\bar a}=0$ Eq.(\ref{VI18}) implies ${}^3{\hat 
{\tilde \pi}}^r_{(a)}$ proportional to $\pi_{\bar a}$; the condition $\Sigma
_{\tau}=R^3$ implies ${}^3K_{rs}=0$ and then $\pi_{\bar a}=0$]. 

Therefore, it is consistent with Einstein equations to
add by hand the two pairs of second class constraints 
$r_{\bar a}(\tau ,\vec \sigma )\approx 0$, $\pi_{\bar a}(\tau ,\vec \sigma )
\approx 0$, to the Dirac Hamiltonian with arbitrary multipliers,\hfill\break
\hfill\break 
$H^{'}_{(D)ADM,R}=H_{(D)ADM,R}+\int d^3\sigma [
\sum_{\bar a}(\mu_{\bar a}r_{\bar a}+\nu_{\bar a}\pi_{\bar a})](\tau ,\vec 
\sigma )$.\hfill\break
\hfill\break
 The time constancy of these second class constraints
determines the multipliers \hfill\break
\hfill\break
$\partial
_{\tau}r_{\bar a}(\tau ,\vec \sigma )\, {\buildrel \circ \over =}\, \nu_{\bar a}
(\tau ,\vec \sigma )+\int d^3\sigma_1 n(\tau ,{\vec \sigma}_1)
\lbrace r_{\bar a}(\tau ,\vec \sigma ),{\hat {\cal H}}_R(\tau ,{\vec \sigma}_1)
\rbrace {}^{*}\approx 0$,\hfill\break
 $\partial_{\tau} \pi_{\bar a}(\tau ,\vec \sigma )\,
{\buildrel \circ \over =}\, -\mu_{\bar a}(\tau ,\vec \sigma )+\int d^3\sigma_1
n(\tau ,{\vec \sigma}_1) \lbrace \pi_{\bar a}(\tau ,\vec \sigma ),
{\hat {\cal H}}_R(\tau ,{\vec \sigma}_1)\rbrace {}^{*}\approx 0$. 
\hfill\break
\hfill\break
By going to 
new Dirac brackets, we remain with the only conjugate pair $q(\tau ,\vec 
\sigma )$, $\rho(\tau ,\vec \sigma )$, constrained by the first class
constraint ${\hat {\cal H}}_R(\tau ,\vec \sigma ){|}_{r_{\bar a}=\pi_{\bar a}
=0} \approx 0$. In this way we
get the description of a family of gauge equivalent
spacetimes $M^4$ without gravitational field (see Ref.\cite{russo3}), which 
could be called a ``void spacetime", with 3-orthogonal coordinates for
$\Sigma_{\tau}$. They turn out to be ``3-conformally flat" because ${}^3{\hat
g}_{rs}=e^q\, \delta_{rs}$. 
Now, the last of Eqs.(\ref{VII8}) [with $r_{\bar a}
=\pi_{\bar a}=0$] is an integral equation to get $\rho$ in terms of ${}^3{\hat
{\tilde \Pi}}\,$ [or ${}^3\hat K$] and $q={1\over 6}\, ln\, {}^3\hat g$.

If we add the extra gauge-fixing $\rho (\tau ,\vec 
\sigma )\approx 0$, we get the 3-Euclidean metric $\delta_{rs}$ on $\Sigma
_{\tau}$, since the superhamiltonian constraint has $q(\tau ,\vec \sigma )
\approx 0$ [$\phi (\tau ,\vec \sigma )\approx 1$]
as a solution in absence of matter. 
The time constancy of $\rho (\tau ,\vec \sigma )\approx 0$ implies 
$n(\tau ,\vec \sigma ) \approx 0$. 
Indeed, for the reduction to Minkowski spacetime, besides the solution $q(\tau 
,\vec \sigma )\approx 0$ of the superhamiltonian constraint [vanishing of the
so called internal intrinsic (many-fingered) time \cite{kku}], 
we also need the gauge-fixings 
$N_{(as)}(\tau ,\vec \sigma )\approx \epsilon$, $N_{(as)r}(\tau ,\vec \sigma )
\approx 0$, $n_r=0$. Many members of the equivalence class
of void spacetimes represent flat Minkowski spacetimes in the most arbitrary 
coordinates compatible with Einstein theory with the associated inertial 
effects as in Newtonian gravity in noninertial Galilean frames. Therefore, 
they seem to represent the most general ``pure acceleration effects without 
gravitational field (i.e. without tidal effects) but with a control on the
boundary conditions" compatible with Einstein's general relativity for
globally hyperbolic, asymptotically flat at spatial infinity spacetimes [see
also the discussion on general covariance and on the various formulations of
the equivalence principle (homogeneous gravitational fields = absence of
tidal effects) in Norton's papers \cite{norton}].

Void spacetimes can be characterized by adding to the ADM action
the Cotton-York 3-conformal tensor with Lagrange multiplier [see Appendix D and
Eq.(\ref{z3})], but this will be studied elsewhere.

See the future papers \cite{russo3,russo4} for the use of this reduced 
symplectic structure for a solution of the deparametrization problem in 
general relativity in presence of matter.

\vfill\eject

\section{Conclusions and
interpretational Problems: Dirac's Observables versus General 
Covariance and Bergmann's Spacetime Coordinates.}

In this second paper dealing with a new formulation of tetrad gravity on 
globally hyperbolic, asymptotically flat at spatial infinity, spacetimes with
Cauchy 3-surfaces $\Sigma_{\tau}$ diffeomorphic to $R^3$ (so that they admit
global coordinate systems), we analyzed the Hamiltonian group of gauge 
transformations whose generators are the 14 first class constraints of the 
model. After introducing a new parametrization of the lapse and shift 
functions, suited for spacetimes asymptotically flat at spatial infinity, we 
studied in detail the subgroup of gauge transformations associated with 
rotations and pseudodiffeomorphisms of the cotriads, namely the automorphism
group of the coframe SO(3) principal bundle. We pointed out the necessity of
using weighted Sobolev spaces to avoid the presence of stability subgroups of
gauge transformations generating the cone over cone structure of
singularities on the constraint manifold and creating an obstruction to the
canonical reduction [Gribov ambiguity of the spin connection and isometries of
the 3-metric]. This description will be valid, in a variational sense, for a
finite interval $\triangle \tau$, after which conjugate points for the 
3-geometry of $\Sigma_{\tau}$ and/or 4-dimensional singularities will develop
due to Einstein equations.

Then we defined and solved the multitemporal equations for cotriads and their 
conjugate momenta on $\Sigma_{\tau}$ associated with rotations and spatial 
pseudodiffeomorphisms. This required a proposal for the parametrization of the 
group manifold of these gauge transformations. Also the corresponding six 
first class constraints have been solved and Abelianized. The final outcome
was the explicit dependence of cotriads and their momenta on the three rotation
angles and on the three pseudodiffeomorphisms parameters. The Dirac observables 
with respect to these gauge transformations are reduced cotriads depending only 
on three arbitrary functions [the reduced momenta also depend on the momenta 
conjugate to these functions]. We have shown that the choice of the coordinate
system on $\Sigma_{\tau}$ is equivalent to the choice of how to parametrize the
reduced cotriads in terms of the three arbitrary functions, and this also gives 
a parametrization of the superspace of 3-geometries ($Riem\, \Sigma_{\tau}
/Diff\, \Sigma_{\tau}$).

By choosing a parametrization corresponding to global 3-orthogonal coordinate
systems on $\Sigma_{\tau}\approx R^3$, we were able to perform a global 
(at least at a heuristic level)
quasi-Shanmugadhasan canonical transformation to a canonical basis in which
13 first class constraints are Abelianized. Next we defined the Dirac brackets 
corresponding to 3-orthogonal gauges (choice of 3-orthogonal coordinates and
of the origin of angles; there is a residual gauge freedom corresponding to
rotating 3-orthogonal gauges; the choice of a congruence of timelike observers
now depends only on the choice of the 3 boost parameters $\varphi_{(a)}$), 
we made a further canonical transformation to more
transparent canonical variables and reexpressed all 3- and 4-tensors in this 
final basis. Besides lapse and shift functions, the final configuration
variables for the superspace of 3-geometries are the conformal factor of the
3-metric $\phi =e^{q/2}=\gamma^{1/12}$ and 
two functions $r_{\bar a}$ (the genuine degrees of 
freedom of the gravitational field) parametrizing the diagonal elements of the 
3-metric. Moreover, there are the two momenta $\pi_{\bar a}$ of the 
gravitational field and the momentum $\rho$ [$\pi_{\phi}=2\phi^{-1}\rho$]
conjugate to the conformal factor $q$ [$\phi$]. 
The momentum $\rho$, containing a nonlocal information on the extrinsic
curvature of $\Sigma_{\tau}$, and not ${}^3K$ (which depends non locally upon 
$\rho$) is the last gauge variable of tetrad gravity.
The only left first class constraint is the reduced superhamiltonian
one, which becomes an integrodifferential equation for the conformal factor [in
metric gravity it would correspond to the Lichnerowicz equation after having
put into it the solution of the supermomentum constraints] as it will be
justified in Ref.\cite{russo3}. A
comparison has been made with the conformal approach of Lichnerowicz
and York.

In future papers\cite{russo3,russo4} there will be the study of the
superhamiltonian constraint (with the refusal of its quantum version, the
Wheeler-DeWitt equation, as an evolution equation),
of the asymptotic Poincar\'e charges, of the ADM
energy as the physical Hamiltonian (and of the related problem of time), of
the deparametrization of tetrad gravity in presence of matter (scalar particles)
. In this way, we will see that the 3-orthogonal 
gauges are the equivalent of the Coulomb gauge in classical
electrodynamics (like the harmonic gauge is the equivalent of the Lorentz
gauge): this will allow to show explicitly the action-at-a-distance 
(Newton-like and gravitomagnetic) potentials among particles hidden in tetrad
gravity (like the instantaneous
Coulomb potential is hidden in the electromagnetic gauge
potential). Spinning particles will be needed to study precessional effects from
gravitomagnetism. Also a reformulation of the canonical reduction done in this 
paper in local normal coordinates on $\Sigma_{\tau}$  will be needed as a
first step towards the study of normal coordinates in $M^4$, necessary to define
local nonrotating inertial observers and to study the geodetic deviation 
equation.

Our approach breaks the general covariance of general relativity completely by
going to the special 3-orthogonal gauge with $\rho (\tau ,\vec \sigma )
\approx 0$. But this is done in a way
naturally associated with presymplectic theories (i.e. theories with first class
constraints like all formulations of general relativity and the standard model
of elementary particles with or without supersymmetry): the global
Shanmugadhasan canonical transformations (when they exist; for instance they
do not exist when the configuration space is compact
like in closed spacetimes) correspond to privileged
Darboux charts for presymplectic manifolds. Therefore, the gauges identified by 
these canonical transformations should have a special (till now unexplored) 
role also in generally covariant theories, in which traditionally one 
looks for observables invariant under 
spacetime diffeomorphisms (but no complete basis is
known for them in general relativity) and not for (not generally covariant) 
Dirac observables. While in electromagnetism and in Yang-Mills theories the
physical interpretation of Dirac observables is clear, in generally covariant
theories there is a lot of interpretational problems and ambiguities.

Therefore, let us finish with some considerations on interpretational problems,
whose relevance has been clearly pointed out in Ref.\cite{stachel}.

First of all, let us interpret metric and tetrad gravity according to
Dirac-Bergmann theory of constraints (the presymplectic approach). Given a
mathematical noncompact, topologically trivial, manifold $M^4$ with a maximal 
$C^{\infty}$-atlas A, its diffeomorphisms in $Diff\, M^4$ are interpreted in 
passive sense (pseudodiffeomorphisms): chosen a reference atlas (contained in A)
of $M^4$, each pseudodiffeomorphism identifies another possible atlas contained 
in A. The pseudodiffeomorphisms are assumed to tend to the identity at spatial 
infinity in a way which will be discussed in the next paper \cite{russo3}. 
Then we add an arbitrary $C^{\infty}$
metric structure on $M^4$, we assume that $(M^4,{}^4g)$ is globally hyperbolic
and asymptotically flat at spatial infinity and we arrive at a family of 
Lorentzian spacetimes $(M^4,{}^4g)$ over $M^4$. On $(M^4,{}^4g)$ one usually
defines \cite{mtw,mw}
the standards of length and time, by using some material bodies, with
the help of mathematical structures like the line element $ds^2$, timelike
geodesics (trajectories of test particles) and null geodesics (trajectories of 
photons), without any reference to Einstein's equations [see the conformal,
projective, affine and metric structures hidden in $(M^4, {}^4g)$ according to
Ref.\cite{pirani}, which replace at the mathematical level the ``material
reference frame" concept \cite{rov,brown,wittt} with its `test' objects]; 
only the equivalence
principle (statement about test particles in an external given gravitational
field) is used to emphasize the relevance of geodesics. Let $\tilde Diff\,
M^4$ be the extension of $Diff\, M^4$ to the space of tensors over $M^4$. Since
the Hilbert action of metric gravity is invariant under the combined action
of $Diff\, M^4$ and $\tilde Diff\, M^4$, one says that the relevant object
in gravity is the set of all 4-geometries over $M^4$ [$(M^4,{}^4g)$ modulo
$Diff\, M^4$ or $Riem\, M^4/Diff\, M^4$] 
and that the relevant quantities (generally covariant observables)
associated with it are the invariants under diffeomorphisms like the curvature
scalars. From the point of view of  dynamics, one has to select those special
4-geometries whose representatives $(M^4,{}^4g)$ satisfy Einstein's equations,
which are invariant in form under diffeomorphisms (general covariance). 
The variation of a solution ${}^4g_{\mu\nu}(x)$ of Einstein's equations under
infinitesimal spacetime diffeomorphisms, namely ${\cal L}_{\xi^{\rho}\partial
_{\rho}}\, {}^4g_{\mu\nu}(x)$, satisfies the Jacobi equations associated with 
Einstein's equations or linearized Einstein equations [see Refs.
\cite{fmm,monc,cfm}; with our assumptions we are in the noncompact case (like
Ref.\cite{cfm}) without Killing vectors: in this case it is known that near
Minkowski spacetime the Einstein empty space equations are linearization 
stable]: therefore these Noether (gauge) symmetries of the Hilbert action are 
also dynamical symmetries of Einstein equations.

One can say that a ``kinematical gravitational field" is a 4-geometry (an
element of $Riem\, M^4/Diff\, M^4$), namely an equivalence
class of 4-metrics modulo $Diff\, M^4$, and that an ``Einstein 
or dynamical gravitational 
field" (or Einstein 4-geometry or equivalence class of Einstein spacetimes) is
a kinematical gravitational field which satisfies Einstein's equations.

However, the fact that the ten Einstein equations are not a hyperbolic system
of differential equations and cannot be put in normal form [this is evident if 
one starts with the ADM action, because the ADM Lagrangian is singular] is only 
considered in connection with the initial data problem. Instead, the ADM
action (needed as the starting point to define the canonical formalism since
it has a well posed variational problem) contains the extra input of a 3+1
splitting of $M^4$: this allows the identification of the surface term
containing the second time derivatives of the 4-metric to be discarded from the
Hilbert action.
As a consequence the ADM action is quasi-invariant under the pullback
of the Hamiltonian group of gauge transformations generated by the first class 
constraints (as every singular Lagrangian) and this group is not $Diff\, M^4$
plus its extension ${\tilde Diff}\, M^4$, as it will be shown in Ref.
\cite{russo3}. In particular, the ADM action is not invariant under 
diffeomorphisms in $Diff\, M^4$ skew with respect to the foliation of $M^4$
associated to the chosen 3+1 splitting, even if the ADM theory is independent 
from the choice of the 3+1 splitting. However, the ADM action generates the same
equations of motion, i.e. Einstein's equations,  so that
the space of the dynamical symmetries of these equations is the same as in the
description base on the Hilbert action: but now not every dynamical symmetry
of Einstein's equations is a Noether symmetry of the ADM action.

Regarding the 10 Einstein equations,
the Bianchi identities imply that four equations are linearly dependent on the 
other six ones and their gradients. Moreover, the four combinations of
Einstein's equations projectable to phase space (where they become the
secondary first class superhamitonian and supermomentum constraints of canonical
metric and tetrad gravity) are independent from the accelerations and are only
restrictions on the Cauchy data. As a consequence,
the Einstein equations have solutions, in which the ten
components ${}^4g_{\mu\nu}$ of the 4-metric depend on only two dynamical
(but not tensorial)
degrees of freedom (defining the physical gravitational field) and on eight
undetermined degrees of freedom [more exactly the four components of the 
4-metric corresponding to the lapse and shift functions and on the four
functions depending on the gradients of the 4-metric 
(generalized velocities) corresponding, through
the first half of Hamilton equations, to the four arbitrary Dirac multipliers
in front of the primary constraints (vanishing of the momenta conjugate to
lapse and shift functions) in the Dirac Hamiltonian].

This transition from the ten components ${}^4g
_{\mu\nu}$ of the tensor ${}^4g$ in some atlas of $M^4$ to the 2 
(deterministic)+8 (undetermined) degrees of freedom breaks general covariance,
because these quantities are neither tensors nor invariants under spacetime
diffeomorphisms (their functional form is atlas dependent in a way dictated
by the 3+1 splittings of $M^4$ needed for defining the canonical formalism).
This is manifest in the canonical approach (we discuss metric gravity, but
nothing changes in tetrad gravity except that there are six more undetermined
degrees of freedom):\hfill\break
i) choose an atlas for $M^4$, a 3+1 splitting $M^{3+1}$
of $M^4$ (with leaves $\Sigma_{\tau}$ of the foliation assumed diffeomorphic
to $R^3$), go to coordinates adapted to the 3+1 splitting
[atlas for $M^{3+1}$ with coordinate charts $(\sigma^A)=(\tau ,\vec \sigma )$,
connected to the $M^4$ atlas by the transition functions $b^{\mu}_A(\tau 
,\vec \sigma )$ of Section II of I ] and replace $Diff\, M^4$ with $Diff\, 
M^{3+1}$ (the diffeomorphisms respecting the 3+1 splitting); \hfill\break
ii) the ten components
${}^4g_{AB}$ of the 4-metric in the  adapted coordinates are non
covariantly replaced with $N$, $N^r$, ${}^3g_{rs}$, whose conjugate momenta
are ${\tilde \pi}_N$, ${\tilde \pi}^{\vec N}_r$, ${}^3{\tilde \Pi}^{rs}$;
\hfill\break
iii) there are four primary [${\tilde \pi}_N\approx 0$, ${\tilde \pi}^{\vec N}
_r\approx 0$] and four secondary [${\tilde {\cal H}}\approx 0$, ${\tilde {\cal 
H}}^r\approx 0$] first class constraints;\hfill\break
iv) therefore, the twenty canonical
variables have to be replaced (with a Shanmugadhasan canonical transformation)
with two pairs of genuine physical degrees of freedom (Dirac's observables),
with eight gauge variables and with eight abelianized first class constraints;
\hfill\break
v) this separation is dictated by the Hamiltonian group ${\bar {\cal G}}$ of
gauge transformations which has eight generators and is not connected with
${\tilde Diff}\, M^{3+1}$ [except for spatial diffeomorphisms 
$Diff\, \Sigma_{\tau} \subset Diff\, M^{3+1}$], 
which has only four generators and whose invariants 
are not Dirac observables [the so called time-diffeomorphisms are replaced by
the 5 gauge transformations generated by ${\tilde \pi}_N$, ${\tilde \pi}_r
^{\vec N}$, and the superhamiltonian constraint];\hfill\break 
vi) as already said at the end of Section V of I, the eight 
gauge variables should be fixed by giving only four gauge fixings for the 
secondary constraints (the same number of conditions needed to fix a 
diffeomorphisms), because their time constancy determines the four secondary
gauge fixings for the primary constraints [and, then, their time constancy
determines the Dirac multipliers (four velocity functions not determined by
Einstein equations) in front of the primary constraints in the Dirac 
Hamiltonian].

Since no one has solved the metric gravity secondary constraints till now, it
is not clear what is undetermined inside ${}^3g_{rs}$ (see Appendix C for what
is known from the conformal approach) and, therefore, which is the physical
meaning (with respect to the arbitrary determination of the standards of length
and time) of the first four gauge-fixings. Instead, the secondary four 
gauge-fixings determine the lapse and shift functions, namely they determine
how the leaves $\Sigma_{\tau}$ are packed in the foliation (the gauge nature 
of the shift functions, i.e. of ${}^4g_{oi}$, is connected with the
conventionality of simultaneity \cite{simul}). Let us remark that the 
invariants under spacetime diffeomorphisms are in general
not Dirac observables, because they depend on the eight gauge variables not 
determined by Einstein's equations. Therefore, all the curvature scalars are 
gauge quantities at least at the kinematical level, as can
be deduced from the expression of 4-tensors given in Appendix B of I and in 
Appendix E.

In this paper we have clarified the situation in the case of tetrad gravity,
and, as a consequence, also for metric gravity since we started from the ADM 
action. The original 32 canonical variables $N$, $N_{(a)}$, $\varphi_{(a)}$,
${}^3e_{(a)r}$, ${\tilde \pi}_N$, ${\tilde \pi}^{\vec N}_{(a)}$, ${\tilde
\pi}^{\vec \varphi}_{(a)}$, ${}^3{\tilde \pi}^r_{(a)}$ (we disregard the
asymptotic part of the lapse and shift functions for this discussion) have
been replaced, in the case of 3-orthogonal coordinates $\vec \sigma$ on 
$\Sigma_{\tau}$ and therefore in the associated coordinates $(\tau ,\vec \sigma
)$ of an atlas of $M^{3+1}$, by the Dirac's observables $r_{\bar a}$, $\pi
_{\bar a}$ [the gravitational field], 
by 14 first class constraints [13 have been abelianized] and by 14
gauge variables: $N$, $N_{(a)}$, $\varphi_{(a)}$, $\alpha_{(a)}$, $\xi^r$,
$\rho$ [the momentum conjugate to the conformal factor $q$ of the 3-metric;
$q$ is determined by the superhamiltonian constraint or Lichnerowicz
equation]. Now we have to add 10 
primary gauge-fixings:\hfill\break
i) 6 gauge-fixings, determining  $\varphi_{(a)}$ and 
$\alpha_{(a)}$, for the primary constraints ${\tilde \pi}^{\vec \varphi}_{(a)}
\approx 0$, ${}^3{\tilde M}_{(a)}\approx 0$ [which do not generate secondary 
constraints]: they fix the orientation of the tetrads 
${}^4E^{(\alpha )}_A$ in every point [the gauge
fixings on the $\varphi_{(a)}$'s are equivalent to choose the (in general
nongeodesic) congruence of timelike worldlines with 4-velocity field $u^A=
{}^4E^A_{(o)}$ corresponding to local observers
either at rest or Lorentz boosted; the gauge fixings on the $\alpha_{(a)}$'s
are equivalent to the fixation of the standard of non rotation of the local
observer];  \hfill\break
ii) 3 gauge-fixings for the parameters $\xi^r$ of the spatial 
pseudodiffeomorphisms
generated by the secondary constraints ${}^3{\tilde \Theta}_r\approx 0$:
they correspond to the choice of an atlas of coordinates on $\Sigma_{\tau}$
[chosen as conventional origin of pseudodiffeomorphisms and influencing the
parametrization of the angles $\alpha_{(a)}$] and, therefore, by adding the
parameter $\tau$, labelling the leaves of the foliation, of an atlas 
on $M^{3+1}$. The gauge-fixings on
$\xi^r$, whose time constancy produces the gauge-fixings for the shift functions
and, therefore, a choice of simultaneity convention in $M^4$ (the choice of how
to synchronize clocks),
can be interpreted as a fixation of 3 standards of length by means of the 
choice of a coordinate system on $\Sigma_{\tau}$; \hfill\break
iii) a gauge-fixing for $\rho$, which, being a momentum, carries an 
information about the extrinsic curvature of $\Sigma_{\tau}$ embedded in $M^4$
[it replaces the York extrinsic time ${}^3K$ of the Lichnerowicz-York conformal
approch] for the superhamiltonian constraint. The
gauge-fixing on $\rho$ has nothing to do with a standard of time (the evolution
is parametrized by the parameter $\tau$ of the induced coordinate system 
$(\tau ,\vec \sigma )$ on $M^4$; see also Ref.\cite{russo3}), but it is a
statement about the extrinsic curvature of a $\Sigma_{\tau}$ embedded in $M^4$
[the Poisson algebra of the superhamiltonian and supermomentum constraints 
reflects the embeddability properties of $\Sigma_{\tau}$; the superhamiltonian
constraint generates the deformations normal to $\Sigma_{\tau}$, which partially
`replace' the $\tau$-diffeomorphisms] and is one of the sources of the gauge 
dependence at the kinematical level 
of the curvature scalars of $M^4$ [the other 
sources are the lapse and shift functions and their gradients]. The natural
interpretation of the gauge transformations generated by the superhamiltonian
constraint is to change the 3+1 splitting of $M^4$ by varying the gauge variable
$\rho (\tau ,\vec \sigma )$ [i.e. something in the extrinsic curvature of the
leaves $\Sigma_{\tau}$ of the associated foliation], so to make the theory 
independent from the choice of the original 3+1 splitting of $M^4$, as it 
happens with parametrized Minkowski theories. However, since the
time constancy of the gauge-fixing on $\rho$ determines the gauge-fixing for the
lapse function [which says how the $\Sigma_{\tau}$ are packed in $M^4$], 
there is a connection with the choice of the standard of local
proper time. Let us remark that only the gauge-fixing $\rho (\tau ,\vec \sigma 
)\approx 0$ [implying ${}^3{\vec K}(\tau ,\vec \sigma )\approx 0$ only in 
absence of matter and of gravitational field] leaves the Dirac observables 
$r_{\bar a}$, $\pi_{\bar a}$, canonical; with other gauge-fixings the canonical
 degrees of freedom of the gravitational field have to be redefined.

Therefore, according to constraint theory, given an atlas on a 3+1 splitting
$M^{3+1}$ of $M^4$, the phase space content of the 8 nondynamical Einstein 
equations is equivalent to the determination of the Dirac observables
(namely a kinematical 
gravitational field not yet solution of the 2 dynamical Einstein 
equations, i.e. of the final Hamilton equations with the ADM energy as 
Hamiltonian, see Ref.\cite{russo3}), 
whose functional form in terms of the original
variables depends on choice of the atlas on $M^{3+1}$ and on a certain part of 
the extrinsic curvature of $\Sigma_{\tau}$. 

Let us define  a ``Hamiltonian kinematical gravitational field"  as the 
quotient of the set of Lorentzian spacetimes $(M^{3+1},{}^4g)$ with a 3+1 
splitting with respect to the Hamiltonian
gauge group ${\tilde {\cal G}}$ with 14 (8 in metric gravity) generators
[$Riem\, M^{3+1}/{\tilde {\cal G}}$]: while
space diffeomorphisms in $Diff\, M^{3+1}$ coincide with those in $Diff\, \Sigma
_{\tau}$, the ``$\tau$-diffeomorphisms" in $Diff\, M^{3+1}$ are replaced by
the 5 gauge freedoms associated with $\rho$, $N$ and $N_{(a)}$. 

A representative of a ``Hamiltonian kinematical gravitational field" in a
given gauge equivalence class is
parametrized by $r_{\bar a}$, $\pi_{\bar a}$ and is an element of a
gauge orbit $\Gamma$ spanned by the gauge variables $\varphi_{(a)}$, $\alpha
_{(a)}$, $\xi^r$, $\rho$, $N$, $N_{(a)}$. Let us consider the reduced gauge
orbit $\Gamma^{'}$ obtained from $\Gamma$ by going to the quotient with
respect to $\varphi_{(a)}$, $\alpha_{(a)}$, $\xi^r$. The solution $\phi =
e^{q/2}$ of the reduced Lichnerowicz equation is $\rho$-dependent, so that the 
gauge orbit $\Gamma^{'}$ contains one conformal 3-geometry (conformal gauge
orbit; see the end of Appendix 
C), or a family of conformal 3-metrics if the $\rho$-dependence of the solution
$\phi$ does not span all the Weyl rescalings. In addition $\Gamma^{'}$
contains the lapse and shift functions. Now, each 3-metric in the conformal
gauge orbit has a different 3-Riemann tensor and different 3-curvature scalars.
Since 4-tensors and  4-curvature scalars depend : i) on the
lapse and shift functions (and their gradients); ii) on $\rho$ both explicitly 
and implicitly through the solution of the Lichnerowicz equation, as shown in
Appendices A and B of I and in Appendix E in the 3-orthogonal gauges (with the
corresponding 3-tensors given in Appendix D), and this influences the 
3-curvature scalars, most of these objects are in general gauge variables from
the Hamiltonian point of view at least at the kinematical level. 
The simplest relevant scalars 
of $Diff\, M^4$, where to visualize these effects, 
are Komar-Bergmann's individuating fields (see later on) and/or the
bilinears ${}^4R_{\mu\nu\rho\sigma}\, {}^4R^{\mu\nu\rho\sigma}$, ${}^4R
_{\mu\nu\rho\sigma}\, \epsilon^{\mu\nu\alpha\beta}\, {}^4R_{\alpha\beta}{}
^{\rho\sigma}$.
Therefore, generically the elements of the gauge
orbit $\Gamma^{'}$ are, from the point of view of $M^4$ based on the Hilbert 
action, associated with different 4-metrics belonging to different 4-geometries
(the standard ``kinematical gravitational fields").

According to the gauge interpretation based on constraint theory, a
``Hamiltonian kinematical 
gravitational field" is an equivalence class of 4-metrics modulo 
the pullback of the Hamiltonian group of gauge transformations, which contains 
all the 4-geometries connected by them and a well defined conformal
3-geometry. This is a consequence of the different 
invariance properties of the ADM and Hilbert actions, even if they generate
the same equation of motion.

Let us define an ` Hamiltonian Einstein or dynamical
gravitational field" as a Hamiltonian kinematical
gravitational field which satisfies the final Hamilton equations with the ADM
energy as Hamiltonian (equivalent to the two dynamical equations hidden in the
Einstein equations). 

These Hamiltonian dynamical gravitational fields correspond to special gauge
equivalence classes, which contain only one 4-geometry whose representative
4-metrics satisfy Einstein's equations, so that they ``coincide" with the
standard dynamical gravitational fields. This highly nontrivial statement is
contained in the results of Refs.\cite{fmm,cfm,monc} (in particular see
Ref.\cite{cfm} for the noncompact asymptotically free at spatial infinity case).
The implication of this fact is that on the space of the solutions of the
Hamilton-Dirac equations (which, together with the first class constraints, are
equivalent to Einstein's equations) the 
kinematical Hamiltonian gauge transformations are
restricted to be dynamical symmetries (maps of solutions onto solutions; with 
them there is not necessarily an associated constant of the motion like with the
Noether symmetries of an action) of Einstein's equations in the ADM presentation
and this implies that the allowed Hamiltonian gauge transformations must be 
equivalent to or contained in
the spacetime pseudodiffeomorphisms of $M^4$ (which are dynamical
symmetries of Einstein's equations as already said). The allowed infinitesimal
Hamiltonian gauge transformations on the space of solutions of the
Hamilton-Dirac equations must be solutions of the Jacobi equations (the 
linearized constraints and the linearized evolution equations; see Refs.
\cite{monc} for their explicit expression) and this excludes most of the
kinematically possible Hamiltonian gauge transformations (all those generating 
a transition from a 4-geometry to another one). In the allowed Hamiltonian 
gauge transformations the gauge parameters $N$, $N_{(a)}$, $\rho$,... are not
independent but restricted by the condition that the resulting gauge 
transformation must be a spacetime pseudodiffeomorphisms.
However, since the infinitesimal
spacetime pseudodiffeomorphisms of a 4-metric solution of Einstein's equations
(i.e. ${\cal L}_{\xi^{\rho}\partial_{\rho}}\, {}^4g_{\mu\nu}(x)$) are
solutions to the Jacobi equations in the Hilbert form, it turns out that among 
the dynamical symmetries of Einstein's equations there are both allowed
strictly Hamiltonian gauge trasformations, under which the ADM action is
quasi-invariant, and generalized transformations under which the ADM action is
not invariant (see Appendix A of the next paper \cite{russo3}). This derives
from the fact that the Noether symmetries of an action and the dynamical
symmetries of its Euler-Lagrange equations have an overlap but do not coincide.
This is the way in which on the space of solutions of Einstein's equations
spacetime diffeomorphisms are rebuild starting from the allowed Hamiltonian 
gauge transformations adapted to the 3+1 splittings of the ADM formalism.
The kinematical freedom of the 8 independent types of Hamiltonian gauge
transformations of metric gravity is reduced to 4 dynamical types like for 
$Diff\, M^4$; partially, this was anticipated at the kinematical level by the 
fact that in the original Dirac Hamiltonian there are only 4 arbitrary Dirac 
multipliers, and that the gauge-fixing procedure starts with the gauge
fixings of the secondary constraints, which generate those for the primary ones
, which in turn lead to the determination of the Dirac multipliers.

This state of affairs implies also that the Dirac observables (namely the 
invariants under the kinematical Hamiltonian gauge transformations and without
any a priori tensorial characted under $Diff\, M^4$) restricted to the 
solutions of the final Hamilton-Dirac equations (and therefore of the original 
Einstein's equations) must be expressible in some way in terms of quantities
scalar under $Diff\, M^4$ when $M^4$ is an Einstein spacetime. A step in this
direction would be to find the connection of our
Dirac observables $r_{\bar a}(\tau ,\vec \sigma )$ in the 3-orthogonal gauges
with the symmetric traceless 2-tensors on 2-planes, which are the independent
gravitational degrees of freedom according to Christodoulou and Klainermann
\cite{chris}, and with the in some way connected Newman-Penrose formalism.

In generally covariant theories (without background fields)
the interpretational difference with respect 
to the Dirac observables of Yang-Mills theories, is that one has to make a
complete gauge-fixing to give a meaning to ``space and time" (in the above 
sense) before being able to identify the functional form of the Dirac 
observables for the gravitational field and moreover we have to formulate the 
problem only for the solutions of Einstein's equations (this is not necessary
for Yang-Mills theory).

This deep difference between the interpretations based on constraint theory
and on general covariance respectively is reflected in 
the two viewpoints about what is observable in general relativity (and, as a 
consequence, in all generally covariant theories) as one can clearly see
in Ref.\cite{rov} and in its  bibliography:\hfill\break
i) The ``non-local point of view" of Dirac \cite{dirac}, according to which
determinism implies that only gauge-invariant quantities (Dirac's observables;
they do not exist globally for compact spacetimes) can be measured. The ``hole
argument" of Einstein\cite{einst} (see Refs.\cite{rov,stachel} for its
modern treatment) supports this viewpoint: points of spacetime are not a priori
distinguishable (their individuality is washed out by general covariance, i.e.
by the invariance under spacetime diffeomorphisms), so that, for instance,
${}^4R(\tau ,\vec \sigma )$ [a scalar under diffeomorphisms, but not a Dirac 
observable at the kinematical level] 
is not an observable quantity. Even if ${}^4R(\tau ,\vec \sigma )\,
{\buildrel \circ \over =}\, 0$ in absence of matter, the other curvature 
scalars are non vanishing after having used Einstein equations and, due to the 
lack of known solutions without Killing vectors, it is not possible to say 
which is their connection with Dirac observables.
More in general, the 4-metric tensor ${}^4g_{\mu\nu}$ is a not observable gauge 
variable. As said in Ref.\cite{stachel} an Einstein  spacetime manifold with a 
metric corresponds to a dynamical gravitational field, but a dynamical
gravitational field corresponds to an equivalence class of
spacetimes. The metrical structure forms part of the set of dynamical variables,
which must be determined before the points of spacetime have any physical 
properties. Therefore, one cannot assume in general relativity what is valid 
in special relativity, namely that the individuation of the points of
Minkowski spacetime is established by a framework of rigid rods and clocks.

In Appendix E this is clearly shown at the kinematical level (i.e. before the
restriction to the solutions of Einstein's equations)
in the 3-orthogonal gauges: there is the 
explicit dependence of the 4-tensors on $(M^4,{}^4g)$ on the residual gauge
variables (to be fixed to have a reconstruction of $M^4$ and, therefore, a
coordinate system on it) $N=N_{(as)}+n$, $N_{(a)}={}^3{\hat e}^r_{(a)} 
[N_{(as)r}+n_r]$, $\rho$,
and on the conformal factor $q$ of the 3-metric, which has to be 
determined by the superhamiltonian constraint (and this will introduce an
extra dependence on the last gauge variable, its conjugate momentum $\rho$;
this is the only gauge freedom of the 3-tensors on $(\Sigma_{\tau},{}^3g)$
given in Appendix D). Instead in the Appendices A and B of I there is shown the
general gauge dependence of 4-tensors on all the gauge variables before 
choosing a coordinate system.

Fixing the gauge freedoms in general relativity means to determine the 
functional form of the 4-metric tensor ${}^4g_{\mu\nu}$: this is a definition 
of the angle and distance properties of the material bodies, which form the 
reference system (rods and clocks). At the kinematical level
the standard procedures of defining 
measures of length and time \cite{ll,mtw} are gauge dependent, because the 
line element $ds^2$ is gauge dependent and determined only after a complete 
gauge-fixing and after the restriction to the solutions of Einstein's
equations (note that in textbooks these procedures are always defined without
any reference to Einstein's equations):
only now the curvature scalars of $M^4$ become measurable, like
the electromagnetic vector potential in the Coulomb gauge. 
The measuring apparatuses should  also be described by the gauge 
invariant Dirac observables associated with the given gauge (namely identified
by the Shanmugadhasan canonical transformation associated with that gauge) as
we shall try to show in Refs.\cite{russo3,russo4}, after the introduction
of matter, since an experimental laboratory corresponds by definition to a
completely fixed gauge.

See also Ref.\cite{but} for the relevance of the ``hole argument" in the
discussions on the nature of spacetime and for the attempts to formulate
quantum gravity. Even if the standard canonical (either metric or tetrad) 
gravity approach presents serious problems in quantization due to the 
intractable Lichnerowicz equation (so that research turned towards either 
Ashtekar's approach or superstring theory with its bigger general covariance 
group), still the problem of what is 
observable at the classical level in generally covariant theories is open.

ii) The `local point of view', according to which the spacetime manifold $M^4$
is the manifold of physically determined `events' (like in special relativity),
namely spacetime points are physically distinguishable, because any
measurement is performed in the frame of a given reference system. The gauge 
freedom of generally covariant theories reflects the freedom of choosing
coordinate systems, i.e. reference systems. Therefore, the evolution is not
uniquely determined (since the reference systems are freely chosen)
and, for instance, ${}^4R(\tau ,\vec \sigma )$ is an
observable quantity, like the 4-metric tensor ${}^4g_{\mu\nu}$. See Ref.
\cite{feng} for a refusal of Dirac's observables in general relativity based on
the local point of view. 

In Ref.\cite{rov} the non-local point of view is accepted and there is a 
proposal for using some special kind of matter to define a ``material
reference system" (not to be confused with a coordinate system) to localize
points in $M^4$, so to recover the local point of view in some approximate way
[the main approximations are: 1) to neglect, in Einstein equations, the
energy-momentum tensor of the matter forming the material reference system
(it's similar to what happens for test particles); 2) to neglect, in the 
system of dynamical equations, the entire set of equations determining the 
motion of the matter of the reference system (this introduces some indeterminism
in the evolution of the entire system)], since in the analysis of
classical experiments both approaches tend to lead to the same conclusions.
See also Refs.\cite{brown,ish,kuchar1} for a complete review of material clocks
and reference fluids. However, we think that one has to consider the use of
``test objects" as an idealization for the attempt to approximate with realistic
dynamical objects the conformal, projective, affine and metric structures
\cite{pirani} of Lorentzian manifolds, which are used to define the ``ideal
geodesic clocks"\cite{mtw} and the basis of the theory of measurement.

Let us remark that in applications, for instance in the search of gravitational 
waves, one is always selecting a background reference metric and the associated
(Minkowski like) theory of measurement: the conceptual framework becomes the 
same as in special relativity. The same happens for every string theory due to 
necessity (till now) of a background metric in their formulation.

Since in Refs.\cite{russo3,russo4} we shall present a different solution for the
time problem (in a scheme in which a ``mathematical time" is identified before 
quantization and never quantized),
we delay the discussion of these problems to these papers. In them the 
deterministic evolution of general relativity, in the mathematical parameter
$\tau$ labelling the leaves $\Sigma_{\tau}$ of the foliation (to be locally
correlated to some physical time), is generated by the ADM energy 
(see also Ref.\cite{cfm}) and there is
a decoupled ``point particle clock" measuring $\tau$. Let us remark that the
refuse of internal either intrinsic or extrinsic times 
implies that the superhamiltonian constraint has to be interpreted as a 
generator of gauge transformations [so that the momentum $\rho$, conjugate to
the conformal factor q of the 3-metric, is a gauge variable] 
and not as a generator of time evolution, contrary to the
commonly accepted wiewpoint for compact spacetimes 
(see Kuchar in Ref.\cite{ku1}) .

Instead, we accept the proposal of Komar and Bergmann\cite{komar,be} of 
identifying the physical points of a spacetime $(M^4, {}^4g)$
without Killing vectors, solution of the Einstein's 
equations, only a posteriori in a way 
invariant under spacetime diffeomorphisms extended to 4-tensors, 
by using four invariants bilinear and trilinear in the
Weyl tensors [as shown in Ref.\cite{gede} there are 14 algebraically independent
curvature scalars for $M^4$, which are reduced to four when Einstein equations
without matter are used], called ``individuating fields'', which do not depend 
on the lapse and shift functions. These individuating fields depend on 
$r_{\bar a}$, $\pi_{\bar a}$ and on the gauge parameters $\xi^r$ (choice of 
3-coordinates on $\Sigma_{\tau}$) and $\rho$ (replacing York's internal 
extrinsic time ${}^3K$): note the difference from the proposal of 
Refs.\cite{ish,dc11} of using $\xi^r$ and $q$ for this aim.
The 4-metric in this ``physical 4-coordinate grid", obtained from ${}^4g_{AB}$
by making a coordinate transformation from the adapted coordinates $\sigma^A=
(\tau ,\vec \sigma )$, depends on the same variables and also on the lapse 
and shift functions.

By using Appendices 
A, B of I and E, one can see that these individuating fields are not Dirac
observables at the kinematical level. They must not be Dirac observables also
when restricted to the solutions of Einstein's equations, because the freedom 
in the choice of the mathematical coordinates $\sigma^A$ is replaced by the
gauge freedom in the choice of $\xi^r$ and $\rho$.
However, in every complete gauge (choice of the coordinate systems 
on $\Sigma_{\tau}$ and on $M^{3+1}$) they describe a special 
gauge-dependent coordinate system for $M^4$, in which the 
dynamical gravitational field degrees of freedom in that
gauge can be used (at least in some finite region) to
characterize distinct points of $M^4$, as also remarked by
Stachel\cite{stachel} in connection with Einstein's hole argument [but
without taking into account constraint theory]. In this way we get a physical
4-coordinate grid on the mathematical 4-manifold $M^4$ dynamically determined
by tensors over $M^4$ itself with a rule which is invariant under $Diff\, M^4$
but with the functional form of the map ``$\sigma^A=(\tau ,\vec \sigma ) \mapsto
\, physical\, 4-coordinates$ depending on the chosen complete gauge:
the ``local point of view" is justified a posteriori in every 
completely fixed gauge.

Finally, let us remember that Bergmann\cite{be} made the following critique of 
general covariance: it would be desirable to restrict the group of
coordinate transformations (spacetime diffeomorphisms) in such a way that it
could contain an invariant subgroup describing the coordinate transformations 
that change the frame of reference of an outside observer (these 
transformations could be called Lorentz transformations; see also the
comments in Ref.\cite{ll} on the asymptotic behaviour of coordinate
transformations); the remaining 
coordinate transformations would be like the gauge transformations of
electromagnetism. This is what we began to do in Section II with the
redefinition of lapse and shift functions and which will be completely
accomplished in the next papers \cite{russo3,russo4} on Poincar\'e charges
and on the deparametrization of tetrad gravity in presence of matter. 
In this way ``preferred" asymptotic 
coordinate systems will emerge [see the preferred congruence of asymptotic 
timelike observers in Ref.\cite{russo3}], which, as said by Bergmann, are
not ``flat": while the inertial coordinates are determined experimentally by 
the observation of trajectories of force-free bodies, these intrinsic 
coordinates can be determined only by much more elaborate experiments
(for instance precessional effects on gyroscopes), since
they depend, at least, on the inhomogeneities of the ambient gravitational 
fields.

See also Ref.\cite{elma} for other critics to general covariance: very often
to get physical results one uses preferred coordinates not merely for
calculational convenience, but also for understanding. In Ref.\cite{zala} this
fact has been formalized as the ``principle of restricted covariance". In our
case the choice of the gauge-fixings has been dictated by the Shanmugadhasan
canonical transformations, which produce generalized Coulomb gauges, in which 
one can put in normal form the Hamilton equations for the canonical
variables of the gravitational field [and, therefore, they also produce a normal
form of the two associated combinations of the Einstein equations which
depend on the accelerations].

This discussion points towards the necessity of finding suitable weighted 
Sobolev spaces such that: i) there are no isometries of the metric (Gribov
ambiguities of the spin connection); ii) there are no supertranslations;
iii) Poincar\'e charges at spatial infinity are well defined; iv) there is a
well defined Hamiltonian group ${\bar {\cal G}}$
of gauge transformations which preserves
properties i), ii) and iii). It is hoped that its pullback ${\tilde {\cal G}}$,
 acting on tensors on $M^{3+1}$, will contain asymptotic Poincar\'e 
transformations as an invariant subgroup (implying the existence of Bergmann's
``preferred" coordinate systems). These problems will be studied in the next 
paper\cite{russo3}.

ACKNOWLEDGEMENTS

One of us (L.L.) thanks Prof.C.Isham for pointing him the relevance of the
automorphism group of the orthogonal coframe principal bundle in the study of
the Hamiltonian group of gauge transformations of tetrad gravity, Prof.
M.Pauri for very clariflying discussions on the interpretational problems,
Prof.K.Kuchar, Prof. C.Rovelli and Prof. M.J.Gotay for constructive
criticism at various stages of this work.

\vfill\eject

\appendix

\section{Special Systems of Coordinates.}

The gauge freedom of general relativity, due to its invariance under general
coordinate transformations or spacetime diffeomorphisms, reflects the 
arbitrariness in the choice of how to describe space and time since coordinates 
have no intrinsic meaning. The choice of a local coordinate system is 
equivalent in standard presentations to the
definition of an observer with his ideal clocks and rods and the principle
of general covariance states that the laws of physics are independent from
this choice. However in trying to solve Einstein partial differential
equations or to find local canonical adapted bases with the Shanmugadhasan
canonical transformation one has to look for those coordinate systems (if
any) which separate the equations. Therefore, the choice of adapted bases
and probably also some future definition of elementary particle in general
relativity (so that the standard Wigner definition will re-emerge in the
limit of flat Minkowski spacetime)  require a breaking of general covariance.
At least locally one has to choose ``physical coordinate systems adapted to
the physical systems under investigation", study there the equations of
motion and then use general covariance in a passive way as mathematical
coordinate transformations, which possibly can transform localized concepts
in spacetime delocalized ones.

While the weak (or Galilei) form of the equivalence principle (implying the
equality of inertial and gravitational masses) is common to Newton and
Einstein gravity [the laws of motion of free particles in a local, freely
falling, nonrotating frame are identical to Newton's laws of motion expressed
in a gravity-free Galilean frame: they will produce straight worldlines in a
local Lorentz frame (i.e. in a freely falling nonrotating frame) as in
special relativity, in absence of electric charge, for small angular momentum, 
for gravitational binding energies much less of rest masses and in a
sufficiently small neighbourhood such that the effects of the geodesic
deviation equation are negligible], the Einstein medium-strong and strong
forms assert the existence of local Lorentz frames for all 
the nongravitational laws and
all the laws of physics respectively. In particular, the strong form implies
that there are no gravitational effects in a local freely falling
nonrotating frame in a sufficiently small spacetime neighbourhood
\cite{mtw,ciuf} in which tidal effects are negligible.

``Ideal" rods and clocks are defined as being ones which measure proper length
$\triangle s=\sqrt{-\epsilon {}^4g_{\mu\nu} \triangle^{\mu}\triangle^{\nu}}$
($\triangle^{\mu}$ spacelike) or proper time $\triangle \tau =\sqrt{\epsilon
{}^4g_{\mu\nu} \triangle^{\mu}\triangle^{\nu}}$ ($\triangle^{\mu}$ timelike); 
one must then determine the accuracy to which a given rod or clock is ideal
under given circumstances by using laws of physics to analyze its
behaviour\cite{mtw} (see the Conclusions for the interpretational problems).

See Refs.\cite{pet,mtw} for a review of relevant  coordinate systems [Ref.
\cite{pet} uses $\epsilon =+1$].

We shall add only some informations about geodesic coordinates, harmonic 
coordinates and holonomic versus nonholonomic coordinates. See the previous
references and Appendix A of I
for coordinates 1) geodesic along a specified curve [which include
Fermi-Walker and Fermi transport of tetrads (gyroscopes) and Fermi normal
coordinates]; 2) semigeodesic [which include Gaussian normal (or synchronous)
coordinates]].

A)`` Coordinates $x^{\mu}_1$ geodesic 
(or locally inertial) at a point $p\in M^4$" chosen as origin
$x_1^{\mu}{|}_p=0$. They are such that ${}^4\Gamma^{\alpha}_{\beta\gamma}
(x_1=0) =0$, so that at p the geodesic equation is ${{d^2}\over {d\tau^2}}
x_1^{\mu} =0$.

Aa)`` Local Lorentzian (or Minkowskian or inertial
or comoving) frames". An observer falling
freely in $M^4$ makes measurements in his local Lorentz frame, defined by

\begin{equation}
{}^4g_{\mu\nu}(x_1=0) = {}^4\eta_{(\mu )(\nu )},\quad\quad \partial_{\alpha}
{}^4g_{\mu\nu}{|}_{x_1=0} =0,\Rightarrow {}^4\Gamma^{\alpha}_{\beta\gamma}
(x_1=0)=0.
\label{b1}
\end{equation}

The observer is at rest in his local Lorentz frame, i.e. his worldline is
$\lbrace x_1^o\, arbitrary,\, x_1^k=0\rbrace$ : his velocity is $u^{\mu}_1=
{{dx_1^{\mu}}\over {d\tau}} {|}_{x_1^k=0} = {}^4\nabla_{u_1}\, u^{\mu}_1
{|}_{x_1^k=0} =({}^4\nabla_{u_1^o} u^o_1 {|}_{x_1^k=0}, \vec 0) 
=u_1^o\, {}^4\Gamma
^{\mu}_{oo}\, u_1^o {|}_{x_1^k=0} =0$. Therefore the observer is freely falling
since he moves along a geodesic [${}^4\nabla_{u_1}u^{\mu}_1=0$; in the local
Lorentz frame the geodesic is an extremal of proper time, $d\tau =
\sqrt{\epsilon {}^4\eta_{(\alpha )(\beta )} dx_1^{(\alpha )}dx_1^{(\beta )}}$
and has no acceleration, $a_1^{\mu} ={d\over {d\tau}}
u_1^{\mu} =0$]. Local Lorentz frames remains geodesic at $p\in M^4$ under linear
transformations of coordinates, because the Christoffel symbols behave like 
tensors under such transformations.

Ab) ``Riemann coordinates $x_2^{\mu}$ geodesic at $p\in M^4$", where $x_2^{\mu}
{|}_p=0$. Their distinctive feature is that geodesics passing through $p\in M^4$
satisfy the same equations for straight lines passing through p as in Euclidean 
geometry with Cartesian coordinates: if $\tau$ is an affine parameter along
anyone of these geodesics ($\tau =0$ in p), the geodesics through p have the 
form $x_2^{\mu}(\tau )=\xi^{\mu} \tau$, where $\xi^{\mu}={{dx^{\mu}_2(\tau )}
\over {d\tau}}{|}_p$ is the tangent to the geodesic at p (it is constant along
the geodesic). Riemann coordinate systems at $p\in M^4$ are related by linear
homogeneous transformations, which preserve the form of the geodesics. They 
exists in a neighbourhood V of $p\in M^4$, in which the geodesics emanating 
from it do not cross each other \cite{wald}
inside V, so that p can be joined to each point of V by means of them (V is
geodesically complete). The singularity theorems tend to say that every 
pseudo-Riemannian $M^4$ cannot be geodesically complete; instead Riemannian 
3-manifolds like $\Sigma_{\tau}$ (only spacelike geodesics) may be geodesically 
complete.

Necessary and sufficient criteria defining Riemannian coordinates at $p\in
M^4$ are:

($\alpha )$ The geodesics through $p\in M^4$ have the form $x_2^{\mu}(\tau )=
\xi^{\mu}\tau$, with $\xi^{\mu}=const.$.

($\beta$) The Christoffel symbols satisfy ${}^4\Gamma^{\alpha}_{\mu\nu}(x_2)\,
=0$ (i.e. Riemann coordinates are geodesic at p),
so that the equation for the geodesics through p is
${{d^2x^{\mu}_2(\tau )}\over {d\tau^2}}=0$.

($\gamma$) If ${}^4g_{\mu\nu}$ are analytic functions in 
a neighbourhood of p, the following symmetrized derivative of the
Christoffel symbols vanish at p: $\partial_{(\mu_1}\partial_{\mu_2}\cdots
\partial_{\mu_r}\, {}^4\Gamma^{\alpha}_{\nu_1\nu_2)}(x_2=0)=0$, and  one has
[similar expansions hold for every tensor]:

\begin{eqnarray}
{}^4g_{\mu\nu}(x_2)&=& {}^4g_{\mu\nu}(0) -{1\over 3} {}^4R_{\mu\alpha\nu\beta}
(0) x^{\alpha}_2x^{\beta}_2 -{1\over {3!}} \partial_{\rho}\, {}^4R_{\mu\alpha
\nu\beta}{|}_{x_2=0} x^{\rho}_2x^{\alpha}_2x^{\beta}_2+\nonumber \\
&+&{1\over {5!}} [-6 \partial_{\rho}\partial_{\sigma}\, {}^4R_{\mu\alpha\nu
\beta}{|}_{x_2=0} +{{16}\over 3} {}^4R_{\rho\nu\sigma}{}^{\epsilon}(0)\,
{}^4R_{\beta\mu\alpha\epsilon}(0)] x^{\rho}_2x^{\sigma}_2x^{\alpha}_2
x^{\beta}_2 +\cdots .
\label{b2}
\end{eqnarray}

Ab1) ``Riemann normal (or simply normal) coordinates $y^{\mu}$". It is a special
system of Riemann coordinates for which one has ${}^4g_{\mu\nu}(0)={}^4\eta
_{(\mu )(\nu )}$; in it one has

\begin{eqnarray}
{}^4g&=& |det\, ({}^4g_{\mu\nu})|= \lbrace 1-{1\over 3} {}^4R_{\mu\nu}(0)
x^{\mu}_2x^{\nu}_2 -{1\over {3!}} \partial_{\rho}\, {}^4R_{\mu\nu}{|}_{x_2=0}
x^{\rho}_2x^{\mu}_2x^{\nu}_2-\nonumber \\
&-&{1\over {4!}} [{4\over 3} {}^4R_{\mu\nu}(0)\, {}^4R_{\alpha\beta}(0) +{4\over
{15}} {}^4R_{\mu\rho\nu}{}^{\lambda}(0)\, {}^4R_{\alpha\lambda\beta}{}^{\rho}
(0) +\nonumber \\
&+&{6\over 5} \partial_{\mu}\partial_{\nu}\, {}^4R_{\alpha\beta}{|}_{x_2=0}]
x^{\mu}_2x^{\nu}_2x^{\alpha}_2x^{\beta}_2+.. \rbrace ,
\label{b3}
\end{eqnarray}

\noindent so that, if ${}^4R_{\mu\nu}(0)=0$, then ${}^4g=\epsilon [1+{1\over 
{90}} {}^4R_{\mu\rho\nu}{}^{\lambda}(0)\, {}^4R_{\alpha\lambda\beta}{}^{\rho}
(0) x^{\mu}_2x^{\nu}_2x^{\alpha}_2x^{\beta}_2+\cdots ]$.
A normal coordinate system at $p\in M^4$ is defined to within a linear Lorentz
transformation.

Normal coordinates exploit to the full the locally Minkowskian properties of
pseudo-Riemannian 4-manifolds: an observer who assigns coordinates in the
neighbourhood of a given event $p\in M^4$ by theodolite measurements at p and
interval measurements from p as if spacetime were flat, will assign normal
coordinates; i.e.
the observer at p [where an inertial observer in using a frame (tetrad) 
${}^4E_{1(\alpha )} ={}^4E^{\mu}_{1(\alpha )} \partial /\partial x_1^{\mu}$]
fills spacetime near p with geodesics radiating out from p, with each geodesic
(with a suitable choice of affine parameter) determined by its tangent vector
at p.

Cartan\cite{cart,spiv} showed that, given Riemann normal coordinates $y^{\mu}$
at $p\in M^4$ [$y^{\mu}{|}_p=0$], one can choose adapted orthonormal frames
and coframes ${}^4E^{(N)}_{(\alpha )}={}^4E^{(N)\mu}_{(\alpha )}(y) \partial
/\partial y^{\mu}$, ${}^4\theta^{(N)(\alpha )}={}^4E^{(N)(\alpha )}_{\mu}(y) 
dy^{\mu}$, obtained from ${}^4E^{(N)}_{(\alpha )}{|}_p=\delta^{\mu}_{(\alpha )}
\partial /\partial y^{\mu}$, ${}^4\theta^{(N)(\alpha )}{|}_p=\delta^{(\alpha )}
_{\mu} dy^{\mu}$, by parallel transport along the geodesic arcs originating at 
p. Then one has the following properties

\begin{eqnarray}
{}^4E^{(N)(\alpha )}_{\mu}(y)\, y^{\mu} &=& \delta^{(\alpha )}_{\mu} y^{\mu}
\nonumber \\
{}^4\theta^{(N)(\alpha )}&=& \delta^{(\alpha )}_{\mu} [dy^{\mu}+
y^{\rho}y^{\sigma}\, N^{\mu}{}_{\rho\sigma\lambda}(y)\, 
dy^{\lambda}],\nonumber \\
&&N_{\mu\rho\sigma\lambda}=-N_{\rho\mu\sigma\lambda}=-N_{\mu\rho\lambda\sigma}.
\label{b5}
\end{eqnarray}

Since normal coordinates are the most natural from a differential geometric
point of view, let us look for a parametrization, in this system of
coordinates, of the Dirac observables ${}^3{\hat e}_{(a)r}(\tau ,\vec \sigma )$
on $\Sigma_{\tau}$
in terms of 3 real functions ${\hat Q}_r(\tau ,\vec \sigma )$, whose conjugate
momenta will be denoted ${\hat {\tilde \Pi}}^r(\tau ,\vec \sigma )$.
Eqs.(\ref{b5}) give the Cartan definition of orthonormal tetrads adapted to
normal coordinates for Lorentzian 4-manifolds. This suggest that for
Riemannian 3-manifolds like $\Sigma_{\tau}$, the reduced cotriads ${}^3{\hat
e}_{(a)r}(\tau ,\vec \sigma )$ may be parametrized as follows

\begin{eqnarray}
{}^3{\hat e}_{(a)r}(\tau ,\vec \sigma )&=&\delta^s_{(a)}[\delta_{rs}+\sum_n
\epsilon_{run}\epsilon_{svn} \sigma^u\sigma^v {\hat Q}_n(\tau ,\vec \sigma )]
\nonumber \\
&\Rightarrow& {}^3{\hat e}_{(a)r}(\tau ,\vec \sigma )\, \sigma^r=\delta
_{(a)r}\, \sigma^r,
\label{w1}
\end{eqnarray}

\noindent with $N_{surv}(\tau ,\vec \sigma )=\sum_n \epsilon_{sun}\epsilon
_{rvn}{\hat Q}_n(\tau ,\vec \sigma )=-N_{usrv}(\tau ,\vec \sigma )=-
N_{suvr}(\tau ,\vec \sigma )=N_{rvsu}(\tau ,\vec \sigma )$. Then one gets

\begin{eqnarray}
{}^3{\hat g}_{rs}(\tau ,\vec \sigma )&=&{}^3{\hat e}_{(a)r}(\tau ,\vec \sigma )
\, {}^3{\hat e}_{(a)s}(\tau ,\vec \sigma )=\delta_{rs}+\nonumber \\
&+&\sigma^u\sigma^v [\sum_n\epsilon_{run}\epsilon_{svn}(2+{\vec \sigma}^2\, 
{\hat Q}_n(\tau ,\vec \sigma )){\hat Q}_n(\tau ,\vec \sigma )-\sum_{nm}
\epsilon_{run}\epsilon_{svm}\sigma^n\sigma^m{\hat Q}_n(\tau ,\vec \sigma )
{\hat Q}_m(\tau ,\vec \sigma )],\nonumber \\
&&{}
\label{w2}
\end{eqnarray}

\noindent to be compared with Eq.(\ref{b2}).

A special case of normal coordinates, $y^{\mu}=\xi^{\mu}$, 
(with a special orientation of the tetrads by means of SO(3,1) rotations) is 
realized, even when torsion is present,  with the ``radial gauge"
\cite{moda} (see Ref.\cite{moda1,moda2} for a review). To get it, one imposes 
the gauge conditions

\begin{eqnarray}
&&\xi^{\mu}\, {}^4\omega^{(\alpha )}_{\mu (\beta )}(\xi )=0\quad
\Rightarrow {}^4\omega^{(\alpha )}_{\mu (\beta )}(0)=0,\nonumber \\
&&\xi^{\mu}\, {}^4E^{(\alpha )}_{\mu}(\xi )=\delta^{(\alpha )}_{\mu}\xi^{\mu}
\quad \Rightarrow {}^4E^{(\alpha )}_{\mu}(0)=\delta^{(\alpha )}_{\mu},
\nonumber \\
&&{}\nonumber \\
&&{}^4\Gamma^{\rho}_{\mu\nu}={}^4E^{\rho}_{(\alpha )}\, {}^4E^{(\beta )}_{\mu}\,
{}^4\omega^{(\alpha )}_{\nu (\beta )}+{}^4E^{\rho}_{(\alpha )} \partial_{\nu}
\, {}^4E^{(\alpha )}_{\mu} \Rightarrow \nonumber \\
&&\Rightarrow \xi^{\nu}\, {}^4\Gamma^{\rho}_{\mu\nu}={}^4E^{\rho}_{(\alpha )}
\, \xi^{\nu}\partial_{\nu}\, {}^4E^{(\alpha )}_{\mu}\quad \Rightarrow
\xi^{\mu}\xi^{\nu}\, {}^4\Gamma^{\rho}_{\mu\nu}=0.
\label{b6}
\end{eqnarray}

\noindent The first condition means that the tetrads are parallel transported 
from the origin $\xi^{\mu}=0$ along the straight lines $\xi^{\mu}(s)=sv^{\mu}$,
$0 \leq s$ and $v^{\mu}=const.$, while the second one means that these lines 
are autoparallel, so that in absence of torsion they are geodesics. 
These equations determine
univocally both the coordinate system $\xi^{\mu}$ and the tetrad field
${}^4E^{\mu}_{(\alpha )}$ in a neighbourhood of the origin [the point with
coordinates $\xi^{\mu}=0$], so that the gauge conditions are locally
attainable and complete. There is a residual global gauge freedom due to
the arbitrariness of the choice of the origin and to the possibility of
making a Lorentz transformation of the tetrad in the origin. 

In $\Sigma_{\tau}$-adapted coordinates $\sigma^A=(\tau ,\vec \sigma )$ and with
the tetrads ${}^4E^{(\alpha )}_A$ of Eqs. (45), (46) of I, the second condition
becomes $\sigma^A\, {}^4E^{(\alpha )}_A (\sigma^d)= \delta^{(\alpha )}_A 
\sigma^A$, ${}^4E^{(\alpha )}_A(0)=\delta^{(\alpha )}_A$: together with the
first condition on the spin 4-connection they determine the gauge variables of
tetrad gravity without using explicitly the 3+1 decomposition of the
4-metric.

The radial gauge conditions can be regarded, in a sense, as an operational
prescription which permits one to locate the measuring instruments in a 
neighbourhood of the observer, who lies at the origin $\xi^{\mu}=0$. In fact,
a simple way to explore this neighbourhood is to send from the origin many
``space-probes" carrying clocks, gyroscopes and measuring instruments. A
space-probe will be launched with 4-velocity $v^{(\alpha )}$ with respect to the
given tetrad ${}^4E^{\mu}_{(\alpha )}$ and, if $\tau$ is the proper time 
measured by the clock, $\xi^{\mu}=\tau v^{\mu}$ are the normal coordinates
(in the absence of torsion). Of course, in Minkowski spacetime only the
interior of the future cone can be explored in this way.

B)``Harmonic coordinates $x^{\mu}_3$". Given an arbitrary system of coordinates 
$y^{\mu}$, let us consider the wave equation in $M^4$, ${}^4\Box \varphi (y)=
{}^4\nabla_{\mu}\, {}^4\nabla^{\mu}\, \varphi (y)= {1\over {\sqrt{{}^4g}}}
\partial_{\alpha} (\sqrt{{}^4g}\, {}^4g^{\alpha\beta}\, \partial_{\beta})
\varphi (y)=0$. The harmonic coordinates are defined as $x^{\mu}_3=\varphi^{\mu}
(y)$ with $\varphi^{\mu}(y)$ four independent solutions of the wave
equation, so that the harmonic coordinate condition can be written as ${1\over
{\sqrt{{}^4g}}} \partial_{\alpha}(\sqrt{{}^4g}\, {}^4g^{\alpha\beta})=0$. See 
Ref.\cite{wald} for a discussion of these coordinates, which have to be used 
for the study of the Cauchy problem for Einstein equations [maximal Cauchy
developments, Cauchy stability,...]. If harmonic coordinates hold on an
initial data slice, they give a ``reduced" form of the Einstein equations
that is hyperbolic and preserves both the constraints and the harmonic 
condition in the evolution.

C) ``Nonholonomic coordinates". Given anyone of the previous coordinate systems,
one can define ``coordinate hypersurfaces" $x^{\mu}=const.$ and ``coordinate
lines" on which only one of the $x^{\mu}$ is not fixed. Moreover, since one 
has local coordinate bases $\partial /\partial x^{\mu}$ and $dx^{\mu}$ for
$TM^4$ and $T^{*}M^4$ respectively, it turns out that the tangent vectors to the
$x^{\mu}$ coordinate line are $l_{(\mu )}=\delta^{\nu}_{(\mu )} \partial
/\partial x^{\nu}$ [we put the index $\mu$ inside round brackets to emphasize
that it numbers the tangent vectors] with controvariant components $l^{\nu}
_{(\mu )}=\delta^{\nu}_{(\mu )}=\partial x^{\nu}/\partial x^{\mu}$ and that
their duals  are $\theta^{(\mu )}=\delta^{(\mu )}_{\nu} dx^{\nu}$ with
covariant components $\theta^{(\mu )}_{\nu}=\delta^{(\mu )}_{\nu}= \partial
x^{\mu}/\partial x^{\nu}$. Therefore, $\theta^{(\mu )}$ are a system of 4
gradient vectors : $\theta^{(\mu )}_{\nu}=\partial_{\nu} x^{\mu}$, $\partial
_{\rho} \theta^{(\mu )}_{\nu} -\partial_{\nu} \theta^{(\mu )}_{\rho}=0$.

A system of coordinates is said ``holonomic" if the basic covariant vectors are
a field of gradient vectors. When this does not happen, the coordinate
system is said ``nonholonomic" 
(it can only be defined in a region of $M^4$ which 
can be shrunk to a point): in this case one has local dual noncoordinate
bases [tetrads and cotetrads, frames and coframes,..; see Section II of I]
${}^4{\hat E}_{(\mu )}={}^4{\hat E}^{\nu}_{(\mu )}(x)\partial /\partial
x^{\nu}$ and ${}^4{\hat \theta}^{(\mu )}={}^4{\hat E}^{(\mu )}_{\nu}(x) 
dx^{\nu}$ with${}^4{\hat E}^{\nu}_{(\alpha )}\, {}^4{\hat E}^{(\beta )}_{\nu}=
\delta^{(\beta )}_{(\alpha )}$. Nonholonomic noncoordinate bases may be chosen 
orthogonal ${}^4{\hat E}^{(\alpha )}_{\mu}\, {}^4{\hat E}_{(\alpha )}^{\nu}=
\delta^{\nu}_{\mu}$.
The basic covariant vectors ${}^4{\hat \theta}^{(\mu )}$ of a nonholonomic
coordinate system are a nonintegrable field of tangents to 4 coordinate
lines, which define coordinates $x^{(\mu )}$ [$dx^{(\mu )} \not= {}^4{\hat 
\theta}^{(\mu )}$ since ${}^4{\hat E}^{(\mu )}_{\nu}$ are not gradients].
In Ref.\cite{rob} it is shown how to construct a local nonholonomic coordinate
system $x^{(\mu )}$ from local holonomic ones $y^{\mu}$ defined in a
neighbourhood of a point $p\in M^4$ [chosen as origin: $y^{\mu}{|}_p=0$],
with associated orthonormal frames and coframes ${}^4E_{(\alpha )}={}^4E^{\mu}
_{(\alpha )}(y)\partial /\partial y^{\mu}$, ${}^4\theta^{(\alpha )}={}^4E
^{(\alpha )}_{\mu}(y) dy^{\mu}$. In the neighbourhood of p, where the coframe
matrix ${}^4E^{(\alpha )}_{\mu}$ is regular, the new coordinates are
$x^{(\alpha )}={}^4E^{(\alpha )}_{\mu}(y) y^{\mu}$, so that $y^{\mu}=
{}^4E^{\mu}_{(\alpha )}(x(y)) x^{(\alpha )}$.

\vfill\eject

\section{Isometries and Conformal Transformations.}

See Section II of I for the notations and Refs.\cite{naka,wald}.

A diffeomorphism $\phi \in Diff\, M^4$ is an ``isometry" of a pseudo-Riemannian
manifold $(M^4,{}^4g)$ if ${}^4g^{'}_{\mu\nu}(x^{'}(x))={}^4g_{\mu\nu}(x)$
with $x^{{'}\mu}=\phi^{\mu}(x)$ [it is called isometry since it preserves the
length of a vector; $\phi$ can be interpreted (in the active sense) as a rigid
motion]. The isometries of a given 4-manifold form a group. The infinitesimal
isometries $x^{{'}\mu}(x)=x^{\mu}+\xi^{\mu}(x)=x^{\mu}+\delta_ox^{\mu}=
x^{\mu}+\epsilon X_{\phi}x^{\mu}$ [$\epsilon$ is an infinitesimal parameter]
are generated by a vector field $X_{\phi}$ called a ``Killing vector field",
satisfying the equation \hfill\break
\hfill\break
$[{\cal L}_{X_{\phi}} {}^4g ]_{\mu\nu}= X^{\rho}_{\phi}
\partial_{\rho}\, {}^4g_{\mu\nu} + \partial_{\mu}X^{\rho}_{\phi}\, {}^4g
_{\rho\nu}+\partial_{\nu}X^{\rho}_{\phi}\, {}^4g_{\mu\rho}=0$, \hfill\break
\hfill\break
which becomes \hfill\break
\hfill\break
$({}^4\nabla
_{\mu} X_{\phi})_{\nu}+({}^4\nabla_{\nu} X_{\phi})_{\mu}=\partial_{\mu} X_{\phi
\, \nu}+\partial_{\nu} X_{\phi \, \mu} -2{}^4\Gamma^{\lambda}_{\mu\nu} 
X_{\phi \, \lambda}=0$ \hfill\break
\hfill\break
[Killing equation]
with the Levi-Civita connection. The Killing vector
fields of a 4-manifold span the Lie algebra of the isometry group.

A diffeomorphism of $(M^4,{}^4g)$ such that ${}^4g^{'}_{\mu\nu}(x^{'}(x))=
e^{2\sigma (x)} {}^4g_{\mu\nu}(x)$ is a ``conformal isometry" of the 4-manifold,
namely a particular conformal transformation [it changes the scale but not the 
shape;  the angles are preserved]. In general, conformal transformations 
[$x^{{'}\mu}(x)$ such that ${}^4g^{'}_{\mu\nu}(x^{'}(x))=
e^{2\sigma (x)} {}^4g_{\mu\nu}(x)$] are not  spacetime diffeomorphisms; 
the conformal transformations form a group, $Conf\, M^4$, and 
the conformal isometries are conformal transformations in $Diff\, M^4 \cap 
Conf\, M^4$ , the only conformal transformations 
under which Einstein metric gravity is invariant. The conformal transformation
preserve the time-, light- (or null-) and space-like character of the objects.
One has the following effects of conformal transformations

\begin{eqnarray}
{}^4g_{\mu\nu}(x)&\mapsto& {}^4g^{'}_{\mu\nu}(x^{'}(x))=e^{2\sigma (x)}\, 
{}^4g_{\mu\nu}(x),\nonumber \\
{}^4\Gamma^{\mu}_{\alpha\beta}(x)&\mapsto& {}^4\Gamma^{{'}\mu}_{\alpha\beta}
(x^{'}(x))={}^4\Gamma^{\mu}_{\alpha\beta}(x) +\delta^{\mu}_{\beta}\partial
_{\alpha}\sigma (x)+\delta^{\mu}_{\alpha}\partial_{\beta}\sigma (x)-{}^4g
_{\alpha\beta}(x)\, {}^4g^{\mu\nu}(x)\partial_{\nu}\sigma (x),\nonumber \\
{}^4R^{\mu}{}_{\alpha\nu\beta}(x)&\mapsto& {}^4R^{{'}\mu}{}_{\alpha\nu\beta}
(x^{'}(x))={}^4R^{\mu}{}_{\alpha\nu\beta}(x)-{}^4g_{\alpha\beta}(x)\, {}^4B
^{\mu}_{\nu}(x)-{}^4g_{\alpha\nu}(x)\, {}^4B^{\mu}_{\beta}(x)+\nonumber \\
&+&{}^4g_{\alpha\rho}(x) [{}^4B^{\rho}_{\nu}(x) \delta^{\mu}_{\beta}-
{}^4B^{\rho}_{\beta}(x) \delta^{\mu}_{\nu}],\nonumber \\
{}^4R_{\mu\nu}(x)&\mapsto& {}^4R^{'}_{\mu\nu}(x^{'}(x))={}^4R_{\mu\nu}(x)-
{}^4g_{\mu\nu}(x)\, {}^4B^{\rho}_{\rho}(x)-2\, {}^4B_{\mu\nu}(x),\nonumber \\
{}^4R(x)&\mapsto& {}^4R^{'}(x^{'}(x))=e^{-2\sigma (x)} [{}^4R(x)-6\, {}^4B
^{\rho}_{\rho}(x)],\nonumber \\
{}^4C_{\mu\nu\alpha\beta}(x)&\mapsto& {}^4C^{'}_{\mu\nu\alpha\beta}(x^{'}(x))=
{}^4C_{\mu\nu\alpha\beta}(x),
\label{a21}
\end{eqnarray}

\noindent where ${}^4B^{\mu}_{\nu}=-\partial_{\nu}\sigma \, {}^4g^{\mu\rho}\, 
\partial_{\rho}\sigma +{}^4g^{\mu\rho} (\partial_{\nu}\partial_{\rho}\sigma -
{}^4\Gamma^{\gamma}_{\nu\rho} \partial_{\gamma}\sigma )+{1\over 2}
\delta^{\mu}_{\nu} {}^4g^{\alpha\beta} \partial_{\alpha}\sigma \partial
_{\beta}\sigma$, ${}^4B_{\mu\nu}={}^4B_{\nu\mu}$.

Instead for 3-manifolds one has [${}^3{\nabla}_r$ is the covariant 
derivative associated with the 3-metric ${}^3{g}_{rs}$]: \hfill\break
\hfill\break
${}^3g_{rs} \mapsto {}^3{\hat g}_{rs}=\phi^4\, {}^3g_{rs}$, \hfill\break
${}^3\Gamma^u_{rs} 
\mapsto {}^3{\hat \Gamma}^u_{rs}={}^3\Gamma^u_{rs}+2 \phi^{-1} 
(\delta^u_r\, {}^3{\nabla}_s\phi+\delta^u_s\, {}^3{\nabla}_r\phi-
{}^3{g}_{rs}\, {}^3{g}^{uv}\, {}^3{\nabla}_v\phi )$,\hfill\break
${}^3R \mapsto {}^3{\hat R}=\phi^{-4}\, {}^3R -8 \phi^{-5} ({}^3{g}^{uv}\,
{}^3{\nabla}_u\, {}^3{\nabla}_v)\phi$.\hfill\break

Given two 4-metrics ${}^4g$ and ${}^4\bar g$ on the same 4-manifold $M^4$, they 
are called ``conformally related" if ${}^4{\bar g}_{\mu\nu}(x)=e^{2\sigma (x)}
\, {}^4g_{\mu\nu}(x)$; 
the equivalence class of the conformally related 4-metrics
on a 4-manifold $M^4$ is called a ``conformal structure". The transformation
${}^4g\mapsto e^{2\sigma}\, {}^4g$ is called a ``Weyl rescaling" and the set of
Weyl rescalings on $M^4$ is a group $Weyl\, M^4$.

If every point $x^{\mu}\in M^4$ of the pseudo-Riemannian manifold $(M^4,{}^4g)$
lies in a coordinate chart where ${}^4g_{\mu\nu}=e^{2\sigma}\, 
{}^4\eta_{\mu\nu}$,
then $(M^4,{}^4g)$ is said to be ``conformally flat"; the vanishing of the
Weyl tensor ${}^4C_{\mu\nu\alpha\beta}$ is the necessary and sufficient
condition for conformal flatness.

An infinitesimal conformal isometry is generated by a so called ``conformal
Killing vector field" $X$, which satisfies the conformal Killing equation
\hfill\break
\hfill\break
$X^{\rho}\partial_{\rho} {}^4g_{\mu\nu}+\partial_{\mu}X^{\rho} {}^4g_{\rho\nu}+
\partial_{\nu}X^{\rho} {}^4g_{\mu\rho}={1\over 4} {}^4g_{\mu\nu}
[X^{\rho} {}^4g^{\alpha\beta}
\partial_{\rho} {}^4g_{\alpha\beta} +2\partial_{\rho} X^{\rho}]$. \hfill\break
\hfill\break
The dilatation
vector field $D=x^{\mu}\partial_{\mu}$ is a 
conformal vector field of Minkowski spacetime. Since $X_{\mu}n^{\mu}$, 
with $n^{\mu}$ tangent to a geodesic $\gamma$, is constant
only for null geodesics [$n^{\mu}n_{\mu}=0$], conformal Killing
vector fields give rise to constants of motion for light rays.

\vfill\eject

\section{The Lichnerowicz-York Conformal Approach to the 
Canonical Reduction of Metric Gravity.}

To give an idea  of the
Lichnerowicz-York conformal approach to canonical reduction
\cite{conf,york} [see Refs.\cite{cho,yoyo,ciuf} for reviews], 
we need some preliminary concepts. See Section V of I for a review of ADM 
canonical metric gravity.

A)
Since the hypersurfaces ${\cal T}=-{4\over 3}\epsilon k\, {}^3K=const.$ 
of constant mean extrinsic curvature [CMC slices] play
an important role for the reduction of Hamiltonian constraints in the conformal
approach [${}^3K(\tau ,\vec \sigma )-const. \approx 0$ is the gauge-fixing
for the superhamiltonian constraint, which is interpreted as an elliptic
equation for the conformal factor of the 3-metric], 
for numerical solutions of Einstein equations and in the proof of
the positive gravitational energy conjecture, let us give some results about
these hypersurfaces \cite{cho,brill} [see Ref.\cite{brill1} for solutions of
the Einstein's equations with matter, which do not admit constant mean extrinsic
curvature hypersurfaces].

In Refs.\cite{yo2,yoyo,york} it is shown that ${}^3K$ defines the chosen rate
of volume expansion of an initial slice $\Sigma_{\tau}$ relative to local
proper time. In fact, an element of proper volume $\sqrt{\gamma} d^3\sigma$ 
[$\gamma ={}^3g$] on a spacelike hypersurface $\Sigma_{\tau}$ undergoes in the 
next unit interval of proper time, as measured normal to $\Sigma_{\tau}$, a 
fractional increase of proper volume given by \hfill\break
\hfill\break
$-{}^3K=l^{\mu}{}_{;\mu}=-{}^3g
^{\mu\nu}\, {}^3K_{\mu\nu}={1\over 2}{}^3g^{\mu\nu} {\cal L}_l\, {}^3g_{\mu\nu}
={1\over {\sqrt{\gamma}}}{\cal L}_l \sqrt{\gamma}={\cal L}_l (ln\, 
\sqrt{\gamma})=-\Theta ({}^4g,l)$, \hfill\break
\hfill\break
where $\Theta$ is the ``expansion or 
dilatation" of $l^{\mu}$ [so that ${}^3K$ is equal to the ``convergence of the 
normals $l^{\mu}\,$" in $(M^4,{}^4g)$; see Appendix A of I]. 
For the volume to be extremal this quantity 
must vanish at every point of $\Sigma_{\tau}$ [this is satisfied in a Friedmann
closed universe and in a Taub closed universe at that value of the natural
time-coordinate t at which the universe switches from expansion to 
recontraction, so that the sign of ${}^3K$ could be used to distinguish the
expansion and contraction epochs; see also Refs.\cite{kku}]. 
Moreover, it can be shown\cite{york} that
the rate of change of ${}^3K$ in timelike directions tends to be positive as a
consequence of the equations of motion [for a freely falling observer (${}^3a
^{\mu}=0$) one has ${\cal L}_l\, {}^3K\geq 0$, i.e. ${}^3K$ increases with
respect to the local standard of proper time; ${}^3K$ is essentially the
volume Hubble parameter], and that ${}^3K$ defines a definite class of 
foliations. See Ref.\cite{qadir} for York cosmic time ${\cal T}$
versus proper time.

Instead, Misner's choice\cite{misner} of
the internal intrinsic time $\Omega =-{1\over 3}ln\, \sqrt{\gamma}=-q=
-2 ln\, \phi$ [the
logarithm of the volume] is acceptable for open always expanding universes;
for closed universes, $\Omega$ stops its forward flow at a moment of maximum
expansion and begins to run backward. The other problem with $\Omega$ is that
it is ``not a scalar" and thus has utility only in the presence of a definite 
choice of 3-dimensional coordinates. Moreover, $\Omega$ contains the conformal
factor of the 3-metric, which is the natural variable in which to solve the
superhamiltonian constraint. Since ${\cal L}_l\, \Omega = \kappa \, {}^3K$,
both these variables define the same family of hypersurfaces in homogeneous 
models. The use of ${}^3K$ as time (and of the associated foliations ) does not
depend on any assumptions of homogeneity, nor does it restrict in any way
the anisotropy of the universe. Finally, with the choice of ${}^3K$ as ``time",
its conjugate variable [the natural Hamiltonian to be introduced as an 
``energy" after having solved the constraints in this approach] 
is the scale factor $\sqrt{\gamma}$, so that the ``energy" becomes equal to 
the volume of the universe.
See Refs.\cite{kku} for the connection of ${}^3K$ with the ``many-fingered
time approach".

A number of theorems regarding the mean extrinsic curvature exist, based on the
linearization of the map ${\cal T}$.

a) Compact Slices- See the results in Ref.\cite{choq2}

b) Noncompact Slices. More difficult to study because of a lack of sufficient
knowledge of the invertibility of the corresponding Laplace operator. There is
one important particular case: $\Sigma_{\tau}$ diffeomorphic to $R^3$ with 
${}^4g$ asymptotic to the Minkowski 4-metric at spatial infinity.
Theorem\cite{choq2,cfmoy}: Every Lorentzian manifold $(M^4,
{}^4g)$ in a neighbourhood of Minkowski spacetime $(R^4,{}^4\eta )$ admits a
``maximal" (i.e. with ${\cal T}=Tr\, {}^3K=0$) spacelike submanifold.
It has been conjectured that all spacetimes satisfying the strong energy
condition, and that can be continuously deformed into Minkowski spacetime,
admit a maximal hypersurface. It is known that there exists a maximal
submanifold of Minkowski spacetime passing through a bounded regularly
spacelike 2-dimensional boundary [see Ref.\cite{cho} for references].
One knows also that ``maximal slicing" exists in many cases when $\Sigma
_{\tau}$ is ``noncompact" but ``not" diffeomorphic to $R^3$, for example, 
in the Schwarzschild,
Reissner-Nordstr\"om and Kerr spacetimes as well as general vacuum static and
stationary spacetimes\cite{cho,sma}.

One may use ${\cal T}=-{4\over 3}\epsilon
 {}^3K=const.\not= 0$ slices when $\Sigma_{\tau}$ is
``noncompact"; here the slices are analogous to the ``mass hyperboloids" of
Minkowski spacetime and are asymptotically null \cite{bri,sma}.
In Minkowski spacetime $(R^4,{}^4\eta )$ the most interesting spacelike slices
\cite{yoyo} are the (future and past) ``mass hyperboloids" [asymptotic to the
(future and past) null cones] with ${}^3K=const.$ and the standard
$x^o=const.$ hyperplanes with ${}^3K=0$.

In Ref.\cite{cho} it is reported that when, in the asymptotically flat case, 
one uses
weighted Sobolev spaces, which prevent the existence of stability subgroups of
gauge transformations (and therefore the Gribov ambiguity) for the spin
connection, then there are no ``conformal Killing vectors" for the Riemann
manifold $(\Sigma_{\tau},{}^3g)$ [the equation $\xi^u\partial_u\, {}^3g_{rs}+
\partial_r\xi^u\, {}^3g_{us}+\partial_s\xi^u\, {}^3g_{ru}={1\over 3} {}^3g_{rs}
[\xi^u\, {}^3g^{mn}\partial_u\, {}^3g_{mn}+2\partial_u\xi^u]$ has no solution
$\xi^u\partial_u$].
In these weighted Sobolev spaces one can also show \cite{cho,choq1}
that the Laplace-Beltrami
operator $\triangle_{{}^3g}$ is an isomorphism: this implies that 
$(\Sigma_{\tau},{}^3g)$ has no isometries [i.e. Killing
vectors $\xi^u\partial_u$ satisfying $\xi^u\partial_u\, {}^3g_{rs}+\partial
_r\xi^u\, {}^3g_{us}+\partial_s\xi^u\, {}^3g_{ru}=0$], because no such vector 
can tend to zero to infinity.

B)
Let us now consider two kinds of decompositions of symmetric 3-tensors $T^{rs}=
T^{sr}={}^3T^{rs}$ defined on a Riemannian 3-manifold $(\Sigma_{\tau}, {}^3g
_{rs})$, whose validity in the noncompact case 
requires the use of weighted Sobolev spaces. A closed manifold means a compact
manifold without boundary. One can show that in closed
manifolds $\Sigma_{\tau}$ every vector field on $\Sigma_{\tau}$ is the sum
of a Killing vector field for ${}^3g$ and the divergence of a symmetric
3-tensor field [all vector fields can be written globally on $\Sigma_{\tau}$
as such divergences if and only if ${}^3g$ has no Killing field].

1) ``Transverse decomposition" \cite{yo1}. Following Ref.\cite{yo2}, it is 
defined as

\begin{eqnarray}
T^{rs}&=&T^{rs}_t+T^{rs}_l,\nonumber \\
&&{}\nonumber \\
T^{rs}_l&=&(KX)^{rs}\equiv
{}^3\nabla^rX^s+{}^3\nabla^sX^r=(KX)^{sr},\nonumber \\
{}^3\nabla_s\, T^{rs}_t&=&T^{rs}{}_{|s}={}^3\nabla_s\, T^{rs}-{}^3\nabla_s(KX)
^{rs}=0,\nonumber \\
{}&\Rightarrow& \nonumber \\
{}^3\nabla_s(KX)^{rs}&\equiv&
({}^3\triangle_KX)^r={}^3\triangle X^r+{}^3\nabla^r
({}^3\nabla_sX^s)+{}^3R^r{}_sX^s={}^3\nabla_s\, T^{rs}.
\label{tt1}
\end{eqnarray}

\noindent Here the longitudinal part $T^{rs}_l=(KX)^{rs}$ is the 
``Killing form" of
$X^r$ [$(KX)_{rs}={\cal L}_X\, {}^3g_{rs}$ and $(KX)_{rs}=0$ is the Killing 
equation for determining the infinitesimal isometries of the 3-metric ${}^3g
_{rs}$]. While ${}^3\triangle$ is the ordinary Laplacian for the 3-metric,
${}^3\triangle_K$ is a linear second-order vector operator, the K-Laplacian.

The trace-free part of $T^{rs}_t$, i.e. $T^{rs}_t-{1\over 3}\, {}^3g^{rs} T_t$
[$T_t={}^3g_{rs}T^{rs}_t=T-2\, {}^3\nabla_rX^r$, $T={}^3g_{rs}T^{rs}$], is
no longer transverse, because, in general, ${}^3\nabla_r T_t={}^3\nabla_r(T-2\, 
{}^3\nabla_sX^s)\not= 0$.

If we make the transverse decomposition of this trace-free part of $T^{rs}$,
we get \hfill\break
\hfill\break
$T^{rs}-{1\over 3}\, {}^3g^{rs} T=(T^{rs}-{1\over 3}\, 
{}^3g^{rs} T)_t+(KS)^{rs}$ \hfill\break
\hfill\break
for some $S^r$. 
Now, $(T^{rs}-{1\over 3}\, {}^3g^{rs} T)_t$ is transverse,
but not trace-free: ${}^3g_{rs} (T^{rs}-{1\over 3}\, {}^3g^{rs} T)_t=2\,
{}^3\nabla_sS^s \not= 0$ in general.

2) ``Transverse-Traceless decomposition" \cite{yo2} [see Ref.\cite{beig0} for
more mathematical properties on the solution of the elliptic equation
for $Y^r$, which is connected to the linearization of the Cotton-York tensor]. 
It is defined as

\begin{eqnarray}
T^{rs}&=&T^{rs}_{TT}+T^{rs}_L+T^{rs}_{Tr},\nonumber \\
&&{}\nonumber \\
T^{rs}_{Tr}&=&{1\over 3} {}^3g^{rs} T,\quad\quad T={}^3g_{rs}T^{rs},\nonumber \\
&&{}\nonumber \\
{}^3g_{rs}T^{rs}_{TT}&=&\, {}^3\nabla_s\, T^{rs}_{TT}=0,\nonumber \\
&&{}\nonumber \\
T^{rs}_L&=&(LY)^{rs}\equiv {}^3\nabla^rY^s+{}^3\nabla^sY^r-{2\over 3} {}^3g
^{rs}\, {}^3\nabla_uY^u=\nonumber \\
&=&(KY)^{rs}-{2\over 3} {}^3g^{rs}\, {}^3\nabla_uY^u,\nonumber \\
{}^3g_{rs}T^{rs}_L&=&0,\nonumber \\
{}^3\nabla_s(LY)^{rs}&\equiv& ({}^3\triangle_LY)^r={}^3\triangle Y^r+{1\over 3}
{}^3\nabla^r({}^3\nabla_sY^s)+{}^3R^r{}_sY^s=\nonumber \\
&=&{}^3\nabla_s(T^{rs}-{1\over 3} {}^3g^{rs} T).
\label{tt2}
\end{eqnarray}

Now, $(LY)^{rs}$ is the ``conformal Killing form" of $Y^r$, because, if ${}^3
{\tilde g}_{rs}=\gamma^{-1/3}\, {}^3g_{rs}$ is the ``conformal metric"
(independent of arbitrary overall scale changes: if ${}^3g_{rs}\mapsto \phi \, 
{}^3g_{rs}$ then ${}^3{\tilde g}_{rs}\mapsto {}^3{\tilde g}_{rs}$), then
${\cal L}_Y\, {}^3{\tilde g}_{rs}=\gamma^{-1/3}\, (LY)_{rs}$ (this describes
the action of infinitesimal coordinate transformations on the conformal
metric) and ${\cal L}_Y\, {}^3{\tilde g}_{rs}=0$ is the conformal Killing 
equation, determining the conformal Killing vectors (if any) of ${}^3{\tilde
g}_{rs}$.

The TT-decomposition gives 
a unique result. It turns out that Tr-, TT- and L-tensors are mutually
orthogonal. This is the content of York's splitting theorem\cite{cho}.

It can be shown\cite{yo2} that one has

\begin{eqnarray}
T^{rs}_{TT}&=&(T^{rs}_t)_{TT}=(T^{rs}_{TT})_t,\nonumber \\
(T^{rs}_{TT})_l&=&(KV)^{rs}=0,\nonumber \\
(T^{rs}_l)_{TT}&=&0,\nonumber \\
(T^{rs}_t)_L&=&(LM)^{rs}=(LY)^{rs}-(LX)^{rs}=\nonumber \\
&=&T^{rs}_L-[(KX)^{rs}-{2\over 3}{}^3g^{rs}\, {}^3\nabla_uX^u]=T^{rs}_L-T^{rs}
_l+{1\over 3}{}^3g^{rs} T_l,\nonumber \\
&&{}\nonumber \\
T^{rs}_t&=&(T^{rs}_t)_{TT}+(T^{rs}_t)_L+(T^{rs}_t)_{Tr}=\nonumber \\
&=&T^{rs}_{TT}+[L(Y-X)]^{rs}+{1\over 3}{}^3g^{rs}T_l=\nonumber \\
&=&T^{rs}_{TT}+T^{rs}_L-T^{rs}_l+{1\over 3}{}^3g^{rs} T_l,\nonumber \\
T^{rs}_l&=&(T^{rs}_l)_{TT}+(T^{rs}_l)_L+(T^{rs}_l)_{Tr}=\nonumber \\
&=&(LZ)^{rs}+{1\over 3}{}^3g^{rs} T_l,\quad for\quad some\quad Z^r,\nonumber \\
T&=&T_l+T_t,\quad\quad T_l=2\, {}^3\nabla_rX^r,\nonumber \\
&&{}\nonumber \\
T^{rs}_{TT}&=&(T^{rs}_{TT})_t+(T^{rs}_{TT})_l=(T^{rs}_{TT})_t,\nonumber \\
T^{rs}_L&=&(T^{rs}_L)_t+(T^{rs}_L)_l=(T^{rs}_t)_L+T^{rs}_l-{1\over 3}{}^3g^{rs}
T_l.
\label{tt3}
\end{eqnarray}

For closed $\Sigma_{\tau}$ one has the theorem\cite{yo2}: Let $Y^r$ be a 
harmonic function of ${}^3\triangle_L$ with nowhere vanishing norm on a closed
manifold M; then, there always exists a manifold $\tilde M$ conformally
related to M for which $Y^r$ is a harmonic function of ${}^3{\tilde \triangle}
_K$.

Every transverse symmetric tensor $T^{rs}_t$ on $(\Sigma_{\tau},{}^3g_{rs})$ can
be split uniquely and orthogonally into a sum of a ``transverse tensor with 
vanishing trace" and a ``transverse tensor with nonvanishing trace". From
$(T^{rs}_t)_{TT}=T^{rs}_{TT}$ and $T^{rs}_t=T^{rs}_{TT}+(LM)^{rs}+{1\over 3}
{}^3g^{rs} T_t$ with $M^r=Y^r-X^r$, we get

\begin{eqnarray}
T^{rs}_t&=&T^{rs}_{TT}+T^{rs}_{Tr,t},\nonumber \\
&&{}\nonumber \\
T^{rs}_{Tr,t}&=&(LM)^{rs}+{1\over 3}{}^3g^{rs} T_t,\nonumber \\
{}^3\nabla_s T^{rs}_{Tr,t}&=&{}^3\nabla_s(LM)^{rs}+{1\over 3} {}^3\nabla^r\, 
T_t={}^3\nabla_s(T^{rs}_t-{1\over 3}{}^3g^{rs} T_t)+{1\over 3}{}^3\nabla^r T_t
=0,\nonumber \\
{}^3g_{rs}T^{rs}_{Tr,t}&=&{}^3g_{rs}[(LM)^{rs}+{1\over 3}{}^3g^{rs} T_t]=T_t.
\label{tt4}
\end{eqnarray}

It follows that the gradient of the trace of a transverse tensor is always
globally orthogonal to conformal Killing vectors, when $\Sigma_{\tau}$ is 
closed.

Therefore, $T^{rs}_t$ contains a TT part, $T^{rs}_{TT}$, plus another tensor
$T^{rs}_{Tr,t}$ which can be expressed as a functional only of $T_t$ due to
the equations ${}^3\nabla_s(LM)^{rs}=-{1\over 3}{}^3\nabla^r T_t$. Since the
supermomentum constraints for metric gravity, 
i.e. 3 of the Einstein equations, imply that \hfill\break
\hfill\break
${}^3K^{rs}\, 
{\buildrel \circ \over =}\, {}^3K^{rs}_t={}^3K^{rs}_{TT}+{}^3K^{rs}
_{Tr,t}$ with ${}^3K^{rs}_{Tr,t}=(LM)^{rs}+{1\over 3}{}^3g^{rs}\, {}^3K$, 
\hfill\break
\hfill\break
one can say that ${}^3K^{rs}_{TT}$ contains the ``wave" part [purely
gravitational spin-two TT-tensor], while ${}^3K=Tr\, {}^3K$ is a kinematical
function defining an essentially arbitrary ``gauge" degree of freedom.

C) In the conformal approach it is assumed that the 
superhamiltonian constraint becomes the ``scale or Lichnerowicz equation" 
\cite{cho}, which is a quasilinear elliptic equation for\hfill\break
\hfill\break

$\phi (\tau ,\vec \sigma )=e^{q(\tau ,\vec \sigma )}=
[\gamma (\tau ,\vec \sigma )]^{1/12}=
{}^3e^{1/6}=[-{\cal P}_{\cal T}(\tau ,\vec \sigma )]^{1/6}$.\hfill\break
\hfill\break
In Ref.\cite{cho} it is shown that
$\phi$ and a 3-vector $X^r$ (the vector part of the TT-decomposition) can be
interpreted physically as generalizations of the single potential function 
that satisfies Poisson's equation in Newtonian gravity: one has for $r=|\vec
\sigma | \rightarrow \infty$ the following results \hfill\break
\hfill\break
$\phi =1+E/16\pi r+...$,\hfill\break
$X^r={1\over {32\pi r^3}} P^s (7r^2\delta^r_s+\sigma^r\sigma^s)+...$,
\hfill\break
\hfill\break
where E and $P^r$ are the asymptotic Poincar\'e translation charges [modulo
the supertranslations connected to asymptotic gauge transformations].

The conformal approach can be formulated either in terms of ${}^3g_{rs}$, 
${}^3K_{rs}$ or in terms of the ADM canonical variables ${}^3g_{rs}$, 
${}^3{\tilde \Pi}^{rs}$ (see Section V of I). The ADM
supermomentum constraints for the tensor density ${}^3{\tilde \Pi}^{rs}$ are
proportional to the Einstein equations \hfill\break
\hfill\break
${}^4{\bar G}_{lr}=\epsilon
[\sqrt{\gamma} ({}^3K_r{}^s-\delta^s_r\, {}^3K)]_{|s}\, {\buildrel \circ \over 
=}\, 0\,$, i.e. \hfill\break
$\, 0\approx {}^3{\tilde \Pi}^{rs}_{|s}=\epsilon k 
[\sqrt{\gamma} ({}^3K^{rs}-
{}^3g^{rs}\, {}^3K)]_{|s}=\epsilon k \sqrt{\gamma}\, {}^4{\bar G}_{l}{}^r\,
{\buildrel \circ \over =}\, 0$. \hfill\break
\hfill\break
One has ${}^3{\tilde \Pi}=-2\epsilon k
\sqrt{\gamma}\, {}^3K$ with ${}^3K={}^3g^{rs}\, {}^3K_{rs}= Tr\, {}^3K$.

The idea is to make a suitable separation between physical and unphysical 
degrees of freedom to identify four candidates for the variables in which either
the Einstein equations ${}^4{\bar G}_{ll}\, {\buildrel \circ \over =}\, 0$,
${}^4{\bar G}_{lr}\, {\buildrel \circ \over =}\, 0$, or the ADM constraints,
have to be solved [in the ADM approach a separation is made based on a 
TT-decomposition referred to a flat background spacetime; this is suited for 
weak fields and linearized theory]. 

First of all, one makes a ``conformal transformation" on the 3-metric,
\hfill\break
\hfill\break
${}^3g_{rs}=\phi^4 \, {}^3{\check g}_{rs}$ \hfill\break
\hfill\break
[${}^3g^{rs}=\phi^{-4}\, {}^3{\check
g}^{rs}$; ${}^3R=\phi^{-4}\, {}^3\check R-8\phi^{-5}\, {}^3{\check \triangle}
\phi$ with ${}^3\check \triangle$ the Laplacian for the 3-metric ${}^3{\check 
g}_{rs}$] with $\phi$ an arbitrary definite positive function (positivity is
crucial for the study of the existence and uniqueness of solutions of the
Lichnerowicz equation \cite{cho}); essentially, one uses \hfill\break
\hfill\break
$\phi =e^{q/2}=\gamma^{1/12}={}^3e^{1/6}$, \hfill\break
\hfill\break
so that ${}^3{\check g}_{rs}={}^3{\tilde g}_{rs}$, where ${}^3
{\tilde g}_{rs}$ is the ``conformal metric" with $det\, {}^3{\tilde g}_{rs}=1$
[at each point it gives only the ratio between any two ``local" distances;
the absolute distances are fixed by the scale factor $\phi$]. 

Secondly, one defines the trace-free part ${}^3A^{rs}$ (also called the
``distortion tensor"), ${}^3g_{rs}\, {}^3A
^{rs}=0$, of ${}^3K^{rs}$: \hfill\break
\hfill\break
${}^3K^{rs}={}^3A^{rs}+{1\over 3}\, {}^3g^{rs}\, 
{}^3K$ or \hfill\break
${}^3{\tilde \Pi}^{rs}=\epsilon k\sqrt{\gamma} ({}^3A^{rs}-{2\over 3}
\, {}^3g^{rs}\, {}^3K)={}^3{\tilde \Pi}^{rs}_A+{1\over 3}\, {}^3g^{rs}\, 
{}^3{\tilde\Pi}$, ${}^3g_{rs}\, {}^3{\tilde \Pi}^{rs}_A=0$. \hfill\break
\hfill\break
Now the supermomentum constraints become \hfill\break
\hfill\break
either ${}^3A^{rs}{}_{|s}-{2\over 3}\,
{}^3\nabla^r\, {}^3K\,
{\buildrel \circ \over =}\, 0$ or ${}^3{\tilde \Pi}_A^{rs}{}_{|s}+{1\over 3}\,
{}^3\nabla^r\, {}^3{\tilde \Pi}\approx 0$. \hfill\break
\hfill\break
Then, in the simplest version of
the approach\cite{cho,yoyo}, one makes the 
``conformal rescaling" ${}^3A^{rs}=
\phi^{-10}\, {}^3{\check A}^{rs}$ [${}^3{\tilde \Pi}_A^{rs}=\phi^{-10}\,
{}^3{\check {\tilde \Pi}}_A^{rs}$], so that \hfill\break
\hfill\break
${}^3A^{rs}{}_{|s}={}^3\nabla_s\, 
{}^3A^{rs}=\phi^{-10}\, {}^3{\check \nabla}_s\, {}^3{\check A}^{rs}$, 
${}^3\nabla^r\, {}^3K=\phi^{-4}\, {}^3{\check \nabla}^r\, {}^3K$\hfill\break
\hfill\break
 [${}^3\nabla
_s\, {}^3{\tilde \Pi}^{rs}_A=\phi^{-10}\, {}^3{\check \nabla}_s\, {}^3{\check 
{\tilde \Pi}}^{rs}_A$, ${}^3\nabla^r\, {}^3{\tilde \Pi}=\phi^{-4}\, {}^3{\check
\nabla}^r\, {}^3{\tilde \Pi}$].\hfill\break
\hfill\break
The next step is to make a TT-decomposition  of 
${}^3{\check A}^{rs}$ [${}^3{\check {\tilde \Pi}}_A^{rs}$]: ${}^3{\check A}
^{rs}={}^3{\check A}^{rs}_{TT}+{}^3{\check A}^{rs}_L$ with ${}^3{\check A}^{rs}
_L=(L\check W)^{rs}$ and \hfill\break
\hfill\break
${}^3{\check \nabla}_s\, {}^3{\check A}^{rs}_L=({}^3
{\check \triangle}_L\check W)^r={}^3{\check \nabla}_s\, {}^3{\check A}^{rs}\,
{\buildrel \circ \over =}\, {2\over 3}\phi^6\, {}^3{\check \nabla}^r\, {}^3K$ 
\hfill\break
\hfill\break
[${}^3{\check {\tilde \Pi}}^{rs}_A={}^3{\check {\tilde \Pi}}^{rs}_{A,TT}+{}^3
{\check {\tilde \Pi}}^{rs}_{A,L}$ with ${}^3{\check {\tilde \Pi}}^{rs}_{A,L}=
(L{\check W}_{\pi})^{rs}$ and ${}^3{\check \nabla}_s\, {}^3{\check {\tilde
\Pi}}^{rs}_{A,L}=({}^3{\check \triangle}_L{\check W}_{\pi})^r={}^3{\check 
\nabla}_s\, {}^3{\check {\tilde \Pi}}_A^{rs}\approx -{1\over 3}\phi^6\,
{}^3{\check \nabla}^r\, {}^3{\tilde \Pi}$]. The equation ${}^4{\bar G}_{ll}\,
{\buildrel \circ \over =}\, 0$ becomes the scale or Lichnerowicz equation

\begin{equation}
8\, {}^3{\check \triangle} \phi -{}^3\check R \phi +({}^3{\check A}^{rs}_{TT}+
{}^3{\check A}^{rs}_L)({}^3{\check A}_{TT,rs}+{}^3{\check A}_{L,rs}) \phi^{-7}
-{2\over 3}({}^3K)^2 \phi^5\, {\buildrel \circ \over =}\, 0,
\label{c5}
\end{equation}

\noindent [the same happens for the superhamiltonian constraint], which is
in general coupled with the equations ${}^4{\bar G}_{lr}\, {\buildrel \circ
\over =}\, 0$

\begin{equation}
{}^3{\check \nabla}_s\, {}^3{\check A}^{rs}_L=({}^3
{\check \triangle}_L\check W)^r={}^3{\check \nabla}_s\, {}^3{\check A}^{rs}\,
{\buildrel \circ \over =}\, {2\over 3}\phi^6\, {}^3{\check \nabla}^r\, {}^3K.
\label{c6}
\end{equation}

The four Einstein equations [ADM constraints] have to be solved in $\phi$ 
[the ``conformal or scale factor"] and in 
the longitudinal part ${}^3{\check A}^{rs}_L$ [${}^4{\check {\tilde \Pi}}^{rs}
_{A,L}$] of ${}^3{\check A}^{rs}$ [${}^3{\check {\tilde \Pi}}^{rs}_A$],
i.e. in the vector ${\check W}^r$ [${\check W}^r_{\pi}$] [see the previous
subsection B)], which is named the ``gravitomagnetic vector potential";
these four functions become functionals of ${}^3{\check g}_{rs}$ 
and of the parts ${}^3{\check A}^{rs}_{TT}$ and ${}^3K$ of ${}^3K_{rs}$
[${}^3{\check {\tilde \Pi}}^{rs}_{A,TT}$ and ${}^3{\tilde \Pi}$ of ${}^3{\tilde
\Pi}^{rs}$]. While ${}^3K$ [${}^3{\tilde \Pi}$], the mean extrinsic curvature, 
is interpreted as the internal extrinsic time conjugated to the momentum $\phi$,
3 components of ${}^3{\check g}_{rs}$ have to be interpreted as conjugate to
the vector ${\check W}^r$; ${}^3{\check A}^{rs}_{TT}$ [the free part of the 
conformally rescaled ``distorsion tensor"] and the remaining two
degrees of freedom hidden in ${}^3{\check g}_{rs}$ are the genuine 
(gravitational wave) physical degrees of freedom in this reduction. For 
constant or vanishing (maximal slicing) ${}^3K$, the supermomentum constraints
decouple from the Lichnerowicz equation. See Ref.\cite{ciuf} for a review on the
existence and unicity of the solutions of Eqs.(\ref{c5}), (\ref{c6}) when
${}^3K=0\, or\, const.$ and Refs.\cite{cho,i1,i2} for the
classification of the known solutions of this equation. The Yamabe theorem is
a fundamental tool in this classification\cite{yama}.

In presence of matter in a closed universe, the conformal current ${}^3j^r$ of
mass-energy has to be orthogonal to the conformal Killing vectors of the
conformal 3-metric (if any). This is the ``condition of confinability" for the 
gravitomagnetic vector potential ${\check W}^r$ \cite{ciuf,my} [it is like in
electrostatic, where, in a closed space, the Poisson equation $\triangle \phi 
=-4\pi \rho$ implies $\int d^3x \sqrt{\gamma} \rho =0$ (i.e. the vanishing of 
the total source charge)\cite{ciuf}].

The previous decomposition suggests to use the variables ${\cal T}=-{4\over 3}
\epsilon k\, {}^3K={2\over {3\sqrt{\gamma}}} \, {}^3{\tilde \Pi}$, 
${\cal P}_{\cal T}=-\sqrt{\gamma}$, ${}^3\sigma_{rs}=
{}^3g_{rs}/ \gamma^{1/3}$, ${}^3{\tilde \Pi}_A^{rs}={}^3{\tilde \Pi}^{rs}-
{1\over 3} {}^3g^{rs}\, {}^3{\tilde \Pi}$, which satisfy the Poisson brackets

\begin{eqnarray}
&&\lbrace {\cal T}(\tau ,\vec \sigma ),{\cal P}_{\cal T}(\tau ,{\vec \sigma}
^{'}\rbrace =-\delta^3(\vec \sigma ,{\vec \sigma}),\nonumber \\
&&\lbrace {}^3\sigma_{rs}(\tau ,\vec \sigma ),{}^3{\tilde \Pi}_A
^{uv}(\tau ,{\vec \sigma}
^{'})\rbrace =[{1\over 2}(\delta^u_r\delta^v_s-\delta^v_r\delta^u_s)-{1\over 3}
\, {}^3\sigma^{uv}\, {}^3\sigma_{rs}](\tau ,\vec \sigma ) \delta^3(\vec \sigma ,
{\vec \sigma}),\nonumber \\
&&\lbrace {}^3{\tilde \Pi}_A^{rs}(\tau ,\vec \sigma ),{}^3{\tilde \Pi}_A
^{uv}(\tau ,{\vec \sigma}^{'})\rbrace ={1\over 3} ({}^3\sigma^{uv}\, 
{}^3{\tilde \Pi}_A^{rs}-{}^3\sigma^{rs}\, {}^3{\tilde \Pi}_A^{uv})
(\tau ,\vec \sigma ) \delta^3(\vec \sigma ,{\vec \sigma}).
\label{c7}
\end{eqnarray}

In Ref.\cite{ise} it is shown that there exists a canonical
basis [${}^3\sigma_{rs}$, ${}^3{\tilde \Pi}^{rs}_{TT}$] hidden in the variables
${}^3\sigma_{rs}$, ${}^3{\tilde \Pi}_A^{rs}$ [but it has never been found
explicitly] and that one can define the 
reduced phase space (the conformal superspace)
${\tilde {\cal S}}$, in which one has gone to the quotient with respect to
the space diffeomorphisms and to the conformal rescalings. It is also shown 
that one can define a ``York map" from this reduced phase space to the subset 
of the standard phase superspace (quotient of the ADM phase space with respect 
to the space diffeomorphisms plus the gauge transformations generated by the
superhamiltonian constraint; it is the phase space of the superspace,
the configuration space obtained from the 3-metrics going to the quotient
with respect to the space- and `time'- diffeomorphisms of the ADM formalism) 
defined by the condition ${}^3K=const.$. 

The ``conformal superspace" ${\tilde {\cal S}}$ may be defined as the space of 
conformal 3-geometries on ``closed" 
manifolds and can be identified in a natural 
way with the space of conformal 3-metrics modulo space diffeomorphisms, or,
equivalently, with the space of Riemannian 3-metrics modulo space 
diffeomorphisms and conformal transformations of the form ${}^3g_{rs}\mapsto
\phi^4\, {}^3g_{rs}$, $\phi > 0$. Instead, the ``ordinary superspace" 
${\cal S}$ is the space of Lorentzian 4-metrics modulo spacetime 
diffeomorphisms. In this way a bridge is built towards the phase superspace, 
which is mathematically connected with the Moncrief splitting
theorem\cite{mon1,cho} valid for closed $\Sigma_{\tau}$ [see however
Ref.\cite{cho} for what is known in the asymptotically flat case by using
weighted Sobolev spaces].

\vfill\eject

\section{3-tensors in the Final Canonical Basis.}

By using the definitions given in I, Eqs.(\ref{VIIa}) imply the following
expressions for the field strengths and curvature tensors of $(\Sigma_{\tau},
{}^3g)$ [we also give their limits for $r_{\bar a}\, \rightarrow \, 0$ and
$q\, \rightarrow \, 0$]

\begin{eqnarray}
{}^3{\hat \Omega}_{rs(a)}&=&
\epsilon_{(a)(b)(c)}\sum_u\delta_{(c)u}\nonumber \\
&&\Big( \delta_{(b)s}e^{{1\over {\sqrt{3}}}\sum_{\bar a}(\gamma_{\bar as}-\gamma
_{\bar au})r_{\bar a}}
\Big[ {1\over {\sqrt{3}}}(\partial_uq+{1\over {\sqrt{3}}}\sum
_{\bar b}\gamma_{\bar bs}\partial_ur_{\bar b})\sum_{\bar c}(\gamma_{\bar cs}-
\gamma_{\bar cu})\partial_rr_{\bar c}+\nonumber \\
&&+\partial_u\partial_rq+{1\over {\sqrt{3}}}\sum_{\bar b}\gamma_{\bar bs}
\partial_u\partial_rr_{\bar b}\Big] -\nonumber \\
&&-\delta_{(b)r}e^{{1\over {\sqrt{3}}}\sum_{\bar a}(\gamma_{\bar ar}-\gamma
_{\bar au})r_{\bar a}}
\Big[ {1\over {\sqrt{3}}}(\partial_uq+{1\over {\sqrt{3}}}\sum
_{\bar b}\gamma_{\bar br}\partial_ur_{\bar b})\sum_{\bar c}(\gamma_{\bar cr}-
\gamma_{\bar cu})\partial_sr_{\bar c}+\nonumber \\
&&+\partial_u\partial_sq+{1\over {\sqrt{3}}}\sum_{\bar b}\gamma_{\bar br}
\partial_u\partial_sr_{\bar b}\Big] \Big) +\nonumber \\
&&+{1\over 2}\sum_{uv}
\Big[ \delta_{(a)(b)}\epsilon_{(c)(d)(e)}-\delta_{(a)(c)}\epsilon
_{(b)(d)(e)}+\delta_{(a)(d)}\epsilon_{(e)(c)(b)}-\delta_{(a)(e)}\epsilon
_{(d)(c)(b)}\Big]\nonumber \\
&&e^{{1\over {\sqrt{3}}}\sum_{\bar a}(\gamma_{\bar ar}+\gamma_{\bar as}-\gamma
_{\bar au}-\gamma_{\bar av})r_{\bar a}}\Big( \partial_uq+{1\over {\sqrt{3}}}\sum
_{\bar b}\gamma_{\bar br}\partial_ur_{\bar b}\Big) \Big( \partial_vq+{1\over 
{\sqrt{3}}}\sum_{\bar c}\gamma_{\bar cs}\partial_vr_{\bar c}\Big)\nonumber \\
&&{}\nonumber \\
&&{\rightarrow}_{r_{\bar a}\rightarrow 0}\, \epsilon_{(a)(b)(c)}\sum_u\delta
_{(c)u}\Big[ \delta_{(b)s} \partial_u\partial_rq-\delta_{(b)r} \partial_u
\partial_sq\Big] +\nonumber \\
&&+{1\over 2}\Big[ \delta_{(a)(b)}\epsilon_{(c)(d)(e)}-\delta_{(a)(c)}\epsilon
_{(b)(d)(e)}+\delta_{(a)(d)}\epsilon_{(e)(c)(b)}-\delta_{(a)(e)}\epsilon
_{(d)(c)(b)}\Big] \partial_uq\partial_vq\, {\rightarrow}_{q\, \rightarrow 0}\, 
0,\nonumber \\
&&{\rightarrow}_{q\, \rightarrow 0}\, {1\over {\sqrt{3}}}\epsilon_{(a)(b)(c)}
\sum_u\delta_{(c)u}\Big( \delta_{(b)s}e^{{1\over {\sqrt{3}}}\sum_{\bar a}(\gamma
_{\bar as}-\gamma_{\bar au})r_{\bar a}}\cdot \nonumber \\
&&\sum_{\bar b}\gamma_{\bar bs}\Big[ \partial_r
\partial_ur_{\bar b}+{1\over {\sqrt{3}}}\partial_ur_{\bar b}\sum_{\bar c}(\gamma
_{\bar cs}-\gamma_{\bar cu})\partial_rr_{\bar c}\Big] -\nonumber \\
&&-\delta_{(b)r}e^{{1\over {\sqrt{3}}}\sum_{\bar a}(\gamma_{\bar ar}-\gamma
_{\bar au})r_{\bar a}}\sum_{\bar b}\gamma_{\bar br}\Big[ \partial_s\partial_ur
_{\bar b}+{1\over {\sqrt{3}}}\partial_ur_{\bar b}\sum_{\bar c}(\gamma
_{\bar cr}-\gamma_{\bar cu})\partial_sr_{\bar c}\Big] \Big) +\nonumber \\
&&+{1\over 6}\sum_{uv}\Big[ \delta_{(a)(b)}\epsilon_{(c)(d)(e)}-\delta_{(a)(c)}
\epsilon_{(b)(d)(e)}+\delta_{(a)(d)}\epsilon_{(e)(c)(b)}-\delta_{(a)(e)}\epsilon
_{(d)(c)(b)}\Big]\nonumber \\
&&e^{{1\over {\sqrt{3}}}\sum_{\bar a}(\gamma_{\bar ar}+\gamma_{\bar as}-\gamma
_{\bar au}+\gamma_{\bar av})r_{\bar a}}\sum_{\bar b}\gamma_{\bar br}\partial_ur
_{\bar b}\sum_{\bar c}\gamma_{\bar cs}\partial_vr_{\bar c},\nonumber \\
&&{}\nonumber \\
{}^3{\hat R}_{rusv}&=&
(\delta_{rv}\delta_{su}-\delta_{rs}\delta_{uv})e^{2(2q+{1\over {\sqrt{3}}}
\sum_{\bar c}(\gamma_{\bar cr}+\gamma_{\bar cu})r_{\bar c})}\nonumber \\
&&\sum_n\Big( \partial_nq
+{1\over {\sqrt{3}}}\sum_{\bar a}\gamma_{\bar ar}\partial_nr_{\bar a}\Big)
\Big( \partial_nq+{1\over {\sqrt{3}}}\sum_{\bar b}\gamma_{\bar bu}\partial_nr
_{\bar b}\Big) +\nonumber \\
&&+e^{2(q+{1\over {\sqrt{3}}}\sum_{\bar c}\gamma_{\bar cr}r_{\bar c})} \Big( 
\delta_{rv}\Big[ \partial_s\partial_uq+{1\over {\sqrt{3}}}\sum_{\bar a}\gamma
_{\bar ar}\partial_s\partial_ur_{\bar a}+\nonumber \\
&&+{1\over {\sqrt{3}}}\Big( \partial_uq+{1\over 
{\sqrt{3}}}\sum_{\bar a}\gamma_{\bar ar}\partial_ur_{\bar a}\Big) \sum_{\bar b}
(\gamma_{\bar br}-\gamma_{\bar bu})\partial_sr_{\bar b}-\nonumber \\
&&-\Big( \partial_uq+{1\over {\sqrt{3}}}\sum_{\bar a}\gamma_{\bar as}\partial_ur
_{\bar a}\Big) \Big( \partial_sq+{1\over {\sqrt{3}}}\sum_{\bar b}\gamma_{\bar 
br}\partial_sr_{\bar b}\Big) \Big]-\delta_{rs}\Big[ \partial_v\partial_uq+{1
\over {\sqrt{3}}}\sum_{\bar ar} \partial_v\partial_ur_{\bar a}+\nonumber \\
&&+{1\over {\sqrt{3}}}\Big( \partial_uq+{1\over {\sqrt{3}}}\sum_{\bar a}\gamma
_{\bar ar}\partial_ur_{\bar a}\Big) \sum_{\bar b}(\gamma_{\bar br}-\gamma
_{\bar bu})\partial_vr_{\bar b}-\nonumber \\
&&-\Big( \partial_uq+{1\over {\sqrt{3}}}\sum_{\bar a}
\gamma_{\bar av}\partial_ur_{\bar a}\Big) \Big( \partial_vq+{1\over {\sqrt{3}}}
\sum_{\bar b}\gamma_{\bar br}\partial_vr_{\bar b}\Big) \Big] \Big) +\nonumber \\
&&+e^{2(q+{1\over {\sqrt{3}}}\sum_{\bar c}\gamma_{\bar cu}r_{\bar c})}\Big(
\delta_{su}\Big[ \partial_v\partial_rq+{1\over {\sqrt{3}}}\sum_{\bar a}\gamma
_{\bar au}\partial_v\partial_rr_{\bar a}+{1\over {\sqrt{3}}}\Big( \partial_rq+
\nonumber \\
&&+{1\over {\sqrt{3}}}\sum_{\bar a}\gamma_{\bar au}\partial_rr_{\bar a}\Big)
\sum_{\bar b}(\gamma_{\bar bu}-\gamma_{\bar br})\partial_vr_{\bar b}-
\nonumber \\
&&-\Big( \partial_rq+{1\over {\sqrt{3}}}\sum_{\bar a}\gamma_{\bar ar}\partial
_rr_{\bar a}\Big) \Big( \partial_vq+{1\over {\sqrt{3}}}\sum_{\bar b}\gamma
_{\bar bu}\partial_vr_{\bar b}\Big) \Big] -\nonumber \\
&-&\delta_{uv}\Big[ \partial_s\partial
_rq+{1\over {\sqrt{3}}}
\sum_{\bar a}\gamma_{\bar au}\partial_s\partial_rr_{\bar a}+\nonumber \\
&&+{1\over {\sqrt{3}}}\Big( \partial_rq+{1\over {\sqrt{3}}}\sum_{\bar a}\gamma
_{\bar au}\partial_rr_{\bar a}\Big) \sum_{\bar b}(\gamma_{\bar bu}-\gamma
_{\bar br})\partial_sr_{\bar b}-\nonumber \\
&&-\Big( \partial_rq+{1\over {\sqrt{3}}}\sum_{\bar a}\gamma
_{\bar as}\partial_rr_{\bar a}\Big) \Big( \partial_sq+{1\over {\sqrt{3}}}\sum
_{\bar b}\gamma_{\bar bu}\partial_sr_{\bar b}\Big) \Big] \Big) \nonumber \\
&&{}\nonumber \\
&&{\rightarrow}_{r_{\bar a}\, \rightarrow 0}\, (\delta_{rv}\delta_{su}-\delta
_{rs}\delta_{uv}) e^{4q} \sum_n (\partial_nq)^2+\nonumber \\
&&+e^{2q}\Big( \delta_{rv}[\partial_s\partial_uq-\partial_sq \partial_uq]-
\delta_{rs}[\partial_v\partial_uq-\partial_vq \partial_uq]+\nonumber \\
&&+\delta_{su}[\partial_v\partial_rq-\partial_vq \partial_rq]-\delta_{uv}
[\partial_s\partial_rq-\partial_sq \partial_rq] \Big) {\rightarrow}_{q\,
\rightarrow 0}\, 0,\nonumber \\
&&{\rightarrow}_{q\, \rightarrow 0}\, {1\over 3}(\delta_{rv}\delta_{su}-\delta
_{rs}\delta_{uv}) e^{ {2\over {\sqrt{3}}}\sum_{\bar c}(\gamma_{\bar cr}+\gamma
_{\bar cu})r_{\bar c}} \sum_n\sum_{\bar a\bar b}\gamma_{\bar ar}\gamma_{\bar bu}
\partial_nr_{\bar a}\partial_nr_{\bar b}+\nonumber \\
&&+{1\over {\sqrt{3}}}e^{{2\over {\sqrt{3}}}\sum_{\bar c}\gamma_{\bar cr}
r_{\bar c}} \sum_{\bar a}\gamma_{\bar ar} \Big( \delta_{rv}\Big[ \partial_s
\partial_ur_{\bar a}+\nonumber \\
&&+{1\over {\sqrt{3}}}\sum_{\bar b}(\gamma_{\bar br}-\gamma_{\bar bu})\partial
_ur_{\bar a}\partial_sr_{\bar b}-{1\over {\sqrt{3}}}\sum_{\bar b}\gamma
_{\bar bs}\partial_sr_{\bar a}\partial_ur_{\bar b}\Big] -\nonumber \\
&&-\delta_{rs}\Big[ \partial_v\partial_ur_{\bar a}+{1\over {\sqrt{3}}}\sum
_{\bar b}(\gamma_{\bar br}-\gamma_{bar bu})\partial_ur_{\bar a}\partial_vr
_{\bar b}-{1\over {\sqrt{3}}}\sum_{\bar b}\gamma_{\bar bv}\partial_vr_{\bar a}
\partial_ur_{\bar b}\Big] \Big) +\nonumber \\
&&+{1\over {\sqrt{3}}} e^{{2\over {\sqrt{3}}}\sum_{\bar c}\gamma_{\bar cu}
r_{\bar c}} \sum_{\bar a}\gamma_{\bar au} \Big( \delta_{su} \Big[ \partial_v
\partial_rr_{\bar a}+\nonumber \\
&&+{1\over {\sqrt{3}}}\sum_{\bar b}(\gamma_{\bar bu}-\gamma_{\bar br})
\partial_rr_{\bar a}\partial_vr_{\bar b}-{1\over {\sqrt{3}}}\sum_{\bar b}
\gamma_{\bar bv}\partial_vr_{\bar a}\partial_rr_{\bar b}\Big] -\nonumber \\
&&-\delta_{uv}\Big[ 
\partial_s\partial_rr_{\bar a}+{1\over {\sqrt{3}}}\sum_{\bar b}
(\gamma_{\bar bu}-\gamma_{\bar br})\partial_rr_{\bar a}\partial_sr_{\bar b}-
{1\over {\sqrt{3}}}\sum_{\bar b}\gamma_{\bar bs}\partial_sr_{\bar a}\partial
_rr_{\bar b}\Big] \Big) ,\nonumber \\
&&{}\nonumber \\
{}^3{\hat R}_{uv}&=&
-\partial_u\partial_vq+{1\over {\sqrt{3}}}\sum_{\bar a}(\gamma_{\bar au}+
\gamma_{\bar av})\partial_u\partial_vr_{\bar a}+\nonumber \\
&&+{1\over 2}\Big[ \Big( \partial_uq+{1\over {\sqrt{3}}}\sum_{\bar a}\gamma
_{\bar av}\partial_ur_{\bar a}\Big) \Big( \partial
_vq-{2\over {\sqrt{3}}}\sum_{\bar b}\gamma_{\bar bu}\partial_vr_{\bar b}\Big) +
\nonumber \\
&&+\Big( \partial_vq+{1\over {\sqrt{3}}}\sum_{\bar a}\gamma_{\bar u}\partial_vr
_{\bar a}\Big) \Big( \partial_uq-{2\over {\sqrt{3}}}\sum_{\bar b}\gamma_{\bar 
bv}r_{\bar b}\Big) \Big] -\nonumber \\
&&-{1\over {2\sqrt{3}}}\sum_n\Big[ \Big( \partial_uq+{1\over {\sqrt{3}}}\sum
_{\bar a}\gamma_{\bar ar}\partial_yr_{\bar a}\Big) \sum_{\bar b}(\gamma
_{\bar bn}-\gamma_{bar bu})\partial_vr_{\bar b}+\nonumber \\
&&+\Big( \partial_vq+{1\over {\sqrt{3}}}\sum_{\bar a}\gamma_{\bar ar}\partial_vr
_{\bar a}\big) \sum_{\bar b}(\gamma_{\bar bn}-\gamma_{\bar bv})\partial_ur
_{\bar b}\Big] -\delta_{uv} e^{2(q+{1\over {\sqrt{3}}}\sum_{\bar c}\gamma
_{\bar cu}r_{\bar c})}\nonumber \\
&&\sum_n \Big( (\partial_nq+{1\over {\sqrt{3}}}\sum_{\bar a}\gamma_{\bar au}
\partial_nr_{\bar a})(2\partial_nq-{1\over {\sqrt{3}}}\sum_{\bar b}\gamma
_{\bar bu}\partial_nr_{\bar b})+\nonumber \\
&&+e^{-2(q+{1\over {\sqrt{3}}}\sum_{\bar c}\gamma
_{\bar cn}r_{\bar c})}\Big[ \partial^2_nq+\nonumber \\
&&+{1\over {\sqrt{3}}}\sum_{\bar a}\gamma
_{\bar au}\partial^2_nr_{\bar a}+{1\over {\sqrt{3}}}(\partial_nq+{1\over 
{\sqrt{3}}}\sum_{\bar au}\partial_nr_{\bar a})\sum_{\bar b}(\gamma_{\bar bu}-2
\gamma_{\bar bn})\partial_nr_{\bar b}\Big] \Big) \nonumber \\
&&{}\nonumber \\
&&{\rightarrow}_{r_{\bar a}\, \rightarrow 0} \, -\partial_u\partial_vq+
\partial_uq \partial_vq-\delta_{uv}e^{2q}\sum_n\Big[ 2e^{2q}(\partial_nq)^2+
\partial^2_nq-(\partial_nq)^2\Big]\, {\rightarrow}_{q\, \rightarrow 0}\, 0,
\nonumber \\
&&{\rightarrow}_{q\, \rightarrow 0}\, {1\over {\sqrt{3}}}\sum_{\bar a}
(\gamma_{\bar au}+\gamma_{\bar av})\partial_u\partial_vr_{\bar a}-{2\over 3}
\sum_{\bar a\bar b}\gamma_{\bar au}\gamma_{\bar bv}\partial_vr_{\bar a}
\partial_ur_{\bar b}-\nonumber \\
&&-{1\over 6}\sum_n\sum_{\bar a\bar b}\gamma_{\bar an}\Big[ (\gamma_{\bar br}-
\gamma_{\bar bu})\partial_ur_{\bar a}\partial_vr_{\bar b}+(\gamma_{\bar bn}-
\gamma_{\bar bv})\partial_vr_{\bar a}\partial_ur_{\bar b}\Big] +\nonumber \\
&&+{1\over {\sqrt{3}}}\delta_{uv}e^{{2\over {\sqrt{3}}}\sum_{\bar c}\gamma
_{\bar cu}r_{\bar c}}\sum_n\Big( {1\over {\sqrt{3}}}\sum_{\bar a\bar b}\gamma
_{\bar au}\gamma_{\bar bu}\partial_nr_{\bar a}\partial_nr_{\bar b}-\nonumber \\
&&-e^{-{2\over {\sqrt{3}}}\sum_{\bar c}\gamma_{\bar cn}r_{\bar c}}\sum_{\bar a}
\gamma_{\bar au}\Big[ \partial^2_nr_{\bar a}+{1\over {\sqrt{3}}}\sum_{\bar b}
(\gamma_{\bar bu}-2\gamma_{\bar bn})\partial_nr_{\bar a}\partial_nr_{\bar b}
\Big] \Big) , \nonumber \\
&&{}\nonumber \\
{}^3\hat R&=&
-\sum_{uv}\Big( (\partial_vq+{1\over {\sqrt{3}}}\sum_{\bar a}\gamma_{\bar au}
\partial_vr_{\bar a})(2\partial_vq-{1\over {\sqrt{3}}}\sum_{\bar b}\gamma
_{\bar bu}\partial_vr_{\bar b})+\nonumber \\
&&+e^{-2(q+{1\over {\sqrt{3}}}\sum_{\bar c}\gamma_{\bar cv}r_{\bar c})}
\Big[ \partial^2_vq+{1\over {\sqrt{3}}}\sum_{\bar a}\gamma_{\bar au}\partial_v
^2r_{\bar a}+\nonumber \\
&&+{2\over {\sqrt{3}}}(\partial_vq+{1\over {\sqrt{3}}}\sum_{\bar a}\gamma
_{\bar au}\partial_vr_{\bar a})\sum_{\bar b}(\gamma_{\bar bu}-\gamma_{\bar bv})
\partial_vr_{\bar b}-\nonumber \\
&&-(\partial_vq+{1\over {\sqrt{3}}}\sum_{\bar a}\gamma_{\bar av}\partial_vr
_{\bar a})(\partial_vq+{1\over {\sqrt{3}}}\sum_{\bar b}\gamma_{\bar bu}
\partial_vr_{\bar b}\Big] \Big) +\nonumber \\
&&+\sum_ue^{-2(q+{1\over {\sqrt{3}}}\sum_{\bar c}\gamma_{\bar cu}r_{\bar c})}
\Big[ -\partial^2_uq+{2\over {\sqrt{3}}}\sum_{\bar a}\gamma_{\bar au} \partial
^2_ur_{\bar a}+\nonumber \\
&&+(\partial_uq+{1\over {\sqrt{3}}}\sum_{\bar a}\gamma_{\bar au}\partial_ur
_{\bar a})(\partial_uq-{2\over {\sqrt{3}}}\sum_{\bar b}\gamma_{\bar bu}
\partial_ur_{\bar b})\Big]\nonumber \\
&&{}\nonumber \\
&&{\rightarrow}_{r_{\bar a}\, \rightarrow 0}\, -6\sum_u(\partial_uq)^2-
4e^{-2q}\sum_u\Big[ \partial_u^2q-(\partial_uq)^2\Big]\, {\rightarrow}_{q\,
\rightarrow 0}\, 0,\nonumber \\
&&{\rightarrow}_{q\, \rightarrow 0}\, -{1\over {\sqrt{3}}}\sum_{uv}\Big(
-{1\over {\sqrt{3}}}\sum_{\bar a\bar b}\gamma_{\bar au}\gamma_{\bar bu}
\partial_vr_{\bar a}\partial_vr_{\bar b}+e^{-{2\over {\sqrt{3}}}\sum_{\bar c}
\gamma_{\bar cv}r_{\bar c}}\sum_{\bar a}\gamma_{\bar au}\cdot \nonumber \\
&&\Big[ \partial^2_vr_{\bar a}+{2\over {\sqrt{3}}}\sum_{\bar b}(\gamma_{\bar 
bu}-\gamma_{\bar bv})\partial_vr_{\bar a}\partial_vr_{\bar b}-{1\over 
{\sqrt{3}}}\sum_{\bar b}\gamma_{\bar bv}\partial_vr_{\bar a}\partial_vr_{\bar 
b}\Big] \Big) +\nonumber \\
&&+{2\over {\sqrt{3}}}\sum_ue^{-{2\over {\sqrt{3}}}\sum_{\bar c}\gamma
_{\bar cu}t_{\bar c}}\sum_{\bar a}\gamma_{\bar au}\Big[ \partial^2_ur_{\bar a}+
{1\over {\sqrt{3}}}\sum_{\bar b}\gamma_{\bar bu}\partial_ur_{\bar a}
\partial_ur_{\bar b}\Big] .
\label{1}
\end{eqnarray}

The Weyl-Schouten tensor \hfill\break
\hfill\break
${}^3C_{rsu}={}^3\nabla_u\, {}^3R_{rs}-{}^3\nabla_s\, 
{}^3R_{ru}-{1\over 4}({}^3g_{rs}\partial_u\, {}^3R-{}^3g_{ru}\partial_s\, 
{}^3R)$ \hfill\break
\hfill\break
of the 3-manifold $(\Sigma_{\tau},{}^3g)$
[see after Eq.(9) of I and Ref.\cite{naka}]
satisfies ${}^3C^r{}_{ru}=0$, ${}^3C_{rsu}=-{}^3C
_{rus}$, ${}^3C_{rsu}+{}^3C_{urs}+{}^3C_{sur}=0$
and has 5 independent components. The related York's conformal 
tensor\cite{york,mtw}  \hfill\break
\hfill\break
${}^3Y^{rs}=\gamma^{1/3}\, \epsilon^{ruv}({}^3R_v{}^s-
{1\over 4} \delta^s_v\, {}^3R){}_{|u}=-{1\over 2} \gamma^{1/3} \epsilon^{ruv}\,
{}^3g^{sm}\, {}^3C_{muv}$ \hfill\break
\hfill\break
[it is a tensor density of weight 5/3 and involves
the third derivatives of the metric]
is symmetric [${}^3Y^{rs}={}^3Y^{sr}$],
traceless [${}^3Y^r{}_r=0$] and transverse [${}^3Y^{rs}{}_{|s}=0$] besides
being invariant under 3-conformal transformations; therefore, it has only 2
independent components [$Y^{rs}=Y^{rs}_{TT}$ according to York's decomposition
of Appendix C] and provides what York calls the pure spin-two
representation of the 3-geometry intrinsic to $\Sigma_{\tau}$. Its explicitly 
symmetric form is the Cotton-York tensor  given by\hfill\break
\hfill\break
${}^3{\cal Y}^{rs}={1\over 2} ({}^3Y^{rs}+{}^3Y^{sr})={1\over 2} \gamma^{1/3} 
(\epsilon^{ruv}\, {}^3g^{sc}+\epsilon^{suv}\, {}^3g^{rc}) {}^3R_{vc|u}=
-{1\over 4}\gamma^{1/3}(\epsilon^{ruv}\, {}^3g^{sm}+\epsilon^{suv}\, {}^3g^{rm})
{}^3C_{muv}$. \hfill\break
\hfill\break
A 3-manifold is conformally flat if and only if either the Weyl-Schouten or the
Cotton-York tensor vanishes \cite{mtw,york,naka}. We have

\begin{eqnarray}
{}^3C_{rsu}&=&{}^3\nabla_u\, {}^3R_{rs}-{}^3\nabla_s\, {}^3R_{ru}-{1\over 4}
({}^3g_{rs}\partial_u\, {}^3R-{}^3g_{ru}\partial_s\, {}^3R)
\mapsto \nonumber \\
&&\mapsto {}^3{\hat C}_{rsu}={}^3{\hat R}_{rs|u}-{}^3{\hat R}_{ru|s}-{1\over 4}
e^{2q_r}(\delta_{rs}\partial_u\, {}^3\hat R-\delta_{ru}\partial_s\, {}^3\hat R)
,\nonumber \\
&&{}\nonumber \\
{}^3{\cal Y}_{mn} &=&{1\over 2}\gamma^{1/3}\sum_{rsu} 
(\epsilon_{mur}\, {}^3g_{ns}+
\epsilon_{nur}\, {}^3g_{ms}) {}^3R_{rs|u} \mapsto \nonumber \\
&&\mapsto {}^3{\hat {\cal Y}}_{mn}={1\over 2}e^{{2\over 3}\sum_vq_v}
\sum_{rsu}e^{-2q_s}(\epsilon_{mur}\delta_{ns}+\epsilon_{nur}\delta_{ms})
{}^3{\hat R}_{rs|u}=\nonumber \\
&&={1\over 2}e^{2q}\sum_{rsu}e^{-2q_s}
(\epsilon_{mur}\delta_{ns}+\epsilon_{nur}\delta_{ms})\cdot \nonumber \\
&&\Big( \, \partial_u\partial_r\partial_s(q_r+q_s-\sum_tq_t)-
\partial_u(q_r+q_s)\partial_r\partial_s(q_r+q_s-\sum_tq_t)-\nonumber \\
&&-\partial_rq_u \partial_u\partial_s(q_u+q_s-\sum_tq_t)-\partial_sq_u 
\partial_u\partial_r(q_u+q_r-\sum_tq_t)-\nonumber \\
&&-{1\over 2}\partial_u[\partial_rq_s\partial_s(2q_r-\sum_tq_t)+\partial_sq_r
\partial_r(2q_s-\sum_tq_t)]-\nonumber \\
&&-{1\over 2}\sum_n \partial_u[\partial_rq_n\partial_s(q_n-q_r)+\partial_sq_n
\partial_r(q_n-q_s)]+\nonumber \\
&&+{1\over 2}\partial_u(q_r+q_s) [\partial_rq_s\partial_s(2q_r-\sum_tq_t)+
\partial_sq_r \partial_r(2q_s-\sum_tq_t)]+\nonumber \\
&&+{1\over 2}\partial_rq_u [\partial_uq_s \partial_s(2q_u-\sum_tq_t)+
\partial_sq_u \partial_u(2q_s-\sum_tq_t)]+\nonumber \\
&&+{1\over 2}\partial_sq_u [\partial_uq_r \partial_r(2q_u-\sum_tq_t)+
\partial_rq_u \partial_u(2q_r-\sum_tq_t)]+\nonumber \\
&&+{1\over 2}\partial_u(q_r+q_s) \sum_n[\partial_rq_n \partial_s(q_n-q_r)+
\partial_sq_n \partial_r(q_n-q_s)]+\nonumber \\
&&+{1\over 2}\partial_rq_u \sum_n[\partial_uq_n \partial_s(q_n-q_u)+
\partial_sq_n \partial_u(q_n-q_s)]+\nonumber \\
&&+{1\over 2}\partial_sq_u \sum_n[\partial_uq_n \partial_r(q_n-q_u)+
\partial_rq_n \partial_u(q_n-q_r)]+\nonumber \\
&&+\delta_{rs}\, e^{2q_r}\sum_n 
\Big[ 2\partial_uq_r[\partial_nq_r\partial_n(q_r-
\sum_tq_t)-e^{-2q_n}(\partial^2_nq_r+\partial_nq_r\partial_n(q_r-2q_n))]+
\nonumber \\
&&+\partial_u[\partial_nq_r\partial_n(q_r-\sum_tq_t)]+e^{-2q_n}[2\partial_uq_n
(\partial_n^2q_r+\partial_nq_r\partial_n(q_r-2q_n))-\nonumber \\
&&-\partial_u(\partial_n^2q_r+\partial_nq_r\partial_n(q_r-2q_n))]-\nonumber \\
&&-2\partial_uq_r[\partial_nq_r\partial_n(q_r-\sum_tq_t)-e^{-2q_n}(\partial^2_n
q_r+\partial_nq_r\partial_n(q_r-2q_n))] \Big] +\nonumber \\
&&+\delta_{ru}\, e^{2q_u}\Big[ \sum_ve^{-2q_v} \partial_vq_u \Big( \partial_v
\partial_s(q_v+q_s-\sum_tq_t)-\nonumber \\
&&-{1\over 2}[\partial_vq_s\partial_s(2q_v-\sum_tq_t)+\partial_sq_v\partial_v
(2q_s-\sum_tq_t)]-\nonumber \\
&&-{1\over 2}\sum_n[\partial_vq_n\partial_s(q_n-q_v)+
\partial_sq_n\partial_v(q_n-q_s)] \Big) +\nonumber \\
&&+\sum_n (\partial_sq_u [\partial_nq_s\partial_n(q_s-\sum_tq_t)-
e^{-2q_n}(\partial_n^2q_s+\partial_nq_s\partial_n(q_s-2q_n))]-\nonumber \\
&&-\partial_sq_u[\partial_nq_r\partial_n(q_r-\sum_tq_t)-e^{-2q_n}(\partial_n
^2q_r+\partial_nq_r\partial_n(q_r-2q_n))] ) \Big] +\nonumber \\
&&+\delta_{su}\, e^{2q_u} \Big[ \sum_ve^{-2q_v} \partial_vq_u 
\Big( \partial_v\partial_r(q_v+q_r-\sum_tq_t)-\nonumber \\
&&-{1\over 2}[\partial_vq_r\partial_r(2q_v-\sum_tq_t)+\partial_rq_v\partial_v
(2q_r-\sum_tq_t)]-\nonumber \\
&&-{1\over 2}\sum_n[\partial_vq_n\partial_r(q_n-q_v)+
\partial_rq_n\partial_v(q_n-q_r)] \Big) +\nonumber \\
&&+\sum_n (\partial_rq_u [\partial_nq_r\partial_n(q_r-\sum_tq_t)-
e^{-2q_n}(\partial_n^2q_r+\partial_nq_r\partial_n(q_r-2q_n)]-\nonumber \\
&&-\partial_rq_u[\partial_nq_s\partial_n(q_s-\sum_tq_t)-e^{-2q_n}(\partial_n
^2q_s+\partial_nq_s\partial_n(q_s-2q_n))] ) \Big] \, \Big) \nonumber \\
&&{}\nonumber \\
&&{\rightarrow}_{r_{\bar a}\, \rightarrow 0}\, {1\over 2}\sum_{rsu}(\epsilon
_{mur}\delta_{ns}+\epsilon_{nur}\delta_{ms})\cdot \nonumber \\
&&-\partial_u\partial_r\partial_sq
+2(\partial_uq \partial_r\partial_sq+\partial_rq \partial_s\partial_uq+
\partial_sq \partial_u\partial_rq)-4\partial_uq \partial_rq \partial_sq+
\nonumber \\
&&+\delta_{rs}\sum_n \Big(
\partial_u[\partial_n^2q-(\partial_nq)^2]+2\partial_uq
[\partial^2_nq-(\partial_nq)^2]-2e^{2q}\partial_u(\partial_nq)^2 \Big) -
\nonumber \\
&&-\delta_{ru}\sum_v\partial_vq (\partial_v\partial_sq-\partial_vq \partial_sq)-
\delta_{su}\sum_v\partial_vq (\partial_v\partial_rq-\partial_vq \partial_rq)=
0,\nonumber \\
&&{}\nonumber \\
&&{\rightarrow}_{q\, \rightarrow 0}\, {1\over 2}e^{2q}\sum_{rsu}e^{-2q_s}
(\epsilon_{mur}\delta_{ns}+\epsilon_{nur}\delta_{ms})\cdot \nonumber \\
&&\Big( \, {1\over {\sqrt{3}}}\sum_{\bar a}(\gamma_{\bar ar}+\gamma_{\bar as})
\partial_u\partial_r\partial_sr_{\bar a}-{1\over 3}\sum_{\bar a\bar b}(\gamma
_{\bar ar}+\gamma_{\bar as})(\gamma_{\bar br}+\gamma_{\bar bs})\partial_ur
_{\bar a}\partial_r\partial_sr_{\bar b}-\nonumber \\
&&-{1\over 3}\sum_{\bar a\bar b}\gamma_{\bar au}[(\gamma_{\bar bu}+\gamma
_{\bar bs})\partial_rr_{\bar a}\partial_u\partial_sr_{\bar b}+(\gamma_{\bar bu}+
\gamma_{\bar br})\partial_sr_{\bar a}\partial_u\partial_rr_{\bar b}]-
\nonumber \\
&&-{2\over 3}\sum_{\bar a\bar b}\gamma_{\bar as}\gamma_{\bar br}\partial_u
[\partial_rr_{\bar a}\partial_sr_{\bar b}]-\nonumber \\
&&-{1\over 6}\sum_{\bar a\bar b}\sum_n\gamma_{\bar an} \partial_u
[(\gamma_{\bar bn}-\gamma_{\bar br})\partial_rr_{\bar a}\partial_sr_{\bar b}+
(\gamma_{\bar bn}-\gamma_{\bar bs})\partial_sr_{\bar a}\partial_rr_{\bar b}]+
\nonumber \\
&&+{2\over {3\sqrt{3}}}\sum_{\bar a\bar b\bar c}(\gamma_{\bar ar}+\gamma
_{\bar s})(\gamma_{\bar bu}\gamma_{\bar cu}+\gamma_{\bar bs}\gamma_{\bar cr})
\partial_ur_{\bar a}\partial_rr_{\bar b}\partial_sr_{\bar c}+\nonumber \\
&&+{1\over {6\sqrt{3}}}\sum_{\bar a\bar b\bar c}\sum_n\gamma_{\bar bn} \Big[
(\gamma_{\bar ar}+\gamma_{\bar as})\partial_ur_{\bar a}[(\gamma_{\bar cn}-
\gamma_{\bar cr})\partial_rr_{\bar b}\partial_sr_{\bar c}+(\gamma_{\bar cn}-
\gamma_{\bar cs})\partial_sr_{\bar b}\partial_rr_{\bar c}]+\nonumber \\
&&+\gamma_{\bar au} (\, \partial_rr_{\bar a}[(\gamma_{\bar cn}-\gamma
_{\bar cu})\partial_ur_{\bar b}\partial_sr_{\bar c}+(\gamma_{\bar cn}-\gamma
_{\bar cs})\partial_sr_{\bar b}\partial_ur_{\bar c}]+\nonumber \\
&&+\partial_sr_{\bar a}[(\gamma_{\bar cn}-\gamma_{\bar cu})\partial_ur_{\bar b}
\partial_rr_{\bar c}+(\gamma_{\bar cn}-\gamma_{\bar cr})\partial_rr_{\bar b}
\partial_ur_{\bar c}]\, )\, \Big] +\nonumber \\
&&+{1\over {3\sqrt{3}}}\delta_{rs}e^{{2\over {\sqrt{3}}}\sum_{\bar e}\gamma
_{\bar er}r_{\bar e}} \sum_n\sum_{\bar a} \Big( 2\gamma_{\bar ar}\partial_ur
_{\bar a}\sum_{\bar b\bar c}\gamma_{\bar br}\gamma_{\bar cr}\partial_nr_{\bar b}
\partial_nr_{\bar c}-\nonumber \\
&&-e^{-{2\over {\sqrt{3}}}\sum_{\bar d}\gamma_{\bar dn}r_{\bar d}}\gamma
_{\bar ar}[\, \sqrt{3} \partial^2_nr_{\bar a}+(\gamma_{\bar br}-2\gamma
_{\bar bn}\partial_nr_{\bar a}\partial_nr_{\bar b})]+\nonumber \\
&&+\sqrt{3}\sum_{\bar b}
\gamma_{\bar ar}\gamma_{\bar br}\partial_u[\partial_nr_{\bar a}\partial_nr
_{\bar b}]+\nonumber \\
&&+e^{-{2\over {\sqrt{3}}}\sum_{\bar d}\gamma_{\bar dn}r_{\bar d}}[2\gamma
_{\bar an}\partial_ur_{\bar a}\sum_{\bar b}\gamma_{\bar br}(\sqrt{3}
\partial_n^2r_{\bar B}+\nonumber \\
&&+\sum_{\bar c}(\gamma_{\bar cr}-2\gamma_{\bar cn})\partial_nr_{\bar b}
\partial_nr_{\bar c})-\gamma_{\bar ar}\partial_u(\sqrt{3}\partial_n^2r_{\bar a}
+\sum_{\bar b}(\gamma_{\bar br}-2\gamma_{\bar bn})\partial_nr_{\bar a}
\partial_nr_{\bar b})]-\nonumber \\
&&-2\gamma_{\bar ar}\partial_ur_{\bar a} \sum_{\bar b}[\sum_{\bar c}\gamma
_{\bar br}\gamma_{\bar cr}\partial_nr_{\bar b}\partial_nr_{\bar c}-\nonumber \\
&&-e^{-{2\over {\sqrt{3}}}\sum_{\bar d}\gamma_{\bar dn}r_{\bar d}}\gamma
_{\bar br}(\sqrt{3}\partial_n^2r_{\bar b}+\sum_{\bar c}(\gamma_{\bar cr}-
2\gamma_{\bar cn})\partial_nr_{\bar b}\partial_nr_{\bar c})]\, \Big) 
+\nonumber \\
&&+{1\over 3}\delta_{ru}e^{{2\over {\sqrt{3}}}\sum_{\bar e}\gamma_{\bar eu}r
_{\bar e}}\, \sum_{\bar a} \Big( \sum_ve^{-{2\over {\sqrt{3}}}\sum_{\bar d}
\gamma_{\bar dv}r_{\bar d}}\sum_{\bar b}\gamma_{\bar au}\partial_vr_{\bar a}
\cdot \nonumber \\
&&\Big[ 
(\gamma_{\bar bv}+\gamma_{\bar bs})\partial_v\partial_sr_{\bar b}-{2\over
{\sqrt{3}}}\sum_{\bar c}\gamma_{\bar bs}\gamma_{\bar cv}\partial_vr_{\bar b}
\partial_sr_{\bar c}-\nonumber \\
&&-{1\over {2\sqrt{3}}}\sum_n\gamma_{\bar bn}\sum_{\bar c}[(\gamma_{\bar cn}-
\gamma_{\bar cv})\partial_vr_{\bar b}\partial_sr_{\bar c}+(\gamma_{\bar cn}-
\gamma_{\bar cs})\partial_sr_{\bar b}\partial_vr_{\bar c}] \Big] +\nonumber \\
&&+{1\over {\sqrt{3}}}\sum_n( \gamma_{\bar au}\partial_sr_{\bar a}[\sum
_{\bar b\bar c}\gamma_{\bar bs}\gamma_{\bar cs}\partial_nr_{\bar b}\partial_nr
_{\bar c}-e^{-{2\over {\sqrt{3}}}\sum_{\bar d}\gamma_{\bar dn}r_{\bar d}}
\sum_{\bar b}\gamma_{\bar bs}(\sqrt{3}\partial_n^2r_{\bar b}+\nonumber \\
&&+\sum_{\bar c}(\gamma_{\bar cs}-2\gamma_{\bar cn})\partial_nr_{\bar b}
\partial_nr_{\bar c})]-\gamma_{\bar au}\partial_sr_{\bar a}\sum_{\bar b}
[\sum_{\bar c}\gamma_{\bar br}\gamma_{\bar cr}\partial_nr_{\bar b}\partial_nr
_{\bar c}-\nonumber \\
&&-e^{-{2\over {\sqrt{3}}}\sum_{\bar d}\gamma_{\bar dn}r_{\bar d}}\gamma
_{\bar br}(\sqrt{3}\partial_n^2r_{\bar b}+\sum_{\bar c}(\gamma_{\bar cr}-
2\gamma_{\bar cn})\partial_nr_{\bar b}\partial_nr_{\bar c})]) \} +
\nonumber \\
&&+{1\over 3}\delta_{su}e^{{2\over {\sqrt{3}}}\sum_{\bar e}\gamma_{\bar eu}r
_{\bar e}}\, \sum_{\bar a} \Big( \sum_ve^{-{2\over {\sqrt{3}}}\sum_{\bar d}
\gamma_{\bar dv}r_{\bar d}}\sum_{\bar b}\gamma_{\bar au}\partial_vr_{\bar a}
\cdot \nonumber \\
&&\Big[ 
(\gamma_{\bar bv}+\gamma_{\bar br})\partial_v\partial_rr_{\bar b}-{2\over
{\sqrt{3}}}\sum_{\bar c}\gamma_{\bar br}\gamma_{\bar cv}\partial_vr_{\bar b}
\partial_sr_{\bar c}-\nonumber \\
&&-{1\over {2\sqrt{3}}}\sum_n\gamma_{\bar bn}\sum_{\bar c}[(\gamma_{\bar cn}-
\gamma_{\bar cv})\partial_vr_{\bar b}\partial_rr_{\bar c}+(\gamma_{\bar cn}-
\gamma_{\bar cr})\partial_rr_{\bar b}\partial_vr_{\bar c}] \Big] +\nonumber \\
&&+{1\over {\sqrt{3}}}\sum_n( \gamma_{\bar au}\partial_rr_{\bar a}[\sum
_{\bar b\bar c}\gamma_{\bar br}\gamma_{\bar cr}\partial_nr_{\bar b}\partial_nr
_{\bar c}-e^{-{2\over {\sqrt{3}}}\sum_{\bar d}\gamma_{\bar dn}r_{\bar d}}
\sum_{\bar b}\gamma_{\bar br}(\sqrt{3}\partial_n^2r_{\bar b}+\nonumber \\
&&+\sum_{\bar c}(\gamma_{\bar cr}-2\gamma_{\bar cn})\partial_nr_{\bar b}
\partial_nr_{\bar c})]-\gamma_{\bar au}\partial_rr_{\bar a}\sum_{\bar b}
[\sum_{\bar c}\gamma_{\bar bs}\gamma_{\bar cs}\partial_nr_{\bar b}\partial_nr
_{\bar c}-\nonumber \\
&&-e^{-{2\over {\sqrt{3}}}\sum_{\bar d}\gamma_{\bar dn}r_{\bar d}}\gamma
_{\bar bs}(\sqrt{3}\partial_n^2r_{\bar b}+\sum_{\bar c}(\gamma_{\bar cs}-
2\gamma_{\bar cn})\partial_nr_{\bar b}\partial_nr_{\bar c})]) \Big) 
\,\, \Big) .
\label{z3}
\end{eqnarray}

\noindent Since the condition $r_{\bar a}=0$ corresponds to conformally flat 
3-manifolds $\Sigma_{\tau}$, the Cotton-York conformal tensor vanishes
in the limit $r_a \rightarrow 0$.

\vfill\eject

\section{4-tensors in the Final Canonical Basis.}

From the results of Appendix B of I and Eq.(\ref{VII8}) for ${}^3{\hat {\tilde 
\pi}}^r_{(a)}$, we get the following expressions for the reconstruction of
4-tensors on $M^4$ [here $N=N_{(as)}+n$ and $N_{(a)}={}^3{\hat e}^r_{(a)}N_r=
{}^3{\hat e}^r_{(a)} [N_{(as) r}+n_r]=e^{-q}\sum_r\delta^r_{(a)} e^{-{1\over 
{\sqrt{3}}} \sum_{\bar a}\gamma_{\bar ar}r_{\bar a}}[N_{(as)r}+n_r]$ 
[$n_r=0$ as after Eq.(\ref{VII1}) will no more hold after the addition of 
surface terms to the Dirac Hamiltonian to make it differentiable, see
Ref.\cite{russo3}; when $n_r\not= 0$, one has to replace $N_{(as)r}$ with
$N_{(as)r}+n_r$ in the following formulas]
are the total lapse and shift functions; only in the
Christoffel symbols we shall put the explicit expression of $N_{(a)}$]:

\begin{eqnarray}
{}^4{\hat \Gamma}^{\tau}_{\tau\tau}\, &=&{1\over N} \Big[ \partial_{\tau}N+
e^{-2q}\sum_re^{-{2\over {\sqrt{3}}}\sum_{\bar a}\gamma_{\bar ar}r_{\bar a}}
N_{(as)r} \partial_rN-\nonumber \\
&-&{{\epsilon}\over {4k}}e^{-4q}\, {}^3G_{o(a)(b)(c)(d)} \sum_{mn} e^{-{1\over
{\sqrt{3}}}\sum_{\bar a}(\gamma_{\bar am}+\gamma_{\bar an})r_{\bar a}}
\nonumber \\
&&\delta_{(a)m}\delta_{(b)n}N_{(as)m}N_{(as)n} \sum_u\delta_{(c)u}e^{{1\over 
{\sqrt{3}}}\sum_{\bar a}\gamma_{\bar au}r_{\bar a}}\, {}^3{\hat {\tilde \pi}}^u
_{(d)}\Big] =\nonumber \\
&&=_{N_{(a)}=0\quad\quad} {1\over {N}} \partial_{\tau}N ,\nonumber \\
{}^4{\hat \Gamma}^{\tau}_{r\tau}&=&{}^4{\hat \Gamma}^{\tau}_{\tau r}={1\over N}
\Big[ \partial_rN-{{\epsilon}\over {4k}}\, {}^3G_{o(a)(b)(c)(d)} \delta_{(a)r}
\nonumber \\
&&\sum_{su} e^{{1\over {\sqrt{3}}}\sum_{\bar a}(\gamma_{\bar ar}-\gamma_{\bar 
as}+\gamma_{\bar au})r_{\bar a}} \delta_{(b)s}N_{(as)s}\delta_{(c)u} \,
{}^3{\hat {\tilde \pi}}^u_{(d)} \Big] =\nonumber \\
&&=_{N_{(a)}=0\quad\quad}{1\over {N}} \partial_rN,\nonumber \\
{}^4{\hat \Gamma}^{\tau}_{rs}&=&{}^4{\hat \Gamma}^{\tau}_{sr}=-{{\epsilon}\over
{4kN}} \sum_ue^{{1\over {\sqrt{3}}}\sum_{\bar a}(\gamma_{\bar ar}+\gamma
_{\bar as}+\gamma_{\bar au})r_{\bar a}} {}^3G_{o(a)(b)(c)(d)}\delta_{(a)r}
\delta_{(b)s}\delta_{(c)u}\, {}^3{\hat {\tilde \pi}}^u_{(d)},\nonumber \\
{}^4{\hat \Gamma}^u_{\tau\tau}\, &=&\Big[ \partial_{\tau}\Big( e^{-q-{1\over
{\sqrt{3}}}\sum_{\bar a}\gamma_{\bar au}r_{\bar a}}N_{(as)u}\Big)-\nonumber \\
&-&{{\partial
_{\tau}N}\over N}e^{-q-{1\over {\sqrt{3}}}\sum_{\bar a}\gamma_{\bar au}r_{\bar 
a}} N_{(as)u}\Big] e^{-q-{1\over {\sqrt{3}}}\sum_{\bar a}\gamma_{\bar au}
r_{\bar a}}+\nonumber \\
&+&Ne^{-2q}\Big( \delta_{(a)(b)}-e^{-2q}\sum_{rs}e^{-{1\over {\sqrt{3}}}\sum
_{\bar a}(\gamma_{\bar ar}+\gamma_{\bar as})r_{\bar a}} \delta_{(a)r}\delta
_{(b)s}{{N_{(as)r}N_{(as)s}}\over {N^2}}\Big) \nonumber \\
&&\sum_ve^{-{1\over {\sqrt{3}}}
\sum_{\bar a}(\gamma_{\bar au}+\gamma_{\bar av})r_{\bar a}} \delta^u_{(a)}
\delta^v_{(b)}\partial_vN+\nonumber \\
&+&e^{-2q} \sum_v e^{-{2\over {\sqrt{3}}}\sum_{\bar a}\gamma_{\bar av}r_{\bar 
a}} N_{(as)v} (e^{-2q-{2\over {\sqrt{3}}}\sum_{\bar a}\gamma_{\bar au}r_{\bar
a}} N_{(as)u}\, )_{|v}-\nonumber \\
&-&{{\epsilon N}\over {2k}}e^{-4q}\, {}^3G_{o(a)(b)(c)(d)}\sum_w e^{-{1\over
{\sqrt{3}}}\sum_{\bar a}\gamma_{\bar aw}r_{\bar a}}\delta_{(a)w}N_{(as)w}
(\delta_{(a)(b)}-\nonumber \\
&-&e^{-2q}\sum_{rs}e^{-{1\over {\sqrt{3}}}\sum_{\bar a}(\gamma_{\bar ar}+\gamma
_{\bar as})r_{\bar a}}{{N_{(as)r}N_{(as)s}}\over {2N^2}}) \sum_v\delta^u_{(a)}
e^{-{1\over {\sqrt{3}}}\sum_{\bar a}(\gamma_{\bar au}-\gamma_{\bar av})r_{\bar 
a}} \delta_{(c)v}\, {}^3{\hat {\tilde \pi}}^v_{(d)}-\nonumber \\
&-&e^{-2q}\sum_re^{-{1\over {\sqrt{3}}}\sum_{\bar a}(\gamma_{\bar ar}+\gamma
_{\bar au})r_{\bar a}}N_{(as)r} \nonumber \\
&&\Big[ {{\epsilon N}\over {4k}}e^{-2q}\delta
_{(a)r}\delta^u_{(b)}\, {}^3G_{o(a)(b)(c)(d)}\sum_se^{{1\over {\sqrt{3}}}
\sum_{\bar a}\gamma_{\bar as}r_{\bar a}}\delta_{(c)s}\, {}^3{\hat {\tilde \pi}}
^s_{(d)}+\nonumber \\
&+&e^{-q-{1\over {\sqrt{3}}}\sum_{\bar a}\gamma_{\bar ar}r_{\bar a}} \Big(
e^{-2q}\sum_we^{-{2\over {\sqrt{3}}}\sum_{\bar a}\gamma_{\bar aw}r_{\bar a}}
N_{(as)w}\partial_w[e^{q+{1\over {\sqrt{3}}}\sum_{\bar a}\gamma_{\bar ar}r
_{\bar a}}]+\nonumber \\
&+&e^{q+{1\over {\sqrt{3}}}\sum_{\bar a}\gamma_{\bar au}r_{\bar a}} \partial_r
[e^{-2q-{2\over {\sqrt{3}}}\sum_{\bar a}\gamma_{\bar au}r_{\bar a}}N_{(as)u}]
\Big) \Big] =\nonumber \\
&&=_{N_{(a)}=0\quad\quad}
N e^{-2(q+{1\over {\sqrt{3}}}\sum_{\bar a}\gamma_{\bar au}r_{\bar a})}
\partial_uN,\nonumber \\
{}^4{\hat \Gamma}^u_{r\tau}&=&{}^4{\hat \Gamma}^u_{\tau r}=
e^{-q-{1\over {\sqrt{3}}}\sum_{\bar a}\gamma_{\bar au}r_{\bar a}} \Big[ (e^{-q}
\sum_se^{-{1\over {\sqrt{3}}}\sum_{\bar a}\gamma_{\bar as}r_{\bar a}}\delta
_{(a)s}N_{(as)s} )_{| r} -\nonumber \\
&-&e^{-q}\sum_s e^{-{1\over {\sqrt{3}}}\sum_{\bar a}\gamma_{\bar as}r_{\bar a}}
\delta_{(a)s}N_{(as)s} {{\partial_rN}\over N} \Big] -\nonumber \\
&-&{{\epsilon N}\over {4k}}e^{-2q}(\delta_{(a)(b)}-e^{-2q}\sum_{rs}e^{-{1\over
{\sqrt{3}}}\sum_{\bar a}(\gamma_{\bar ar}+\gamma_{\bar as})r_{\bar a}}\delta
_{(a)r}\delta_{(b)s}{{N_{(as)r}N_{(as)s}}\over {N^2}} )\nonumber \\
&&e^{-{1\over {\sqrt{3}}}\sum_{\bar a}\gamma_{\bar au}r_{\bar a}}\delta^u_{(a)}
\, {}^3G_{o(b)(c)(d)(e)}\sum_w e^{{1\over {\sqrt{3}}}\sum_{\bar a}(\gamma
_{\bar ar}+\gamma_{\bar aw})r_{\bar a}}\delta_{(c)r}\delta_{(d)w}\, {}^3{\hat 
{\tilde \pi}}^w_{(e)}=\nonumber \\
&&=_{N_{(a)}=0\quad\quad}-
{{\epsilon N}\over {4k}}\sum_{uv}e^{-2q+{1\over {\sqrt{3}}}\sum_{\bar a}
(\gamma_{\bar ar}+\gamma_{\bar av}+\gamma_{\bar au})r_{\bar a}}
{}^3G_{o(a)(b)(c)(d)}\delta^u_{(a)}\delta
_{(b)r}\delta_{(c)v}\, {}^3{\hat {\tilde \pi}}^v_{(d)},\nonumber \\
{}^4{\hat \Gamma}^u_{rs}&=&{}^3{\hat \Gamma}^u_{rs}+{{\epsilon}\over {4k}}
e^{-2q-{2\over {\sqrt{3}}}\sum_{\bar a}\gamma_{\bar au}r_{\bar a}}N_{(as)u}\,
{}^3G_{o(a)(b)(c)(d)}\nonumber \\
&&\sum_v e^{{1\over {\sqrt{3}}}\sum_{\bar a}(\gamma_{\bar ar}+\gamma_{\bar as}
+\gamma_{\bar av})r_{\bar a}}\delta_{(a)r}\delta_{(b)s}\delta_{(c)v}\,
{}^3{\hat {\tilde \pi}}^v_{(d)}=\nonumber \\
&&=_{N_{(a)}=0\quad\quad}{}^3{\hat \Gamma}^u_{rs},\nonumber \\
&&{}\nonumber \\
{}^4{\hat {\buildrel \circ \over {\omega}}}_{\tau (o)(a)}&=&-{}^4{\hat
{\buildrel \circ \over {\omega}}}_{\tau (a)(o)}=-\epsilon \sum_r\delta_{(a)r}
e^{-(q+{1\over {\sqrt{3}}}\sum_{\bar a}\gamma_{\bar ar}r_{\bar a})}
\partial_rN-\nonumber \\
&-&{{e^{-2q}}\over {4k}} {}^3G_{o(a)(b)(c)(d)}N_{(b)} \sum_u e^{{1\over 
{\sqrt{3}}}\sum_{\bar a}\gamma_{\bar au}r_{\bar a}}\delta_{(c)u}\, {}^3{\hat 
{\tilde \pi}}^u_{(d)}=\nonumber \\
&&=_{N_{(a)}=0\quad\quad} -\epsilon \sum_r\delta_{(a)r}
e^{-(q+{1\over {\sqrt{3}}}\sum_{\bar a}\gamma_{\bar ar}r_{\bar a})}
\partial_rN,\nonumber \\ 
{}^4{\hat {\buildrel \circ \over {\omega}}}_{\tau (a)(b)}&=&-{}^4{\hat
{\buildrel \circ \over {\omega}}}_{\tau (b)(a)}\,
{\buildrel \circ \over =}\,-\epsilon \sum_r e^{-q}\, {}^3{\hat \omega}
_{r(a)(b)}  e^{-{1\over {\sqrt{3}}}\sum_{\bar a}\gamma_{\bar ar}r_{\bar a}}
\delta^r_{(c)} N_{(c)} =\nonumber \\
&&=_{N_{(a)}=0\quad\quad} 0,\nonumber \\
{}^4{\hat {\buildrel \circ \over {\omega}}}_{r(o)(a)}&=&-{}^4{\hat
{\buildrel \circ \over {\omega}}}_{r(a)(o)}=-{1\over {4k}} \sum_u
e^{-q+{1\over {\sqrt{3}}}\sum_{\bar a}(\gamma_{\bar ar}+\gamma_{\bar au})r
_{\bar a}} {}^3G_{o(a)(b)(c)(d)}\delta_{(b)r}\delta_{(c)u}\, {}^3{\hat
{\tilde \pi}}^u_{(d)},\nonumber \\
{}^4{\hat {\buildrel \circ \over {\omega}}}_{r(a)(b)}&=&-{}^4{\hat
{\buildrel \circ \over {\omega}}}_{r(b)(a)}=-\epsilon
{}^3{\hat \omega}_{r(a)(b)},\nonumber \\
&&{}\nonumber \\
{}^4{\hat {\buildrel \circ \over {\Omega}}}_{rs(a)(b)}&=&-\epsilon 
\, \Big[ {}^3{\hat \Omega}_{rs(a)(b)}+
{{e^{-2q}}\over {4k}}\, {}^3G_{o(a)(c)(d)(e)}\, {}^3G_{o(b)(f)(g)(h)}\cdot
\nonumber \\
&&\sum_{uv}
e^{{1\over {\sqrt{3}}}\sum_{\bar a}(\gamma_{\bar ar}+\gamma_{\bar as}+\gamma
_{\bar au}+\gamma_{\bar av})r_{\bar a}}
(\delta_{(c)r}\delta_{(f)s}-\delta_{(c)s}\delta_{(f)r}) \delta_{(d)u}
\, {}^3{\hat {\tilde \pi}}^u_{(e)}\, \delta_{(g)v}\, {}^3{\hat {\tilde \pi}}^v
_{(h)}\Big] ,\nonumber \\
{}^4{\hat {\buildrel \circ \over {\Omega}}}_{rs(o)(a)}&=&{1\over N}\sum_v
e^{-(q+{1\over {\sqrt{3}}}\sum_{\bar a}\gamma_{\bar av}r_{\bar a})} \delta
_{(a)v}\Big( {}^4{\hat R}_{\tau vrs}-N_{(b)}\sum_ue^{-(q+{1\over {\sqrt{3}}}\sum
_{\bar b}\gamma_{\bar bu}r_{\bar b}}\delta_{(b)u}\, {}^4{\hat R}_{uvrs}\Big) =
\nonumber \\
&=&{1\over {4k}} \sum_ue^{-(q+{1\over {\sqrt{3}}}\sum_{\bar a}\gamma_{\bar au}r
_{\bar a})}\delta_{(a)u}\nonumber \\
&&\Big[ \Big( e^{{1\over {\sqrt{3}}}\sum_{\bar b}(\gamma_{\bar br}
+\gamma_{\bar bu}+\gamma_{\bar bv})r_{\bar b}}\, {}^3G_{o(b)(c)(d)(e)}\,
\delta_{(b)r}\delta_{(c)u}\delta_{(d)v}\, {}^3{\hat {\tilde \pi}}^v_{(e)}
\Big)_{|s}-\nonumber \\
&-&\Big( e^{{1\over {\sqrt{3}}}\sum_{\bar b}(\gamma_{\bar bs}
+\gamma_{\bar bu}+\gamma_{\bar bv})r_{\bar b}}\, {}^3G_{o(b)(c)(d)(e)}\, 
\delta_{(b)s}\delta_{(c)u}\delta_{(d)v}\, {}^3{\hat {\tilde \pi}}^v_{(e)}\Big)
_{|r}\Big] ,\nonumber \\
{}^4{\hat {\buildrel \circ \over {\Omega}}}_{\tau r(a)(b)}\, &=&\sum_{uv}
e^{-(2q+{1\over {\sqrt{3}}}\sum_{\bar a}(\gamma_{\bar au}+\gamma_{\bar av})
r_{\bar a})}\delta_{(a)u}\delta_{(b)v)}\, {}^4{\hat R}_{uv\tau r}
{\buildrel \circ \over =}\, \nonumber \\
&&{\buildrel \circ \over =}\, -\epsilon \Big( 
\partial_{\tau}\, {}^3{\hat \omega}
_{r(a)(b)}+{1\over 2}\Big[ \epsilon_{(a)(b)(c)}\epsilon_{(d)(e)(f)}-\epsilon
_{(a)(b)(d)}\epsilon_{(c)(e)(f)}\Big] \cdot \nonumber \\
&&\sum_s e^{-(q+{1\over {\sqrt{3}}}\sum_{\bar a}\gamma_{\bar as}r_{\bar a})}
\delta_{(c)s}\nonumber \\
&&\Big[
{{\epsilon N}\over {4k}} \sum_ve^{-q+{1\over {\sqrt{3}}}\sum_{\bar b}(\gamma
_{\bar bs}+\gamma_{\bar bv})r_{\bar b}}\, {}^3G_{o(d)(l)(m)(n)}\,
\delta_{(l)s}\delta_{(m)v}\, {}^3{\hat {\tilde \pi}}^v_{(n)}+\nonumber \\
&+&N_{(l)}\sum_ue^{(q+{1\over {\sqrt{3}}}\sum_{\bar b}\gamma_{\bar bu}r_{\bar b}
)}\delta_{(l)u}\partial_u\Big( 
\delta_{(d)s}e^{q+{1\over {\sqrt{3}}}\sum_{\bar c}
\gamma_{\bar cs}r_{\bar c}}\Big) +\nonumber \\
&+&\sum_ue^{q+{1\over {\sqrt{3}}}\sum_{\bar b}\gamma_{\bar bu}r_{\bar b}}
\delta_{(d)u}\partial_s\Big( N_{(l)}\delta_{(l)u}e^{-(q+{1\over {\sqrt{3}}}\sum
_{\bar c}\gamma_{\bar cu}r_{\bar c})}\Big) +\nonumber \\
&+&\epsilon_{(d)(m)(n)}{\hat \mu}_{(m)}\delta_{(n)s}e^{q+{1\over {\sqrt{3}}}
\sum_{\bar b}\gamma_{\bar bs}r_{\bar b}}-\nonumber \\
&-&N_{(g)}\sum_ue^{-(q+{1\over {\sqrt{3}}}\sum_{\bar b}\gamma_{\bar bu}r
_{\bar b})}\delta_{(g)u} \partial_u\Big( \delta_{(d)s}e^{q+{1\over {\sqrt{3}}}
\sum_{\bar c}\gamma_{\bar cs}r_{\bar c}}\Big) -\nonumber \\
&-&\sum_ue^{q+{1\over {\sqrt{3}}}\sum_{\bar b}\gamma_{\bar bu}r_{\bar b}}
\delta_{(d)u}\partial_s\Big( N_{(g)}\delta_{(g)u}e^{-(q+{1\over {\sqrt{3}}}\sum
_{\bar c}\gamma_{\bar cu}r_{\bar c})}\Big) \Big] 
{}^3{\hat \omega}_{r(e)(f)}+\nonumber \\
&+&N_{(c)}\delta_{(c)s}e^{-(q+{1\over {\sqrt{3}}}\sum_{\bar b}\gamma_{\bar bs}
r_{\bar b})}\, [{}^3{\hat \omega}_s,{}^3{\hat \omega}_r]_{(a)(b)}+\nonumber \\
&+&{{\epsilon}\over {4k}} \sum_ue^{-2q+{1\over {\sqrt{3}}}\sum_{\bar b}\gamma
_{\bar br}r_{\bar b}}\, {}^3G_{o(c)(d)(e)(f)}\delta_{(c)r}\delta_{(e)u}\, 
{}^3{\hat {\tilde \pi}}^u_{(f)}\nonumber \\
&&\Big( \delta_{(a)(d)}\delta_{(b)u}-\delta_{(b)(d)}\delta_{(a)u}\Big) 
\partial_uN+\nonumber \\
&+&{1\over {(4k)^2}}\Big( \delta_{(a)(l)}\delta_{(b)(d)}-\delta_{(a)(d)}
\delta_{(b)(l)}\Big) {}^3G_{o(d)(e)(f)(g)}\, {}^3G_{o(h)(l)(m)(n)}\cdot
\nonumber \\
&\cdot& \sum_{wv}e^{-3q+{1\over {\sqrt{3}}}\sum_{\bar b}(\gamma_{\bar br}+
\gamma_{\bar bw}+\gamma_{\bar bv})r_{\bar b}}\delta_{(h)r}
N_{(e)}\delta_{(f)w}\, {}^3{\hat {\tilde \pi}}^w_{(g)}\delta_{(m)v}\, 
{}^3{\hat {\tilde \pi}}^v_{(n)} \Big) ,\nonumber \\
{}^4{\hat {\buildrel \circ \over {\Omega}}}_{\tau r(o)(a)}\, &{\buildrel \circ
\over =}\,& {1\over N}\, \sum_ue^{-(q+{1\over {\sqrt{3}}}\sum_{\bar a}\gamma
_{\bar au}r_{\bar a})}\delta_{(a)u}
\Big[ {}^4{\hat R}_{\tau u\tau r}-N_{(b)}\sum_s
e^{-(q+{1\over {\sqrt{3}}}\sum_{\bar b}\gamma_{\bar bs}r_{\bar b})}\delta
_{(b)s}\, {}^4{\hat R}_{su\tau r}\Big]\, {\buildrel \circ \over
=}\nonumber \\
&{\buildrel \circ \over =}&\, -\epsilon \, \sum_se^{-(q+{1\over {\sqrt{3}}}
\sum_{\bar a}\gamma_{\bar as}r_{\bar a})}\delta_{(a)s}\Big[ \partial_{\tau}\, 
{}^3{\hat K}_{rs} + N_{|s|r}-\nonumber \\
&-&{{\epsilon}\over {4k}}\sum_{uw}e^{{1\over {\sqrt{3}}}\sum_{\bar b}(\gamma
_{\bar bu}+\gamma_{\bar bw})r_{\bar b}} \,{}^3G_{o(c)(d)(e)(f)}\delta_{(d)u}
\delta_{(e)w}\, {}^3{\hat {\tilde \pi}}^w_{(f)}\nonumber \\
&& \Big( e^{{1\over {\sqrt{3}}}\sum_{\bar c}\gamma_{\bar cr}r_{\bar c}}\delta
_{(c)r}\Big( N_{(b)}e^{-(q+{1\over {\sqrt{3}}}\sum_{\bar d}\gamma_{\bar du}r
_{\bar d})}\delta_{(b)u}\Big)_{|s} + \nonumber \\
&+&e^{{1\over {\sqrt{3}}}\sum_{\bar c}\gamma_{\bar cs}r_{\bar c}}\delta_{(c)s}
\Big( N_{(b)}e^{-(q+{1\over {\sqrt{3}}}\sum_{\bar d}\gamma_{\bar du}r_{\bar d}}
\delta_{(b)u}\Big)_{|r} \Big) -\nonumber \\
&-&{{\epsilon}\over {4k}}\sum_{usw} N_{(b)}e^{q+{1\over {\sqrt{3}}}\sum_{\bar b}
\gamma_{\bar bu}r_{\bar b}}\delta_{(b)u} \nonumber \\
&&\Big( 
e^{3q+{1\over {\sqrt{3}}}\sum_{\bar c}(\gamma_{\bar cs}+\gamma_{\bar cu}+
\gamma_{\bar cw})r_{\bar c}}\, {}^3G_{o(c)(d)(e)(f)}\, 
\delta_{(c)s}\delta_{(d)u}\delta_{(e)w}\, {}^3{\hat {\tilde \pi}}^w_{(f)}\Big)
_{|r}\, \Big],\nonumber \\
&&{}\nonumber \\
{}^4{\hat R}_{rsuv}&=&\delta_{(a)r}\delta_{(b)s}e^{2q+{1\over {\sqrt{3}}}
\sum_{\bar a}(\gamma_{\bar ar}+\gamma_{\bar as})r_{\bar a}}
\, {}^4{\hat {\buildrel \circ \over {\Omega}}}_{uv(a)(b)}=\nonumber \\
&&=-{}^3{\hat R}_{rsuv}+{{N^2}\over {16k^2}}\sum_{tw}e^{{1\over {\sqrt{3}}}
\sum_{\bar a}(\gamma_{\bar ar}+\gamma_{\bar as}+\gamma_{\bar au}+\gamma
_{\bar av}+\gamma_{\bar at}+\gamma_{\bar aw})r_{\bar a}}\cdot \nonumber \\
&&{}^3G_{o(a)(b)(c)(d)}\, {}^3G_{o(e)(f)(g)(h)}\nonumber \\
&&\cdot \delta_{(a)r}\delta_{(e)s}\Big( \delta_{(b)u}\delta_{(f)v}-\delta_{(b)v}
\delta_{(f)u}\Big)
\delta_{(c)t}\delta_{(g)w}\, {}^3{\hat {\tilde \pi}}^t_{(d)}\,
{}^3{\hat {\tilde \pi}}^w_{(h)},\nonumber \\
{}^4{\hat R}_{\tau ruv}&=&N \delta_{(a)r}e^{q+{1\over {\sqrt{3}}}\sum_{\bar a}
\gamma_{\bar ar}r_{\bar a}}\, {}^4{\hat {\buildrel \circ
\over {\Omega}}}_{uv(o)(a)},\nonumber \\
{}^4{\hat R}_{\tau r\tau s}&=&N \delta_{(a)r}e^{q+{1\over {\sqrt{3}}}\sum_{\bar
a}\gamma_{\bar ar}r_{\bar a}}\, {}^4{\hat {\buildrel
\circ \over {\Omega}}}_{\tau s(o)(a)}, \nonumber \\
&&{}\nonumber \\
{}^4{\hat R}_{\tau\tau}&=&-\epsilon \sum_re^{-2(q+{1\over {\sqrt{3}}}\sum
_{\bar a}\gamma_{\bar ar}r_{\bar a})}\, {}^4{\hat R}
_{r\tau r\tau},\nonumber \\
{}^4{\hat R}_{\tau r}&=&{}^4{\hat R}_{r\tau}=-\epsilon \sum_u
e^{-2(q+{1\over {\sqrt{3}}}\sum_{\bar a}\gamma_{\bar au}r_{\bar a})}
\, {}^4{\hat R}_{u\tau ur},\nonumber \\
{}^4{\hat R}_{rs}&=&{}^4{\hat R}_{sr}={{\epsilon}\over {N^2}}{}^4{\hat R}
_{\tau r\tau s}-\epsilon \sum_ue^{-2(q+{1\over {\sqrt{3}}}\sum_{\bar a}\gamma
_{\bar au}r_{\bar a})}\, {}^4{\hat R}_{urus},\nonumber \\
{}^4{\hat R}&=&{{\epsilon}\over {N^2}}{}^4{\hat R}_{\tau\tau}-\epsilon \sum_r
e^{-2(q+{1\over {\sqrt{3}}}\sum_{\bar a}\gamma_{\bar ar}r_{\bar a})}
\, {}^4{\hat R}_{rr},\nonumber \\
&&{}\nonumber \\
{}^4{\hat C}_{rsuv}&=&{}^4{\hat R}_{rsuv}+{{\epsilon}\over 2}\Big[ e^{2(q+
{1\over {\sqrt{3}}}\sum_{\bar a}\gamma_{\bar ar}r_{\bar a})}(\delta_{rv}
\, {}^4{\hat R}_{su}-\delta_{ru}\, {}^4{\hat R}_{sv})+\nonumber \\
&&+e^{2(q+{1\over {\sqrt{3}}}\sum_{\bar a}\gamma_{\bar as}r_{\bar a})}
(\delta_{su}\,
{}^4{\hat R}_{rv}-\delta_{sv}\, {}^4{\hat R}_{ru})\Big] +\nonumber \\
&&+{1\over 6}e^{2(2q+{1\over {\sqrt{3}}}\sum_{\bar a}(\gamma_{\bar ar}+\gamma
_{\bar as})r_{\bar a})}(\delta_{ru}\delta_{sv}-\delta_{rv}\delta_{su})
{}^4{\hat R},\nonumber \\
{}^4{\hat C}_{\tau ruv}&=&{}^4{\hat R}_{\tau ruv}+{{\epsilon}\over 2}
e^{2(q+{1\over {\sqrt{3}}}\sum_{\bar a}\gamma_{\bar ar}r_{\bar a})}
(\delta_{ru}\, {}^4{\hat R}_{\tau v}-\delta_{rv}\, {}^4{\hat R}_{\tau u}),
\nonumber \\
{}^4{\hat C}_{\tau r\tau s}&=&{}^4{\hat R}_{\tau r\tau s}+{1\over 2}\Big[ N^2\, 
{}^4{\hat R}_{rs}-\epsilon 
e^{2(q+{1\over {\sqrt{3}}}\sum_{\bar a}\gamma_{\bar ar}r_{\bar a})}
\delta_{rs}\, {}^4{\hat R}_{\tau\tau}\Big] -\nonumber \\
&&-{1\over 6}N^2
e^{2(q+{1\over {\sqrt{3}}}\sum_{\bar a}\gamma_{\bar ar}r_{\bar a})}
\delta_{rs}\, {}^4\hat R,
\label{z6}
\end{eqnarray}

\noindent To get $\partial_{\tau}\, {}^3{\hat \omega}_{r(a)(b)}$ and
$\partial_{\tau}\, {}^3{\hat K}_{rs}$ we need Eqs.(65) of I, where in
Appendix B it is noted that $\partial_{\tau}\, {}^3{\hat K}_{rs}$ needs the 
use of the second half of Hamilton equations.

The York almost canonical basis of Appendix C takes the form [${\cal T}(\tau
,\vec \sigma )$ is the ``extrinsic internal time" proportional to the
``mean extrinsic curvature"; ${}^3\sigma_{rs}(\tau ,\vec \sigma )$ is the 
``conformal metric" [$det\, ({}^3\sigma_{rs})=1$], 
which is a density of weight=-2/3 like the momentum 
${}^3{\tilde \Pi}^{rs}_A(\tau ,\vec \sigma )$]

\begin{eqnarray}
{\cal T}(\tau ,\vec \sigma )&=&-{4\over 3}\epsilon k\, {}^3K(\tau ,\vec \sigma 
)=[{{2\, {}^3g_{rs}\, {}^3{\tilde \Pi}^{rs}}\over {3\sqrt{\gamma}}}]
(\tau ,\vec \sigma ) \mapsto {\hat {\cal T}}(\tau 
,\vec \sigma )=\nonumber \\
&&={{\epsilon}\over 3} e^{q(\tau ,\vec \sigma )} \sum_r[e^{{1\over {\sqrt{3}}}
\sum_{\bar a}\gamma_{\bar ar}r_{\bar a}}](\tau ,\vec \sigma ) \int d^3\sigma_1 
{\cal K}^r_{(b)s}(\vec \sigma ,{\vec \sigma}_1,\tau |q,r_{\bar a}]\cdot 
\nonumber \\
&\cdot&(e^{-q-{1\over 3}\sum_{\bar a}\gamma_{\bar as}r_{\bar a}})(\tau ,{\vec 
\sigma}_1) \Big[ {1\over 3}\rho +\sqrt{3} \sum_{\bar b}
\gamma_{\bar bs} \pi_{\bar b}\Big] (\tau ,{\vec \sigma}_1),\nonumber \\
{\cal P}_{\cal T}(\tau ,\vec \sigma )&=&-\sqrt{\gamma} \mapsto {\hat {\cal P}}
_{\cal T}(\tau ,\vec \sigma )=-e^{3q(\tau ,\vec \sigma )},
\nonumber \\
{}^3\sigma_{rs}(\tau ,\vec \sigma )&=&[{{{}^3g_{rs}}\over {\gamma^{1/3}}}]
(\tau ,\vec \sigma ) 
\mapsto {}^3{\hat \sigma}_{rs}(\tau ,\vec \sigma )={}^3g^Y_{rs}
(\tau ,\vec \sigma )=
e^{{2\over {\sqrt{3}}}\sum_{\bar a}\gamma_{\bar ar}r_{\bar a}(\tau ,\vec 
\sigma )}\, \delta_{rs},\nonumber \\
{}^3{\tilde \Pi}_A^{rs}(\tau ,\vec \sigma )&=&
\Big[ \gamma^{1/3}({}^3{\tilde \Pi}^{rs}-{1\over 3}\, {}^3g^{rs}\, 
{}^3{\tilde \Pi})\Big] (\tau ,\vec \sigma )\mapsto \nonumber \\
&&\mapsto {{\epsilon}\over 4}
\Big( e^{4q}\Big[ e^{-{1\over {\sqrt{3}}}\sum_{\bar a}\gamma
_{\bar ar}r_{\bar a}}\delta^r_{(a)}\, {}^3{\hat {\tilde \pi}}^s_{(a)}+
\nonumber \\
&&+e^{-{1\over {\sqrt{3}}}\sum_{\bar a}\gamma_{\bar as}r_{\bar a}}
\delta^s_{(a)}\, {}^3{\hat {\tilde \pi}}^r_{(a)}-{2\over 3}e^{{1\over 
{\sqrt{3}}}\sum_{\bar a}(\gamma_{\bar au}-2\gamma_{\bar ar})r_{\bar a}}
\delta^{rs}\delta^u_{(a)}\, {}^3{\hat {\tilde \pi}}^u_{(a)}\Big]\, \Big)
(\tau ,\vec \sigma )
\label{z1}
\end{eqnarray}

Using Eqs.(68) and (69) of I, Ashtekar's variables become

\begin{eqnarray}
&&{}^3{\tilde h}^r_{(a)}(\tau ,\vec \sigma )
\mapsto {}^3{\hat {\tilde h}}^r_{(a)}(\tau ,\vec \sigma )=\delta^r_{(a)} 
\Big[ e^{2q} e^{-{1\over {\sqrt{3}}}\sum_{\bar a}\gamma_{\bar ar}r_{\bar a}}
\Big] (\tau ,\vec \sigma ),\nonumber \\
&&{}\nonumber \\
&&{}^3A_{(a)r}(\tau ,\vec \sigma ) 
\mapsto {}^3{\hat A}_{(a)r}(\tau ,\vec \sigma )=
{1\over {2k}}(e^{2q+{2\over 
{\sqrt{3}}}\sum_{\bar a}\gamma_{\bar ar}r_{\bar a}})(\tau ,\vec \sigma )
\nonumber \\
&&\int d^3\sigma_1\, {\cal K}^r_{(a)s}(\vec \sigma ,{\vec \sigma}_1,\tau |
q,r_{\bar c}]\, \Big[ e^{-q-{1\over {\sqrt{3}}}
\sum_{\bar d}\gamma_{\bar ds}r_{\bar d}}({{\rho}\over 3}+\sqrt{3}\sum_{\bar b}
\gamma_{\bar bs}\pi_{\bar b})\Big] (\tau ,{\vec \sigma}_1)+\nonumber \\
&+&i\epsilon_{(a)(b)(c)}\delta_{(b)r}\delta_{(c)u} \Big[ e^{ {2\over {\sqrt{3}}}
\sum_{\bar a}(\gamma_{\bar ar}-\gamma_{\bar au})r_{\bar a}}(\partial_uq+
{1\over {\sqrt{3}}}\sum_{\bar b}\gamma_{\bar br} \partial_ur_{\bar b})\Big]
(\tau ,\vec \sigma ).
\label{z4}
\end{eqnarray}

With the further gauge fixing ${\tilde \lambda}_r(\tau )=0$ 
[i.e. when $n_r=0$ and $N_{(as)(a)}=0$]
the 4-geodesic and 4-geodesic deviation equations given at the end of
Appendix A of I become respectively in the 3-orthogonal gauges

\begin{eqnarray}
{{d^2\tau (s)}\over {ds^2}}&+& {{\partial_{\tau}N}\over N}+2{{\partial_rN}\over 
N} {{d\tau (s)}\over {ds}}{{d\sigma^r(s)}\over {ds}}-\nonumber \\
&-&{{\epsilon}\over {4kN}}\sum_ue^{{1\over {\sqrt{3}}}\sum_{\bar a}(\gamma
_{\bar ar}+\gamma_{\bar as}+\gamma_{\bar au})r_{\bar a}}\, {}^3G_{o(a)(b)(c)(d)}
\delta_{(a)r}\delta_{(b)s}\delta_{(c)u}\, {}^3{\hat {\tilde \pi}}^u_{(d)}
{{d\sigma^r(s)}\over {ds}}{{d\sigma^s(s)}\over {ds}}=0,\nonumber \\
{{d^2\sigma^u(s)}\over {ds^2}}&+&N\phi^{-4}e^{-{2\over {\sqrt{3}}}\sum_{\bar a}
\gamma_{\bar au}r_{\bar a}} \partial_uN ({{d\tau (s)}\over {ds}})^2-\nonumber \\
&-&{{\epsilon N}\over {2k}}\phi^{-4}\sum_{mn}e^{{1\over {\sqrt{3}}}\sum_{\bar 
a}(\gamma_{\bar ar}+\gamma_{\bar am}+\gamma_{\bar an})r_{\bar a}}\, {}^3G
_{o(a)(b)(c)(d)}\delta_{(a)m}\delta_{(b)r}\delta_{(c)n}\, {}^3{\hat {\tilde 
\pi}}^u_{(d)} {{d\tau (s)}\over {ds}}{{d\sigma^r(s)}\over {ds}}+\nonumber \\
&+& {}^3{\hat \Gamma}^u_{rs} {{d\sigma^r(s)}\over {ds}}{{d\sigma^s(s)}\over 
{ds}}=0,\nonumber \\
&&{}\nonumber \\
a^{\tau}&=&-{{\epsilon}\over {N^2}}\Big( {}^4{\hat R}_{\tau m\tau n}{{d\sigma
^m}\over {ds}}{{d\sigma^n}\over {ds}} \triangle x^{\tau} -\nonumber \\
&-&\Big[ {}^4{\hat R}_{\tau m\tau s}{{d\tau}\over {ds}}-{}^4{\hat R}_{\tau msn}
{{d\sigma^n}\over {ds}}\Big] {{d\sigma^m}\over {ds}} \triangle x^s \Big) ,
\nonumber \\
a^u&=& -\epsilon \, {}^3{\hat e}^u_{(a)}\, {}^3{\hat e}_{(a)}^r\Big( \Big[
{}^4{\hat R}_{\tau r\tau n}{{d\tau}\over {ds}}-{}^4{\hat R}_{rm\tau n}
{{d\sigma^m}\over {ds}}\Big] {{d\sigma^n}\over {ds}} \triangle x^{\tau}-
\nonumber \\
&-&\Big[ {}^4{\hat R}_{\tau r\tau s}({{d\tau}\over {ds}})^2-({}^4{\hat R}_{\tau
rsm}+{}^4{\hat R}_{rm\tau s}){{d\tau}\over {ds}}{{d\sigma^m}\over {ds}}+
{}^4{\hat R}_{rmsn} {{d\sigma^m}\over {ds}}{{d\sigma^n}\over {ds}} \Big] 
\triangle x^s \Big) .
\label{z5}
\end{eqnarray}

\vfill\eject

\end{document}